\documentstyle[11pt,epsfig]{article}

\newcommand{\beq}{\begin{equation}}
\newcommand{\eeq}{\end{equation}}
\newcommand{\bea}{\begin{eqnarray}}
\newcommand{\eea}{\end{eqnarray}}

\def\laq{~\raise 0.4ex\hbox{$<$}\kern -0.8em\lower 0.62
ex\hbox{$\sim$}~}
\def\gaq{~\raise 0.4ex\hbox{$>$}\kern -0.7em\lower 0.62
ex\hbox{$\sim$}~}

\def \pa {\partial}
\def \ti {\widetilde}
\def \se {{\prime\prime}}
\def \ra {\rightarrow}
\def \la {\lambda}
\def \La {\Lambda}
\def \Da {\Delta}
\def \b {\beta}
\def \a {\alpha}
\def \ap {\alpha^{\prime}}
\def \Ga {\Gamma}
\def \ga {\gamma}
\def \sg {\sigma}
\def \da {\delta}
\def \ep {\epsilon}
\def \r {\rho}
\def \om {\omega}
\def \Om {\Omega}

\def \bp {\dot{\beta}}

\def \fpu {\dot{\phi}}
\def \fpp {\ddot{\phi}}
\def \hp {\dot{h}}
\def \hpp {\ddot{h}}

\def \fb {\overline \phi}
\def \rb {\overline \rho}
\def \pb {\overline p}
\def \fbp {\dot{\fb}}
\def \ls {\lambda_{\rm s}}

\begin{document}
\begin{titlepage}

\begin{flushright}
BA-TH/02-437\\
CERN-TH/2002-104\\
hep-th/0207130
\end{flushright}

\vspace*{0.5in}

\begin{center}
\huge
{\bf The Pre-Big Bang Scenario \\ in String Cosmology}

\vspace{.5in}

\large{
M. Gasperini$^1$ and G. Veneziano$^2$}

\normalsize
\vspace{.2in}

{\sl $^1$Dipartimento di Fisica , Universit\`a di Bari, \\ 
Via G. Amendola 173, 70126 Bari, Italy\\
and \\
Istituto Nazionale di Fisica Nucleare, Sezione di Bari\\
Via G. Amendola 173, 70126 Bari, Italy\\
\vspace{0.3cm}
E-mail: {\tt maurizio.gasperini@ba.infn.it}}

\vspace{.2in}

{\sl $^2$Theory Division, CERN, \\
CH-1211 Geneva 23, Switzerland \\
\vspace{0.3cm}
E-mail: {\tt gabriele.veneziano@cern.ch}}

\vspace{.7in}

\begin{abstract}
We review physical  motivations, phenomenological consequences,
and open problems of the so-called pre-big bang scenario in
superstring cosmology. 
\end{abstract}

\end{center}
\end{titlepage}

\tableofcontents

\newpage

\parskip 0.2cm

\section{Introduction}
\label{Sec1}
\setcounter{equation}{0}
\setcounter{figure}{0}

During the past thirty years, mainly thanks to accelerator
experiments of higher and higher energy and precision,
 the standard model of particle physics  has established itself as
the uncontested winner in the race for a consistent description
of electroweak and strong interaction phenomena at distances above
$10^{-15}$ cm or so. There are, nonetheless, good
reasons  (in particular the increasing evidence for non-vanishing
neutrino masses \cite{KaTo01,SNO,SNOa})  to believe that the standard
model is not the end of the story. The surprising validity of this model at
energies below $100$ GeV, as well as the (in)famous Higgs mass
fine-tuning problem, suggest some supersymmetric extension of the
standard model  (for a review see \cite{Nilles}) as the most likely
improved description of non-gravitational phenomena over a few more
decades in the ladder of scales. It is however quite likely that other
questions that are
 left unanswered by the standard model, such as the peculiarities
 of fermionic masses and mixings, the family pattern, C, P, CP, B 
violation, etc., will only find their answers at --or around-- the  much
higher energies at which all gauge interactions appear to unify
\cite{Ama91}.   This energy scale appears to be embarrassingly
close (on a logarithmic scale) to the so-called Planck mass, $M_{\rm P} 
\sim 10^{19}$ GeV, the scale at which gravity becomes strong and needs
to be quantized.

The situation with gravitational phenomena is completely different.  
Even the good old Newton law is known to be valid only down to the $1$
mm scale \cite{Hoyle01}, so that much interest has been  devoted  to the
possibility of large modifications of gravity below that distance,  either
from new forces mediated by light scalars such as the dilaton of string
theory \cite{Tay88}, or from the existence of large extra dimensions felt
exclusively  by gravity  \cite{Arkani98,RS2}. General relativity is well
tested at large scales; nevertheless; present evidence for a (small)
vacuum energy density \cite{Riess98,Perlmutter99} suggests that, even
on cosmologically  large distances, the  strict Einstein theory might turn
out to be inadequate. Evidently, the construction of a standard model for
gravity  and cosmology lags much behind its particle physics counterpart.

The hot big bang model (see for instance \cite{Weinberg}), originally
thought of as another great success of general relativity, was later
discovered  to suffer from huge fine-tuning problems. Some of these
conceptual  problems are solved by the standard inflationary paradigm
(see \cite{Lin90,KT90} for a review), yet inflation remains a generic idea
 in search of a theory that will embody it naturally.
Furthermore, the classical theory of inflation does not really address
the problem of how the initial conditions needed for a successful
inflation came about. The answer to  this question is certainly related to
even more fundamental issues, such as: How did it all start? What caused
the big bang? Has there been a singularity at $t=0$?
Unfortunately, these questions lie deeply inside the short-distance,
high-curvature regime of gravity where quantum corrections cannot be
neglected. Attempts at answering these questions using quantum
cosmology based on Einstein's theory has resulted in a lot of heated
discussions  \cite{Linde98,TuHa98}, with no firm conclusions.

It is very likely that both a standard model for gravity and cosmology 
and a full understanding of the standard model of particle physics will 
require our understanding of physics down to the shortest scale,
the Planck length $\la_{\rm P}\sim 10^{-33}$ cm.
Until the Green--Schwarz revolution of 1984 \cite{GSW87},
the above conclusion would have meant postponing indefinitely those
kinds of questions. Since then, however, particle theorists have studied
and developed superstring  theory (see \cite{Pol98}
for a   recent review, as well as \cite{Greene99} for a non-specialized
introduction), which appears to represent a consistent framework not
only for addressing  (and possibly answering)  those questions, but even
for unifying our understanding of gravitational and non-gravitational
phenomena, and therefore for relating the two classes of questions. 

The so-called ``pre-big bang" scenario described in this report  has to
be seen in the above perspective as a possible example, even just as a 
toy model, of what cosmology can look like if we assume that the
sought for standard model  of gravity and cosmology is based on (some
particular version of) superstring theory. Although most string theorists
would certainly  agree on the importance of studying the cosmological
consequences of string theory, it is a priori far from obvious that
the ``state of the art" in this field can provide an
unambiguous answer to this question. Indeed, most of our understanding
of superstring theory is still based on  perturbative expansions,  while 
 most of the recent progress in non-perturbative string theory has been
achieved in the context of   ``vacua" (i.e.
classical solutions to the field equations) that respect a large
number of supersymmetries \cite{Pol98}.
By contrast, our understanding of string theory at large curvatures
 and couplings, especially in the absence of supersymmetry,
is still largely incomplete.  A cosmological background, and a  
fortiori one that evolves rapidly in time, breaks (albeit spontaneously)  
supersymmetry. This is why the Planckian regime of cosmology appears
to be intractable for the   time being.

It is very fortunate, in this respect, that in the 
 pre-big bang  scenario  the Universe is supposed to emerge from a 
highly perturbative initial state {\it preceding} the big bang.
Therefore,  early enough before (and late enough after) the big bang, 
 we may presume to know the effective
theory to be solved. The difficult part to be dealt with
non-perturbatively remains the transition from the pre- to the post-big
bang regime, through a high-curvature (and/or possibly a large-coupling)
phase. Thus, from a more phenomenological standpoint, the relevant
question becomes: Are the predictions of the pre-big bang scenario 
robust with respect to the details of  the non-perturbative
phase?

It is difficult of course to give a clear-cut answer to this question, but
an analogy with QCD and the physics of strong interactions may be 
helpful. Because of asymptotic freedom, QCD can be treated
perturbatively at short distance (high momentum transfers). However,
even ``hard" processes such as $e^+ e^- \rightarrow {\rm hadrons}$ are
not fully within perturbative control.  Some soft non-perturbative
physics  always gets mixed in at some level, e.g. when partons
eventually turn into hadrons. The reason why certain sufficiently
inclusive quantities are believed to be calculable is that  large- and
short-distance   physics ``decouple", so that, for instance, the 
hadronization process does not   affect certain ``infrared-safe"
quantities, computed at the quark--gluon level.

In the case of string  cosmology the situation should be similar, 
although somehow reversed \cite{GVGvsG95}.  For gravity, in fact, the
large-distance, small-curvature regime is easy to deal with, while the
short-distance, high-curvature is hard. Yet, we shall argue that some
consequences of string cosmology, those concerning length scales that
were very large with respect to the string scale (or the horizon) in the
high-curvature regime, should not be affected (other than by a trivial
kinematical redshift)   by the details of the pre- to post-big bang
transition. The above reasoning does not imply, of course, that string
theorists should not address the hard, non-perturbative questions {\it
now}. On the contrary, the ``easy" part of the   game will provide 
precious information about what the relevant hard questions are,  and
on how to formulate them.

Finally, possible reservations on a ``top--down" string cosmology
approach may naturally arise from a cosmology community accustomed
to a  data-driven, ``bottom--up" approach. We do  believe ourselves 
that a good model of cosmology is unlikely to emerge from theoretical
considerations alone. Input from the data will be
essential in the selection among various theoretical alternatives.  We
also believe, however, that a balanced combination of theoretical and
experimental imput should be the best guarantee for an eventual
success. 

Insisting on the soundness of the underlying theory (e.g. on its 
renormalizability)  was indeed  essential in the progressive
construction of the standard model, just as were  the quantity and the
quality of  experimental data. Cosmology today  resembles the particle
physics of the  sixties: there is no shortage of data, and these are 
becoming more and more precise but also more and more challenging 
while, theoretically,  we are still playing with very phenomenological
(even if undoubtedly successful) models, lacking a clear connection to
other branches of fundamental physics, and therefore remaining largely
unconstrained.

\subsection{Coping with a beginning of time}
\label{Sec1.1}

Both the standard Friedmann--Robertson--Walker (FRW) cosmological
scenario \cite{Weinberg} and  the standard inflationary scenario
\cite{Gut81,KT90,Lin90}  assume that time had a beginning.
 Many of the problems with the former model simply stem  from the
fact that, at the start of the classical era, so little time had elapsed
since the beginning. Indeed, in the FRW framework, the  proper size of
the (now observable) Universe was about $10^{-3}$ cm across at the
start of  the classical era, say at a time of the order of a few Planck
times, $t_{\rm P} \sim 10^{-43}$ s.  This is of course a very tiny Universe 
with respect to  its present size ($\sim 10^{28}~ \rm{cm}$), yet it is
huge with  respect to the horizon  (the distance travelled by light) at
that time,  $\la_{\rm P}  = c t_{\rm P} \sim 10^{-33}$ cm. 

In other words, a few Planck times after the big bang, our observable
Universe consisted of about $(10^{30})^3 = 10^{90}$ Planckian-size,
causally disconnected  regions. Simply not enough time had elapsed
since the beginning for the Universe to become homogeneous (e.g. to
thermalize) over its entire size. Furthermore, soon after $t=t_{\rm P}$, 
the Universe must have been characterized by a huge
hierarchy between its Hubble radius, on the one hand, and its
spatial-curvature radius,  on the other. The relative factor of (at least)
$10^{30}$ appears as an incredible amount of fine-tuning on the initial
state of the Universe, corresponding to a huge asymmetry between 
space and time derivatives, or, in more abstract terms,
between intrinsic and extrinsic curvature.   Was this asymmetry
really there? And, if so, can it be explained in any, more natural way?

The conventional answer to the difficulties of the standard scenario is 
to wash out inhomogeneities and spatial curvature by introducing, in 
the history of the Universe, a long period of accelerated expansion, 
called inflation \cite{Gut81,KT90,Lin90}. It has been pointed out,
however, that standard inflation cannot be  ``past-eternal"
\cite{BoGuVil01} (and cannot avoid the initial singularity
\cite{Vil92,BoVil94}), so that the question of what preceded inflation is
very relevant. Insisting on the assumption that the Universe (and time
itself) started at the big bang leaves only the possibility of having
post-big bang  inflation mend an insufficiently smooth and flat Universe
arising from the big bang. 

Unfortunately, that solution has its own problems, for instance those of
fine-tuned initial conditions for the inflaton field and its potential.
A consistent quantum cosmology approach giving birth
to a Universe in the ``right" initial state is still  much under debate
\cite{Haw84,Vil84,Linde84,Zeldo84,Rub84}. Furthermore, the inflaton is
introduced {\em ad hoc} and inflation is not part  of  a grander theory of
elementary particles and fundamental interactions such as superstring
theory.  In spite of its possible importance, and of repeated motivated
attempts \cite{EENQ86,MP86,BG86},  a conventional realization of an
inflationary phase in a string theory context is in fact problematic
\cite{CAO90}, in particular because the dilaton  --the fundamental 
string theory scalar-- cannot be (at least trivially) identified with the
inflaton  --the fundamental scalar of the standard inflationary scenario
\cite{BS93}.

Here we shall argue that, instead, superstring theory gives strong hints
in favour of a totally different approach to solving the problems of the
standard cosmological scenario. This new possibility arises if we  assume
that, in string theory, the big bang singularity is fictitious
and that it makes therefore sense to ``continue" time to the   past of
the big bang itself.

\subsection{Inflation before the big bang}
\label{Sec1.2}

If the history of the Universe can be continued backward in time past
the big bang, new possibilities arise for a causal evolution to have {\it
produced}  a big bang with the desired characteristics. The actual pre-big
bang scenario  presented  in this report is just one possible realization
of the above general idea. Since, as we shall see, it is easy to generate a
phase of pre-big bang inflation  driven by the kinetic energy of the 
dilaton (somewhat in analogy with kinetic-inflation ideas \cite{Levin}),
we will discuss, as the simplest possibility,  a minimal
cosmological scenario, which avoids making use of standard (i.e.
potential-energy-driven)  post-big bang inflation.

This does not, though, that pre- and post-big bang inflation are
mutually exclusive or incompatible. Should near-future
high-precision experiments definitely
indicate that an inflation that is exclusively of the pre-big
bang type is disfavoured with respect to conventional, post-big bang,
``slow-roll" inflation, one should ask whether a pre-big bang 
phase can naturally lead to ``initial" conditions suitable for
igniting an inflationary epoch of the slow-roll type, rather than a
standard, non-inflationary, FRW cosmology.

One model-independent feature of pre-big bang cosmology is clear:  by
its very definition, the pre-big bang phase should be  an evolution
towards --rather than away from-- a high-curvature regime. As we
shall see in Section \ref{Sec2}, this is precisely what the symmetries of
the string cosmology equations suggest, an unconventional realization
of the inflationary scenario, in which the phase of accelerated
cosmological evolution  occurs while the
Universe is approaching --rather than getting away from-- the
high-curvature, Planckian regime.

\begin{figure}[t]
\centerline{\epsfig{file=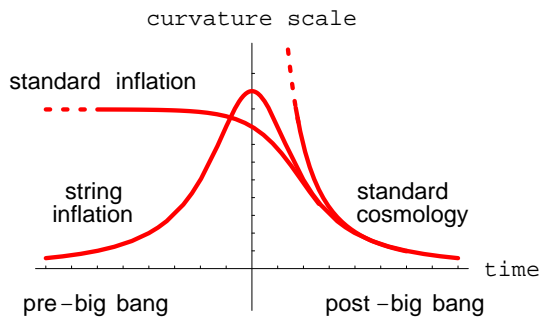,width=72mm}}
\vskip 5mm
\caption{\sl Qualitative evolution of the curvature scale in the standard
cosmological model, in conventional inflationary models and in 
string-cosmology models.}
\label{f11}
\end{figure}

The main difference between the string cosmology and the standard
inflationary scenarios can therefore be underlined through  the opposite
behaviour of the curvature scale as a function of time, as  shown in Fig.
\ref{f11}. As we go backward in time, instead of a monotonic growth
(predicted by the standard scenario), or of a ``de-Sitter-like" phase of
nearly constant curvature (as in the conventional inflationary picture),
the curvature grows,  reaches a maximum controlled by the string scale 
$M_{\rm s}=\la_{\rm s}^{-1}$,  and then starts decreasing towards an
asymptotically flat state, the string perturbative vacuum. The big bang
singularity is regularized by a ``stringy" phase of high but finite
curvature, occurring at the end of the initial inflationary evolution.

We should warn the reader, from the very beginning of this
review, that this scenario is far from being complete and
understood in all of its aspects, and that many important problems
are to be solved still. Nevertheless, the results obtained up to now have
been encouraging, in the sense that it now seems possible to formulate
models for the pre-big bang evolution of our Universe that fit
consistently in a string theory context, and which are compatible with
various phenomenological and theoretical bounds. Not only: the
parameter space of such models seems to be accessible to direct
observations in a  relativiely near future and, at present, it is already
indirectly constrained by various astrophysical, cosmological,
 and particle physics data.

To close this subsection we should mention, as a historical note, that the
idea of a phase of growing curvature preceding that of standard 
decelerated expansion, is neither   new  in cosmology, nor  peculiar to
string theory. Indeed, if the growth of the curvature corresponds to a
contraction, it is reminiscent of Tolman's cyclic Universe \cite{Tol}, in
which the birth of our  present Universe is preceded by a phase of
gravitational collapse (see also \cite{DurLau96,BP91,Novello93}). Also,
and more conventionally, the growth of the curvature may be
implemented as a phase of Kaluza--Klein superinflation
\cite{Sha84,Abb84,Kolb84}, in which the accelerated expansion of our
three-dimensional space is sustained by the  contraction of the internal
dimensions and/or by some exotic source, with the appropriate equation
of state (in particular, strings \cite{GSV91} and extended objects). 

In the context of general relativity, however, the problem is how to
avoid the curvature singularity appearing at the end of the phase of
growing curvature. This is in general impossible, for both contraction 
and superinflationary expansion, unless one accepts rather drastic
modifications of the classical gravitational theory. In the contracting
case, for instance, the damping of the curvature and a smooth
transition to the phase of decreasing curvature can be arranged
through the introduction of a non-minimal and gauge-non-invariant
coupling of gravity to a cosmic vector \cite{Novello78} or scalar 
\cite{SatSi86,Barrow93} field,  with a (phenomenological) modification
of the equation of state in the Planckian curvature regime
\cite{Rosen85,Wesson85}, or with the use of a non-metric,
Weyl-integrable connection \cite{Novello93}. In the case of
superinflation,  a smooth transition can be arranged through a breaking
of the local Lorentz symmetry of general relativity
\cite{Gas85,Gas98a}, a geometric contribution of the spin of the
fermionic sources \cite{Gas86}, or the embedding of the space-time
geometry into a more fundamental quantum phase-space dynamics
\cite{Caia91,Gas91}. In the more exotic context of topological
transitions, a smooth evolution from contraction to expansion, through 
a state of minimal size, is also obtained with the adiabatic compression
and the dimensional transmutation of the de Sitter vacuum
\cite{GasUm91}. 

In the context of string theory, on the contrary, the growth of the
curvature is naturally associated to the growth of the dilaton and of the
coupling constants (see for instance Section \ref{Sec2}). This effect, on
the one hand,  sustains the phase of superinflationary expansion, with
no  need of matter sources or extra dimensions. On the other hand, it
necessarily leads the Universe to a regime in which not only the
curvature but also the couplings become strong, so that typical 
``stringy" effects   become important and  are expected to smooth 
out the curvature singularity. This means that there is 
no need to look for more or less {\em ad hoc}  modifications of the
theory,  as string theory itself is expected to provide the appropriate
tools for a complete and self-consistent cosmological scenario.

\subsection{Pre-big bang inflation and conformal frames}
\label{Sec1.3}

While  postponing to the next section the issue of
physical motivations, it is important to classify the various inflationary
possibilities just from their kinematical properties. To be more precise,
let us consider the so-called flatness problem (the arguments  are
similar, and the conclusions the same, for the horizon problem 
mentioned in Subsection \ref{Sec1.1}). We shall assume, on the basis of
the approximate isotropy observed at large scale, that our present
cosmological phase can be correctly described by the ordinary
Einstein--Friedmann equations.  In that case, the gravitational part of
the equations contains two  contributions from the metric: $k/a^2$,
coming from the spatial (or intrinsic) curvature, and $H^{2}$, coming
from the gravitational kinetic  energy (or extrinsic curvature). Present
observations imply that the spatial curvature term, if not
negligible, is at least non-dominant, i.e.
\beq
 r^2={k\over a^2H^2} \laq 1. 
\label{11}
\eeq

On the other hand, during a phase of standard, decelerated expansion,
 the ratio $r$ grows with time. Indeed, if $a \sim t^\b$, 
\beq
r \sim \dot a ^{-1} \sim t^{1-\b} ,
\label{12}
\eeq
so that $r$ is growing both in the matter-dominated ($\b=2/3$)
and in the radiation-dominated ($\b=1/2$) era. Thus, as we go back in 
time, $r$ becomes smaller and smaller. If, for instance,  we wish to
impose initial conditions at the Planck scale,  we must require a
fine-tuning suppressing by $30$ orders of magnitude the spatial
curvature term with respect to the other terms of the cosmological
equations. Even if initial conditions are given at a lower scale (say the
GUT scale) the amount of fine-tuning is still nearly as bad.

This problem can be solved by introducing an early phase during which
the value of $r$, initially of order $1$, decreases so much in time that
its subsequent growth  during FRW evolution keeps it still below $1$
today. It is evident that, on pure kinematic grounds, this requirement
can be implemented in two classes of backgrounds.

\begin{itemize}
\item[{\bf (I)}] : $a \sim t^\b, ~\b >1,  ~t \ra +\infty$. This class of
background corresponds to what is conventionally called ``power
inflation" \cite{Luc85}, describing accelerated expansion and decreasing
curvature scale,  $\dot a >0$, $\ddot a >0$, $\dot H <0$. It contains,  as
the  limiting case ($\b \ra \infty$), exponential de Sitter inflation,  $a
\sim
e^{Ht}$, $\dot H =0$, describing accelerated expansion with constant
curvature.
 \item[{\bf(II)}] : $a \sim (-t)^\b,~ \b <1, ~ t \ra 0_-$. This case
contains
two subclasses.
\begin{itemize}
\item[{\bf(IIa)}] : $\b<0$, corresponding to ``superinflation" or ``pole
inflation" \cite{Sha84,Abb84,Kolb84}, and describing  accelerated
expansion with growing curvature scale,  $\dot a >0$, $\ddot a >0$,
$\dot H >0$;
\item[{\bf(IIb)}] : $0<\b<1$, describing accelerated contraction and
growing curvature scale \cite{GasVe93b},   $\dot a <0$, $\ddot a <0$,
$\dot H <0$. 
\end{itemize}
\end{itemize}
In the first class of backgrounds, corresponding to post-big bang
inflation, the Universe is driven {\em away} from the
singularity/high-curvature regime, while in the second class inflation
drives the Universe {\em towards} it, with the typical pre-big bang
behaviour illustrated in Fig. \ref{f11}.  We may thus immediately note  a
very important ``phenomenological" difference between post- and
pre-big bang inflation. In the former case the Planck era  lies very far in
the past, and its physics  remains screened from present observations,
since the scales that probed Planckian physics are still far from
re-entering. By contrast, in the pre-big bang case, the Planck/string
regimes are closer to us (assuming that no or little inflation occurs
after the big bang itself). Scales that probe Planckian physics are now
the first to re-enter, and to leave an imprint for our observations (see,
for instance, the case of a stochastic background of relic
gravitational waves, discussed in Section \ref{Sec5}).

The inflationary character of {Class IIa} backgrounds is well known,
and recognized since the earlier studies of the inflationary scenario
\cite{Luc85}. The inflationary character of {Class IIb}  is
more unconventional --a sort of ``inflation without inflation"
\cite{Gas94a}, if we insist on looking at inflation as accelerated
expansion--  and was first pointed out only much later \cite{GasVe93b}.
It is amusing to observe that, in the pre-big bang scenario, both
subclasses {IIa} and  {IIb} occur. However, as discussed in detail in
Subsection \ref{Sec2.5}, they do not correspond to different models of
pre-big bang inflation, but simply  to different kinematical
representations of the  {\em same} scenario in two {\em different}
conformal frames.

In order to illustrate this point, which is important also for  our
subsequent arguments, we shall proceed in two
steps. First we will show that, through a field redefinition $g=g(\ti g,\ti
\phi)$, $\phi= \phi (\ti \phi)$, it is always possible to move from
the string frame (S-frame), in which the lowest order gravidilaton 
effective action takes the form
\beq
S(g,\phi)= - \int d^{d+1}x \sqrt{|g|}~
e^{-\phi}\left[R+ g^{\mu\nu}\pa_\mu\phi \pa_\nu \phi\right],
\label{13}
\eeq
to the Einstein frame (E-frame), in which the dilaton  $\phi$ is minimally
coupled to the metric and has a canonical  kinetic term:
\beq
S(\ti g,\ti \phi)= - \int d^{d+1}x \sqrt{|\ti g|}~
\left[\ti R-{1\over 2} \ti g^{\mu\nu}\pa_\mu\ti \phi \pa_\nu \ti
\phi\right]
\label{14}
\eeq
(see Subsection \ref{Sec1.4} for notations and conventions).
Secondly, we will show that, by applying such a redefinition, a
superinflationary solution obtained in the S-frame becomes an
accelerated contraction in the E-frame, and vice versa.

We shall consider, for simplicity, an isotropic, spatially flat background
with $d$ spatial dimensions, and  set:
\beq
g_{\mu\nu}= {\rm diag} \left( N^2, -a^2 \da_{ij}\right), ~~~~~~~
\phi=\phi(t),
\label{15}
\eeq
where $g_{00}=N^2$ is to be fixed by an arbitrary gauge choice.  For
this background  the S-frame action (\ref{13}) becomes,  modulo a total
derivative, \beq
S(g,\phi)= - \int d^{d+1}x {a^d
e^{-\phi}\over N}\left[\dot \phi^2-2dH\dot\phi +d(d-1)H^2 \right],
\label{16}
\eeq
where, as expected, $N$ has no kinetic term and plays the role of a
Lagrange multiplier.
In the E-frame the variables are $\ti N, \ti a, \ti \phi$, and the
action(\ref{14}), after integration by parts, takes the canonical form
\beq
S(\ti g, \ti \phi)= - \int d^{d+1}x {\ti a^d
\over \ti N}\left[-{1\over 2}\dot {\ti \phi}^2 +d(d-1)H^2\right]. 
\label{17}
\eeq
A quick comparison with Eq. (\ref{16}) finally leads  to the field
redefinition (not a coordinate transformation!) connecting the Einstein
and String frames:
\beq
\ti a = a e^{-\phi/(d-1)}, ~~~~~~~~~
\ti N = N e^{-\phi/(d-1)}, ~~~~~~~~~
\ti \phi =\phi \sqrt{2\over d-1}.
\label {18}
\eeq

Consider now  an isotropic, $d$-dimensional vacuum solution of the
action (\ref{16}), describing a superinflationary, pre-big bang expansion
driven by the dilaton (see Section \ref{Sec2}) \cite{Mul90,GV91}, of 
Class IIa, with  $\dot a >0, \ddot a >0, \dot H>0, \dot \phi >0$:
\beq
 a= (-t)^{-1/\sqrt{d}}, ~~~~~~~~~
e^\phi=  (-t)^{-(\sqrt{d}+1)}, ~~~~~~~~~
t<0, ~~~~~~~~~~t \ra 0_-, 
\label{19}
\eeq
 and look for the
corresponding E-frame solution. Since the above solution is valid in the
synchronous gauge, $N=1$,  we can choose, for instance, the
synchronous gauge also in the E-frame, and  fix $\ti N$ by the
condition:
\beq
\ti N dt \equiv N e^{-\phi/(d-1)} dt = d \ti t,
\label{110}
\eeq
which defines the E-frame cosmic time $\ti t$ as:
 \beq
d \ti t = e^{-\phi/(d-1)} dt .
\label{111}
\eeq
After integration
\beq
t \sim \ti t ^{d-1\over d +\sqrt{d}};
\label{112}
\eeq
the transformed solution takes the form:
\beq
\ti a= (-\ti t)^{1/{d}}, ~~~~~~~~~
e^{\ti \phi}=  (-\ti t)^{-\sqrt{2(d-1)\over d}}, ~~~~~~~~~
\ti t<0, ~~~~~~~~~ \ti t \ra 0_- .
\label{113}
\eeq
It can easily be checked that this solution describes accelerated
contraction of {Class IIb}, with growing dilaton and growing
curvature scale: \beq
{d\ti a\over d \ti t}<0,~~~~~~~
{d^2\ti a\over d \ti t^2}<0,~~~~~~~
{d\ti H\over d \ti t}<0,~~~~~~~
{d\ti \phi\over d \ti t}>0.
\label{114}
\eeq
The same result applies if we transform other isotropic solutions from
the String to the Einstein frame, for instance the superinflationary
solutions with perfect fluid sources \cite{GasVe93b}, presented in
Section \ref{Sec2}.

To conclude this section, and for later use, let us stress that the
main dynamical difference between post-big bang inflation, 
Class I metrics,   and pre-big bang inflation, {Class II} metrics,  can
also be conveniently illustrated in terms of the  proper size of the event
horizon, relative to a given comoving observer.

Consider in fact the proper distance $d_e(t)$ of the event
horizon from a comoving observer, at rest in an isotropic,
conformally flat background \cite{Rin56}: 
\beq
d_e(t)= a(t)\int _t^{t_M} dt' a^{-1} (t'). 
\label{115}
\eeq
Here $t_M$ is the maximal allowed extension, towards the future, of the
cosmic time coordinate for the given background manifold. The above
integral converges for all the above classes of accelerated (expanding or
contracting) scale factors. In the case of {Class I} metrics we have, in
particular,
\beq
d_e(t)= t^\beta\int _t^{\infty} dt' t'^{-\b}= {t\over \b-1} =
{\b\over \b-1} H^{-1}(t)
\label{116}
\eeq
for power-law inflation ($\b >1, t>0$), and
\beq
d_e(t)= e^{Ht}\int _t^{\infty} dt' e^{-Ht'}=  H^{-1}
\label{117}
\eeq
for de Sitter inflation. For {\sl Class II} metrics ($\b <1, t<0$) we have
instead
\beq
d_e(t)= (-t)^\beta\int _t^{0} dt' (-t')^{-\b}= {(-t)\over 1-\b} 
={\b\over \b-1} H^{-1}(t). 
\label{118}
\eeq
In all cases the proper size $d_e(t)$ evolves in time like the so-called
Hubble horizon (i.e. the inverse of the modulus of the Hubble parameter),
and then like the inverse of the curvature scale. The size of the  horizon
is thus constant or growing in standard inflation ({Class I}),
decreasing in pre-big bang inflation ({Class II}), both in the S-frame
and in the E-frame.

\begin{figure}[t]
\centerline{\epsfig{file=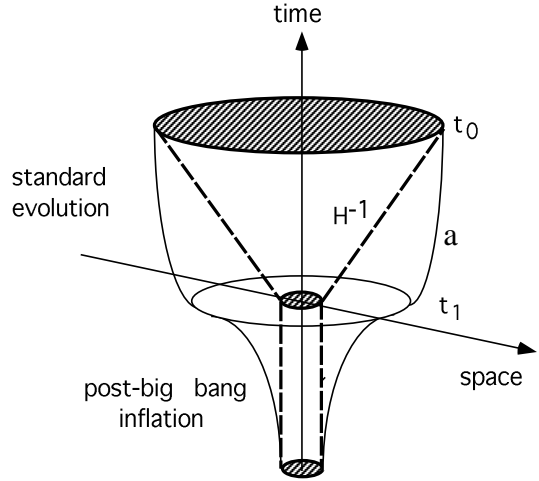,width=72mm}}
\vskip 5mm
\caption{\sl Qualitative evolution of the Hubble horizon (dashed curve)
and of the scale factor (solid curve) in the standard, post-big bang
inflationary scenario.}
\label{f12}
\end{figure}

\begin{figure}[t]
\centerline{\epsfig{file=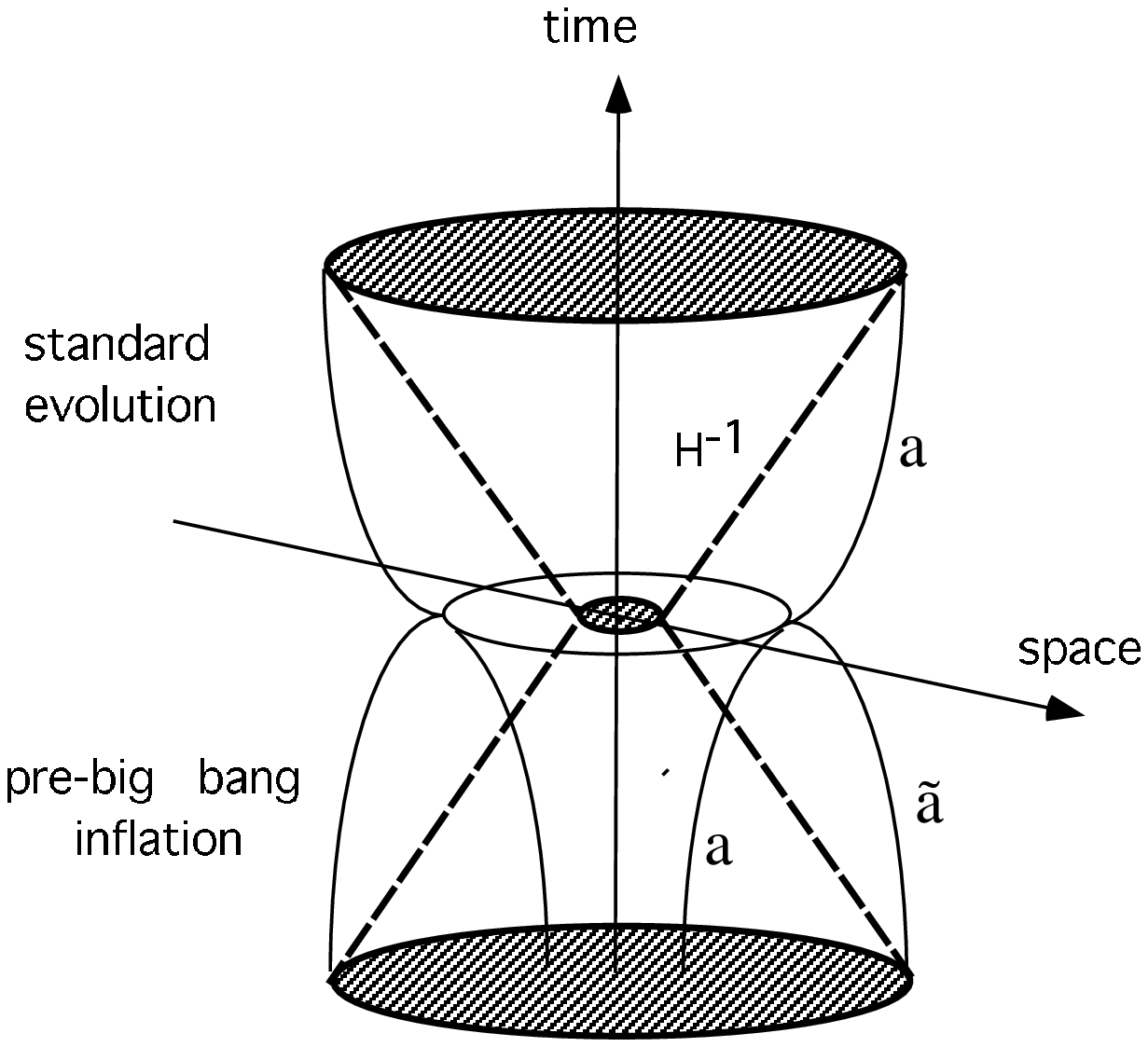,width=72mm}}
\vskip 5mm
\caption{\sl Qualitative evolution of the Hubble horizon (dashed curve)
and of the scale factor (solid curve) in the pre-big bang inflationary
scenario, in the S-frame, $a(t)$, and in the E-frame, $\ti a(t)$.}
\label{f13}
\end{figure}

Such an important difference is clearly illustrated in Figs.  \ref{f12} and
\ref{f13}, where the dashed lines represent the evolution of the
horizon and the solid curves the evolution of the scale factor.  The
shaded area at time $t_0$ represents the portion of Universe inside our
present Hubble radius. As we go back in time, according to the standard
scenario, the horizon shrinks linearly ($H^{-1} \sim t$); however, the
decrease of the scale factor is slower, so that, at the beginning of the
phase of standard evolution ($t=t_1$), we end up with a causal horizon
much smaller than the  portion of Universe that we presently observe.
This is the ``horizon problem" already mentioned at the beginning of this
section.

In Fig. \ref{f12} the phase of standard evolution is preceded in  time
by a phase of standard, post-big bang (in particular de Sitter) inflation.
Going back in time, for $t<t_1$, the scale factor keeps shrinking, and our
portion of  Universe ``re-enters" the Hubble radius during a phase
of constant (or slightly growing in time) horizon.

In Fig. \ref{f13} the standard evolution is preceded in time by a phase
of pre-big bang inflation, with growing curvature. The  Universe
``re-enters"  the Hubble radius during a phase of shrinking  horizon. To
emphasize the difference, we have plotted the evolution of the scale
factor both as expanding in the S-frame, $a(t)$, and as contracting
in the E-frame, $\ti a (t)$. Unlike in post-big bang inflation, the proper
size of the initial portion of the Universe may be very large in strings (or
Planck) units, {\em but not larger than the initial horizon} itself
\cite{Gas00}, as emphasized in the picture, and as will be discussed in a
more quantitative way in Section \ref{Sec2.4}. The initial horizon
$H_i^{-1}$ is large because the initial curvature scale is small (in string
cosmology, in particular,  $H_i \ll 1/\la_{\rm s}$).

This is a basic consequence of the choice of the initial state; in
the context of the string cosmology scenario, this approaches the flat,
cold and empty string perturbative vacuum, (see the discussion of
Section \ref{Sec3}). This initial state has to be contrasted with the
extremely curved, hot and dense initial state of the standard scenario,
characterizing a  Universe that starts inflating at (or soon after) the
Planck scale, $H_i \sim 1/\la_{\rm P}$ (see also \cite{Gas00a} for a more
detailed comparison and discussion of pre-big bang  versus post-big bang
inflation).

\subsection{Outline, notations and conventions}
\label{Sec1.4}

We give here a general overlook at the material presented
in the various sections of this report. Furthermore, each section will 
begin with
an outline of the content of each of its subsections, and will try to be
as self-contained as possible, in order to help the reader interested  only
in some particular aspects of this field.

In Section {\ref{Sec2}, after a very quick reminder of some
relevant properties of superstring theory,  we review   the
string-theoretic motivations behind the pre-big bang  scenario,   and
outline the main ideas. In Section \ref{Sec3}, after formulating on   the
basis of the previous discussion a postulate of `` asymptotic past
triviality", we discuss, within that framework, the problem of initial
conditions and fine-tuning.

As a preliminary to the discussion of the observational consequences of 
the pre-big bang scenario, Section \ref{Sec4} presents some general
results on the evolution of quantum fluctuations. Section \ref{Sec5}  
will deal  with the specific case of tensor (metric) perturbations and 
Section \ref{Sec6} with  scalar (dilatonic) ones, while Section \ref{Sec7}
will consider gauge (in particular electromagnetic) and axionic
perturbations, and their possible physical relevance to galactic magnetic
fields and large scale structure, respectively.

The last part of this report is devoted to various open problems,
 in particular to the  possibility of a smooth transition from the
pre- to the post-big bang regime, the so-called ``exit problem" (Section
\ref{Sec8}).   Several lines of approach to  (and partial solutions of) this
problem are presented, including a discussion of heat and entropy
production in pre-big bang cosmology.  A possible minisuperspace
approach to the quantum regime  will be illustrated in
Section \ref{Sec9}. Finally,  Section \ref{Sec10} contains the
discussion of a possible dilatonic interpretation of the quintessence, a
short presentation of other open problems, and  an outlook. 

For further study of string cosmology and of the pre-big bang
scenario, we  also refer the interested reader to other 
review papers \cite{Lidsey00,Gas00a,Bar98,Fen00,Bran01}, as well as to
two recent introductory lectures \cite{Gas99a,Ve99a}. A regularly 
updated  collection of selected papers on the pre-big bang scenario is
also available at \cite{GasWeb}. 

We finally report  here our conventions for the metric and the curvature
tensor, together with the definitions of some variables frequently used in
the paper.

We shall always use natural units $\hbar=c=k_B=1$. Unless otherwise
stated, the metric signature is fixed by $g_{00}>0$; the Riemann and Ricci
tensors are defined by 
\beq
R_{\mu\nu\a}\,^\b= \pa_\mu\Ga_{\nu\a}\,^\b+ \Ga_{\mu\r}\,^\b
\Ga_{\nu\a}\,^\r - (\mu \leftrightarrow \nu),
~~~~~~
R_{\nu\a}= R_{\mu\nu\a}\,^\mu.
\label{119}
\eeq
In particular, for a Bianchi-I-type metric, and in the synchronous gauge, 
\beq
g_{\mu\nu}= {\rm diag} \left( 1, -a^2_i(t) \da_{ij}\right),
\label{120}
\eeq
our conventions lead to
\bea
&&
 R_0\,^0= - \sum_i \left(\dot H_i +H_i^2\right), ~~~~~~~~~
R_i\,^j= - \dot H_i\da_i^j -H_i\da_i^j \sum_kH_k,\nonumber \\
&&
R= -\sum_i\left( 2 \dot H_i + H_i^2\right)- \left(\sum_i H_i\right)^2,
\label{121}
\eea
where $H_i= d \ln a_i/dt$. 
In a $D=d+1$ space-time manifold, Greek indices run from $0$ to $d$,
while Latin indices run from $1$ to $d$.

The duality invariant dilatonic variable, the ``shifted dilaton"  $\fb$, is
referred to a $d$-dimensional spatial section of finite volume,
$(\int d^dx\sqrt{|g|})_{t={\rm const}}<\infty$, and is defined by
\beq
e^{-\fb}= \int{ d^{d}x \over \la_{\rm s}^d}\sqrt{|g_d|}
e^{-\phi}.
\label{122}
\eeq
In a Bianchi-I-type metric background, in particular, we shall absorb
into $\phi$ the constant shift $-\ln(\la_{\rm s}^{-d}\int  d^dx)$ (required to
secure the scalar behaviour of $\fb$ under coordinate
reparametrizations), and we shall set
\beq
\fb= \phi- \sum_i \ln a_i .
\label{123}
\eeq

Finally, $\la_{\rm s}$  is the fundamental length scale of string theory,
related to the string mass $M_{\rm s}$  and to the string tension $T$
(the mass per unit length) by
\beq
\la_{\rm s}^2= M_{\rm s}^{-2}T^{-1}\equiv 2\pi \ap.
\label{124}
\eeq
At the tree level (i.e. at lowest order) in the string coupling   $g_{\rm
s}$, the string length is related to the Planck length $\la_{\rm P}$,  and
to the gravitational constant $G_D$ in $d+1$ dimensions, by
\beq
8\pi G_D=\la_{\rm P}^{d-1}= \la_{\rm s} ^{d-1}e^{\phi}.
\label{125}
\eeq
In $d=3$, in particular, the relation between the string and the Planck
mass $M_{\rm P}=\la_{\rm P}^{-1}$ reads
\beq
(\la_{\rm P}/\la_{\rm s})^2=(M_{\rm s}/M_{\rm P})^2= e^{\phi}.
\label{126}
\eeq
We shall often work in units such that $2 \la_{\rm s} ^{d-1}=1$,  i.e.
$16 \pi G_D=1$, in which $\exp (\phi)$ parametrizes, in the String frame,
the (dimensionless) strength of the gravitational coupling.

\section{String theory motivations}
\label{Sec2}
\setcounter{equation}{0}
\setcounter{figure}{0}

A very important concept in string theory, as well as in field theory, is
that of moduli space, the space of vacua. In field theory, and in the
absence of gravity, the coordinates of moduli space
label the possible ground states of the theory.
It is very important to immediately distinguish  classical moduli space
from its exact, quantum counterpart. The two are generally
different, since a classical-level ground state can fail to be a  true
ground state when perturbative or non-perturbative quantum
corrections are added (consider for instance  dynamical symmetry
breaking \`a la Coleman--Weinberg, or the double-well
potential in quantum mechanics).

When gravity is added to the picture (and this is always the case in s
tring theory) the concept of a lowest-energy state becomes less well
defined, since total energy is always zero in a general-covariant theory.
It is therefore better, in string theory, to extend the definition of moduli
space to include all string backgrounds that allow a consistent string
propagation, i.e. those consistent with world-sheet conformal
invariance \cite{GSW87}.  Such backgrounds  correspond to the vanishing
of the two-dimensional $\sigma$-model $\beta$-functions
\cite{Love84} and, at the same time, they can also be shown to satisfy
the field equations of an effective action living in ordinary space-time.
The most famous example of a consistent background is, for
superstrings, $D=10$ Minkowski space-time with trivial (i.e. constant)
dilaton and antisymmetric tensor potentials. Unfortunately, even if
quantum (i.e. string-loop) corrections are neglected, our knowledge of
moduli space is very limited. Basically,  apart from a handful of exact
conformal field theories, such as the Wess--Zumino--Witten models,
only low-energy solutions
 (i.e. classical solutions of the effective low-energy
field theory) are known.

The known solutions are, at the same time, too many and too few.  They
are too many because they typically leave the vacuum expectation 
values of a few scalar fields completely undetermined. Such fields
correspond to gravitationally coupled  massless scalar
particles that mediate dangerous long-range forces, badly violating the
well-tested equivalence principle. The way out of this problem is clear:
these flat directions should be lifted by quantum corrections, typically 
(in the supersymmetric case)  at the non-perturbative level.
The known solutions are also too few, because some of them, which we
would like to see appearing, are missing: notably those describing
gravitational collapse or cosmological backgrounds, which evolve as a
function of time from a regime of low curvature and/or coupling to one
of high curvature and/or coupling  (and vice versa). These solutions are
very hard to analyse for two reasons: first, because time evolution
spontaneously breaks supersymmetry, rendering the solutions unstable
to radiative corrections;  secondly, because the solutions go out of
theoretical control, as they enter the non-perturbative regime.

For the purpose of this section, the important property of superstring's
moduli space is that it exhibits duality symmetries. Generically, this
means that points in moduli space that seem to describe different
theories actually describe the same theory (up to some irrelevant
relabelling of the fields). Let us illustrate this in the simplest example of
$T$-duality (for a review see, for instance, \cite{GiPoRa94}).

Consider a theory of closed strings moving in a space endowed with
some compact dimensions, say, for the sake of simplicity, with one 
extra dimension having the topology of a circle. Let us denote by $R_c$
the radius of this circle. In point-particle theory, momentum along the
compact dimension is quantized, in units of $\hbar/R_c$. This is also true
in string theory, as far as the motion of the string's centre of mass (a
non-oscillatory ``zero mode") is concerned. However, for closed strings
moving on a compact space, like our circle, there is a second zero mode:
the string can simply wind around the circle an integer number of times.
By doing so it acquires winding energy, which is quantized in units  of
$(2 \pi T) R_c$, if $T=(2 \pi \ap)^{-1}$ is the string tension.

Something remarkable  does happen if we replace $R_c$ by $\ti{R_c}
\equiv \hbar/TR_c \equiv \lambda_{\rm s}^2/R_c$. A point  particle
would certainly notice the difference between $R_c$ and $\ti{R_c}$
(unless $R_c=\ti{R_c}=\lambda_{\rm s}$), since the new momenta will
be different from the old ones. A closed string, instead, does not feel
the change of $R$ since the role of the momenta in the original theory
will mow be played by the winding modes, and vice versa. This
symmetry of closed string theory has been called $T$-duality and is
believed to be exact, at least to all orders of perturbation theory,
provided a suitable transformation of the dilaton accompanies the one
on the radius (it also has interesting extensions to discrete groups of 
the $O(d,d;Z)$ type  \cite{GiPoRa94}).  It is important to stress, in our
context, that $T$-duality actually implies that there is a physical lower
limit to the dimensions of a compact space, controlled by the string
length $\lambda_{\rm s}$ itself.  

When applied to open strings, $T$-duality leads to the concept of
Dirichlet strings, or $D$-strings.  In other words, while closed strings are
self-dual, open strings with Neumann boundary conditions are dual to
open strings with Dirichlet boundary conditions, and vice versa. These 
developments \cite{Pol96} have led to the study of $D$-branes, the
manifolds on which the end-points of open $D$-strings are confined;
they  play a major role in establishing the basic unity of all five known
types of $10$-dimensional superstring theories (Type I, Type IIA,
Type IIB, HETSO(32), HETE8) as different limits of a single, more
fundamental  ``M-theory" \cite{Wit95}. 

In order to briefly illustrate this point we start recalling that all
superstring theories are actually defined through a {\it double}
perturbative expansion in {\it two} dimensionless parameters: the first,
the string coupling expansion, can be seen as the analogue of the loop
expansion in quantum field theory, except that the coupling constant 
gets promoted to a scalar field, the dilaton. Consequently, the range of
validity of the loop expansion depends on the value of the dilaton, and
can break down in certain regions of space-time if the dilaton is not
constant. The second expansion, which has no quantum field theory
analogue, is an expansion in derivatives,  the dimensionless parameter
being $\lambda_{\rm s}^2 \partial^2$, with $\lambda_{\rm s}$ the fundamental
length scale of string theory (see Subsection 2.1). Obviously, the validity
of this second expansion breaks down when curvature or field 
space-time derivatives become of order $1$ in string-length units.

One of the most amazing recent developments in string theory  
\cite{Wit95} is the recognition that the above five theories, rather than
forming isolated islands in moduli space, are connected to one another
via a web of duality transformations. In the huge moduli space, they
represent ``corners" where the above-mentioned perturbative
expansions are, qualitatively at least, correct. A sixth corner actually
should be added, corresponding to $11$-dimensional supergravity
\cite{Tow95}.  The mysterious theory approaching these six known
theories in appropriate limits was given the name of M-theory. It
might seem curious, at first sight, that one is able to flow
continuously from $10$ to $11$ dimensions, within a single theory. The
puzzle was solved after it was realized that the $D=11$ supergravity
theory  has no dilaton, hence no free coupling. In other words,  the
dilaton of the five $10$-dimensional superstring theories becomes (at
large coupling) an extra dimension of space \cite{Wit95,Hor96}.  At weak
coupling, this extra dimension is so small that one may safely describe
physics in $10$ dimensions.

In spite of the beauty of all this, the previous discussion shows that
the moduli-space diagram connecting the five superstring theories can
be quite misleading. Since each point in the diagram is supposed to
represent a possible solution of the $\beta$-function constraints,  and
the diagram itself is supposed to show how apparently different 
theories are actually connected by moving in coupling constant (or other
moduli-) space, it necessarily includes the flat directions we have been
arguing against. At the same time, cosmological solutions of the
above-mentioned type, i.e. evolving in different regions of coupling
constant/curvature, cannot be ``localized" in the diagram. A single
cosmology, for instance, may indeed correspond to an initial
Universe, well described by heterotic string theory, ending  in
another Universe better described by perturbative Type I theory. Yet, 
such a cosmological solution should be only a point, not a curve, in
moduli space. As we shall argue below, thanks to some gravitational
instability, solutions of the above type are rather the rule than the
exception.

It seems appropriate, at this point, to comment on the fact that modern
research in  string/M-theory looks to be strongly biased towards
analysing supersymmetric vacua, or at least solutions that preserve a
large number of supersymmetries \cite{Pol98}. While the
mathematical motivation for that is quite obvious, and physically
interesting results can be rigorously derived in special cases
(concerning, for instance, BPS states and black holes),  it is quite clear
that cosmology, especially inflationary or rapidly evolving solutions,
requires extensive breaking of supersymmetries. Given their
phenomenological importance, we think that more effort should be
devoted to developing new techniques dealing with non-supersymmetric
solutions and, in particular, with their high-curvature and/or 
large-coupling regimes.

In this section we will emphasize the role of the duality symmetry 
for the pre-big bang scenario, starting from the  short- and
large-distance properties of string theory (Subsection \ref{Sec2.1}), and
from the related  invariance of the classical cosmological equations 
under a large group of non-compact  transformations (Subsections
\ref{Sec2.2}, \ref{Sec2.3}). We will discuss the inflationary aspects of
families of solutions related by such groups of transformations in 
Subsection \ref{Sec2.4}, and will end up  with a discussion of the 
relative merits of the so-called Einstein and String frames  (in
Subsection \ref{Sec2.5}),  in spite of their ultimate equivalence.

\subsection{Short- and large-distance motivations}
\label{Sec2.1}

Since the classical (Nambu--Goto)  action $S$ of a string
is proportional to the area $A$ of the space-time surface it sweeps, its
quantization requires the introduction of 
a  fundamental ``quantum" of length $\lambda_{\rm s}$, through the 
rlation: 
\begin{equation}
S/\hbar =  A/\lambda_{\rm s}^2\; .
\end{equation}
As discussed  in \cite{GVFC86,GVFC89}, the appearance of this length in
string theory is so much tied to quantum mechanics
that,  after introducing $\lambda_{\rm s}$, we are actually dispensed 
from introducing $\hbar$ itself, provided we use the natural units of
energy of string theory, i.e. provided we replace $E$ by $\alpha' E$,
the classical length of a string of energy $E$.
For instance, the  Regge trajectory relation between angular
momentum and mass,  $J = \alpha' M^2 + \alpha(0) \hbar$ ($c=1$
throughout), is rewritten in string units as $J = M^2 + \alpha(0)
\lambda_{\rm s}^2$,  where $\hbar$ has neatly disappeared.

One of the most celebrated virtues of string theory is its soft behaviour 
at short distances. This property, which is deeply related to the extended
nature of fundamental particles in the theory, makes gravity and gauge
interactions not just renormalizable, but finite, at least order by  order
in the loop expansion \cite{Pol98}. At the same time, the presence of an
effective short-distance cut-off in loop 
diagrams should have a counterpart in a modification of the theory,
when the external momenta of the diagram approach the cut-off itself. 

An analogy with the standard electroweak theory may be of order here:
the Glashow--Weinberg--Salam model makes Fermi's model softer at
short distances, turning a non-renormalizable effective theory into a
fully-fledged renormalizable gauge theory. At the same time, the naive
predictions of Fermi's model of electroweak interactions become largely
modified when the $W/Z$-mass region is approached. Note that this
mass scale, and not $G_F^{-1/2}$, is where new physics comes into play.

In string theory, something similar is expected to occur, this time in the
gravitational sector, where Einstein's effective theory should be
recovered at low energy, as soon as the string scale (and {\it not}
$G_N^{-1/2} \sim  M_{\rm P}$) is reached. At present, there are not so
many available tests of this idea: one comes from the study of
trans-Planckian energy collisions in string theory \cite{ACV1,ACV2}. 
When one  prepares the initial state in such a way as to expect the
formation of  tiny black holes, having radius (or curvature) respectively
smaller (or larger) than the fundamental scale of string theory, one finds
that such objects are simply not formed. Arguments based on entropy
considerations \cite{Bowick87} also suggest that the late stage of black
hole evaporation ends, in string theory, into a normal non-collapsed
string state of radius equal to the string length, rather than into a
singularity.  Finally, we recall the arguments from $T$-duality for a
minimal compactification scale, given at the beginning of this section.

This  considerable amount of circumstantial evidence leads to the 
conclusion  that $\lambda_{\rm s}$ plays, in string theory, the role of a
minimal observable length, i.e. of an ultraviolet cut-off. Physical
quantities are expected to be bounded (in natural units)  by the
appropriate powers of $\lambda_{\rm s}$, e.g. the curvature scale $R
\sim H^2 \sim G\rho \leq \lambda_{\rm s}^{-2}$, the temperature $T\leq
\lambda_{\rm s}^{-1}$, the compactification radius $R_c \geq \la_{\rm
s} $, and so on. It follows, in particular, that space-time singularities are
expected to be avoided (or at least reinterpreted) in any geometric
model of gravity which is compatible with string theory.  

In other words, in quantum string theory,  relativistic  quantum
mechanics should solve the singularity problems in much  the same way
as non-relativistic quantum mechanics solves the singularity
problem of the hydrogen atom, by keeping the electron and the proton a
finite distance apart. By the same token, string theory gives us a
rationale for asking  daring questions such as: What was there before
the big bang? Even if we do not know the answer to this question in
string theory, in no other currently available framework can such a
question be meaningfully asked.

To answer it, we should keep in mind, however, that even
at large distance (i.e. low energy, small curvatures), superstring theory
does {\it not} automatically give Einstein's general relativity. Rather, it
leads to a scalar--tensor theory (which in some limit reduces to a 
theory of the Jordan--Brans--Dicke variety). In fact, the conformal
invariance of string theory (see for instance \cite{GSW87}) unavoidably
requires a  new scalar particle/field $\phi$, the dilaton, which, as already
mentioned,  gets reinterpreted as the radius of a new dimension of 
space in M-theory \cite{Wit95,Hor96}. 

By supersymmetry, the dilaton is massless to all orders in perturbation
theory (i.e. as long as supersymmetry remains unbroken) andcan thus
give rise to dangerous violations of the equivalence principle. This is
just one example of the general problem with classical moduli space
that we mentioned at the beginning of this section. Furthermore,  $\phi$
controls the strength of all forces \cite{Wit84}, gravitational and gauge
alike, by fixing the grand-unification coupling through the (tree-level)
relation \cite{Kap85}:  
\begin{equation}
\a_{\rm GUT} \simeq \left( \la_{\rm P}/ \la_{\rm s}\right)^2 \simeq  \exp (\phi)
\; , 
\label{21}
\end{equation}
showing the basic unification of all forces in string theory and the
fact that, in our conventions, the weak-coupling region coincides with
$\phi \ll -1$.

The bottom line is that, in order not to contradict precision tests of the
equivalence principle and of the constancy  (today and in the ``recent"
past) of the gauge and  gravitational couplings, we
require \cite{Tay88} the dilaton to have a mass  and to be frozen at the
bottom of  its own potential {\it today}  (see, however,
\cite{DP94a,DP94b} for an ingenious alternative,  whose possible
consequences will be discussed in Sections. \ref{Sec6} and \ref{Sec10}).
This does not exclude, however, the possibility of the dilaton having
evolved cosmologically in the past (after all, the metric did!), within the
weak  coupling region ($\phi \ra -\infty$) where it was practically
massless.  The amazing (yet simple) observation \cite{GV91} is that, by
so doing, the dilaton may have inflated the Universe. 

A simplified argument, which, although not completely accurate,
captures the essential physical point, consists in writing the Friedmann
equation $3 H^2 = 8 \pi G \rho$,  and in noticing that a growing dilaton
(meaning through Eq. (\ref{21}) a growing $G$) can drive the growth of
$H$ even if the energy density of standard matter decreases (as
typically expected in an expanding Universe). This particular type of
superinflation, characterized by growing $H$ and $\phi$,  has been
termed dilaton-driven inflation. 

The basic idea of pre-big bang cosmology 
\cite{GV91,GasVe93a,GasVe93b,GasVe94a} can then be illustrated as  in
Fig. \ref{f21}, where the dilaton  starts at very large negative values,
rolling up a potential which, at the beginning, is practically zero  (at
weak coupling,  the potential is  known to be instantonically suppressed
\cite{BG86},  $V(\phi) \sim \exp [-\exp (-\phi)]$). The dilaton grows and
inflates the Universe, until the potential develops  some
non-perturbative structure, which can eventually  damp and trap the
dilaton (possibly after some oscillations). The whole process may be 
seen as the slow, but eventually explosive, decay of the string
perturbative vacuum (the flat and interaction-free asymptotic initial
state of the pre-big bang scenario), into a final, radiation-dominated
state typical of standard cosmology.  Incidentally, as shown in Fig.
\ref{f21}, the dilaton of string theory can easily roll-up --rather than
down-- potential hills, as a consequence of its non-standard coupling to
gravity.

A phase of accelerated evolution, sustained by the kinetic energy of a
growing dilaton \cite{GV91} (and possibly by other antisymmetric tensor
fields \cite{GMV91,Cop94}, in more complicated backgrounds)  is not just
possible: it does necessarily occur in a class of (lowest-order)
cosmological solutions based on a cosmological variant of the
(previously mentioned) $T$-duality  symmetry 
\cite{GV91,Tse91,MeiVe91,MeiVe91a,Sen91,Has92,Tse92,TseVa92,GasVe92}. 
In such a way duality provides a strong motivation for (and becomes a
basic ingredient of) the pre-big bang scenario first introduced in
\cite{GV91} and whose developments are the subject of this report. 

\begin{figure}[t]
\centerline{\epsfig{file=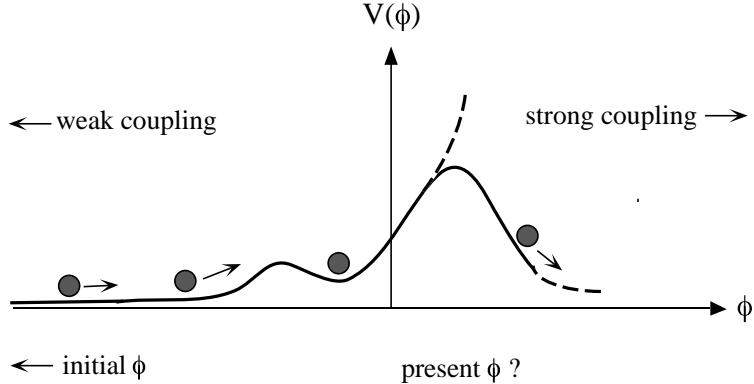,width=100mm}}
\vskip 5mm
\caption{\sl According to the pre-big bang scenario, the dilaton starts
in the asymptotic past of our Universe at very large negative values,
and grows through a flat potential towards the strong-coupling regime.
At present, it is either trapped at a minimum of the potential, or 
keeps growing monotonically towards $+\infty$ (see Section 
\ref{Sec10}).} 
\label{f21}
\end{figure}

The importance of duality
for a string-motivated cosmology was indeed pointed out already in
some pioneer papers \cite{Alvarez85,Leblanc88,AlvaOs89,BraVa89} based
on superstring thermodynamics; the fact that our standard 
Universe could emerge after a phase of inflation with ``dual" 
cosmological properties was also  independently suggested by the study
of string motion in curved backgrounds \cite{GSV91}. 
For future applications (see Section \ref{Sec8}) we wish to recall here,
in particular, the $T$-duality approach to a superstring Universe
discussed in \cite{BraVa89} (see also \cite{Hotta97}), in which all nine
spatial dimensions are compact (with similar radius), and the presence
of winding strings wrapped around the tori prevents the expansion of the
primordial Universe, unless such winding modes disappear by mutual
annihilation.  

The probability of annihilation, however,  depends
on the number of dimensions. In $d=9$ it is so small that strings prevent
the nine dimensions from expanding. If $n<6$ dimensions contract to the 
string size, string annihilation in the other $9-n$ is still so small that
even they cannot expand. Only for $n\ge 6$ is the annihilation 
probability in the remaining dimensions large enough, and the
expansion becomes possible:  this would give $3$ as the maximal number
of large space dimensions  (such a mechanism has recently  been 
extended also to a more general brane--gas context, see Subsection
\ref{Sec8.5}). 

This as well as the other, early attempts were based on Einstein's
cosmological equations, i.e. on gravitational equations at fixed dilaton.
Taking into account the large-distance modifications of general
relativity required by string theory, and including a dynamical dilaton, 
the target-space duality typical of closed  strings moving in compact
spaces can be extended (in a somewhat modified version)  even to 
non-compact cosmological backgrounds
\cite{GV91,MeiVe91a,Tse91,Tse92,TseVa92}. Consider in fact a generic
solution of the field equations of string theory (hence a point in our
moduli space), which possesses a certain number $n$ of Abelian
isometries (the generalization to non-Abelian isometries is subtle, see
\cite{Ossa93}). Working in an adapted coordinate system, in which
the  fields appearing in the solution are independent of
$n$-coordinates, it can then be argued \cite{MeiVe91} that there is an
$O(n,n;R)$ group that, acting on the solution, generates new ones (in
other words, this group has a representation in that part of moduli space
that possesses the said isometries). 

Note that, unlike strict $T$-duality, this 
continuous $O(d,d;R)$ extension is not a true symmetry of the theory, 
but only a  symmetry of the classical field equations. The corresponding
transformations can be used to generate, from a given solution, other,
generally inequivalent ones, and this is possible even in the absence of
compactification. In the next subsections we will show in detail that
such transformations, applied to a decelerated cosmological solution 
(and combined with a time-reversal transformation) lead in general to
inflation, and we shall present various (low-energy) exact inflationary
solutions, with and without sources, which may represent possible
models of pre-big bang evolution. We shall consider, in particular, both 
scale-factor \cite{GV91,Tse91,TseVa92} and $O(d,d)$
\cite{MeiVe91,MeiVe91a,Sen91,Has92,GasVe92} duality tranformations,
and we will discuss some peculiar kinematic aspects of such pre-big
bang solutions.

\subsection{Scale-factor duality without and with sources}
\label{Sec2.2}

We start by recalling that, in general relativity, the Einstein action is
invariant under time reflections. It follows that, if we consider an
isotropic, spatially flat metric parametrized by the scale factor $a(t)$,
\beq
ds^2=dt^2- a^2(t) dx_i^2,
\label{22}
\eeq
and if $a(t)$ is a solution of the Einstein equations, then $a(-t)$ is also
a solution. On the other hand, when $t \ra -t$ the Hubble parameter
flips sign; 
\beq
a(t) \rightarrow a(-t), ~~~~~~~~~~~~~~~~
H=\dot a/a \rightarrow -H.
\label{23}
\eeq
Thus, to any standard cosmological solution $H(t)$, describing
decelerated expansion and decreasing curvature ($H>0$, $\dot H <0$),
time reversal associates a ``reflected" solution, $H(-t)$,
describing a contracting Universe.

In string theory, the reparametrization and gauge invariance of
conventional field theories are expected to be only a tiny subset of a
much larger symmetry group, which should characterize the effective
action even at lowest order. The string effective action that we shall
use in this section, in particular, is determined by the usual
requirement that the string motion is conformally invariant at the
quantum level \cite{GSW87}. The starting point is the 
(non-linear) sigma model describing the coupling of a closed 
string to external metric ($g_{\mu\nu}$), scalar ($\phi$), and
antisymmetric tensor ($B_{\mu\nu}$) fields. In the bosonic sector the
action reads: 
\beq
S= -{1\over 4 \pi \ap} \int d^2 \xi \left[\sqrt{-\ga}
\ga^{ij}\pa_ix^\mu\pa_jx^\nu g_{\mu\nu}(x) +
\ep^{ij}\pa_ix^\mu\pa_jx^\nu B_{\mu\nu}(x)+{\ap\over 2}\sqrt{-\ga}
R^{(2)} \phi(x)\right].
\label{24}
\eeq
Here $2\pi \ap=\la_{\rm s}^2$, $\pa_i \equiv \pa/\pa \xi^i$, and  $\xi^i$ are
the coordinates spanning the two-dimensional string world-sheet
($i,j=1,2$), whose induced metric is $\ga_{ik}(\xi)$.  The coordinates
$x^\mu=x^\mu(\xi)$ are the fields determining the embedding of the
string world-sheet in the external (also called ``target") space,
$\ep_{ij}$ is the two-dimensional Levi-Civita tensor density, and
$R^{(2)}(\ga) $ is the scalar curvature for the world-sheet metric $\ga$.
We have included the interaction of the string with all  three
massless states (the graviton, the dilaton and the 
antisymmetric tensor) appearing in the lowest energy level of the
spectrum of quantum string excitations (the unphysical tachyon is
removed by supersymmetry \cite{GSW87}). We note, for further use,
that the antisymmetric field $B_{\mu\nu}$ is often called the 
Neveu--Schwarz/Neveu--Schwarz (NS-NS) two-form.  

If we quantize the above action for the self-coupled fields
$x^\mu(\xi)$,  we can expand the loop corrections of the corresponding
non-linear quantum field theory in powers of the curvature (i.e. in
higher derivatives of the metric and of the other background fields)
\cite{GSW87}. Such a higher derivative expansion is typical of an
extended object like a string, and is indeed controlled by the powers of
$\ap$, i.e. of the string length parameter $\la_{\rm s}$. At each order in
$\ap$, however, there is an additional higher genus expansion in the
topology of the world-sheet metric, which corresponds, in the quantum
field theory limit, to the usual loop expansion, controlled by the
effective coupling parameter $g_{\rm s}^2= \exp(\phi)$.

The conformal invariance of the classical string motion in an external
background can then be imposed, at the quantum level, at any loop
order: we obtain, in this way, a set of differential conditions to be
satisfied by the background fields for the absence of conformal
anomalies, order by order. At tree level in $g_{\rm s}$, and to lowest
order in $\ap$, such differential equations (in a critical number of
dimensions) are \cite{GSW87}:
\bea
&&
R_{\mu\nu} + \nabla_{\mu}\nabla_{\nu} \phi
-{1\over 4}
H_{\mu\a\b}H_\nu~^{\a\b}=0,
\label{25} \\
&&
R+ 2 \nabla^2 \phi -\left(\nabla_\mu \phi\right)^2 - {1\over 12}
H_{\mu\nu\a}^2=0,
\label{26}
\eea
where $H_{\mu\nu\a}=\pa_\mu B_{\nu\a}+$ cyclic permutations. By
introducing the Einstein tensor, $G_{\mu\nu}=
R_{\mu\nu}-g_{\mu\nu}(R/2)$, the first equation can be rewritten in a
more ``Einsteinian form" as
\beq
G_{\mu\nu} + \nabla_\mu\nabla_\nu \phi +{1\over 2}g_{\mu\nu}
\left[\left(\nabla \phi\right)^2 -2 \nabla^2 \phi  +{1\over 12}
H_{\a\b\ga}^2\right] -{1\over 4} H_{\mu\a\b}H_\nu\,^{\a\b}=0, 
\label{27}
\eeq
and it can be easily checked that Eqs. (\ref{26}), (\ref{27}) can be
derived by the $(d+1)$-dimensional effective action
\beq
S= -{1\over 2\la_{\rm s}^{d-1}} \int d^{d+1}x \sqrt{|g|}~
e^{-\phi}\left[R+ \left(\nabla \phi\right)^2- {1\over 12}
H_{\mu\nu\a}^2 \right] .
\label{28}
\eeq
This action (possibly supplemented by a non-perturbative dilaton
potential, and/or by a cosmological term in non-critical dimensions
\cite{GSW87}) is the starting point for the formulation of a
string-theory-compatible cosmology in the small-curvature and 
weak-coupling regime, $\ap R \ll1$, $g_{\rm s}^2 \ll1$ (see Section
\ref{Sec8} for higher-order corrections).

For an illustration of scale factor duality it will now be sufficient  to
restrict our attention to the gravidilaton sector of the action
(\ref{28}) (setting e.g. $H_{\mu\nu\a}=0$). We will consider an
anisotropic Bianchi-I-type metric background, with homogeneous 
dilaton $\phi=\phi(t)$, and we will parametrize the action in terms of 
the ``shifted" scalar field $\fb$ (see Section \ref{Sec1.4} for the 
notation). The field equations (\ref{26}), (\ref{27}) then provide a
system of $d+2$ equations for the $d+1$ variables $\{a_i, \fb\}$:
\bea
&&
\fbp^2 - \sum_i H_i^2=0,
\label{29}\\
&&
\dot H_i -H_i \fbp=0,
\label{210}\\
&&
2 \ddot {\fb} -\fbp^2 -\sum_i H_i^2=0.
\label{211}
\eea
In the absence of sources only $d+1$ equations are
independent (see for instance \cite{GasMaVe97}; Eq. (\ref{29}), in
particular, represents a constraint on the set of initial data, which is
preserved by the evolution).

The above string cosmology equations are invariant not only under a
time-reversal transformation,
\beq
t \ra -t, ~~~~~~~~~~ H \ra -H,  ~~~~~~~~~~\fbp \ra -\fbp,
\label{212}
\eeq
but also under a transformation that inverts any one of the
scale factors, preserving the shifted dilaton,
\beq
a_i \ra \ti a_i = a_i^{-1}, ~~~~~~~~~~~~~~
\fb \ra \fb,
\label{213}
\eeq
and which represents a so-called scale-factor duality transformation
\cite{GV91,Tse91}. Note that the dilaton $\phi$ is not invariant under 
this transformation: if we invert, for instance, the first $ k \leq d$
scale factors, the transformed dilaton $\ti \phi$ is determined by the
condition
\beq
\fb = \phi - \sum_{i=1}^d \ln a_i = \ti \phi -\sum_{i=1}^k \ln \ti a_i-
\sum_{i=k+1}^d \ln a_i.
\label{214}
\eeq
Thanks to scale-factor duality, given any exact solution of Eqs. 
(\ref{29})--(\ref{211}), represented by the set of variables
\beq
\{a_1, ...  , a_d, ~\phi\},
\label{215}
\eeq
the inversion of $k \leq d$ scale factors then defines a new exact
solution, represented by the set of variables
\beq
\left\{a_1^{-1}, ...  , a_k^{-1}, a_{k+1}, ... , a_d, ~\phi-2 \sum_{i=1}^k
\ln  a_i \right\}.
\label{216}
\eeq

Consider in particular the isotropic case $a_i=a$, where all the scale
factors get inverted, and the duality transformation takes the form:
\beq
a \rightarrow \tilde a = a^{-1}, ~~~~~~~~~~~~~~~~
\phi \rightarrow \tilde \phi = \phi- 2 d \ln a.
\label{217}
\eeq
When $a \ra a^{-1}$ the Hubble parameter $H= d(\ln a)/dt$  goes
into $-H$ so that, to each  of the two solutions related by time
reversal, $H(t)$ and $H(-t)$, is also associated a dual solution,
$\tilde H(t)$ and $\tilde H(-t)$, respectively (see Fig. \ref{f22}).
The space of solutions, in a string cosmology context, 
is thus richer 
than in the standard Einstein cosmology,  because of the
combined invariance under duality and time-reversal transformations.
In string cosmology, a  solution has in general four
branches: $a(t), a(-t), a^{-1}(t), a^{-1}(-t)$. Two branches describe
expansion ($H>0$), the other two branches describe contraction ($H<0$).
Also, as illustrated in Fig. \ref{f22}, for two branches the curvature scale
($\sim H^2$) grows  in time, so that they describe a Universe that
evolves {\em towards} a singularity, with a typical ``pre-big bang"
behaviour; for the other two branches the  curvature scale decreases, 
so that the corresponding Universe emerges {\em from} a singularity,
with a typical ``post-big bang" behaviour.

What is important, in our context, is that to any given decelerated, 
expanding solution, $H(t)>0$, with decreasing curvature, $\dot H(t)<0$
(typical of the standard cosmological scenario), is always associated an
inflationary ``dual partner" describing accelerated expansion, $\ti
H(-t)>0$,  and growing curvature, $\dot{\ti H}(-t)>0$. This pairing of
solutions (which has no analogue  in the
context of the Einstein cosmology, where there is no dilaton, and  the
duality symmetry cannot be implemented) naturally suggests a
``self-dual" completion of standard cosmology, in which the Universe
smoothly evolves from the inflationary pre-big bang branch
$\ti H(-t)$ to the post-big bang branch $ H(t)$ (after an appropriate
regularization of the curvature singularity appearing in the 
lowest-order solutions).

\begin{figure}[t]
\centerline{\epsfig{file=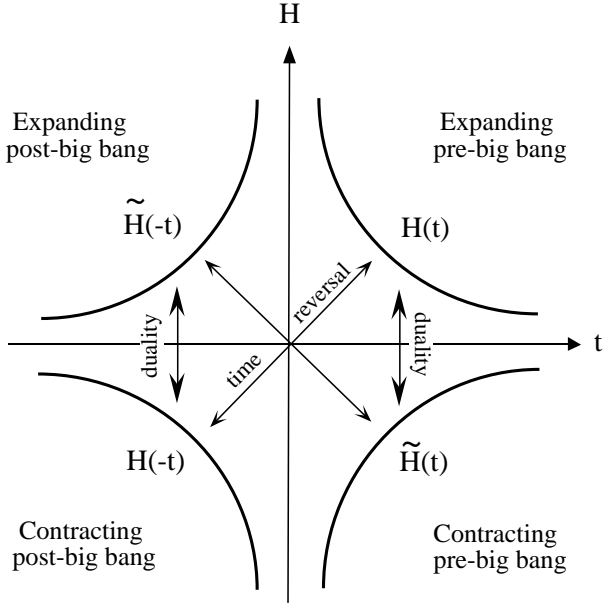,width=80mm}}
\vskip 5mm
\caption{\sl The four  branches of the low-energy string cosmology
backgrounds.}
\label{f22}
\end{figure}

As a simple example of the four cosmological branches,  we may
consider here the particular isotropic solution defined in the positive
range of the time coordinate,
\beq
a= (t/t_0)^{1/\sqrt d}, ~~~~~~~~~~
\fb = -\ln (t/t_0), ~~~~~~~~~~ t>0,
\label{218}
\eeq
which is singular at $t=0$ and  satisfies identically the set of
equations (\ref{29}) -- (\ref{211}). By applying a duality {\em and} a
time-reversal transformation we obtain the four inequivalent  
solutions
\bea
&&
 a_{\pm} (t) = t^{\pm 1/\sqrt d},~~~~~~~~~~~~~~\fb (t)= -\ln t ,
~~~~~~~~~~~~~ t>0,
\nonumber \\
&&
 a_{\pm} (-t) = (-t)^{\pm 1/\sqrt d}, ~~~~~~~\fb (-t)= -\ln (-t) ,
~~~~~~ t<0,
\label{219}
\eea
corresponding to the four branches illustrated in Fig. \ref{f22}. These
solutions are separated  by a curvature singularity at $t=0$; they 
describe decelerated expansion $a_+(t)$,  decelerated contraction 
$a_-(t)$, accelerated contraction $a_+(-t)$,  accelerated expansion 
$a_-(-t)$ (the solution is accelerated or decelerated according to
whether $\dot a$ and $\ddot a$ have the same or  opposite signs,
respectively). The curvature is growing for $ a_{\pm} (-t)$, decreasing
for $ a_{\pm} (t)$. Note that the transformation connecting two different
branches represents in this case not really a symmetry, but rather a
group acting    on the space of solutions, transforming
non-equivalent conformal backgrounds into each other, like in the case
of the Narain transformations \cite{Nara86,Nara87}.

For further applications, it is also important to consider the dilaton
evolution in the various branches of Eq. (\ref{219}). Using the definition
(\ref{123}),
\beq
\phi_\pm (\pm t) =\fb (\pm t) + d \ln a_\pm (\pm t) =
\left( \pm \sqrt d -1\right) \ln (\pm t).
\label{220}
\eeq
It follows that, in a phase of growing curvature ($t<0, t \ra 0_-$), the
dilaton is growing only for an expanding metric, $a_-(-t)$. This means
that, in the isotropic case, the expanding inflationary
solutions describe a cosmological evolution away from the string
perturbative vacuum ($ H = 0, \phi = -\infty$), i.e. are solutions
characterized by a growing string coupling, $\dot g_{\rm s} =
(\exp \phi/2)\dot{}>0$. The string perturbative vacuum thus naturally
emerges as the initial state for a state of pre-big bang inflationary
evolution. This is to be contrasted with the recently proposed
``ekpyrotic" scenario \cite{KOST1}, where the ``pre-big bang"
configuration (i.e. the phase of growing curvature preceding the brane
collision that simulates the big bang explosion) is contracting even in
the S-frame \cite{KOSST} and indeed corresponds  to a phase of
decreasing dilaton.

Note that in a more general, anisotropic case, and in the presence of
contracting dimensions, a growing-curvature solution is associated to a
growing dilaton only for a large enough number of expanding
dimensions. To make this point more precise, consider the particular,
exact solution of Eqs.   (\ref{29})--(\ref{211}), with $d$ expanding and
$n$ contracting dimensions, and scale factors $a(t)$ and $b(t)$,
respectively: \beq
a = (-t)^{-1/\sqrt{d+n}}, ~~~~~
b= (-t)^{1/\sqrt{d+n}}, ~~~~~\fb =-\ln (-t), ~~~~ t \ra 0_-.
\label{221}
\eeq
This gives
\beq
\phi= \fb + d \ln a +n\ln b ={n-d-\sqrt{d+n}\over \sqrt{d+n}} \ln (-t),
\label{222}
\eeq
so that the dilaton is growing if
\beq
d+\sqrt{d+n}>n.
\label{223}
\eeq

We note, incidentally, that this result may have interesting implications
for a possible ``hierarchy" of the present size of extra dimensions, if
we assume that our Universe starts evolving from the string
perturbative vacuum (i.e. with initial $\dot \phi >0$). Indeed, in a
superstring theory context ($d+n=10$), it follows that the initial
number of expanding dimensions is $d>3$, while only $n<6$ may be
contracting. A subsequent freezing, or late-time contraction (possibly
induced by quantum effects \cite{BuoGas97}), of $d-3$ dimensions will
eventually leave only three expanding dimensions, but with a possible
huge asymmetry in the spatial sections of our Universe, even in the
sector of the ``internal" dimensions (see  Refs. \cite{BraVa89,ABE00},
and the discussion of Section \ref{Sec8.5}, for a possible mechanism 
selecting $d=3$ as the maximal, final number of expanding dimensions in
a string- or brane-dominated Universe).

The invariance of the gravidilaton equations  (\ref{29})--(\ref{211}) is
in general broken by a dilaton potential $V(\phi)$, unless $V$ is just a
function of $\fb$. The invariance, however, is still valid in the presence
of matter sources, provided they transform in a way that is compatible
with the string equations of motion in the given background
\cite{GasVe92} (see the next subsection). In the perfect-fluid
approximation, in particular, a scale-factor duality transformation is
associated to a ``reflection" of the equation of state \cite{GV91}.

Consider in fact the addition to the action (\ref{28}) of a matter action
$S_m$, minimally coupled to the S-frame metric (but uncoupled to the
dilaton), and describing an anisotropic fluid with diagonal stress tensor,
\beq
T_{\mu\nu}= {2\over \sqrt{-g}}{\da S_m\over \da g_{\mu\nu}},
~~~~T_\mu\,^\nu=  {\rm diag} (\rho, -p_i~ \da_i^j),
~~~~~ p_i/\rho=\ga_i= {\rm const},  ~~~~ \r=\r(t).
\label{224}
\eeq
The field equations (\ref{27}) are now completed by a source term
\beq
G_{\mu\nu} + \nabla_\mu\nabla_\nu \phi +{1\over 2}g_{\mu\nu}
\left[\left(\nabla \phi\right)^2 -2 \nabla^2 \phi +{1\over 12}
H_{\a\b\ga}^2\right] -{1\over 4} H_{\mu\a\b}H_\nu\,^{\a\b}
={1\over 2}e^\phi T_{\mu\nu}
\label{225}
\eeq
(in units in which $2 \la_{\rm s}^{d-1}=1$), and the cosmological
equations (\ref{29})--(\ref{211}) become
\bea
&&
\fbp^2 -\sum_iH_i^2= e^{\fb} \rb, \label{226}\\
&&
\dot H_i -H_i \fbp={1\over 2} e^{\fb} \pb_i, \label{227}\\
&&
2 \ddot{\fb} -\fbp^2 -\sum_iH_i^2=0,
\label{228}
\eea
where we have introduced the ``shifted" density and pressure
\beq
\rb=  \r \sqrt{-g}= \r \prod_i a_i, ~~~~~~~~~~~~~
\pb= p \sqrt{-g}= p\prod_i a_i.
\label{229}
\eeq
They are a system of $d+2$ independent equations for the $d+2$
variables $\{a_i, \phi, \r\}$. Their combination gives
\beq
\dot {\rb} +\sum_i H_i \pb_i=0,
\label{230}
\eeq
which represents the usual covariant conservation of the source
energy density. The above equations with sources are invariant under
time reflection and under the duality transformation \cite{GV91}
\beq
a_i \ra a_i^{-1}, ~~~~~~~ \fb \ra \fb, ~~~~~~~
\rb \ra \rb, ~~~~~~~ \pb_i \ra -\pb_i,
\label{231}
\eeq
which preserves $\rb$ but changes $\r$ in a non-trivial way,
and``reflects" the barotropic equation of state, $\ga_i \ra -\ga_i$. Thus,
a cosmological solution is still characterized by four distinct branches.

We will present here a simple isotropic example, corresponding to the
power-law evolution
\beq
a \sim t^\a, ~~~~~~~~~ \fb \sim -\b \ln t, ~~~~~~~~~p=\ga \r
\label{232}
\eeq
(see the next subsection for more general solutions).
We use (\ref{226}), (\ref{228}), (\ref{230}) as independent equations.
The integration of Eq. (\ref{230})  immediately gives 
\beq
\rb =\r_0a^{-d\ga};
\label{c233}
\eeq
Eq. (\ref{226}) is then satisfied, provided
\beq
d\ga \a +\b=2.
\label{c234}
\eeq
Finally, Eq. (\ref{228}) leads to the constraint
\beq
2\b -\b^2 -d \a^2 =0.
\label{c235}
\eeq
We  then have  a (quadratic) system of two equations for the two
parameters  $\a,\b$ (note that, if $\a$ is a solution for a given $\ga$,
then also $-\a$ is a solution, associated to $-\ga$). We have in general
two solutions. The  flat-space solution,  $\b=2, \a=0$, corresponds to a
non-trivial dilaton evolving in a frozen pseudo-Euclidean background,
sustained by the energy density of dust matter ($\ga=0$),  according to
Eq. (\ref{227}).  For $\ga \not= 0$
we obtain instead
\beq
\a = {2 \ga\over 1+ d\ga^2}, ~~~~~~~~~~~
\b = {2 \over 1+ d\ga^2},
\label{c236}
\eeq
which fixes the time evolution of $a$ and $\fb$:
\beq
a \sim t^{2\ga\over 1+ d\ga^2}, ~~~~~~~~~~
\fb = - {2 \over 1+ d\ga^2}\ln t,
\label{237}
\eeq
and also of the more conventional variables $\r, \phi$:
\beq
\r = \rb a^{-d} = \r_0 a^{-d(1+\ga)}, ~~~~~~~~~~
\phi = \fb +d \ln a=  {2(d\ga -1) \over 1+ d\ga^2}\ln t.
\label{238}
\eeq

This particular solution reproduces the small-curvature limit of  the
general  solution with perfect fluid sources sufficiently far from the
singularity, as we shall see in the next subsection.  As in the vacuum
solution (\ref{218}), there are four branches, related by time-reversal
and by the duality transformation (\ref{231}), and characterized by the
scale factors
\beq
a_{\pm} (\pm t) \sim (\pm t) ^{\pm 2 \ga/(1+d\ga^2)}.
\label{239}
\eeq

For $d=3$, $\ga=1/3$ and $t>0$ we recover in particular the standard, 
radiation-dominated solution with constant dilaton:
\beq
a \sim t^{1/2}, ~~~~~~~~~
\r=3p \sim a^{-4}, ~~~~~~~~~
\phi ={\rm const}, ~~~~~~~~~t \ra +\infty,
\label{240}
\eeq
describing decelerated expansion and decreasing curvature:
\beq
\dot a >0, ~~~~~~~~~ \ddot a <0,
~~~~~~~~~\dot H <0, ~~~~~~~~~\dot\phi=0 ,
\label{241}
\eeq
typical of the post-big bang radiation era. Through a duality and
time-reversal transformation we obtain the ``dual" complement: 
\beq
a \sim (- t)^{-1/2}, ~~~~~ \phi \sim -3 \ln (-t), ~~~~~
\r=-3p \sim a^{-2}, ~~~~~t\ra -\infty,
\label{242}
\eeq
which is still an exact solution of the string cosmology equations,
and  describes  accelerated (i.e.
inflationary) expansion, with growing dilaton and growing curvature:  
\beq 
\dot a >0, ~~~~~~~~~ \ddot a >0,
~~~~~~~~~\dot H>0, ~~~~~~~~~\dot\phi>0
\label{243}
\eeq
(the unconventional equation of state, $\ga=-1/3$, is typical of a gas of
stretched strings, see \cite{GSV91a,GSV91} and the next subsection).

This confirms that,  if we start with our present cosmological phase in
which the dilaton is constant (in order to guarantee a constant strength
of  gravitational and gauge interactions), and we postulate  for
the early Universe a dual complement preceding the big bang explosion,
then string theory requires for the pre-big bang phase not only {\em
growing curvature}, but also {\em growing dilaton}.

In other words, string theory naturally suggests to identify the initial
configuration of our Universe with a state asymptotically approaching
the flat, cold and empty string perturbative vacuum, $H^2 \ra 0$, $\exp
(\phi) \ra 0$. As a consequence, the initial cosmological evolution
occurs in the small curvature ($H^2/M_{\rm s}^2 \ll 1$) and weak coupling
($g_{\rm s} \ll 1$) regime, and can be appropriately described by the lowest
order effective action (\ref{28}) (see Fig. \ref{f23}). The solutions
(\ref{240}) and (\ref{242}) provide a particular, explicit representation
of the scenario represented in Fig. \ref{f23}, for the two asymptotic
regimes of $t$ large and positive, Eq. (\ref{240}), and $t$ large and
negative, Eq. (\ref{242}). 

\begin{figure}[t]
\centerline{\epsfig{file=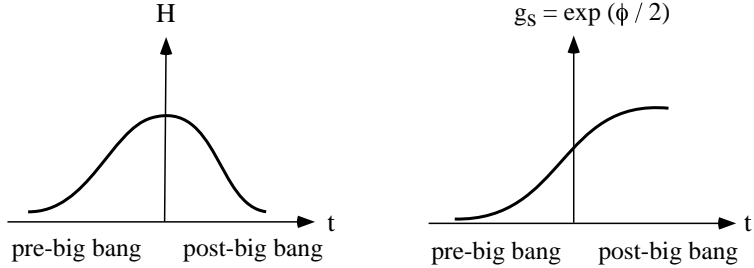,width=100mm}}
\vskip 5mm
\caption{\sl Qualitative time evolution of the curvature scale and of the
string coupling for a typical self-dual solution of the string-cosmology
equations.}
\label{f23}
\end{figure}

It should be mentioned, to conclude this subsection, that the invariance
under the discrete symmetry group $Z_2^d$, generated by the
inversion of $d$ scale factors, can be generalized  so as to be extended
to spatially flat solutions of more general scalar--tensor theories
\cite{CapDeRi93,Lid95}, with the generic Brans--Dicke parameter $\om$,
described by the action
\beq
S=-\int d^{d+1}x \sqrt {-g}~e^{-\phi}~\left[R-\om (\nabla \phi)^2 -2
\La\right].
\label{143a}
\eeq
The case $\om =-1$ corresponds to the string effective action. For
$\om \not= -(d+1)/d$, the equations of an isotropic and spatially flat
background (with scale factor $a(t)$) are invariant under the
transformation  $\a \ra \ti \a$, $\phi \ra \ti \phi$, where $\a =\ln a$,
and \cite{Lid95}
\beq
\a= {d-1+d\om\over d+1+d\om} ~\ti \a -{2(1+\om)\over d+1+d\om} ~\ti
\phi, ~~~~~
\phi=-{2d\over d+1+d\om} ~\ti \a + {d-1+d\om\over d+1+d\om} ~ \ti \phi
\label{143b}
\eeq
(when $\om=-1$ one recovers the transformation (\ref{217})). Similar
symmetries are also present in a restricted class of homogeneous, 
Bianchi-type models \cite{ClaLiTa98} (but the presence of spatial
curvature tends to break the scale-factor duality symmetry); for the
Bianchi-I-type metrics, in addition, the discrete transformation
(\ref{143b}) can be embedded in a continuous $O(3)$ symmetry group.

It should be stressed, however, that when $\om \not=-1$ such
generalized transformations {\em do not} necessarily associate, to any
decelerated solution of the standard scenario, an inflationary solution
with growing dilaton. In this sense, a self-dual cosmological scenario in
which inflation emerges naturally from the perturbative vacuum seems
just to be a peculiar prediction of string theory, in its low energy limit.

In the next subsection we will extend the discussion of this section to
more general models of background and sources.

\subsection{$O(d,d)$-covariance of the cosmological equations}
\label{Sec2.3}

The target-space duality introduced in the previous subsection is not
restricted to the gravidilaton sector and to cosmological backgrounds,
but is expected to be a general property of the solutions of the string
effective action (possibly valid at all orders \cite{Sen91,Has92}, with
the appropriate generalizations \cite{Meis97,KalMeis97}).

Already at the lowest order, in fact, the inversion of the scale factor is
only a special case of a more general transformation of the global
$O(d,d)$ group which leaves invariant the action (\ref{28}) for all
background characterized by $d$ Abelian isometries, and which mixes in 
a non-trivial way the components of the metric and of the antisymmetric
tensor $B_{\mu\nu}$ (see \cite{GiPoRa94} for a general review).

Such an invariance property of the action can also be extended to
non-Abelian isometries \cite{Ossa93}, but then there are problems for
``non-semisimple" isometry groups \cite{GRV93}. Here we shall restrict 
ourselves to the Abelian case, and we shall consider (for cosmological
applications) a set of background fields $\{\phi, g_{\mu\nu},
B_{\mu\nu}\}$ which is isometric with respect to $d$ spatial
translations and for which there exists a synchronous frame 
where $g_{00}=1, ~g_{0i}=0, ~B_{0\mu}=0$, and all the non-zero
 components of $g_{ij}, B_{ij}$ (as well as the dilaton itself) are only
dependent on time. 

In order to illustrate the invariance properties of such a class of
backgrounds under global $O(d,d)$ transformations, we shall first
rewrite the action (\ref{28})
directly in the synchronous gauge (since, for the moment, we are not
interested in the field equations, but only in the symmetries of the
action). We set $g_{ij}= -\ga_{ij}$ and  find, in this gauge,
\bea
&&
\Ga_{ij}\,^0= {1\over 2} \dot \ga_{ij}, ~~~~~~
\Ga_{0i}\,^j={1\over 2}g^{jk}\dot g_{ik}= {1\over 2} \left(g^{-1}\dot
g\right)_i\,^j= \left(\ga^{-1}\dot \ga\right)_i\,^j
\nonumber \\
&&
R_0\,^0 = -{1\over 4} {\rm Tr }\left(\ga^{-1}\dot \ga\right)^2
-{1\over 2} {\rm Tr }\left(\ga^{-1}\ddot \ga\right)
-{1\over 2} {\rm Tr } \left(\dot \ga^{-1}\dot \ga\right), \nonumber\\
&&
R_i\,^j =  -{1\over 2} \left(\ga^{-1}\ddot \ga\right)_i\,^j
-{1\over 4} \left(\ga^{-1}\dot \ga\right)_i\,^j
{\rm Tr } \left(\ga^{-1}\dot \ga\right)
+{1\over 2} \left(\ga^{-1}\dot \ga \ga^{-1}\dot \ga\right)_i\,^j ,
\label{244}
\eea
where
\beq
 {\rm Tr }\left(\ga^{-1}\dot \ga\right)= \left(\ga^{-1}\right)^{ij}
\dot \ga_{ji}= g^{ij} \dot g_{ji},
\label{245}
\eeq
and so on [note also that    $\dot \ga^{-1}$  means
$d\left(\ga^{-1}\right)/dt$]. Similarly we find, for the antisymmetric
tensor,
\bea
&&
H_{0ij}= \dot B_{ij}, ~~~~~~~~
H^{0ij}= g^{ik}g^{jl} \dot B_{kl} =\left( \ga^{-1} \dot B
\ga^{-1}\right)^{ij},
\nonumber\\
&&
H_{\mu\nu\a}H^{\mu\nu\a}= 3 H_{0ij}H^{0ij}=
- 3{\rm Tr }\left(\ga^{-1}\dot B\right)^2.
\label{246}
\eea
Let us introduce the shifted dilaton, by absorbing the spatial
volume into $\phi$, as in Section \ref{Sec1.4}:
\beq
\sqrt{|\det g_{ij}|} e^{-\phi}= e^{-\fb},
\label{247}
\eeq
from which
\beq
\fbp= \dot \phi -{1\over 2} {d\over dt} \ln \left( \det \ga\right) =
\dot \phi -{1\over 2} {\rm Tr }\left(\ga^{-1}\dot \ga\right).
\label{248}
\eeq
By collecting the various contributions from $\phi, R$ and
$H^2_{\mu\nu\a}$,  the action
(\ref{28}) can  be rewritten as:
\bea
&&\qquad
S=-{\la_{\rm s}\over 2} \int dt e^{-\fb} \Bigg[ \fbp^2 +
{1\over 4} {\rm Tr }\left(\ga^{-1}\dot \ga\right)^2
-{\rm Tr }\left(\ga^{-1}\ddot \ga\right)
\nonumber \\
&&\qquad
-{1\over 2} {\rm Tr } \left(\dot \ga^{-1}\dot \ga\right)
+\fbp {\rm Tr }\left(\ga^{-1}\dot \ga\right)
+{1\over 4} {\rm Tr }\left(\ga^{-1}\dot B\right)^2 \Bigg].
\label{249}
\eea
We can now eliminate the second derivatives, and the mixed terms
containing $\fbp \dot \ga$, by noting that
\beq
{d \over dt} \left[ e^{-\fb} {\rm Tr }\left(\ga^{-1}\dot \ga\right)\right]=
e^{-\fb}\left[{\rm Tr }\left(\ga^{-1}\ddot \ga\right)
+ {\rm Tr } \left(\dot \ga^{-1}\dot \ga\right)
-\fbp {\rm Tr }\left(\ga^{-1}\dot \ga\right)\right].
\label{250}
\eeq
Finally, by using the identity 
$\left(\ga^{-1}\right)\dot{}= -\ga^{-1}\dot \ga \ga^{-1}$,   we can
rewrite the action in quadratic form, modulo a total derivative, as
\beq
S=-{\la_{\rm s}\over 2} \int dt e^{-\fb} \left[ \fbp^2
-{1\over 4} {\rm Tr }\left(\ga^{-1}\dot \ga\right)^2
+{1\over 4} {\rm Tr }\left(\ga^{-1}\dot B\right)^2 \right].
\label{252}
\eeq

This action can be set into a more compact form by using the $2d
\times 2d$ matrix $M$, defined in terms of the spatial components of
the metric and of the antisymmetric tensor,
\bea
&&
M=\pmatrix{G^{-1} & -G^{-1}B \cr
BG^{-1} & G-BG^{-1}B \cr}, \nonumber \\
&&
G= g_{ij}\equiv-\ga_{ij},  ~~~~~ G^{-1}\equiv g^{ij}, ~~~~~ B\equiv
B_{ij},
\label{253}
\eea
and using also the matrix $\eta$, representing the invariant metric of
the $O(d,d)$ group in the off-diagonal representation: 
\beq
\eta=\pmatrix{0 & I \cr I & 0 \cr}
\label{254}
\eeq
($I$ is the unit $d$-dimensional matrix).
By computing $M\eta, \dot M \eta$ and $(\dot M \eta)^2$ we find,  in
fact,
\beq
{\rm Tr } \left(\dot M \eta\right)^2=
2  {\rm Tr }\left[\dot \ga^{-1}\dot\ga+  \left(\ga^{-1}\dot
B\right)^2\right]=
2  {\rm Tr }\left[-\left(\ga^{-1}\dot \ga\right)^2 +\left(\ga^{-1}\dot
B\right)^2\right],
\label{255}
\eeq
so that the action can be rewritten as \cite{MeiVe91,MeiVe91a}
\beq
S=-{\la_{\rm s}\over 2} \int dt e^{-\fb} \left[ \fbp^2
+{1\over 8} {\rm Tr }\left(\dot M\eta\right)^2\right].
\label{256}
\eeq

We may note, at this point, that $M$ itself is  a (symmetric) matrix
element of the pseudo-orthogonal  $O(d,d)$ group, since
\beq
M^T \eta M =\eta, ~~~~~~~~~~ M=M^T,
\label{257}
\eeq
for any $B$ and $G$. Therefore:
\beq
M\eta=\eta M^{-1}, ~~~~~~~~\left(\dot M\eta\right)^2=
\eta \left( M^{-1}\right)\dot {} M \eta,
\label{258}
\eeq
and the action can be finally rewritten in the form
\beq
S=-{\la_{\rm s}\over 2} \int dt e^{-\fb} \left[ \fbp^2
+{1\over 8} {\rm Tr } ~\dot M \left(M^{-1}\right)\dot{}\right],
\label{259}
\eeq
which is explicitly invariant under  global $O(d,d)$
transformations preserving the shifted dilaton $\fb$:
\beq
\fb \ra \fb, ~~~~~~~~ M \ra \ti M= \La^{T} M \La, ~~~~~~~~
\La^T\eta \La =\eta.
\label{260}
\eeq
When $B=0$, the matrix $M$ is block-diagonal, and the
special $O(d,d)$ transformation represented by $\La = \eta$ corresponds
to an inversion of the metric tensor:
\beq
M ={\rm diag} (G^{-1}, G), ~~~~~~~~~~~~
\ti M = \La^{T} M \La= \eta M \eta = {\rm diag} (G, G^{-1}),
\label{262}
\eeq
so that $\ti G = G^{-1}$.
For a diagonal metric, in particular,  $G= a^2 I$, and the invariance under
the scale factor duality  transformation (\ref{217}) is recovered as a
particular case of the global $O(d,d)$ symmetry (as already
anticipated).

This $O(d,d)$ invariance holds even in the presence of sources
representing bulk string matter \cite{GasVe92}, namely sources
evolving consistently with the solutions of the string equations of
motion in the background we are considering. A distribution of
non-interacting strings, minimally coupled to the metric and the
antisymmetric tensor of an $O(d,d)$-covariant background, is in fact
characterized by a stress tensor (source of $g_{\mu\nu}$) and by an
antisymmetric current (source of $B_{\mu\nu}$), which are
automatically $O(d,d)$-covariant).

In order to illustrate this important point we add to Eq. (\ref{28}) an
action $S_m$ describing matter sources coupled to $g$ and to $B$, and
we define
\beq
{1\over 2} \sqrt{-g} T_{\mu\nu}= {\da S_m\over \da g^{\mu\nu}},
~~~~~~~~~~~~~~
{1\over 2} \sqrt{-g} J^{\mu\nu}= {\da S_m\over \da B_{\mu\nu}}.
\label{263}
\eeq
The variation with respect to the dilaton, to $g_{\mu\nu}$ and
$B_{\mu\nu}$ gives then, respectively, Eqs. (\ref{26}), Eq. (\ref{225})
and the additional equation
\beq
\nabla_\mu \left(e^{-\phi} H^{\mu\a\b}\right) = J^{\a\b}.
\label{264}
\eeq
For a background with $d$ spatial isometries, and in the synchronous
gauge, such field equations can be written in matrix form using $M,
~\fb$, and a new set of ``shifted" variables defined as follows:
\beq
\rb = \sqrt{-g} T_0~^0, ~~~~~~~~~
{\overline \theta} = \sqrt{-g} T^{ij}, ~~~~~~~~~
{\overline J}= \sqrt{-g} J^{ij},
\label{265}
\eeq
where ${\overline \theta}$ and ${\overline J}$ are $d \times d$
matrices. In particular, the dilaton equation (\ref{26}) takes the form
\beq
\dot {\fb}^2 -2\ddot {\fb} -{1\over 8}\,{\rm Tr}\, (\dot M\eta)^2 =0,
\label{266}
\eeq
the $(0,0)$ component of Eq. (\ref{225}) gives
\beq
\dot {\fb}^2 +{1\over 8}\,{\rm Tr}\, (\dot M\eta)^2 = \rb e^{\fb},
\label{267}
\eeq
while the spatial part of Eq. (\ref{225}), combined with Eq. (\ref{264}),
can be written in the form
\beq
{d\over dt}(e^{-\fb}M\eta \dot M)=\overline T,
\label{268}
\eeq
where $\overline T$ is a $2d \times 2d$ matrix composed with
${\overline \theta}$ and ${\overline J}$:
\beq
\overline T= \pmatrix{-{\overline J},& - \overline \theta G+{\overline J}B
\cr
G\overline \theta -B{\overline J},&
G\overline J G+B\overline J B -G\overline \theta B -B \overline \theta G
\cr}
\label{269}
\eeq
(see \cite{GasVe92} for an explicit computation). By differentiating Eq.
(\ref{267}), using Eqs. (\ref{266}), (\ref{268}), and the identity
\beq
(M\eta \dot M \eta)^2= -(\dot M \eta)^2 ,
\label{270}
\eeq
we obtain the generalized energy conservation equation, written in
matrix form as
\beq
\dot{\rb}+{1\over 4}\,{\rm Tr}\, (\overline T \eta M\eta \dot
M\eta)=0.
\label{271}
\eeq

The $O(d,d)$ covariance of the string cosmology equations is thus
preserved even in the presence of matter sources, provided
$\overline T$ transforms in the same way as $M$. In that case, the
whole set of equations (\ref{266})--(\ref{268}) is left invariant by the
generalized transformation
\beq
\fb \ra \fb,\,\,\,\,\,\,\,\,~~
\rb \ra \rb, \,\,\,\,\,\,\,\,~~
M \ra \La^T M\La, \,\,\,\,\,\,\,~~ \overline T \ra \La^T \overline T \La,
\label{272}
\eeq
where $\La^T \eta \La =\eta$. This is indeed what happens if the
sources are represented by a gas of non-self-interacting
strings.

Suppose in fact that the matter action is given by the sum over all
components of the string distribution, $S_m=\sum_i S^i_{\rm strings}$,
where we use the action (\ref{24}), and we choose the conformally flat
gauge for the world-sheet metric (i.e. $\ga_{ij}=0$, $R^{(2)}=0$):
\bea
S_{\rm string} &=& -{1\over 4 \pi \ap}\int d^{d+1}x ~\da^{d+1}
(x-X(\sg,\tau)) ~d\sg d\tau   \nonumber\\
&&
\times \left[\left(\dot X^\mu \dot X^\nu- X^{\prime \mu} X^{\prime
\nu} \right)g_{\mu\nu}(x) +\left(\dot X^\mu  X^{\prime \nu}- X^{\prime
\mu} \dot X^{\nu} \right)B_{\mu\nu}(x)
\right].
\label{273}
\eea
Here $\sg$ and $\tau$ are the world-sheet coordinates, $\dot X=
dX/d\tau$ and $X'= dX/d\sg$. The variation with respect to $g$ and $B$
gives the tensors (\ref{265}), which depend on the world-sheet
integral of a form bilinear in $X^\mu (\sg,\tau)$. The variation with
respect to $X^\mu$ gives the string equations of motion, the variation
with respect to $\ga_{ij}$ (before imposing the gauge) gives the
constraints. If we have a solution $X^\mu$ of the equations of motion,
in a given background $M(t)$, and we perform an $O(d,d)$
transormation $M \ra \ti M= \La^T M \La$, the new solution $\ti
X^\mu$ is found to correspond to transformed matrices $\ti
{\overline \theta}$ and $\ti{\overline J}$, which combine to give
$\ti{\overline T}= \La^T \overline T \La$ \cite{GasVe92}. The $O(d,d)$
covariance is thus preserved, provided the matter sources transform
according to the string equations of motion, which can themselves be
written in a fully covariant form. 

Let us now exploit this $O(d,d)$ covariance to find more general
solutions of the string cosmology equations, and more general
examples of duality-related backgrounds corresponding to possible
models for the pre-big bang scenario. Let us introduce a convenient
(dimensionless) time-coordinate $x$, such that
\beq
dx= L \rb ~dt
\label{274}
\eeq
($L$ is a constant length). Assuming that the equation of state can
be written in terms of a given $2d \times 2d$ matrix
$\Ga(x)$, such that 
\beq
\rb~ d\Ga= \overline T dx,
\label{275}
\eeq
we can integrate a first time Eqs. (\ref{266})--(\ref{268}), also with the
help of the identity (\ref{270}). The result is \cite{GasVe94a}
\bea
&&
\fb^{~\prime} =-{2\over D(x)} (x+x_0), \label{276}\\
&&
M\eta M' = {4\Ga (x)\over D(x)},
\label{277}
\eea
where
\beq
D(x)\equiv (x+x_0)^2-{1\over 2}\,{\rm Tr}\,(\Ga \eta)^2=
4 L^2 \rb e^{-\fb}
\label{278}
\eeq
(a prime denotes differentiation with respect to $x$, and $x_0$ is an
integration constant).

By exploiting the fact that $M$ is a symmetric $O(d,d)$ matrix, $M\eta
M=\eta$, and that $M\eta \Ga =-\Ga \eta M$, because of the definition
of $\overline T$ (see \cite{GasVe92}), the above equations can be
formally integrated to give
\bea
&&
\fb (x)= \phi_0 -2\int {dx\over D} (x+x_0), \label{279}\\
&&
M(x)= P_x \exp \left[- 4 \int {dx\over D} \Ga \eta \right] M_0,
\label{280}
\eea
where $\phi_0$ is a constant,  $M_0$ a constant symmetric $O(d,d)$
matrix,  and $P_x$ denotes the ``$x$-ordering" of the exponential.
For any given ``equation of state" $\overline T =\overline T (\rb)$,
providing an integrable relation for $\Ga(x)$, according to Eq. 
(\ref{275}), the general exact solution of the low-energy string
cosmology equations (for space-independent fields and vanishing
dilaton potential) is then represented by Eqs. (\ref{278})--(\ref{280}).

Such solutions  contain in general singularities for the curvature and
the string coupling, in correspondence with the zero of $D(x)$. Near the
singularity (i.e., for $D(x) \ra 0$)  the contribution of the matter sources
becomes negligible with respect to the curvature terms in the field
equations (as,in general relativity, for the Kasner anisotropic solutions
\cite{Lif63}), and one recovers the general vacuum solutions presented
in \cite{MeiVe91}. This can be shown in general for any background
$M(x)$,  and any kind of matter distribution $\Ga(x)$ \cite{GasVe94a};
for further applications, however, it will be sufficient to report here the
case of anisotropic but torsionless ($B=0$) backgrounds, with diagonal
metric $g_{ij}=-a^2_i \da_{ij}$,  and  fluid sources with a barotropic
equation of state, as in Eq. (\ref{224}).

In this case we obtain, from the previous definitions,
\bea
&&
M\eta M' = 2 \pmatrix{0 & {a'_i \over a_i} \da_{ij} \cr
-{a'_i \over a_i} \da_{ij}  & 0 \cr}, \,\,\,\,\,
\overline T = \pmatrix{0 & \overline p_i \da _{ij} \cr
-\overline p_i \da _{ij} & 0 \cr}, \nonumber\\
&&
\Ga= \pmatrix {0 & \Ga_i \da_{ij} \cr
-\Ga_i \da_{ij} & 0 \cr} ,\,\,\,\,\,\,\,
~~~~~~~~\Ga_i= \ga_i x +x_i, \nonumber\\
&&
D= (x+x_0)^2 - \sum_i (\ga_i x +x_i)^2= \a (x-x_+)(x-x_-) ,
\label{281}
\eea
where
\bea
&&
\overline p_i = p_i \sqrt{|g|}, \,\,\,\,\,\,\,\,\,\,\,\,\,
\a = 1-\sum_i \ga_i^2, \nonumber\\
&&
x_{\pm}= {1\over \a} \left\{\sum_i\ga_i x_i -x_0 \pm
\left[\left(\sum_i\ga_ix_i-x_0\right)^2+\a\left(\sum_i
x_i^2-x_0^2\right) \right]^{1/2} \right\}; 
\label{282}
\eea
$x_\pm$ are the zeros of $D$,
and $x_i$, $x_0$ are integration constants. The general solution
(\ref{278})--(\ref{280}) reduces to \cite{GasVe93b}
\bea
&&
a_i= a_{0i}|(x-x_+)(x-x_-)|^{\ga_i/\a} |{x-x_+ \over x-x_-}|^{\a_i},
\label{283}\\
&&
e^{\fb} = e^{\phi_0}
|(x-x_+)(x-x_-)|^{-1/\a} |{x-x_+ \over x-x_-}|^{-\sigma},
\label{284}\\
&&
\rb = {\a \over 4 L^2} e^{\phi_0}
|(x-x_+)(x-x_-)|^{(\a - 1) /\a} |{x-x_+ \over x-x_-}|^{-\sigma},
\label{285}
\eea
where
\beq
\sigma = \sum_i\a_i
\ga_i  , \,\,\,\, ~~~~
\a_i = {\a x_i+\ga_i(\sum_i \ga_i x_i -x_0)\over
\a \left[(\sum_i\ga_ix_i-x_0)^2+\a(\sum_i x_i^2-x_0^2)\right]^{1/2}}, 
\label{286}
\eeq
and $a_{i0}$, $\phi_0$ are additional integration constants.

This solution has two curvature singularities at $x_{\pm}$. Near the
singularity the sources are negligible, and one recovers the anisotropic
vacuum solution of string cosmology in critical dimensions
\cite{GV91,Mul90}. Indeed, for $x \ra x_{\pm}$, one finds 
from Eqs. (\ref{274}) and (\ref{285}) that 
$|x|\sim |t|^{\a/(1\pm \a \sum \a_i \ga_i)}$,
and the solution reduces to 
\beq
a_i(t) \sim |t-t_{\pm}|^{\b_i^{\pm}}~,~~~~~~~~
\,\,\,\,\,\,\, \fb \sim -\ln |t-t_{\pm}|, 
\label{287}
\eeq
where
\beq
\b_i^{\pm}= {x_i \pm \ga_i x_{\pm} \over x_0 + x_{\pm}},~~~~~
~~~~~~~~~~~~\sum_i(\b_i^{\pm})^2 =1.  
\label{288}
\eeq
In the large-$|x|$ (small-curvature) limit, on the contrary, the relation
between $x$ and cosmic time is $|x|\simeq |t|^{\a /(2-\a)}$, and the
solution ({\ref{283})--(\ref{285}) behaves like (for $|x|\ra \pm \infty$)
\bea
&&
a_i(t) \sim |t| ^{2\ga_i /( 1+\sum \ga_i^2)},\,\,\,\,\,\,~~~~~~~~
\fb \sim -{2\over 1+\sum \ga_i^2} \ln |t|, \nonumber\\
&&
\phi \sim 2{\sum \ga_i-1\over 1+\sum \ga_i^2} \ln |t|, \,\,\,\,\,\,
~~~~~~~~~~~\rb \sim |t|^{-2\sum \ga_i^2/( 1+\sum \ga_i^2)}.
\label{289}
\eea

We may note, in general, that the relative importance of the matter
sources with respect to the curvature is measured by the ratio (see Eq. 
(\ref{267}), 
\beq
\Om (x)= -{8\rb e^{\fb} \over (d-1)\,{\rm Tr}\, (\dot M \eta)^2}\equiv
{2\over d-1} {D\over {\rm Tr}\, (\Ga \eta)^2} =
{\r e^\phi \over (d-1)\sum_i H_i^2}
\label{290}
\eeq
(normalized  in such a way that one recovers the usual critical 
energy--density ratio, $\Om =\r/\r_c$, when $\phi$ is constant and the
metric is isotropic). It follows that $\Om \ra 0$ when $x \ra x_\pm$,
while in the opposite limit $|x| \ra \infty$, dominated by the matter
sources, $\Om$ goes to a constant,
\beq
\Om \ra \Om_\infty ={1-\sum_i\ga_i^2 \over (d-1)\sum_i\ga_i^2},
\label{291}
\eeq
which is obviously $\Om_\infty =1$ for the isotropic,
radiation-dominated background with $\ga_i=1/d$ and constant dilaton.

Now it can  easily be checked that, in the isotropic case $a_i=a$,
$\ga_i=\ga$, the vacuum solution (\ref{219}) and the
matter-dominated solutions (\ref{237}), (\ref{238}) presented in the
previous subsection can be directly recovered from Eqs. 
(\ref{287}) and (\ref{289}), respectively. As before, we have invariance
under time reflections and under the duality transformation $a_i \ra
a_i^{-1}$, $\ga_i \ra -\ga_i$. In addition, we can now perform more
complicated $O(d,d)$ transformations to find anisotropic and
non-diagonal pre-big bang backgrounds  \cite{GMV91} (the possibility
of singularity-free solutions, in this context, will be discussed in
Section \ref{Sec8}). It should be stressed, however, that the ``$O(d,d)$
complement" of a standard cosmological solution is in general
associated to the introduction of an effective viscosity in the matter
stress tensor, as pointed out in \cite{GasVe92}.

To conclude this subsection let us finally report the cosmic-time and
conformal-time parametrization of the isotropic solutions, in the
asymptotic limit of large and small curvature, both in the S-frame and in
the E-frame.

The small-curvature limit, dominated by a perfect fluid with
$p/\r=\ga$, has been given in cosmic time in Eqs.  (\ref{237}),
(\ref{238}). In the conformal time gauge, defined by $dt = a d\eta$, we
obtain
\beq
a(\eta)\sim |\eta|^{2\ga/(1-2\ga+d\ga^2)}, ~~~~
\phi \sim {d\ga -1\over \ga} \ln a , ~~~~
\r \sim a^{-d (\ga +1)}.
\label{292}
\eeq
Instead, the vacuum, dilaton-dominated limit (\ref{219}) becomes, in
conformal time, 
\beq
a_{\pm}(\eta)\sim |\eta|^{\pm 1 /(\sqrt d \mp 1)}, ~~~~~~~
\phi_{\pm} \sim \sqrt d (\sqrt d \mp 1) \ln a_\pm.
\label{293}
\eeq

In the E-frame, the scale-factor and the curvature parameters are
obtained through the transformation (\ref{18}), (\ref{111}) (the
conformal time is obviously the same in both frames). The 
small-curvature limit (\ref{237}),
(\ref{238}) becomes, in cosmic time,
\beq
\ti a(\ti t) \sim |\ti t|^\beta,~~~~~~~
\ti \phi\sim \sqrt{{2\over d-1}} {(d-1)(1-d\ga)\over \ga-1} \ln \ti a,
~~~~~~~ \ti \r \sim \ti a^{-2/\b},
\label{294}
\eeq
where
\beq
\ti \r =\r {\sqrt{|g|}\over \sqrt{|\ti g|}} = \r e^{\phi(d+1)/(d-1)}, ~~~~
\b={2(1-\ga)\over (d-1)(1+d\ga^2)-2(d\ga-1)}.
\label{295}
\eeq
In conformal time,
\beq
a_E(\eta)\sim |\eta|^{-2(\ga-1)/(d-1)(1-2\ga+d\ga^2)}.
\label{296}
\eeq
The vacuum, dilaton-dominated limit (\ref{219}) becomes, in cosmic
time,
\beq
\ti a_\pm (\ti t) \sim  |\ti t|^{1/d}, ~~~~~~~~~~~
\ti \phi_{\pm} \sim \pm \sqrt{2d(d-1)} \ln \ti a_\pm ,
\label{297}
\eeq
and
\beq
\ti a_\pm (\eta) \sim |\eta|^{1/(d-1)}
\label{298}
\eeq
in conformal time.

 The above solutions will be used to discuss the
kinematics of the phase of pre-big bang inflation (in the next
subsection), and to compute the corresponding amplification of the
quantum fluctuations (see Section \ref{Sec4}).  For the discussion of
more general homogeneous solutions (see for instance
\cite{BaKu97,BaDa97,BaDa98a}), we refer the reader to the
existing literature, and in particular to the detailed phase-space
analysis of the string cosmology equations performed including
non-trivial two-form potentials ($B_{\mu\nu} \not= 0$), moduli fields,
spatial curvature, anisotropy, and a cosmological constant in both  the
matter and  the gravidilaton parts of the action
\cite{BiCoLid1,BiCoLid2,BiCoLid3}.

\subsection{Kinematics of pre-big bang inflation}
\label{Sec2.4}

The symmetry properties of the string effective action suggest a
``self-dual" cosmological scenario, in which the Universe evolves from
an initial configuration approaching (asymptotically) the string
perturbative vacuum ($H \ra 0$, $\phi \ra -\infty$). Such a state, on the
other hand, is particularly appropriate to provide the initial conditions,
because it is unstable with respect to the decay into an inflationary
phase eventually driven by the dilaton kinetic energy. A small
inhomogeneity of the perturbative vacuum is indeed sufficient to trigger
inflation, as we shall discuss in Section \ref{Sec3}. Or, keeping
homogeneity, a small perturbation represented by a constant non-zero
energy density is also sufficient.

In the first models of pre-big bang evolution
\cite{GasVe93a,GasVe94a}, such a perturbation was generated by a
(sufficiently diluted) gas of fundamental strings, as its presence tends
to enhance the instability of the full system (background fields plus
mattere sources). Suppose in fact that we look for self-consistent,
simultaneous solutions of the gravidilaton background equations
(\ref{226})--(\ref{228}) and of the string equations of motion (in the
same background), obtained from Eq. (\ref{273}) with $B_{\mu\nu}=0$:
\bea
&&
\ddot X^\mu - X^{\prime\prime\mu} +\Ga_{\a\b}^\mu
\left(\dot X^\a + X^{\prime \a}\right)\left (\dot X^\b - X^{\prime
\b}\right) =0, \nonumber\\
&&
g_{\mu\nu}\left(\dot X^\mu \dot X^\nu  +
X^{\prime \mu} X^{\prime \nu}\right)=0,  ~~~~~~~~~
g_{\mu\nu} \dot X^\mu X^{\prime \nu} =0 .
\label{299}
\eea
Given a consistent embedding $X^\mu (\sg, \tau)$, the sum over all the
stress tensors of each individual string,
\beq
\sqrt{-g} T^{\mu\nu}(x)= {1\over 2\pi \ap}\int d\sg d\tau
\left(\dot X^\mu \dot X^\nu  -
X^{\prime \mu} X^{\prime \nu}\right) \da^{d+1} (x- X(\sg,\tau) )
\label{2100}
\eeq
(after an appropriate averaging procedure) provides the effective
sources to be inserted into the background equations.

The above system of string and background equations can be solved in
a perturbative way \cite{GGMV96,GGMV98}, by imposing the string
perturbative vacuum as asymptotic initial conditions at $t \ra
-\infty$.  To zeroth order, the background manifold is flat, and the
string solutions are characterized by $|\dot X^0| \gg |X^{\prime 0}|$,
$|\dot X^i| = |X^{\prime i}|$, corresponding to a pressureless effective
stress tensor, $T_0^0=$ const, $T_i^i=0$ (dust matter, in the
perfect-fluid approximation).

To first order, the background solution thus corresponds to the case
$\ga_i=0$ and to the negative branch ($x<x_-$) of Eqs. 
(\ref{283})--(\ref{285}), namely
\bea
&&
a_i(t)=a_{i0}\left|{t-2t_c\over t}\right|^{-\b_i},\,\,\,\,\,\, ~~~~~
e^{\fb} ={16L^2e^{-\phi_0}\over |t(t-2t_c)|},\,\,\,\,\,\, ~~~~~
\rb = {e^{\phi_0} \over 4L^2}= {\rm const}, \nonumber\\
&&
-\b_i={t_i\over t_c},\,\,\,\,\,\,\,\, ~~~~~
t_c=\left(\sum_it_i^2\right)^{1/2},\,\,\,\,\,\,\,\, ~~~~~ t\leq 0
\label{2101}
\eea
 ($t_i$ are integration constants). Note that $t \sim x$ since $\rb=$
const, and that we have performed a time translation to shift the
singularity from $x=x_-$ to the origin. Note also that, at $ t \ra
-\infty$, the metric is flat and only the dilaton is rolling, $\phi \sim -2
\ln (-t)$, sustained by a constant (pressureless) string energy density
$\r=$ const, $p=0$ (consistently with the chosen initial conditions).

The string equation of state keeps unchanged until $a\simeq$ const,
namely for $ |t| \gg t_c$. When $t$ approaches the critical scale $-t_c$,
however, the scale factors start rolling and evolve towards a final
(expanding or contracting) accelerated configuration
\beq
a_i(t) \sim (-t)^{\b_i},\,\,\,\,\,\,\,~~~~~~~ |\b_i|<1,\,\,\,\,\,\,\,
~~~~~~~\sum_i\b_i^2=1,
\label{2102}
\eeq
reached asymptotically for $|t| \ll t_c$. When $t \sim -t_c$, therefore,
the horizon $|H|^{-1}$ starts shrinking (see Section \ref{Sec1.3}) and
we know, in that regime, that extended objects become ``unstable"
\cite{SV90,MG91,MG92}. In the case of string matter, in particular, we
know that the asymptotic solutions of the string equations of motion,
outside the horizon \cite{GSV91a,GSV91}, are characterized by  $|\dot
X^0| \gg |X^{\prime 0}|$,   $|\dot X^i| \ll |X^{\prime i}|$ for accelerated
expansion ($\b_i<0$), and by $|\dot X^0| \gg |X^{\prime 0}|$,  $|\dot X^i|
\gg |X^{\prime i}|$ for accelerated contraction ($0<\b_i<1$). As a
consequence, the string stress tensor (\ref{2100}) satisfies,
respectively, the conditions \bea
&&
T_0^0 \simeq T_i^i, ~~~~~~~~~~~~~~~~~ \b_i<0,
\nonumber\\
&&
T_0^0 \simeq -T_i^i, ~~~~~~~~~~~~~~~ 0<\b_i<1,
\label{2103}
\eea
corresponding to the radiation equation of state ($\ga =1/d$) for
accelerated contraction, and to its dual ($\ga =-1/d$) for accelerated
expansion. In both cases, the string pressure triggers a regime of
positive feedback, which enhances the instabilitys and pushes the
system (background fields plus sources) even further away from the
string perturbative vacuum \cite{GSV91,GasVe93a,GGMV96,GGMV98}.

When $|t| \ll t_c$, however, the contribution of the matter sources
becomes negligible in the field equations: a first differentiation of Eq. 
(\ref{2101}) gives in fact
\beq
H_i={2t_i \over t(t-2t_c)}, \,\, \,\,\,~~~~\dot {\fb} =-{2(t-t_c)\over
 t(t-2t_c)},\,\,\,\,\, ~~~~
\rb e^{\fb}={4\over t(t-2t_c)}= \r e^\phi
\label{2104}
\eeq
and then $\r e^\phi \ll H^2 \sim {\fbp}^2$ for $ t \ra 0_-$. So, even if
the string pressure tends to deviate from its initial vanishing value, its
dynamical effects eventually become negligible when the background
enters the dilaton-dominated regime. The solution (\ref{2101}) thus
represents a reliable description of the coupled system background
plus string sources, both in the $|t| \gg t_c$ and $|t| \ll t_c$ limits,
namely in the initial and final regimes, sufficiently far from the critical
time scale $t = t_c$ \cite{GGMV98}.

What is important, for the purpose of this section, is that the phase of
pre-big bang inflation, at low energy, is in general characterized by
two possible (connected) branches: an initial (model-dependent) phase
in which matter (like string, or other sources) are possibly important,
and a second ``vacuum" phase dominated by the dilaton kinetic energy.
The validity of the low energy solutions stops of course at the string
scale, $|t| \sim \la_{\rm s}$. At that point the higher order (loop and $\ap$)
quantum corrections come into play, but inflation can nonetheless
continue during a phase of constant curvature and linearly growing
dilaton \cite{GasMaVe97}, appearing as late-time attractor in the
high-curvature regime (see Section \ref{Sec8}).

So, in spite of the fact that the inflationary epoch of the pre-big bang
scenario is often referred to as the vacuum, dilaton-driven phase
appearing in the low energy solutions, it is important to stress that the
{\em total} pre-big bang inflationary period possibly contains also a
phase of ``late inflation" at high curvature (as also stressed in
\cite{MagTu97}), and a phase of ``early inflation" driven by matter
sources (as also stressed in \cite{GasVe99}). To discuss the main
kinematic aspects of pre-big bang inflation, however, it will be
sufficient to concentrate our attention on the dilaton-driven, isotropic
vacuum solutions given by Eq. (\ref{219}) (or by Eq. (\ref{293}), in the
conformal-time parametrization).

For a quantitative estimate of the duration of the inflationary period we
observe that the ratio $r$, determined by the flatness problem as in Eq. 
(\ref{12}), also controls the solution of the horizon problem through the
ratio
\beq
{\rm proper~ size~ horizon~ scale \over
proper ~size~ homogeneous~ region} \sim {H^{-1}(t)\over a(t)}
\sim r(t),
\label{2105}
\eeq
and that, for a power-law metric $a \sim t^\b$, such a ratio scales
linearly with respect to the conformal time coordinate,
\beq
r \sim t^{1-\b} \sim \int a^{-1} dt \sim \eta.
\label{2106}
\eeq
Also, we shall assume that the standard radiation era can be extended
back in time down to the Planck scale, and that it is immediately
preceded by a phase of accelerated expansion. At the beginning of the
radiation era the horizon size is then controlled by the Planck
length $\la_{\rm P}$, while the proper size of the homogeneous and
causally connected region inside our present Hubble radius,
rescaled down according to the decelerated evolution of the standard
scenario, is unnaturally larger than the horizon by the factor $ \sim
10^{30} \la_{\rm P}$ (as already remarked in Subsection {\ref{Sec1.1}).
During the inflationary epoch the ratio $r$ must thus increase by  at
least a factor $10^{30}$, so as to push the homogeneous region outside
the horizon, by he amount required  by the subsequent decelerated
evolution. This determines the condition
\beq
|\eta_f|/|\eta_i| ~\laq ~ 10^{-30},
\label{2107}
\eeq
where $\eta_i$ and $\eta_f$ mark, respectively, the beginning and the
end of the inflationary period.

Let us now suppose, for simplicity (and for a more direct comparison
between the pre-big bang and the standard inflationary scenarios), that
at the end of inflation the dilaton immediately freezes out, and that the
string and Planck lengths have already comparable sizes at
$\eta=\eta_f$ (which means $\exp(\phi_f) \sim 1$, according to Eq.
(\ref{21})).

If we go back in time during a dilaton-dominated phase of isotropic
superinflation,
\beq
a\sim (-t)^{-1/\sqrt d} \sim (-\eta)^{-1/(\sqrt d +1)},
\label{2108}
\eeq
starting from the string/Planck scale ($t_f \sim \la_{\rm s}$), it follows
from (\ref{2107}) that the scale factor is {\em at most} reduced by
the factor $a_i/a_f \sim 10^{-30/(1+\sqrt d)}$, corresponding to a
homogeneous region, which, at the beginning of inflation, is still very
large in string units, i.e. at least
\beq
10^{30\sqrt d/(1+ \sqrt d)}\la_{\rm s}
\label{2109}
\eeq
(i.e. $\sim 10^{19}\la_{\rm s}$ for $d=3$). This is to be contrasted with
the case of standard (although not fully realistic) de Sitter inflation, $a
\sim (-\eta)^{-1}$, where, going back in time, the scale factor during
inflation is reduced by the factor $10^{-30}$, so that the size of the
initial homogeneous region is just  $\la_{\rm P}$, like the horizon, which
stays frozen  during inflation.

The contrast is even more striking if the initial size (\ref{2109}) is
expressed in Planck units, since, at the beginning of inflation, the string
coupling $g_{\rm s}=\exp (\phi/2)$ --and thus the Planck length 
$\la_{\rm P}$-- is reduced with respect to its final value ($\la_{\rm P}
\sim \la_{\rm s}$) by the factor 
\beq
{g_{\rm s} (\eta_i)\over g_{\rm s}(\eta_f)}= {\la_{\rm P}(\eta_i)\over \la_{\rm
P} (\eta_f)}= \left(\eta_f\over \eta_i\right)^{\sqrt d/2}
\laq 10^{-15 \sqrt d}, 
\label{2110}
\eeq
so that the lower bound (\ref{2109}) for the size of the initial region can
be written as
\beq
10^{\sqrt d (45+15 \sqrt d)/(1+\sqrt d)} \la_{\rm P}
\label{2111}
\eeq
(i.e. $\sim 10^{45}\la_{\rm P}$ for $d=3$). This, by the way, is exactly
the initial size evalued in the E-frame where $\la_{\rm P}$ is a constant,
and the pre-big bang phase is represented as a contraction, according to
Eq. (\ref{298}).

One might think that, in such a case, the situation is not much better
than  in the case of non-inflationary cosmology, where one finds that the
initial size of the homogeneous part of  our Universe is greater than
$10^{30}\la_{\rm P}$ \cite{Lin99}. It should be noted, however, that
without inflation the initial  homogeneous region has to be much bigger
than the horizon. In the pre-big bang scenario,  on the contrary, the 
initial homogeneous region is large in Planck or string units,  but not
larger than the horizon itself \cite{Gas00}.  Indeed, during
superinflation, the horizon scale  shrinks linearly in cosmic time. As we
go backwards in time, in the particular example that we are
considering, the horizon increases by the factor
$H^{-1}_i/H^{-1}_f=(t_i/t_f)= (\eta_i/\eta_f)^{\sqrt d/(1+ \sqrt d)}$, so
that, at the beginning of inflation, $H^{-1}_i \sim 10^{30  \sqrt
d/(1+\sqrt d)}\la_{\rm P}$, i.e. the initial horizon size is just the
same as that of the homogeneous region  (\ref{2109}) (as also
illustrated in Fig. \ref{f13}).

In this sense, the initial state of pre-big bang inflation is similar to the
one of the standard inflationary scenario. In one case, however,
the initial horizon is large in Planck  units; in the other case it is of
order $1$, as an obvious consequence  of the different curvature scales
at the beginning of inflation. The question about the naturalness of the
initial conditions (see the next section)  thus seems to concern the initial
unit of length used to measure the size of the initial homogeneous
domain: Should we use the Planck (or string) length, or the classical
radius of the causal horizon?

The Planck length certainly provides  natural units for the size of
the initial homogeneous patches when initial conditions  are imposed on
a cosmological state approaching the high-curvature, quantum-gravity
regime. In the pre-big bang scenario, however, initial conditions are to
be imposed when the Universe is deeply inside the low-curvature,
weak-coupling, classical regime. In that regime, the Universe does not
know about the Planck length, and the causal horizon $H^{-1}$ seems to
provide a natural candidate for the initial homogeneity scale. It should
be noted, to this purpose, that the formation of a large homogeneous
domain is not prevented by the amplification of quantum fluctuations
\cite{GPV98}. However, classical inhomogeneities, not normalized to a
vacuum fluctuation spectrum, may require a constraint on the
duration of inflation stronger than Eq. (\ref{2107}), as recently pointed
out in \cite{BuoDa01}.

Note also that a large (homogeneous) Hubble horizon, if we assume the
saturation of the holographic bound \cite{Thooft93} applied in a
cosmological context \cite{GV99}, implies a large initial entropy $S \sim$
(horizon area in Planck units) and corresponds, in a quantum context, to
a small probability that such configuration be obtained through a
process of quantum tunnelling  (proportional to $\exp [-S]$). In the
context of the pre-big bang scenario, however, quantum cosmology
effects, such as tunnelling or reflection of the Wheeler--De Witt wave
function, are expected to be important towards {\sl the end} of inflation,
and  {\sl not the beginning}, as  they may be effective {\sl to exit},
eventually, from the inflationary regime (see Section \ref{Sec9}), {\sl not
to enter} it, and to explain the origin of the initial state. A large entropy
of  the initial state, in the weakly coupled, highly classical regime, can
only correspond to a large probability of such a configuration 
(proportional to $\exp [+S]$), as expected for classical and macroscopic
configurations.

The naturalness of the initial conditions will be discussed in Section
\ref{Sec3}. Here we note, to conclude this subsection, that an initial
state characterized by a set of large (or small) dimensionless
parameters, determined by  a large initial horizon scale $H^{-1}$, is an
unavoidable aspect of all models in which inflation starts at scales
smaller than Planckian. 
In the standard scenario, where observational tracks of what happened
before the Planck epoch are washed out, one may regard as unnatural
\cite{Lin99} having an initial homogeneity scale of order $H^{-1}$
whenever $H$ is very small in Planck or string units. 
In the context of the pre-big bang scenario, however, the
phenomenological imprint of the Planck epoch is not necessarily washed
out by a long and subsequent inflationary phase. The pre-Planckian
history may become visible, and the sub-Planckian initial conditions are
accessible, in principle, to observational tests, so that their
naturalness can also be analysed with a Bayesan approach, in terms of
``a posteriori" probabilities, as in \cite{BDV99} (see also
\cite{CLT98,CLT99,CFLT99} for a discussion of ``generic" initial
conditions in the pre-big bang scenario).

\subsection{Frame-independence: Which is the ``right" metric?}
\label{Sec2.5}

As already stressed in the previous discussion, and explicitly computed
in Section \ref{Sec1.3}, an S-frame inflationary solution of the string
cosmology equations, when transformed to the E-frame, tends to
describe a phase of accelerated contraction. This is strictly true only
for isotropic backgrounds because, if the metric is anisotropic, and
there is a large enough number of (S-frame) contracting dimensions,
the other dimensions may keep expanding (and accelerated) in both
frames.

Consider, for instance, the S-frame vacuum solutions (\ref{221}),
corresponding to $d$ expanding and $n$ contracting dimensions. When
expressed in terms of the E-frame metric $\ti g$, and in the cosmic
time gauge, through the transformation
\beq
\ti g_{\mu\nu}= g_{\mu\nu} ~e^{-2\phi/(d+n-1)}, ~~~~~~
d\ti t = dt ~ e^{-\phi/(d+n-1)}, ~~~~~~
\ti \phi= \phi \left(2\over d+n-1\right)^{1/2}
\label{2112}
\eeq
(which generalizes to the anisotropic case the transformations of Eqs.
(\ref{18}), (\ref{111})), the solution becomes
\bea
&&
\ti a (\ti t) = \left(-\ti t\right)^{-{2n-1-\sqrt{d+n}\over
\sqrt{d+n}(d+n)+d-n}}, ~~~~~~~~~
\ti b (\ti t) = \left(-\ti t\right)^{{2d-1+\sqrt{d+n}\over
\sqrt{d+n}(d+n)+d-n}},
\nonumber\\
&&
\ti \phi (\ti t) = \left(2\over d+n-1\right)^{1/2}{(n-d-\sqrt{d+n})
(d+n-1) \over \sqrt{d+n}(d+n)+d-n} \ln (-\ti t).
\label{2113}
\eea
The expanding part of the solution thus remains expanding (and
inflationary $\dot {\ti a}>0, ~\ddot {\ti a}>0$) even in the E-frame,
provided $2n>1+\sqrt{d+n}$. Note that, for $d=3$, this condition is
compatible with the condition of growing dilaton, $ n< d+\sqrt{d+n}$,
only for $n=1$.

However, isotropic ($n=0$) metrics in the E-frame are
contracting. This does not represent a difficulty for the pre-big bang
scenario because the contraction is accelerated, $\dot {\ti a}<0, ~\ddot
{\ti a}<0$, and thus corresponds to a phase in which the ratio $r$
decreases in time, as required for the solution of the horizon/flatness
problems (see Sections \ref{Sec1.3} and \ref{Sec2.4}). In particular, if the
phase of accelerated expansion is long enough to solve the problems in
the S-frame, according to Eq. (\ref{2107}), then the problems are also
solved in the E-frame \cite{GasVe93b}, since the conformal time is the
same in both:
\beq
d\ti \eta = {d \ti t\over \ti a (\ti t)}={dt\over a(t)} =d\eta.
\label{2114}
\eeq
Furthermore, as we will see in Section \ref{Sec4.1}, the spectral
distribution of the metric fluctuations amplified by inflation is also the
same in both frames \cite{GasVe93b,GasVe94a}.

Such a frame independence of the inflationary aspects of the pre-big
bang solutions also extends to the string-driven (and, more generally,
matter-dominated) solutions presented in the previous subsection.

Suppose, in fact, that we start, at some initial time $t_i$, with a
distribution of strings characterized in the S-frame by a packing factor
(average distance/average size) of the order of unity, and with
sufficiently small interactions to represent a perfect fluid source for
the superinflationary solution (\ref{237}), (\ref{238}), with $\ga=-1/d$.
In the S-frame, where the metric $a$ is expanding, the perfect-fluid
approximation is also valid at any subsequent time $t>t_i$, as the
number of strings {\em per unit of string volume}  $n(t)$ is diluted in
time by the factor $n/n_i =(a_i/a)^d<1$.
In the E-frame the metric $\ti a$ is contracting, but
the string's proper size $\ti L(t)$ shrinks with time as
$\ti L=(\ti a/a)\la_{\rm s}$. As a
consequence, the number of strings per unit of string volume scales as
$\ti n(t)=(\ti L/\la_{\rm s})^d ~\ti a^{-d}=a^{-d}$, and it is again diluted as
time goes on, exactly by the same amount as in the S-frame
\cite{GasVe94a}.

With similar arguments  it can be shown that the heating up of the
string gas with respect to the radiation, which is easy to understand in
the S-frame where the metric is expanding and the radiation is
redshifted, also occurs in the E-frame, in spite of the fact that the
radiation is blueshifted because of the contraction \cite{GasVe93b}.

Using the standard thermodynamical arguments (see for instance
\cite{Weinberg}) we find indeed, for the perfect-fluid sources with
$p/\r=\ga$, generating the solution (\ref{237}), that their
temperature $T_\ga$ satisfies $T_\ga \sim a^{-d\ga}$. On the other
hand, $\r_\ga \sim a^{-d(1+\ga)}$. In the S-frame the radiation
($\ga=1/d$) is thus supercooled and diluted with respect to the (string)
sources that drive inflation ($\ga<1/d$), \beq
{T_\ga\over T_r}= {\r_\ga\over \r_r}\sim a^{1-d\ga},
\label{2115}
\eeq
since the metric is expanding.

In the E-frame the fluid sources satisfy a modified conservation
equation,
\beq
\dot {\ti \r} + d \ti H (\ti \r +\ti p) -{1\over \sqrt{2(d-1)}}
\dot {\ti \phi}(\ti \r -d\ti p)=0
\label{2116}
\eeq
(the dots refer to the E-frame cosmic time $\ti t$), obtained from Eq. 
(\ref{230}), with $\ti \r$ defined in Eq. (\ref{295}). The radiation, which
has a traceless stress tensor, still evolves adiabatically, but now its
temperature is blueshifted because of the contraction: $\ti T_r
\sim {\ti a}^{-1}$. The effective temperature of the other matter
sources is also blueshifted, however, and using thermodynamical
arguments we obtain \cite{GasVe93b}, from Eq. (\ref{2116}):
\beq
{\ti T_\ga\over \ti T_r}= {\ti \r_\ga\over \ti \r_r}
\sim {\ti a}^{\ga(d-1)(d\ga-1)/(\ga-1)}.
\label{2117}
\eeq

For the sources of the inflationary S-frame expansion we have $\ga
<0$, so that even in the E-frame the above ratio is growing, with a
corresponding cooling down and dilution of the radiation density (with
respect to the density of the inflationary sources). 

There is  thus a complete frame-independence for  the
physical effects of the inflationary solutions of the pre-big bang
scenario. This can also be understood as an invariance of physics under
a local field redefinition. Different frames are just related by a local
(field-dependent) conformal transformation of the metric. No physical
observables should depend on it, and this is what we shall find
whenever we push a calculation to its very end (see also \cite{AlCon01}
for a recent discussion of this point). Further help is provided by the
fact that, at late times, when the dilaton is frozen, all frames actually
coincide.

Yet it is true that, in developing some physical intuition, there are
important kinematical differences in different frames, as also stressed
by the examples illustrated before. Inflation in the S-frame, for
instance, can be represented as gravitational collapse in the
E-frame (see the next section).  We think that the S-frame (whose
metric coincides with the sigma-model metric to which fundamental
strings are directly coupled, see Section \ref{Sec2.2}) is the one offering
the simplest intuitive picture of how things evolve and work. The
following three main points support this attitude. 

The first point concerns the ultraviolet cut-off. The string metric is the
one in which the ultraviolet cut-off of string theory is kept fixed, i.e.
does not depend on the dilaton (at least in perturbation theory). This can
be easily seen from the form of the tree-level action in the S-frame, in
$d=3$,  in which we include a typical higher-derivative correction,
$\ap R^2$:  
\beq
\Gamma^{\rm SF} =  -
{1\over 2\lambda_{\rm s}^{2}}\int\,d^{4}x\,\sqrt{|g|}\,e^{-\phi} \,\left(R +
g^{\mu\nu}\partial_{\mu}\phi\partial_{\nu}\phi + \ap R^2\right). 
\eeq
Obviously, higher-derivative corrections become relevant, in this frame, 
as soon as $R \sim \lambda_{\rm s}^{-2}$ and independently of $\phi$. By
contrast, the equivalent E-frame action would read:
\beq
\Gamma^{\rm EF} =  - {1\over 2\lambda_{\rm
s}^{2}}\int\,d^{4}x\,\sqrt{|\ti g|} \,\left(\ti R -{1\over 2}
\ti g^{\mu\nu}\partial_{\mu}\phi\partial_{\nu}\phi + \ap
e^{-\phi} \ti R^2\right).  \eeq
indicating that higher-derivative corrections become relevant at $\ti R
\sim \lambda_{\rm s}^{-2}e^{\phi}$, which is again the (dilaton-dependent)
string scale expressed in Planck units.

The second point concerns masses. 
Tree-level string masses are dilaton-independent in the S-frame,  but
depend on $\phi$ in the E-frame since the S-frame relation
$g^{\mu\nu} p_{\mu}p_{\nu} = m^2$ becomes $\ti g^{\mu\nu}
p_{\mu}p_{\nu} = m^2 e^{\phi}$ in the E-frame.

The third point concerns couplings. It is only an illusion that, by going to
the E-frame, we make the Newton constant $G_N$ dilaton-independent.
The physical quantity is not so much Newton's constant but  the
gravitational force, say between two heavy strings.
The force will be proportional to $G_N m_1 m_2$. In the S-frame, masses
are dilaton-independent and $G_N \sim e^{\phi}$, while,  in the
E-frame, $G_N$ is fixed but all masses scale like $e^{\phi/2}$. The same
physical result is obtained for the gravitational  force but, clearly, it is
much more economical to attribute the dilaton dependence of Newton's
force to Newton's constant rather than to a universal
dilaton dependence  of all masses. We think that this is a good example
of how physics is frame-independent, while our own physical intuition
may prefer one frame to the other.

For all these reasons we shall  interpret our results mainly  in the
string frame. Nonetheless, we shall often use other frames at
intermediate stages of the calculation, to make it as simple as  possible.

\section{Initial conditions}
\label{Sec3}
\setcounter{equation}{0}
\setcounter{figure}{0}

We have already mentioned that, in standard non-inflationary
cosmology, initial conditions have to be fine-tuned to an
incredible accuracy in the far past (i.e. at $t \sim t_{\rm P} \sim
10^{-43}$ s), or else it is impossible to explain some of the most striking
properties of today's Universe, such as its
homogeneity and flatness.

In the pre-big bang scenario, having extended time to the past of the big
bang event can certainly help solving the fine-tuning problems
of standard cosmology: after all,  most of these  come from a
``shortage  of time" for  things to happen early on after the big bang.
Yet it is important to discuss the conditions under which a pre-big bang
phase can naturally generate a patch that can evolve into  the Universe
we live in. In other words: What does the fine-tuning problem look like
if, accepting hints from scale-factor duality, we assume perturbative,
yet generic, initial conditions?

In this section we shall address this  kind of questions by
considering, for simplicity, a class of  space-times endowed with a
regular (time-like and null) asymptotic past,  and by imposing on these
geometries a condition of asymptotic-past-triviality (APT)  (see
Subsection \ref{Sec3.2}). We are aware of the fact that this is already  a
schematization of a more general set-up where one would simply
assume that the Universe, almost everywhere and almost all the time, is
in a chaotic, highly entropic perturbative state in which the arrow of
time itself is ill-defined. Within such a more general set-up a
mechanism for generating a low-entropy subsystem  has to be found, so
that, within this subsystem,  a time arrow (the direction of increasing
entropy) can  possibly be identified, and the organized Universe we
happen to be living in can develop. 

We shall argue  in Subsection \ref{Sec3.3} that such a mechanism can be
identified in the well-known phenomenon of gravitational collapse, and
that the  low-entropy region corresponding to our own Universe lies in
the interior of the closed trapped surface that emerged from the 
collapse. We shall illustrate in Subsection \ref{Sec3.4}  the
particular cases of spherical and planar symmetry,   comment on
the problem of chaotic oscillations in Subsection \ref{Sec3.5}, and 
finally present our conclusion on the fine-tuning problem in  Subsection
\ref{Sec3.6}. 

In our simplified approach  we will treat one of these regions
as representing, by itself, the full space-time manifold. Such a
simplifying procedure is by no means  new in general relativity:  
asymptotic flatness (asymptotic  spatial triviality, in our terminology) is
often assumed in order to prove rigorous theorems. Yet, causality
guarantees that the collapse of a star within our galaxy can only depend
on local conditions, and this allows for a simplified treatment of the
problem. 

In a similar spirit we shall assume that the dynamics that gave rise to
our Universe can be approximately described by a space-time with an
asymptotically trivial past. Thanks to  gravitational instability, this 
evolved  towards the formation of a closed trapped surface  and of  a
space-like singularity of the big bang type. Assuming that the actual
singularity is avoided, thanks to  string corrections to general
relativity,  this process may give  rise, eventually, to a (post-big bang)
phase resembling the one of standard hot big bang cosmology.

In the next subsection, to approach the question of whether our present
Universe may arise from generic initial conditions, we will relax the
homogeneity assumption adopted in Section \ref{Sec2} and 
consider more general solutions of the string cosmology equations.  

\subsection{Quasi-homogeneous and inhomogeneous solutions }
\label{Sec3.1}

If we want to explain the large-scale homogeneity and isotropy of our
Universe out of a cosmological phase preceding the
big bang, we  have to consider the case of non-homogeneous
cosmologies with the hope to find  homogeneity, isotropy, and flatness
as asymptotic features of late-time attractors. The question, in
particular, is how to obtain  space-time manifolds where spatial
gradients are subdominant with respect to time derivatives, starting
from initial conditions in which they are of the same order. In a rather
commonly used terminology, we would like to see ``velocity-dominated"
cosmologies  emerge, at late times, from non-velocity-dominated initial
conditions. This is what ordinary, slow-roll inflation supposedly does.

If we start from very generic inhomogeneous conditions, it is
conceivable that, as in the case of chaotic inflation \cite{Lin83}, only
some spatial patches  will end up exhibiting the required homogeneity
and flatness to be consistent with our Universe, while inhomogeneity
and anisotropy  will prevail almost everywhere else. In the context of
such a type of background evolution it is clear that, as a necessary
preparatory exercise, we need to understand  the nature of solutions
that are not too inhomogeneous, leaving momentarily aside  
 the problem of showing how
patches of this kind may emerge  at late times.

Such a program was started in \cite{GV97} and further extended in 
\cite{BMUV98b}. The analysis performed in \cite{GV97} was based on the
general equations for the gravidilaton system in $D=4$, together with
the assumption of regularity at past infinity.  It was shown that, on a
constant-time hypersurface (defined in the synchronous gauge),
 sufficiently  isotropic initial patches with $\nabla g \sim \dot g$ and
with a growing dilaton would inevitably expand,
 in the S-frame, and evolve towards homogeneity and strong
coupling. Within those patches, a reliable expansion in spatial gradients
\cite{GE} can be justified, the evolution quickly 
becomes velocity-dominated, and a Kasner-like singularity is 
approached according to the following generalization of
the solution (\ref{287}),  (\ref{288}):
\begin{eqnarray}
&& 
d s ^2 =  dt^2 - \sum_a e_i^a(x)~ e_j^a(x)~ (-t)^{2
\alpha_a(x)} dx^i dx^j  \;,  \nonumber \\ 
&&
\phi = \left[\sum_i
\alpha_i(x)-1\right] {\rm log} (-t) , ~~~~~~~ 1 = \sum_i
\alpha_i^2(x) \;, ~~~~~  t < 0 \; . 
\label{inhKasner}
\end{eqnarray}

If the above equations represented a general
quasi-homogeneous solution, the $Z_2^d$ generalization of 
scale-factor duality would be explicitly respected  in the form
$\alpha_i(x) \rightarrow - \alpha_i(x)$  (for any subset of indices $i$),
while keeping all $e_i^a(x)$ and $\bar{\phi}$ fixed. However, 
the momentum constraints still  have to be imposed on Eq.
(\ref{inhKasner}): they are trivial in the homogeneous case, but quite
non-trivial in the presence of spatial gradients. These constraints
preserve the simplest scale-factor duality [$\alpha_i(x)
\rightarrow - \alpha_i(x)$  for all $i$], but  explicitly break its
$Z_2^d$ generalization  \cite{BMUV98b}. The full symmetry is expected
 to be recovered as the solution approaches the singularity.

The (S-frame) local expansion rate, for the above metric, is 
given by the trace $\Theta$ of the volume expansion tensor 
$\Theta_{\mu\nu}$ \cite{Wald},
\beq
\Theta_{\mu\nu}= h_\mu^\a h_\nu ^\b \nabla _{(\a} n_{\b)},
~~~~~ h_{\mu\nu} = g_{\mu\nu} - n_\mu n_\nu, 
~~~~~ \Theta\equiv h^{\mu\nu}\Theta_{\mu\nu} = {\sum_i
\alpha_i(x) \over t},  ~~ t<0,
\label{expansion}
\eeq
where $n_\mu$ is the unit time-like vector normal to the (space-like)
homogeneity hypersurfaces. Under the Kasner constraint of Eq. 
(\ref{inhKasner}), it is clear   that the maximal rate of expansion is
reached where all $\alpha_i$ are equal and negative \cite{GV97}. Hence
the  most isotropic patches are inflated most during this Kasner-like
phase, suggesting that most of the space becomes fairly homogeneous 
and isotropic. This is true in spite of the fact that, unlike in ordinary 
inflation, the modulus $\sg^2$ of the shear tensor \cite{Wald}, 
\beq
\sg^2= {1\over 2}\sg_{\mu\nu}\sg^{\mu\nu}, ~~~~~~~~~~~~~
\sg_{\mu\nu}= \Theta_{\mu\nu}- {1\over 3} \Theta h_{\mu\nu}
\eeq
is not washed away as  a Kasner-like singularity is approached
\cite{KunDur00} (unless higher-order corrections play an important
role in this respect \cite{Giov99c}). 
 
In a subsequent work \cite{BMUV98b} these arguments were
generalized in several directions. First, the equations were extended to
an  arbitrary number of dimensions, showing that in $D \not= 4$   no
qualitatively new results emerge: in particular there is no critical value
of $D$ below which the phenomenon of BKL oscillations (to be  described
in Subsection \ref{Sec3.5}) takes place. Secondly, quasi-homogeneous
solutions were constructed  with the inclusion of the NS-NS
two-form $B_{\mu\nu}$: again, they look very similar to
the known homogeneous solutions, provided the exponents are allowed
to depend on space.  The only non-trivial point consists, again, in
imposing the momentum constraints that explicitly break
 $O(d,d)$ symmetry. 

The case of a quasi-homogeneous  axion field in $D=4$  was also
considered in \cite{BMUV98b}. Since the axion and $B_{\mu\nu}$ are
related  by a  (Poincar\'e)  duality transformation that mixes space 
and time, this case is genuinely different from the one discussed above. 
 Actually, the case of a homogeneous axion
shows interesting new features, such as the blowing up of the
dilaton 
 both at very late and at very early times. These results,
however, have to be taken with much precaution since some  of the
approximations may break down before such strong-coupling regimes
are entered.

In  \cite{BMUV98b} a conjecture was made that, as one approaches past 
infinity, there is a wide basin of attraction
 towards a trivial Milne geometry, which is nothing but a wedge of
 empty, flat Minkowski space in convenient coordinates. If this
conjecture were true, we would conclude that the asymptotic past of
pre-big bang cosmology is indeed trivial. However, the conjecture was
proved not to be quite correct \cite{ClaFe99,Kunze99}, through the
construction of exact inhomogeneous solutions of various kinds. Only
some of them exhibit a Milne-like behaviour at very early times, while,
more often than not, the asymptotic past consists of weakly interacting
waves, endowed with the symmetries of the solutions, travelling 
 on top of a trivial background \cite{ClaFe99,Kunze99,Yaza01}. 
Examples of this kind were constructed by various techniques.  

In the case of cylindrical symmetry, exact, non-homogeneous and 
anisotropic solutions, containing both a dilaton and an axion fields, were
already discussed in \cite{BaKu97a}. Up to a trivial time-reversal
transformation, those solutions approach,  near the late-time, 
a special case of the background (\ref{inhKasner}), while, at very early
times, an oscillatory behaviour prevails. The authors of \cite{BaKu97a}
have also discussed symmetry properties of the solutions under $O(d,d)$ 
duality transformations.

Inhomogeneous, cylindrical backgrounds for  the bosonic 
massless modes of heterotic and type-IIB superstrings have also been
obtained \cite{FLV97} by applying $T$- and $S$-duality transformations
(as well as the usual Geroch transformations) to vacuum solutions
of the Einstein equations exhibiting a two-dimensional group ($G_2$) of
Abelian isometries. The inhomogeneous, string theory version of the
Mixmaster (Bianchi-IX-type) cosmology, in particular, has been
presented in   \cite{FLV97}.  Other inhomogeneous  Einstein--Rosen
string cosmologies were studied with a similar technique in
\cite{ClaFe99}, by applying the global duality symmetries of  the
superstring effective action to homogeneous models of
Bianchi-$VI_h$-type. Again, in these solutions, the conjecture of an
asymptotic-past Milne metric is not usually satisfied. New
inhomogeneous string cosmologies were also obtained in  \cite{Lid99} by
applying a discrete (mirror-like) symmetry  of the string effective
action to $D=4$ to vacuum solutions of general relativity.

On a more phenomenological ground, an interesting model containing a
spherically symmetric inhomogeneity was studied at the
non-perturbative level in \cite{BaKu97b}. Such a treatment allows the
investigation of primordial black hole formation in string cosmology, an
issue that we shall reconsider in the context of tensor perturbations in
Section \ref{Sec5} (see also \cite{Lidsey00}).  The question of how 
robust  the existence of inflationary solutions is, in the presence of
various axion-like backgrounds, was addressed in \cite{FeVa98}; it
was shown that in some cases the parameter space leading to inflation
shrinks down to the empty set. Such a phenomenon may well be related
to the generic existence of chaos and BKL oscillations, which will be
discussed later, in  Subsection \ref{Sec3.5}. 

The question of the asymptotic-past behaviour of inhomogeneous
string cosmology was addressed in general in \cite{Kunze99} within
$G_2$ space-times, admitting two Abelian space-like Killing vectors.
Once more, the asymptotic past, rather than empty and Milne-like, was
 found to contain weakly interacting waves. The inclusion of a Maxwell
field was finally considered in \cite{Yaza01}: even in this case, while
the behaviour near the singularity is Kasner-like, the asymptotic past is
filled with weakly interacting gravitational, dilatonic, and
electromagnetic waves.

As a general remark  all these special, exact, non-homogeneous
solutions (and general, quasi-homogeneous solutions) can shed
important light on the evolution of perturbations  discussed in Sections
\ref{Sec4}--\ref{Sec7}, in particular when perturbation theory becomes
inadequate. We shall  occasionally make use of this connection later in
this report. The conclusion that we wish to draw, for the time being, is
that  a simple structure in the asymptotic past (although not necessarily
as trivial as Milne's)  can be consistently assumed for the generic
solution. This suggests to replace the assumption of  past regularity by a
somewhat stronger postulate, that of asymptotic past triviality. Such a
concept will be introduced and developed in the rest of this
section.

\subsection{Asymptotic past triviality and symmetries}
\label{Sec3.2}

We start from the general postulate that the pre-collapse, pre-big
bang evolution of the Universe 
can be described in terms of the low-energy, tree-level action
of string theory. Taking a generic closed superstring theory, this reads
(see Section \ref{Sec2}): 
\beq
S = - {1\over 2} 
\lambda_{\rm s}^{1-d}\int\,d^{d+1}x\,\sqrt{|g|}\,e^{-\phi} \,\left[R +
g^{\mu\nu}\partial_{\mu}\phi\partial_{\nu}\phi - \frac{1}{12}
 (dB)^2 - 2~ \Lambda \right] ~~,
\label{31}
\eeq
where $dB$ is the (three-form) field strength associated with the
antisymmetric tensor  $B_{\mu\nu}$. A further simplification comes
from assuming that we are  dealing with so-called critical superstring
theory, the case in which the tree-level (and actually
the all-order-perturbative) cosmological constant $\Lambda$ vanishes.
This requires a total of $D= 10$ space-time dimensions \cite{GSW87}.  If
$D \ne 10$ there will  indeed be an effective cosmological constant
$O(\lambda_{\rm s}^{-2})$ preventing the existence of any
low-curvature solution of the field equations.  

Equation (\ref{31}) receives corrections {\em either} when curvatures
become $O(\lambda_{\rm s}^{-2})$,  {\em or} when the coupling $e^{\phi}$
becomes $O(1)$. If  such corrections are both negligible, it sometimes
becomes  useful to perform a change of variable by going to the
E-frame, already introduced in the previous sections. In general,  
this is done by defining \cite{Syn,Wald}:
\begin{equation}
g_{\mu\nu} =\ti g_{\mu\nu} e^{\frac{2}{d-1}(\phi - \phi_0)} \; .
\label{ESF}
\end{equation}
Using the Einstein metric $\ti g$ (and dropping the tilde), the result for 
the action (\ref{31}) is simply:
\beq
S = -{1\over 2}\la_{\rm P}^{1-d} \int\,d^{d+1}x\,\sqrt{|g^{(E)}|}\;
\,\left[R - \frac{1}{d-1} \partial_{\mu}\phi\partial^{\mu}\phi -
\frac{1}{12}
 e^{- \frac{4}{d-1} \phi} (dB)^2  \right] \; ,
\label{33}
\eeq
where $ \la_{\rm P}^{d-1} = e^{\phi_0} \lambda_{\rm s}^{d-1}$ is  the present
value of the Planck length, if we take $\phi_0$ to coincide with the
present (constant) value of the dilaton. 

Although the use of the E-frame could simplify some calculations,
and we shall see examples of this below, it should be kept in mind that
the form of the corrections is no longer so simple. For instance, as
already stressed at the end of Section \ref{Sec2}, higher-derivative
corrections become important when the E-frame curvature is
$O( \la_{\rm P}^{-2} e^{\frac{2}{d-1} \phi} = \lambda_{\rm s}^{-2} )$, i.e. when
it reaches a dilaton-dependent critical value.  Similarly, having a
constant Newton ``constant" in this frame is a mere illusion, 
because (even tree-level) string masses now depend upon $\phi$ (as
already illustrated in Subsection \ref{Sec2.5}). For these reasons,
although physical results are frame-independent, we shall always
describe them  with reference to the original S-frame metric in
which the string length $\lambda_{\rm s}$ is constant.

Let us finally remark that the two frames  coincide
today if the dilaton is eventually fixed at its present value
$\phi_0$. Similarly, as we shall see, the assumption of APT 
leads to a very large and negative value of the initial dilaton which, if
finite,  would also allow the identification of the two frames in the far
past.  However, the two E-frames that coincide with the S-frame at  $t =
\pm \infty$ differ  from each other by an enormous conformal   factor, i.e.
by a huge blowing-up of all physical scales.

The classical equations that follow from varying (\ref{31}) or
(\ref{33}), besides being generally covariant,
are also invariant under a two-parameter group of (global)
transformations acting as follows:
\beq 
\phi \rightarrow  \phi + c, ~~~~~~~~~~~~~~~~~~~
g_{\mu\nu} \rightarrow  \lambda^2 g_{\mu\nu},
\label{symmetries}
\eeq
where $\la$ and $c$ are constant parameters (using general
covariance, the latter symmetry is also equivalent to an overall rescaling
of all the coordinates). Indeed,  both actions are simply rescaled by a
constant factor under this group. Note that these
two symmetries depend crucially on the validity of the tree-level, 
low-energy approximation, and on the absence of a cosmological
constant. Loop corrections clearly spoil invariance under dilaton shifts,
while lower-derivatives (a cosmological constant) or higher-derivatives 
 ($\alpha'$) corrections spoil
invariance under a rescaling of the metric. The importance of dealing
with critical superstring theory now  becomes evident:  if wewould
consider non-supersymmetric string theories, a cosmological constant
would almost certainly be generated at some finite order of the
loop expansion, which would change completely the large-distance
properties and  would spoil the symmetries of the field equations.

The immediate consequence of the above two symmetries is that
solutions of the field equations depend on two free parameters, a
dimensionless one related to the overall value of $\phi$, and a
dimensionful one corresponding to an overall length scale. As a
consequence, neither  the value of the fundamental string length nor 
that of the Planck length are of relevance to the classical
solutions. The importance of this point on the issue of fine-tuning will
become clear later in this section. 

Only if (and when) the
solutions are driven into the high-curvature regime will
higher-derivative corrections (involving the string length) will become
relevant and break scale invariance.  The analysis presented in this 
section  can only be trusted up to this point, or until the strong coupling
regime is reached. Which of the two limitations will be reached first
depends, in turn, on the two above-mentioned classical moduli. For
sufficiently small initial coupling the high-curvature regime is
encountered first, while, for sufficiently small initial curvature, the 
opposite is true. Both possibilities will be discussed in the last part of
this report (see Section \ref{Sec8}).

As already mentioned, the way we now introduce (and
exploit) the concept of APT in string
cosmology parallels the familiar use of  ``asymptotic flatness"
in general relativity \cite{Bon62,Sac62,Pen65}. In both cases 
some assumption of asymptotic simplicity is made, the asymptotic past
in our case, spatial infinity in the more familiar case. It seems indeed
physically (and philosophically) satisfactory to identify the beginning
with simplicity. However, simplicity should not be confused with
complete triviality:  a rigorously empty and flat Universe,
besides being uninteresting, is also very special,  i.e.
non-generic. By contrast, asymptotically trivial
 Universes, though initially simple, are also generic
 in a precise mathematical sense that we shall now discuss.

In the weak-coupling, small-curvature, low-energy regime under
consideration,  all heavy string modes can be integrated out, and one is
left with an effective theory of the massless fields. Some of these
fields, the graviton, the dilaton, and the NS-NS two-form, are always
present. Others (e.g. gauge fields and higher-order forms)
 may vary, depending on the particular superstring theory under study
or, in an M-theory perspective, on the particular M-theory  vacuum
around which the generic solution is looked for.

For the moment, let us restrict our attention to those three
universal massless fields whose low-energy small-coupling dynamics is
controlled by the effective action (\ref{31}). The general solution of the
field equations, in this regime, will consist of an arbitrary linear
superposition of (long-wavelength) gravitational, dilatonic and axionic
waves. From the point of view of target space (taken here, for the sake
of illustration, to be $(3+1)$-dimensional), 
the generic solution depends upon four arbitrary functions of three
coordinates \cite{LL62}
related to the metric, plus two more each for the dilaton and for the
axion field associated to $B_{\mu\nu}$. We will now see how these
arbitrary functions appear   in the asymptotic
expansion of our fields.

Let us  assume, in accordance with APT, the existence, in the far past,
 of Bondi--Sachs-like coordinates $r, v, \theta, \varphi$ in which  our
fields, as  $r \rightarrow \infty$, take the asymptotic forms
\cite{Bon62,Sac62} 
\beq
g_{\mu\nu} =  \eta_{\mu \nu} + \frac{f_{\mu \nu} (v,\theta ,
\varphi)}{r} + ... \; ,
\label{APTdata}
\eeq
and let us count the number of arbitrary functions needed to
specify our initial data up to gauge transformations.
This problem has been widely discussed in the general relativistic
literature and the result (up to functions of less than three coordinates)
is the one we have mentioned: a total of  $4+2+2=8$ functions of three
coordinates are needed (see for instance \cite{LL62}). 

Interestingly, there is an exact correspondence between this
``target-space" counting and a ``world-sheet" counting. In the latter,
those eight arbitrary functions correspond to eight arbitrary functions of
three-momentum entering the most general physical (i.e. on shell)
vertex operator describing gravitons, dilatons, and the antisymmetric
tensor field (which, in four dimensions, is equivalent to a pseudoscalar,
also called Kalb--Ramond axion). It is amusing to compare this number
with that of marginal operators,  able to deform the trivial conformal 
field theory on a flat target space into another conformal theory. These
are associated with physical vertex operators for the graviton, the dilaton
and the NS-NS two-form (see for instance \cite{GSW87}). It is easy to
verify that such (1,1) vertex operators depend on the same number
of arbitrary functions as our APT data. 

For instance, a physical graviton
is associated with a vertex operator depending on two
physical-polarization vectors,  each of which depends arbitrarily upon
the three-momentum  of the on-shell graviton \cite{GSW87}:
\beq
V^i_{\rm grav} = \epsilon^{i}(k)_{\mu\nu} \partial X^{\mu} \bar{\partial}
X^{\nu} e^{i k\cdot X}, ~~~~~~~~~~~~ i=1,2.
\label{gravitonvertex}
\eeq
Similar considerations apply to the dilaton and two-form vertex
operators. They lead to a total number of   eight functions of space,
needed to define the most general deformation of the conformal field
theory around its flat space-time limit.

From this point of view we see that APT conditions are generic, in a
technical sense. Yet APT is not a general property of initial conditions.
For instance, it looks as if, by choosing APT, we have eliminated the
possibility of ``white holes", objects that can emit but not absorb
matter and  that could possibly violate the second law of
thermodynamics. In a sense, by our APT postulate,  we are enforcing
the second law, at least in that part of space-time that will give rise to
our Universe.

\subsection{Pre-big bang inflation as gravitational collapse}
\label{Sec3.3}

 For simplicity, we will only illustrate here  the simplest case of
gravidilaton 
 system already compactified to four space-time dimensions.  Through
the field redefinition (\ref{ESF}), our problem is  reduced to the study
of a massless scalar field minimally coupled to gravity.
It is well known that such a form of matter cannot give inflation (since
it has positive pressure). Instead, it can easily lead to gravitational
collapse. Thus, in the E-frame, the problem becomes that of finding out 
under which conditions  gravitational collapse  occurs, if
asymptotically trivial initial data are assigned. Gravitational collapse 
usually means that the (Einstein) metric (hence the volume of 3-space) 
shrinks to zero at a space-like singularity. However, typically, the dilaton
blows up at that same singularity. Given the relation (\ref{ESF})
between the Einstein and the (physical) string metric, we can easily
imagine that
the latter blows up near the singularity, as implied by the phase
of dilaton-driven inflation illustrated in Section \ref{Sec2}.

How generically does a gravitational collapse take place? Let us recall
the singularity theorems of Hawking and Penrose \cite{Pen65a,HP70},
which state that, under some
general assumptions, singularities are inescapable in general
relativity. Looking at the validity of those assumptions in the case at
hand, one finds that all but one are automatically satisfied. The only
condition to be imposed is the existence of a closed trapped surface (a
closed surface from which  future-directed light rays converge).
Rigorous results \cite{Chr86} show that this condition cannot  be
waived:  sufficiently weak initial data
do not lead to closed trapped surfaces, to collapse, or to
singularities. Sufficiently strong initial data do. But where is the
borderline? This is not known in general,
but precise criteria do exist for particularly symmetric space-times, e.g.
for those endowed with spherical or plane symmetry  (see Subsection
\ref{Sec3.4}).

However, no matter what the general collapse/singularity criterion will
eventually turn out to
be, we do know, from the classical symmetries  described in the
previous subsection, that such a criterion cannot depend
{\em either}   on an overall additive constant in $\phi$, {\em or}
on an overall multiplicative factor in $g_{\mu\nu}$.

A characterization of APT initial data can be made \cite{BDV99}
following the pioneering work of
Bondi, Sachs, Penrose, and others. Since our initial quanta are
assumed to consist of massless gravitons and dilatons, their past
infinity is null: it is the so-called  ${\cal I}^-$ boundary of
the Penrose diagram (see Fig. \ref{f31}). APT means that dilaton and
metric can be   expanded near ${\cal I}^-$ in inverse powers of $r
\rightarrow \infty$, while   advanced time $v$
and two angular variables, $\theta$ and $\varphi$, are kept fixed. We
shall thus write:
\beq
\phi (x^{\lambda}) = \phi_0 + \frac{f(v,\theta , \varphi)}{r} + {O} 
\left( \frac{1}{r} \right) \, ,
\label{eqn2.4}
\eeq
\beq
g_{\mu \nu} (x^{\lambda}) = \eta_{\mu \nu} + \frac{f_{\mu \nu}
(v,\theta , \varphi)}{r} + { O} \left( \frac{1}{r} \right) \, .
\label{eqn2.5}
\eeq
The null wave data on ${\cal I}^-$ are: the asymptotic dilatonic
waveform $f(v,\theta , \varphi)$, and  two polarization components,
$f_+ (v,\theta , \varphi)$ and $f_{\times} (v,\theta , \varphi)$, of the
asymptotic gravitational waveform $f_{\mu \nu} (v,\theta , \varphi)$,
whose other components can be gauged away. The three functions $f$,
$f_+$, $f_{\times}$ of $v,\theta , \varphi$ are equivalent to six
functions of $r,\theta , \varphi$ with $r \geq 0$, because
the advanced time $v$ ranges over the full line $(-\infty , +\infty)$.
This is how the six arbitrary functions of the generic solution to the
gravidilaton 
system are recovered.

\begin{figure}[t]
\centerline{\epsfig{file=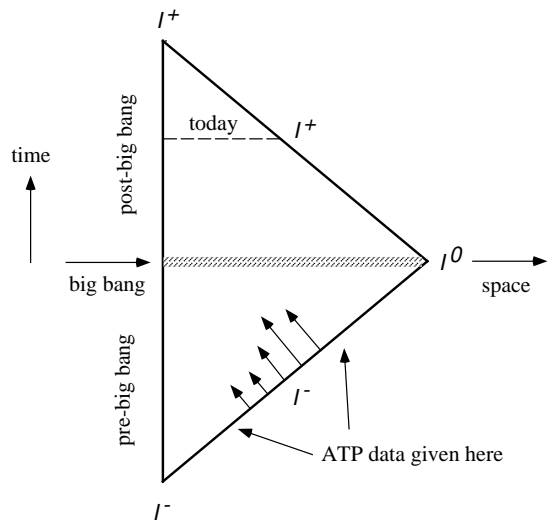,width=72mm}}
\vskip 5mm
\caption{\sl Penrose diagram for a possible model of complete 
string-cosmology scenario.} 
\label{f31}
\end{figure}

Of particular interest here are the so-called News functions, simply
given by
\beq
N (v,\theta , \varphi) \equiv \partial_v \, f (v,\theta , \varphi), 
~~~~~~~ N_+\equiv \partial_v \, f_+, ~~~~~~~
  N_{\times} \equiv \partial_v ~f_{\times}, 
\label{eqn2.8}
\eeq
and the ``Bondi mass'' given by:
\beq
M_-(v) = \frac{1}{4\pi} \int d^2\Omega M_-(v, \theta, \varphi), 
~~~~~
g_{vv} = - \left( 1- {2 M_-(v,\theta , \varphi) \over r}\right) +
{ O} \left( \frac{1}{r} \right).
\label{Bondi}
\eeq
The Bondi mass and the News are connected by
the energy--momentum conservation equation, which tells us that the
advanced-time
derivative of $M_-(v)$
is positive-semidefinite, and related to incoming energy fluxes
controlled by the News:
\beq
{dM_- (v)\over dv}  = \frac{1}{4} \, \int d^2\Omega \left(
N^2 +  \, N_+^2 +  \,
N_{\times}^2 \right)\, .
\label{eqn2.7}
\eeq

The physical meaning of $M_-(v)$ is that it represents the  energy
brought into the system (by massless sources) by advanced time $v$.
In the same spirit, one can define the Bondi mass $M_+(u)$ at future null
infinity ${\cal I}^+$.
It represents the energy still present in the system at  retarded time
$u$.
If only massless sources are present, the so-called ADM mass is  given
by
\begin {equation}
M_-(+\infty) = M_+(-\infty) = M_{\rm ADM} \; ,
\end{equation}
while $M_-(-\infty) = 0$, and  $M_+(+\infty) = M_{\rm C}$  represents
the mass that
 has not been radiated away
even after waiting an infinite time, i.e. the mass that
 underwent gravitational collapse \cite{Chr87}. Collapse (resp.
no-collapse)
 criteria thus aim at establishing
under which initial conditions one expects to find  $M_{\rm C} > 0$ (resp.
$M_{\rm C} =0$).

Since, as we shall see in the particular case of spherical and planar
symmetry,
collapse criteria {\em i}) do not involve any particularly large number,
and {\em ii}) do not contain any intrinsic scale, but just dimensionless
ratios of various classical scales, we expect:
\begin{itemize}
\item  Gravitational collapse to be quite a
generic phenomenon.
\item
  Nothing to be there to fix either the size of the horizon
 or the value of $\phi$ at the onset of collapse.
\end{itemize}
Generically,  randomly, chaotically,  some regions of space at  some
moment in time will undergo gravitational
collapse and  will form horizons and singularities therein. When this is
translated into the S-frame, regions of space-time within the
horizon actually undergo a period of inflation in which the initial values
of both the Hubble parameter and
 $\phi$ are left arbitrary. In  Subsection \ref{Sec3.6} we shall see that
such an arbitrariness
provides an answer to the fine-tuning allegations that
have been moved \cite{TW97,Lin99} to the pre-big bang
scenario. Before dealing with that issue, however, we shall discuss
in the following two subsections
 cases endowed with special isometries, for which
much more is known than for the general case.

\subsection{The case of spherical  and planar symmetry}
\label{Sec3.4}

In the spherically symmetric case, many authors have studied the
problem of gravitational collapse
of a minimally coupled scalar field, both numerically and analytically.
In the former case we should recall the existence of well-known
results \cite{Chp93}  pointing at  mysterious universalities
near critical collapse (i.e. at the  borderline situation in which
the collapse criteria are just barely met). In this case, a very small
black hole forms. This is not the case
we are really interested in, for the reasons we just explained.
We shall thus turn, instead, to what happens when the collapse criteria
are largely fulfilled. For this we make use of the classical results
presented in  \cite{Chr86,Chr87,Chr91,Chr93}. 

In the case of spherical symmetry there are no gravitational waves, so
that null wave data consist  of  just an angle-independent, asymptotic
dilatonic waveform $f(v)$, with the associated
scalar News $N(v) =  f'(v)$. A convenient system of coordinates is the
double null system $(u,v)$, such that
\beq
\phi = \phi (u,v) \, , ~~~~~~
ds^2 = - \Omega^2(u,v)\,du \, dv + r^2(u,v)\,d \omega^2 \,,
\label{eq1.6}
\eeq
where $d\omega^2 = d\theta^2 + \sin^2 \theta \, d\varphi^2$. The field
equations are conveniently re-expressed in terms of the three
functions $\phi
(u,v)$, $r(u,v)$ and $m(u,v)$, where the local mass function $m(u,v)$ is
defined by:
\beq
1 - \frac{2 m}{r} \equiv g^{\mu \nu}\,(\pa_\mu r)\,(\pa_\nu r) =
-\frac{4}{\Omega^2}\, \frac{\pa r}{\pa u}\,
\frac{\pa r}{\pa v}\,.
\label{eq1.7}
\eeq
One then gets the following  set of evolution equations for $m$, $r$ and
$\phi$: 
\bea
\label{eq1.8}
&& 2\frac{\pa r}{\pa u}
\,\frac{\pa m}{\pa u} = \left ( 1 - \frac{2 m}{r} \right )
\,\frac{r^2}{4}\,\left (\frac{\partial \phi}{\pa u} \right )^2, ~~~~
 2 \frac{\pa r}{\pa v}\,
\frac{\pa m}{\pa v} = \left ( 1 - \frac{2 m}{r} \right )
\,\frac{r^2}{4}\,\left (\frac{\partial \phi}{\pa v} \right )^2 
\\
\label{eq1.10}
&& r\,\frac{\pa^2 r}{\pa u \pa v} =
\frac{2 m}{r-2m}\,\frac{\pa r}{\pa u}\,
\frac{\pa r}{\pa v}, ~~~~
r\,\frac{\pa^2 \phi}{\pa u \pa v} +
 \frac{\pa r}{\pa u} ~ \frac{\pa \phi}{\pa v}
+ \frac{\pa r}{\pa v} ~ \frac{\pa \phi}{\pa u}
=0.
\eea

The quantity
\beq
\mu (u,v) \equiv \frac{2 m (u,v)}{r}
\label{eqn3.3}
\eeq
plays a crucial r\^ole in the problem.
 If $\mu$ stays everywhere below 1, the field configuration will
not collapse but will finally disperse at infinity as outgoing waves. By
contrast, if the mass ratio $\mu$ can reach anywhere the value 1, this
signals the formation of an apparent horizon 
${\cal A}$ (a closed trapped surface),  whose location  is indeed defined
by the equation $ \mu (u,v) = 1 $. 
The above statements are substantiated by some rigorous inequalities
\cite{Chr93}, 
stating that:
\beq 
\label{rel1}
\frac{\pa r}{\pa u} < 0\,, ~~~~~~ \frac{\pa m}{\pa v} > 0\,, ~~~~~~
\frac{\pa r}{\pa v}\,
\left ( 1- \mu \right ) > 0\,, ~~~~~~ \frac{\pa m}{\pa u}\,
\left ( 1- \mu \right ) < 0\,.
\eeq
Thus, in weak-field regions ($\mu < 1$), we have $\partial_v
r > 0$, while, as $\mu > 1$, we have $\partial_v r < 0$, meaning that the
outgoing radial null rays (``photons'') emitted by the sphere $r =
{\rm const}$
become convergent, instead of having their usual behaviour.
This is nothing  but the signature of a closed trapped surface.

 The emergence
 of a closed trapped surface implies the existence of a future
singular boundary ${\cal B}$ of space-time where a curvature
singularity occurs \cite{HawEl73,Chr91}. 
Furthermore, the behaviour of various fields near the singularity is
just that of a quasi-homogeneous phase of dilaton-driven inflation
as described by Eqs. (\ref{inhKasner}).  This highly non-trivial result
strongly supports the idea that
pre-big bang inflation in the S-frame
is the counterpart of gravitational collapse in the E-frame.

Reference~\cite{Chr91} gives the following sufficient criterion on
the strength of
characteristic data, considered at some finite retarded time $u$
\beq
\frac{2 \Delta m}{\Delta r }  \geq
\left [ \frac{r_1}{r_2}\,\log \left (\frac{r_1}{2 \Delta r} \right )
+ \frac{6r_1}{r_2} -1 \right ]\,,
\label{eq2.2}
\eeq
where $r_1$ and $r_2$ define two spheres, such that $r_1 \leq r_2$
and $r_2 \leq 3r_1/2$, , $\Delta r = r_2 - r_1$ is the
width of the ``annular'' region between the two
spheres, and $\Delta m = m_2 - m_1 \equiv m(u,r_2) - m (u,r_1)$ is the
mass ``contained'' between the two spheres, i.e. more precisely the
energy flux through the outgoing null cone $u =$ const,  between $r_1$
and $r_2$. Note the absence of any intrinsic scale (in particular of any
short-distance cut-off) in the above criterion.
The theorem proved in \cite{Chr91} is not exhausted in the above
statement. It contains various bounds as well, in particular: {\em i}) 
an upper bound on the retarded time at which the closed trapped
surgace (i.e. a horizon) is formed, {\em ii}) a lower bound on the mass,
i.e. on the radius of  the collapsing region.

The latter quantity is very important for the discussion of the
previous subsection
since it gives, in the equivalent S-frame problem, an upper limit
on the Hubble parameter at the beginning of the phase of inflation. Such
an upper limit depends only on the size of the advanced-time interval
satisfying  the collapse criterion; since the latter is determined by the
scale-invariant condition   (\ref{eq2.2}),
the initial scale of inflation will be classically undetermined.

The above criterion is rigorous, but probably too conservative. It also
has the shortcoming
that it cannot be used directly on ${\cal I}^-$, since $u \rightarrow
-\infty$ on ${\cal I}^-$.
In Ref. \cite{BDV99} a less rigorous (or less general) but simpler
criterion, 
directly expressible in terms of the News (i.e. on ${\cal I}^-$), was
proposed on
the basis of a perturbative study. It has the following attractive form:
\beq
 \sup_{v_1,v_2 \atop v_1 \leq v_2} \, {\rm Var} (N(x))_{x \in
[v_1,v_2]} > C = O(1/4) \, ,
\label{eqn4.16}
\eeq
where:
\beq
{\rm Var} \, (N(x))_{x \in [v_1,v_2]} \equiv  \langle (N(x) -
\langle N \rangle_{[v_1,v_2]})^2 \rangle_{x \in [v_1,v_2]} \, .
\label{eqn4.15}
\eeq
Thus ${\rm Var} \, (g)_{[v_1,v_2]}$ denotes the ``variance'' of the
function
$g(x)$ over the interval $[v_1,v_2]$, i.e. the average squared
deviation from its
mean value.

According to this criterion the largest interval satisfying
(\ref{eqn4.16}) determines the size of the collapsing region and thus,
through the collapse inflation connection, the initial value of the Hubble
parameter. It would be interesting to confirm the validity of the above
criterion,  and to determine more precisely the value of the constant
appearing on its r.h.s., through more analytic or numerical
work. This problem has been studied by two perturbative numerical
analyses \cite{MOV98,Chiba99} and also, more recently, by a numerical
approach that takes into account the full non-linear evolution
\cite{Madden01}, and stresses the possible role of non-linearities in
triggering the collapse of a long train of low-amplitude waves, even
though the criterion (\ref{eqn4.16}) would suggest that they remain
perturbative.

Let us discuss now another toy model example
of gravitational collapse, that of the collision of plane waves, following
closely Refs.  \cite{FKV} and \cite{BV}. The advantage of this case is that 
it can often be solved analytically, thereby providing a  useful laboratory
for investigating  properties that we would like to be shared by more
realistic situations (see some comments at the end of this subsection).
For this reason we shall treat this example in considerable detail. 

In order to include eventually the NS-NS two-form, we will 
start from  the low-energy effective action in a form that already
exhibits the  $O(d, d)$ non-compact symmetry expected from the
$d=D-2$ Abelian isometries of the problem\footnote{In this
subsection, for  simplicity, we have denoted by $d$ what in the
rest of the report would be $d-1$.}.
This is achieved by working with a ``reduced action" living in the
non-trivial (two-dimensional) subspace spanned by time and by the
common direction of propagation of the
two waves.
Following \cite{MahSch} we write:
\begin{equation}
S=\int
{dx^0}{dx^1}\sqrt{-g} e^{-\overline{\phi}} \left[ R+ \partial_ \alpha
\overline{\phi} \partial^\alpha \overline{\phi}
+\frac{1}{8}{\rm Tr}\left(\partial_\alpha M^{-1} \partial^\alpha M
\right) \right].
\label{O(d,d) action}
\end{equation}
Here the  metric $g_{\mu\nu}$ is taken to be
block-diagonal, with blocks  $g_{ij}$ and $g_{\alpha\beta}$
(the Roman
indices ($i,j,\dots$)  span the transverse directions,
 while the Greek indices take the values 0 and 1). We arranged the
components $g_{ij}$ in a $d$-dimensional matrix $G$ and the
non-vanishing components $B_{ij}$ in another matrix $B$. $M$ is then the
usual $2d$-dimensional matrix containing $G$  and $B$ (see Eq. 
(\ref{253})), while $\overline{\phi}=\phi-(1/2)\log \det G + {\rm
constant}$ is the corresponding shifted dilaton (see Subsection
\ref{Sec1.4}). The reduced action (\ref{O(d,d) action})  is manifestly
invariant under the global $O(d,d)$ transformations (\ref{260}).

The corresponding equations of motion are written most  conveniently
by going to the conformal gauge for
the 2-metric, $g_{\alpha\beta} = e^F \eta_{\alpha\beta}$,
and by working with ``light-cone" coordinates, $\sqrt{2} u =
(x^0-x^1) , \sqrt{2} v =
(x^0+x^1)$. They take the explicitly $O(d,d)$-invariant
form
\begin{eqnarray}
&& \partial_u\partial_v {\rm exp}(-
\overline{\phi}) = 0, ~~~~
 \partial_u \left( e^{-\overline{\phi}} M^{-1}
\partial_v M \right) + \partial_v \left( e^{-\overline{\phi}} M^{-1}
\partial_u M \right) = 0 ; \label{phibar}\\
&&   \partial_u^2 \overline{\phi} - \partial_u  F  \partial_u
\overline{\phi} + \frac{1}{8}{\rm Tr}\left(\partial_u M^{-1}
\partial_u M \right) = 0 , ~~{\rm and} ~{\rm the} ~{\rm same} \; {\rm
with} \; u \rightarrow v \label{Vir};\\
 &&   \partial_u
\partial_v \overline{\phi} -  \partial_u \partial_v F +
\frac{1}{8}{\rm Tr}\left(\partial_u M^{-1} \partial_v M \right)=0.
\label{Fevol}
\end{eqnarray}
The two colliding waves are defined as having their fronts at $u=0$ and
$v=0$, and thus to collide at  $u = v = 0$  (i.e. at $x^0 =
x^1 = 0$). The two waves are {\it not} assumed to be impulsive, i.e. their
energy density can have any (finite) support at  positive $u$ and $v$,
respectively. 

Space-time is naturally divided in four regions:
Region I, defined by $u,v<0$, is the space-time in front of the
waves, before any interaction
takes place. It is trivial Minkowski space-time with $B=0$ and a
constant perturbative dilaton
(${\rm exp}(\phi_0) \ll 1$). It will be convenient to fix the constant
in the definition of $\overline{\phi}$  so that, in region I,
$\overline{\phi}=
0$. Region II, defined by $u>0$, $v<0$, represents the wave coming from
the left
before the collision. Metric and dilaton depend only on
$u$:
\begin{equation}
ds_{\rm II}^2=-2dudv+G_{ij}^{\rm II}(u)dx^i dx^j , ~~~~~
B_{ij}=B_{ij}^{\rm II}(u),
~~~~~
\phi = \phi^{\rm II}(u),
\end{equation}
and we can consistently set $F=0$.
Region III can be similarly described up to interchange of $u$ and $v$.
Note that we have not assumed any special shape for $G$, so that
our results, so far,
hold irrespectively of the polarization of the waves.
Finally, in region IV ($u>0$, $v>0$), i.e. in the interaction region, we write
\begin{equation}
ds_{\rm IV}^2=-2e^F dudv+G_{ij}^{\rm IV}dx^i dx^j,
\end{equation}
 and $F$, $G^{\rm IV}$, $\phi^{\rm IV}$ and
$B^{\rm IV}$ are all functions of both $u$ and $v$.
Of course, the metric must be continuous along with its derivative  on
the boundary lines between the four regions. The same must be true
for the dilaton
$\phi$ and the antisymmetric field $B$.

Let us begin by solving the equations in region II.
The only non-trivial equation is
(\ref{Vir}),
which (dropping the subscripts II) reads
\begin{equation}
\ddot{\overline{\phi}}=\frac{1}{8}{\rm{Tr}}\left[\left( M^{-1}
\dot{M}\right)^2\right],\label{Region II1}
\end{equation}
where the dot indicates the derivative with respect to $u$.
It is now useful to change  variables from $u$ to $\tilde{u}$, with
\begin{equation}
\frac{d}{du}=e^{-2\phi/d}\frac{d}{d\tilde{u}}.
\end{equation}
After some algebra we can rewrite (\ref{Region II1}) as:
\begin{equation}
e^{\overline{\phi}/d}\left(e^{-\overline{\phi}/d} \right)''=
-\frac{1}{d^2}\phi'^2-\frac{1}{4d}\left\{{\rm{Tr}}
\left[\left(G^{-1}G'\right)
_{t}^2\right]-{\rm{Tr}}\left[\left(
G^{-1}B'\right)^2\right]\right\},\label{Region II2}
\end{equation}
where $\left(G^{-1}\dot{G}\right)_{t}$ is the traceless part of
$\left(G^{-1}\dot{G}\right)$, and
the prime denotes the derivative with respect to $\tilde{u}$.

It can easily be shown that all  terms on the r.h.s. of Eq. 
(\ref{Region II2}) are negative-definite.
Hence, for any non trivial wave, $e^{-\overline{\phi}/d}$, which
is constant and identically equal to 1 in region I, must acquire a
non-vanishing, negative, and never increasing derivative in region
II. Thus, $e^{-\overline{\phi}/d}$  must vanish at some finite
$\tilde{u} = \tilde{u}^*$. Returning now to the coordinate $u$, we
see that, if the dilaton
 is bounded
(a necessary assumption
 if we want to use the tree-level effective action),
there exists a finite $u = u^*$ where
$e^{-\overline{\phi}/d}$ vanishes. Correspondingly, also $\det G$
vanishes,
and the metric of the transverse space will collapse to zero proper
volume,
thereby producing a (coordinate) singularity.
The same arguments can be repeated in region III, where $\det G$ has
to vanish at some finite  $v=v^*$, with $\overline{\phi} \rightarrow +
\infty$.
These conclusions  generalize a well-known result
in $D=4$ general relativity \cite{Szekeres,Yurtsever}
to any $D$ and to non trivial antisymmetric fields, .

Let us finally analyse region IV (where the interaction between
the two waves occurs) dropping, for simplicity, the subscript IV  from
all functions. We begin by using Eq. (\ref{phibar}),
 which tells us that $e^{-\overline{\phi}}$ is the sum of a function
of $u$ and
a function of $v$.
The unique function of this type that matches the
boundary conditions with region I is
\begin{equation}
e^{-\overline{\phi}(u,v)}=
e^{-\overline{\phi}_{\rm II}(u)}+
e^{-\overline{\phi}_{\rm III}(v)}
-1.
\end{equation}
 We see that $e^{-\overline{\phi}(u,v)}$ must vanish on a
hypersurface that joins  the coordinate singularities in
regions II and III and is contained
 in the region $u\leq u_0$, $v \leq v_0$ within region IV.

Let us now introduce two  new sets of coordinates
that simplify the analysis in region IV.
One set is of the light-cone type:
\begin{equation}
r = r(v) = 2 e^{-\overline{\phi}_{\rm III}(v)  } - 1, ~~~~~~
s = s(u) = 2 e^{-\overline{\phi}_{\rm II}(u)  } - 1, 
\end{equation}
while the second set is of the $t-x$ kind:
 \begin{eqnarray}
&& \xi= \frac {1}{2}  (r + s) = e^{-\overline{\phi}(u,v)}
\sim -t, \\
&& z = \frac {1}{2} (s - r) =
e^{-\overline{\phi}_{\rm II}(u)}-e^{-\overline{\phi}_{\rm III}(v)} .
\end{eqnarray}
Note that the coordinates $r,s$ run from $+1$ to $-1$ in  region IV, with
their sum always positive except on the singular boundary where  $ r+s
= 0$.
Going from the original coordinates to either of the new sets
changes only the conformal factor of the 2-metric.
We may thus write:
\begin{equation}
ds_{\rm IV}^2= - e^f d\xi^2+e^f dz^2 +G_{ij}dx^i dx^j = - 2e^f dr ds
+G_{ij}dx^i dx^j ~~,
\label{metricxi}
\end{equation}
while the shifted dilaton is simply
\begin{equation}
\overline{\phi}=-\log \xi = - \log (r+s)/2  .
\end{equation}

Let us discuss in some detail  the case of $B=0$ and parallel  polarized
waves following \cite{FKV}.
In each region, we take a diagonal $G$ with $G_{ii}=e^{\lambda+\psi_i}$,
where $\sum \psi_i=0$ and $\lambda=(1/d) \log\det G$ goes to
$-\infty$
at $u=u^*$ in region II, and at $v=v^*$ in region III. In the new
coordinates, Eq. (\ref{phibar}) becomes trivial, while
the equations for the $\psi_i$ and
$\lambda$ are decoupled and can be solved separately. They are also
formally the same, so
the solutions have the same structure \cite{Szekeres}:
\begin{eqnarray}
&& - (r+s)^{{\frac{1}{2}}} \psi_i\left(r,s\right) = 
 \int\limits_{\rm s}^1
 ds' \left[(1+s')^{\frac{1}{2}} \psi_{i}(1,s') \right]_{,s'}
 P_{-\frac{1}{2}}\left[1+2\frac{
\left(1-r\right)\left(s'-s \right)}{\left(1+s' \right)
\left(r+s\right)}\right]
 \nonumber \\
&&  + \int\limits_r^1 dr' \left[(1+r')^{\frac{1}{2}} \psi_i(r',1)
\right]_{,r'}
P_{-\frac{1}{2}}\left[1+2\frac{
\left(1-s\right)\left(r'-r \right)}{\left(1+r' \right) \left(r+s\right)}
\right] \, ,
 \label{psiformula}
\end{eqnarray}
where $P_{-1/2}(x)$ are  Legendre functions
written in standard notation, and the commas denote partial
derivatives.  The same expression holds for $\lambda$, with the obvious
replacements. At this point $\phi$ can  easily be recovered
 once $\lambda$ and $\overline{\phi}$ are known.

We see that $\lambda$ and the $\psi_i$ are singular on the
hypersurface $\xi=r+s=0$. So, in a very general way, we  find that
 the collision of two plane
waves leads to a (curvature) singularity in the space-time,  whatever
 the number of
dimensions. Although we do not have, at the moment, such an explicit
solution for the general case, the discussion given in the previous
section
makes us believe that a curvature singularity will always emerge
along the hypersurface $\xi=0$.

For the solutions (\ref{psiformula}), the asymptotic behaviour for
$\xi \rightarrow0$ is easily found by taking
the large-argument limit of the Legendre function
\cite{Szekeres} (see also \cite {Yurtsever}):
\beq
\psi_i\left(\xi,z \right)\sim\epsilon_i\left(z\right)
\log \xi, ~~~~~~
\lambda\left(\xi,z \right)\sim\kappa\left(z\right) \log \xi, ~~~~~~
f\left(\xi,z \right)\sim a\left(z\right) \log \xi.
\eeq
The coefficients multiplying the logarithm are functions of $z$,
whose range, on the singular surface, goes from $-1$ to $1$.
One easily finds
\begin{eqnarray}
&\epsilon_i\left(z\right)=&\frac{1}{\pi\sqrt{1+z}}\int\limits_z^1 ds
 \left[\left(1+s\right)^{\frac{1}{2}}\psi\left(1,s\right) \right]_{,s}
\left(\frac{s+1}{s-z} \right)^{\frac{1}{2}}+ \nonumber \\
&&+ \frac{1}{\pi\sqrt{1-z}}\int\limits_{-z}^1 dr
 \left[\left(1+r\right)^{\frac{1}{2}}\psi\left(r,1\right) \right]_{,r}
\left(\frac{r+1}{r+z} \right)^{\frac{1}{2}} \; ,\label{epsilon} \\
&a\left(z \right)=& \frac{1}{4}\sum\epsilon_i^2
\left(z\right)+\frac{d}{4}\kappa^2\left(z\right)-1 \; ,
\end{eqnarray}
with $\kappa\left(z\right)$ given by the same expression as
$\epsilon_i$, but with $\psi_i$ replaced by $\lambda$. The sum of the
$\epsilon_i$ must be zero
according to the definition of the $\psi_i$.

The asymptotic form of the metric is
\begin{equation}
ds_{\rm IV}^2=-\xi^{a\left(z\right)}d\xi^2+\xi^{a\left(z\right)}dz^2+
\xi^{\kappa(z)}\sum \xi^{
\epsilon_i\left(z\right)} \left(dx^i\right)^2 \; ,
\end{equation}
while
\begin{equation}
\phi \sim - \left(1+ \frac{d}{2}\kappa (z) \right) \log \xi \; .
\end{equation}
Going over to  cosmic time $\xi=t^{\frac{2}{a\left(z\right)+2}}$ 
gives the metric in  Kasner form with exponents
\beq
p_1\left(z\right)=\frac{a\left(z\right)}{a\left(z\right)+2}
\label{p1}, ~~~~~~
p_i\left(z\right)=\frac{\kappa(z)+
\epsilon_i\left(z\right)}{a\left(z\right)+2} \label{pi}.
\eeq
The following relations are immediately verified:
\beq
\phi=\left(\sum\limits_{\alpha=1}^{D-1}
p_\alpha\left(z\right)-1\right)\log t, ~~~~~~
\sum\limits_{\alpha=1}^{D-1} p_\alpha^2\left(z\right)=1.
\eeq

The behaviour of the fields near the singularity is thus of
Kasner type, modified, as usual, by the presence of the dilaton.
Note that, at the two tips of the singularity, $\epsilon_i$ and $\kappa$
diverge in such a way that the Kasner exponents near the tips
are simply $p_1 =1$, $p_i =0$. This corresponds to
 a (contracting) Milne-like metric, 
 which, being non-singular, nicely matches the non-singular
behaviour in regions II
and III. Away from these two points,
 $\kappa$ and $\epsilon_i$ can take any
value: it is easy to verify that the whole Kasner
sphere can be covered by appropriately choosing  the initial data.

It may be interesting to note that, in such a context, it is possible to
estimate the entropy generated by the non-linear interaction of the
incoming waves, before the Universe enters the phase of inflation,
during the intermediate phase dominated by both  the velocity and
spatial gradients \cite{FeKuVa00}. The total produced entropy is of the
order of the product of the focal lengths of the two incoming waves (in
Planck units), and satisfies the entropy bounds that will be discussed in
Section \ref{Sec8.4}. 

 Let us finally perform $O(d,d)$ boosts on our
solutions, in order to introduce a non-trivial $B_{\mu\nu}$ field.
We will refer to \cite{BV} for a full discussion and only summarize here
the final result.
Cosmologies with non-trivial $B_{\mu\nu}$ that can
be obtained, by an $O(d,d)$ transformation, from a metric having
Kasner behaviour near the singularity, generally contract in all
dimensions. In even dimensions, at least one  inflates when the original
metric has just one inflating or just one contracting dimension. In odd
dimensions, if the $O(d,d)$ transformation has $\det Q=0$ and
we start  from a full inflating metric or from a metric with one
inflating dimension, then we are left with one inflating dimension. If
$\det P=0$, then one dimension inflates when we start from a full
contracting metric or from  a metric having just one contracting
dimension. We should also recall that none of the $O(d,d)$
transformations affects the $g_{11}$ component of the metric, which,
therefore, can either inflate or contract.

How do these results generalize to more realistic situations such as the
head-on collision of two plane-fronted waves, with a transverse profile
that extends only over a finite range? Unfortunately such cases cannot
be dealt with analytically, and numerical results are not
available either, as far as we know. There have been claims
\cite{Yurtsever}, however,  that  black holes (and singularities therein)
are inevitably formed, provided certain
conditions are met.  

Consider, an an example, two identical  waves of transverse size $L$,
 moving head on against each other, and carrying, in the
centre-of-mass frame, energy density (per unit area) $\rho$. The
conjecture \cite{Yurtsever} is that a black hole of mass $O(\rho L^2)$
will form,  provided 
\beq
L > c (G\rho)^{-1} \, ,
\label{Yurtsever}
\eeq
where $c$ is a number $O(1)$.

This conjecture appears to be confirmed by applying a recently
 proposed method \cite{Giddings} for determining when
 black holes are formed in point-like particle collisions at finite impact
parameter.
A simple extension of that method \cite{Kohl} to two finite-size
massless waves shows
that Eq. (\ref{Yurtsever}) is indeed  a good criterion for black-hole
formation with
$c \le {1 / 4 \pi}$.

\subsection{Adding $p$-forms and BKL chaotic behaviour}
\label{Sec3.5}

We have already mentioned that  adding a NS-NS two-form
to the background strongly reduces
the parameter space available for inflating the Universe (in string
units). However, very generally, a smooth Kasner-like
behaviour is still generically valid near the singularity.
Spatial gradients become small with respect to time derivatives, 
and solutions become velocity-dominated, thereby solving, at least in
principle, the flatness/homogeneity problems.

This result appears to be in contrast with what is known to happen
in a standard general relativity context. Indeed,
Belinskii,  Khalatnikov and  Lifshitz (BKL)  postulated a long
time ago \cite{BKL} that the approach to a big crunch singularity (the
time-reversal of our situation) is typically chaotic, in contrast with
the naive expectation that,  near the singularity, kinetic terms should
dominate over spatial gradients  since $^{(3)}R \sim a^{-2} \ll R \sim
t^{-2}$, for a scale factor going to zero with a small power of $t$. This
would certainly be true for an isotropic situation. However, because of
the  constraints on the exponents $\a_i =(d \ln a_i/ d\ln t)$ of the
(general relativistic) Kasner solution,
\beq
\label{Kasnerconstr}
\sum \alpha_i = \sum \alpha_i^2 = 1,
\eeq
the only non-trivial vacuum cosmologies are anisotropic. 

As a consequence, there are 
gradient terms that are subdominant and others that, instead, 
grow faster than the extrinsic
curvature itself  (the  expansion contribution to curvature),  and start
dominating at some point. When this happens the Kasner exponents
suddenly change and a new Kasner epoch starts, until another spatial
gradient dominates \dots, and so on indefinitely. One can visualize 
this  behaviour as the motion of a particle in a billiard, where the
Kasner-like phases are the straight motions while the sudden changes in
the Kasner exponents correspond to reflections. Unless the ``billiard" in
question is particularly simple, a chaotic behaviour then follows. This
phenomenon goes under the name of BKL oscillations.

It was known for some time that BKL oscillations occur in general
relativity for  any $D< 11$. However, the phenomenon
was known to disappear if a massless, minimally coupled scalar field is
added. The dilaton of string cosmology is just such a field, and therefore
BKL oscillations do not appear in gravidilaton string cosmology, as
explicitly shown in  \cite{BaDa98}. One way to see this is to realize that,
in the presence of a dilaton, an isotropic solution  is perfectly
allowed, and therefore all spatial gradients can be kept under control.
Thus, possibly after a small number of oscillations, the Kasner exponents
end up in a region (close to the isotropic point) where no more
oscillations occur until the singularity.

Adding the NS-NS two-form does not appear to change the situation 
with respect to BKL oscillations, as  shown in many of the examples
of inhomogeneous string cosmology discussed in Subsection
\ref{Sec3.1}.  The same is true in the presence of various kinds of axion
fields (see e.g.  the case of the Type IIB string cosmology discussed in
\cite{FeVa98}), or in the context of  the gravidilaton--axion--Maxwell
system giving rise to inhomogeneous Gowdy-type solutions 
\cite{Narita00,Yaza01}. On the basis of these examples it was believed
for some time that BKL oscillations were the exception, rather than the
rule, for string cosmology.

The surprise came when it was pointed out \cite{DH00/1,DH00/2} that  
BKL oscillations reappear generically if all the massless bosonic
background fields of superstring  theory (or M-theory) are turned on. 
Note that, even under the assumption of asymptotic past triviality, 
there is no reason to artificially put to zero any of these massless
backgrounds. According to APT, the generic initial state should be an
arbitrary superposition of weakly interacting, low-frequency waves of
all kinds. Setting arbitrarily to zero some of them would amount to
fine-tuning the initial state, and could even be  unstable under quantum
effects.

The emergence of chaos discovered in \cite{DH00/1,DH00/2}
 could have been anticipated, since we can view the
gravidilaton system
 in $D=4$ as coming from dimensional reduction of a pure gravity theory
in more dimensions. Upon dimensional reduction, the higher dimensional
graviton generates both the dilaton and additional gauge fields.
The electric and/or magnetic components of these fields, unless
artificially turned off, will grow and cause BKL behaviour. Let us now go
in some length into this matter, nevertheless referring to the original
articles for full details. 

The starting point is the  low-energy tree-level bosonic action of a 
generic string theory (or of M-theory) which, in the Einstein frame, can
be written in the form: 
\beq
\label{genericaction}
S = -\int d^Dx \sqrt{g} \left[ R(g) + ....+ \sum_p c_p e^{\lambda_p
\varphi}(dA_p)^2 \right],
 \eeq
where the index $p$ labels the various $p$-forms that are present in  a
given superstring model, and $c_p, \la_p$ are numerically coefficients.
There are actually possible additional terms, e.g. Chern-Simons terms,
but these are not relevant to this discussion. One then assumes 
the inhomogeneous-Kasner ansatz of Subsection  \ref{Sec3.1}, 
rewritten in the Einstein frame as: 
\begin{eqnarray}  
&&
d s ^2 =  dt^2 -
\sum_a e_i^a(x)~ e_j^a(x)~ (-t)^{2 p_a(x)} dx^i dx^j , ~~~~~~ \phi =
p_{\varphi} ~ {\rm log} (-t) \, , \nonumber \\ 
&&
  \sum_a p_a  = 1\;, ~~~~~~~~~~~~  p_{\varphi}^2 = 1 - \sum_a p_a^2 ,
~~~~~~~~~~~ t < 0 \; . 
\label{inhKasnerEF}
\end{eqnarray}

In order to check whether such a non-chaotic, non-oscillatory ansatz is
an asymptotic solution one has to check whether all neglected terms
remain parametrically small as $t \rightarrow 0$. Several terms
are possibly dangerous. Dilaton gradients are always harmless, but
 metric gradients are ``down" with respect to the time-derivative terms
by a factor $|t|^{2g_{ijk}}$, where $g_{ijk}$ is the following combination
of Kasner exponents, $g_{ijk} = 1 +p_i - p_j - p_k$. In order for these
terms to be negligible, it is necessary that all $g_{ijk}$ be positive.
Furthermore, the equations of motion for the various forms turn out to
imply the constancy of both the ``magnetic" components,  $F_{j_1\dots
j_{p+1}}$, and of the ``electric" components, $\sqrt{g}
 e^{\lambda_p \varphi} F^{0i_i\dots i_p}$, of the field strength $F_{p+1}
\equiv  dA_p$. When such a result is inserted back  in the complete
equations of motion one finds further constraints on the Kasner
 exponents and on the ``dilaton couplings" $\lambda_p$, which need to
be satisfied in order for  the solution (\ref{inhKasnerEF}) to hold.

The explicit analysis of \cite{DH00/1,DH00/2} shows that in any known 
$D=10$ superstring theory, as well as in $D=11$ supergravity (the
low-energy limit of M-theory), the couplings $\lambda_p$ are just what
is needed to reduce the parameter space for the validity of
(\ref{inhKasnerEF}) to a set of zero measure. The generic case is instead
chaotic, and can be described in terms of motion on a billiard in a
9-dimensional hyperbolic space \cite{DH01}. The reflection rules in the
billiard form a group that turns out to be the Weyl group of the 
hyperbolic Kac--Moody algebras of $E_{10}$ or $BE_{10}$,  depending on
the particular superstring theory under consideration. Similar groups
 were also found to emerge in the case of Kaluza--Klein
compactification of pure gravity in  $D=d+1\ge 4$ \cite{DHJN01}. 
The Weyl group is now related to the Kac--Moody algebra $AE_d$, and
the disappearance of chaos for $D\ge 11$ gets related to the fact that
these algebras are hyperbolic only for $d < 10$.

Let us finally mention a connection between the BKL chaotic behaviour
and the peculiar features of some homogeneous string cosmology
models, in particular those with a non-trivial antisymmetric field
$B_{\mu\nu}$. We have already seen that the presence of such a
backgound  tends to reduce drastically (if not completely) the
phase space left for inflationary pre-big bang behaviour. In Ref.
\cite{DH00/3} it was shown that this can be understood again as motion
in a billiard. Whenever the initial Kasner exponents are such as to give
inflation, some walls are hit a certain (finite) number of times until the
ensuing reflections bring  the S-frame Kasner exponents to be
positive, thereby stopping the inflationary behaviour and turning it into
a contraction. 

From the physical viewpoint the
implications of all these results are not so obvious,  also because
the conditions for BKL behaviour are just met at the tree level and to
lowest order in $\alpha'$.
It is possible that the inclusion of either $\alpha'$ or loop corrections
will bring the oscillations
to a stop (after a relatively small number of them 
\cite{DH00/1,DH00/2}). It is not clear, however, which kind of
string-scale-curvature hypersurface will be generated by the whole
process. It is unlikely that it will be very
simple and homogeneous,  but it is not obvious  that it will not
contain isotropic and homogeneous enough
patches for our Universe to emerge from.

\subsection{Is pre-big bang cosmology fine-tuned?}
\label{Sec3.6}

The two arbitrary  parameters discussed in the previous subsection
are very important, since they
determine the range of validity of our description. In fact, since
both curvature and coupling increase during  the initial phase of
dilaton-driven inflation,  the low-energy and/or tree-level description is
bound to break down at some point. The smaller the   initial Hubble
parameter (i.e. the larger the initial horizon size) and the smaller the
initial coupling, the   longer we can follow the phase of
low-energy inflation through the effective action equations, and the
larger the number of reliable e-folds we shall gain.

This does answer, in our opinion, the objections raised
\cite{TW97} to the pre-big bang scenario according to which it is 
fine-tuned. The situation here actually resembles that of  chaotic
inflation \cite{Lin83}. Given some generic (though APT) initial data, we
should ask which is the distribution of sizes of the collapsing regions 
and of couplings therein.
Then, only the ``tails" of these distributions, i.e. those corresponding
to sufficiently large 
and sufficiently weakly coupled regions will produce Universes like
ours, the rest will not.
The question of how likely a ``good" big bang is to take place is not
very well posed
and can be greatly affected by anthropic considerations \cite{BDV99}.
Furthermore, asking for a  long enough period of
low-energy inflation amounts to setting upper limits on two arbitrary
moduli of the classical solutions.

Figure \ref{f32}  is a $(2+1)$-dimensional sketch of a possible
pre-big bang  Universe: an   original ``sea" of
dilatonic and gravity waves leads to collapsing regions of different
initial size, possibly to a scale-invariant distribution of them. Each one of
these collapses is reinterpreted, in the S-frame, as the process by
which a baby Universe is born after a period of
pre-big bang inflationary ``pregnancy",  the size of each baby Universe
being   determined by the duration of the corresponding pregnancy, i.e.
by the initial size of (and coupling in) the corresponding collapsing
region.  By using, in particular, the result of Eq. (\ref{2109}) one finds
that, in $d=3$,  regions initially larger than $10^{19} \la_{\rm s}$ (namely,
larger than $10^{-13}$ cm if $\la_{\rm s} \sim \la_{\rm P} \sim 10^{-32}$ cm) 
can generate Universes like ours, while smaller ones cannot.

\begin{figure}[t]
\centerline{\epsfig{file=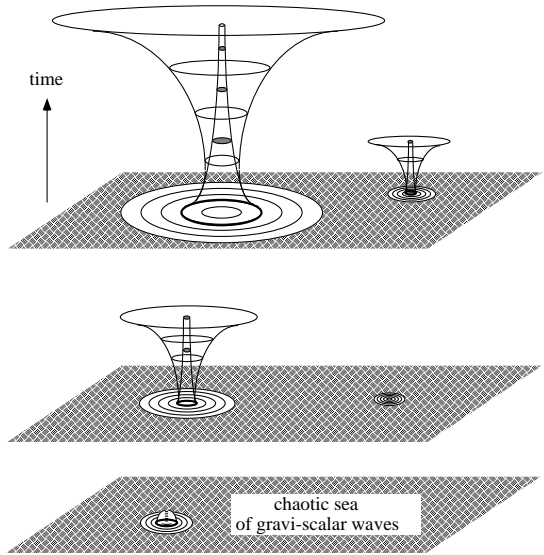,width=72mm}}
\vskip 5mm
\caption{\sl Qualitative space-time illustration of the possible birth of
pre-big bang Universes from a chaotic sea of gravidilaton waves.
Each baby Universe is simultaneously represented in the S-frame (like
an expanding cone) and in the E-frame (like a shrinking hole), and 
it is followed for an increasing time interval as we move upward in the
figure.}
\label{f32}
\end{figure}

A basic difference between the large numbers needed in  
(non-inflationary) FRW cosmology and the large numbers needed in 
pre-big bang cosmology, should be stressed. In the former, the ratio of
two classical scales, e.g. of total   curvature to its
spatial component, which is expected to be $O(1)$,
has to be taken as large as $10^{60}$. In the latter, the above ratio is
initially $O(1)$ in the collapsing/inflating region, and ends up being very
large in that same region, thanks to the inflation. However, the (common)
order of magnitude of these two classical quantities is a free parameter,
and it is taken to be much larger than the classically irrelevant quantum
scale. Indeed, the smallness of quantum corrections (which would
introduce a scale in the problem) was explicitly checked in \cite{GPV98}.

An example of what we have just said is the case of the collision of
two finite-front plane  waves. It is clear from Eq. (\ref{Yurtsever}) that
all that matters is the ratio of two geometrical classical quantities, the
transverse size $L$ and the focal distance $(G\rho)^{-1}$. Neither the
Planck nor the string lengths appear in the collapse criterion.
Even the appearance of the Newton constant in (\ref{Yurtsever}) is
somewhat misleading: as always in the context of classical general
relativity, the Newton constant can be removed by a convenient choice
of units of energy, and then everything reduces to geometrical quantities.

We can visualize analogies and differences between standard and
pre-big bang inflation by looking again at the two cases of Figs. 
\ref{f12} and \ref{f13}. The common feature in the two pictures is
that the fixed comoving scale corresponding to the present horizon was
``inside the horizon" for some time during inflation,  possibly very
deeply inside at its onset.  The difference between the two scenarios
is just in the behaviour of the Hubble radius during inflation:
increasing in standard inflation, decreasing in string cosmology. Thus,
while standard inflation is still facing the initial-singularity question, and
needs a non-adiabatic phenomenon to reheat the
Universe (a kind of small bang), pre-big bang cosmology faces the
singularity   problem later, combining it with the exit and heating
problems (see Section \ref{Sec8}).

A much more relevant objection to this approach to initial conditions is
the one raised in \cite{DH00/1,DH00/3}. If the initial
conditions are to be really generic (although satisfying APT), then all
massless waves should be allowed. Already with antisymmetric tensor 
waves the phase of dilaton-driven inflation is becoming non-generic, 
while, when all massless forms are included, BKL behaviour seems to be
the generic situation, at least before strong curvature and/or strong
coupling is reached. Understanding the fate of such evolutions calls for a
deeper understanding of how
string effects may resolve the singularity problem (to be discussed in
Section \ref{Sec8}).

\section{Amplification of quantum fluctuations}
\label{Sec4}
\setcounter{equation}{0}
\setcounter{figure}{0}

If we accept, at least as a working hypothesis, the idea that the pre-big
bang scenario illustrated in the previous sections might represent a
possible candidate for a self-consistent representation  of the
primordial cosmological evolution,  we are  eventually led to the 
following fundamental questions: Is it possible to find phenomenological
effects that can differentiate the pre-big bang scenario from other,
more conventional  inflationary scenarios, and Are such
differences  observable today, at least in principle? 

In order to answer this question we may note that the accelerated
evolution of the Universe is a process accompanied by the
production of an enormous amount of radiation. 
Almost all types of fields are excited, and particles are copiously
produced, with an efficiency that reaches its maximum during the
explosive ``big bang" regime. By studying the properties of the relic
cosmic backgrounds we may  obtain information about the very
early state of the Universe, just as by studying the quantum numbers
of the particles produced in a  decay process we can  get  information
on the quantum state of the system before the decay. 

From the various mechanisms of radiation production, active on a
cosmological scale, we  will consider here the parametric amplification
of the quantum fluctuations of the vacuum, because it is typical of all
backgrounds with accelerated kinematics, and it is effective
in all inflationary scenarios. It thus represents a good ``indicator" of the
dynamics of the early Universe \cite{Gris88,Gris88a,GriSol91} and an
appropriate starting point to discriminate among different
inflationary models.  

Although in this section we will mainly concentrate on the study of the
metric fluctuations, our discussion can be easily applied to the evolution
of the fluctuations of any background field. The convenient formalism to
be adopted for the study of this process is the theory of cosmological
perturbations (see  \cite{MFB92,Lid93} for  two excellent and
comprehensive reviews), which provides a straightforward and  (in
principle)  simple approach to the problem. According to this approach,
given a set of background field equations     
\beq
 G_{\mu\nu}=T_{\mu\nu},  
\label{41}
\eeq
the standard procedure is to perturb (to first order) the metric and the
matter sources: 
\beq
g_{\mu\nu} = g_{\mu\nu}^{(0)}+ \da^{(1)} g_{\mu\nu},
~~~~~~~~~~~~T_{\mu\nu} = T_{\mu\nu}^{(0)}+ \da^{(1)} T_{\mu\nu},
\label{42}
\eeq
expanding around a given cosmological solution. By using the
unperturbed equations for the bakground fields, $G_{\mu\nu}^{(0)}=
T_{\mu\nu}^{(0)}$, one easily obtains a
system of linear equations describing the evolution of the first-order 
fluctuations of the metric and of the matter fields, 
\beq 
\da^{(1)}G_{\mu\nu}= \da^{(1)} T_{\mu\nu}, 
\label{43}
\eeq
and  can then discuss, on the basis of these equations, the amount of
radiation production.

This perturbative approach is the same for all cosmological scenarios.
There are two important differences, however, that characterize  
pre-big bang models with respect to other inflationary models. 

The first difference concerns the higher-dimensional and scalar-tensor
(i.e. gravidilaton) nature of the string cosmology backgrounds. As a
consequence, there are possible contributions to the amplification of the
fluctuations not only from the inflation of the three-dimensional space,
but also from the dynamics of the extra-dimensions
\cite{Gar89,Am90,Dem90,GG92a}, and from the time evolution of the
coupling constants (i.e. of the dilaton), as first pointed out in \cite{GG93}.
The amplification of the perturbations thus becomes possible also for
fields conformally coupled to the 3-dimensional metric, unlike in the
standard cosmological scenario \cite{Bir82}. 

The second difference concerns the growing character of the
background curvature and of the couplings. In such a background the
quantum fluctuations start evolving from an asymptotically flat,
zero-coupling state, so that they can be naturally normalized to an
initial vacuum fluctuation spectrum. After the amplification their final
configuration thus corresponds to a  ``squeezed vacuum" state
\cite{GriSid89,GriSid90,Gris92}, and not to a ``squeezed-thermal vacuum"
\cite{GGV93}, or to other, more complicated states. In addition, since the
final energy distribution tends to follow (in frequency) the behaviour (in
time) of the curvature scale, one typically obtains in such a context a
growing (or ``blue") spectrum
\cite{Star79,AbbWi84,Gris88,Al88,Sa90,GriSol91}. As a consequence,  the
peak amplitude of the produced  radiation may be high enough to support
a picture in which all the radiation filling our present Universe was
directly produced from the quantum fluctuations of the vacuum
\cite{Ve95,Gas95}.  This effect may facilitate the detection of a cosmic
background of relic gravitational waves \cite{GG92,Gas94}, as will be
discussed in Section \ref{Sec5}, but it complicates the generation of a
scale-invariant spectrum of curvature perturbations, eventually
producing the observed CMB anisotropy (see Section \ref{Sec7}). 

A further consequence of the growth of the curvature scale, however, is
that the comoving amplitude of the perturbations may even grow 
outside the horizon, instead of being frozen as in standard inflationary
models. This effect, implicitly contained in the earlier pioneering studies
of cosmological perturbations \cite{Star79}, was first explicitly pointed
out in \cite{Abb86}, and only later independently rediscovered in a string
cosmology context \cite{GasVe93b} and in the context of scalar-tensor
gravity \cite{Bar93}.  Such a growth of the amplitude may require an
appropriate and non-conventional choice of the gauge \cite{BruMu95} in
order  to restore the validity of the linear approximation,  and to
allow the application of the standard perturbative formalism. 

In the following subsections we will  discuss in some detail
the above effects, typical of the evolution of perturbations in the
context of the pre-big bang scenario. 

\subsection{``Frame" independence}
\label{Sec4.1}

The computation of the spectrum according to the theory of
cosmological perturbations \cite{MFB92} requires a series of formal
steps, the first of which is the choice of the ``frame", i.e.  of
the basic set of fields (metric included) used to parametrize the action.
Any theory is in general characterized by a preferred set of fields, but
one might wonder what happens when there are more than one
preferred choices (this is a typical situation of all scalar-tensor models
of gravity, as in the case of the gravidilaton string effective action). 

In string theory there is a preferred frame (the so-called
 S-frame), in which the coupling to a constant
dilaton is unambiguously fixed (at the tree level) for all fields and to
all orders in the higher-derivative $\ap$ expansion. Also, the metric of
this frame is the one to which fundamental strings are minimally
coupled \cite{SV90,GSV91a} (in other words, the world-paths of strings
freely evolving in a curved background are geodesic surfaces of this
metric).  In the S-frame, and in $d+1$ dimensions, the 
gravidilaton effective action takes the form 
\beq
S = {1\over 2 \la_{\rm s}^{d-1}}\int\,d^{d+1}x\,\sqrt{|g|}\,e^{-\phi}
\,\left[-R+ \om \left(\nabla_\mu\phi\right)^2 \right]
\label{44}
\eeq
(for the sake of generality we have included here the Brans--Dicke
coefficient $\om$, which in our context could  represent a dilaton
self-coupling possibly arising from higher-loop corrections
\cite{GSW87}; in particular, $\om=-1$ for the lowest-order string
effective action). 

For any $\om$, however, this action can always  be
rewritten in the (more conventional)   E-frame, in which the graviton and
dilaton kinetic terms are diagonalized in the standard canonical form, 
and the dilaton is minimally coupled to the metric. 
Consider, in fact, the field redefinition $ g \ra \ti g$ performed with the
help of a new scalar variable $\psi$:
\beq
g_{\mu\nu} =\ti g_{\mu\nu} e^\psi.
\label{45}
\eeq
By expressing the scalar curvature $R$ in terms of $\ti g$ 
\cite{Wald,Syn}, the action (\ref{44}) becomes 
\beq
S = {1\over 2 \la_{\rm s}^{d-1}} \int d^{d+1}x\,\sqrt{|\ti g|}
\,e^{-\phi+{(d-1)\over 2} \psi} \left[-\ti R+ d {\ti \nabla}^2
\psi+{d(d-1)\over 4} \left({\ti \nabla}_\mu\psi\right)^2 +\om
\left({\ti \nabla}_\mu\phi\right)^2 \right], 
\label{46} 
\eeq
where the tilde denotes geometrical quantities computed with respect
to $\ti g$. By setting ($\om=$ const, $\om >-d/(d-1)$):
\beq
(d-1)\psi = 2\phi, ~~~~~~~~~~~
\ti \phi =\phi \left[2d +2\om (d-1) \over d-2\right]^{1/2},
\label{47}
\eeq
and neglecting a total derivative, we are led eventually to the E-frame 
action of general relativity for the new variables $\ti g, \ti \phi$:
\beq
S = {1\over 2 \la_{\rm s}^{d-1}} \int\,d^{d+1}x\,\sqrt{|\ti g|}
\,\left[-\ti R+ {1\over 2}\left(\ti \nabla_\mu \ti \phi\right)^2 \right]. 
\label{48}
\eeq

Please note the ``right" canonical coefficient in front of the
dilaton kinetic term (we remember that we are using the signature 
$g_{00}>0$) and the possible change of sign from the S-frame action
(\ref{44}) (where $\om=-1$, to lowest order). The E-frame is thus
preferred for performing canonical quantization, for identifying the
particle content of the theory, and for defining the effective low-energy
masses and couplings (see also Section \ref{Sec6}). 

However, the field equations for $g, \phi$, obtained by varying the action
(\ref{44}), are  different from the equations for $\ti g, \ti
\phi$,  obtained by varying the action (\ref{48}). As a consequence,
also the perturbation equations (\ref{43}), in the S-frame, are different
from the E-frame perturbation equations, and one might
wonder which is the ``right"  frame to be used,  when computing
cosmological effects to be compared with present observations.  

This is a false problem, however, because the (final) observable
variables, such as the spectral amplitude, the spectral energy density
(see Subsection \ref{Sec4.4}),  are exactly the same in both 
frames \cite{GasVe93b} (an obvious consequence of the fact that 
physical measurable quantities cannot be  changed by field
redefinitions  such as the transformation (\ref{45}); see also
\cite{AlCon01} for a recent discussion of the equivalence between the
String and Einstein frames).   The formal reason of this
frame-independence is that the perturbation equations are indeed
different in the two  frames, but the classical solutions around which the
expansion is performed are also different, and the two differences 
exactly compensate each other. This result can be generally proved by
perturbing the action, better than the equations of motion (see Section 
\ref{Sec4.3}). It is instructive, however, to give also a particular 
example, as will be done here by considering tensor metric
perturbations propagating in a ($d+1$)-dimensional external
background.  

We will start  the computation in the S-frame, where the background
equations obtained from the action (\ref{44}) can be written explicitly
as 
\bea
&&
R_\mu\,^\nu + \nabla_\mu \nabla^\nu \phi + (\om +1) 
\left[ \da_\mu^\nu \left(\nabla_\a \phi\right)^2 -
\da_\mu^\nu \Box \phi -  \nabla_\mu \phi  \nabla^\nu \phi \right]
=0, \nonumber \\
&&
 R+ \om \left(\nabla_\a \phi\right)^2-2 \om \Box \phi =0.
\label{49}
\eea
We shall consider  the transverse,  traceless part of  the metric
perturbations, 
\beq
\da^{(1)}\phi=0, ~~~~
\da^{(1)} g_{\mu\nu} = h_{\mu\nu}, ~~~~~
\da^{(1)} g^{\mu\nu} = -h^{\mu\nu}, ~~~~~
\nabla_\nu h_\mu\,^\nu =0=h_\mu\,^\nu, 
\label{410}
\eeq  
propagating in a higher-dimensional, spatially flat, factorizable 
background with  $d$ ``external" and $n$ ``internal" dimensions, with
scale factors $a(t)$ and $b(t)$, respectively.  Also, we shall assume that
the translations along the internal dimensions are isometries of the full,
perturbed metric (see \cite{Giov97} for a more general situation in which
also the internal gradients of the fluctuations are non-zero, and see
the comment after Eq. (\ref{425}) for backgrounds with
non-factorizable geometry). The  fluctuations of the external space
$h_{ij}(x_i,t)$ are then conveniently described in the synchronous gauge
\cite{LL62},  where ($i,j=1,...,d$; $m,n =d+1, ..., d+n$): 
 \bea && 
g_{00}=1, ~~~~~g_{0i}=0, ~~~~~ g_{ij}=-a^2\da_{ij},
~~~g_{m n}=-b^2\da_{mn}, \nonumber \\
&&
h_{00}=0, ~~~~~h_{0i}=0, ~~~~~ g^{ij}h_{ij}=0, ~~~~~ \pa_j h_i\,^j =0 .
\label{411}
\eea
In this gauge 
\bea
&&
\da^{(1)} \Ga_{0i}\,^j={1\over 2}\hp_i\,^{j}, ~~~~ 
\da^{(1)}\Ga_{ij}\,^0=-{1\over 2}\hp_{ij}, \nonumber \\ 
&&
\da^{(1)}\Ga_{ij}\,^k={1\over 2}\left(\pa_i h_j\,^k+
\pa_j h_i\,^k-\pa^k h_{ij}\right); 
\label{412}
\eea
to first order in $h$, the  perturbation of the
dilaton equation  is found to be trivially satisfied, and
there are no contributions from the terms inside the square brackets  
in (\ref{49}). The perturbation equations are thus (remarkably)
$\om$-independent \cite{GG93}:
\beq
\da^{(1)}R_\mu\,^\nu -\left(\da^{(1)} g^{\nu\a}\Ga_{\mu\a}\, ^0 +
g^{\nu\a}\da^{(1)} \Ga_{\mu\a}\, ^0 \right) \dot \phi=0. 
\label{413}
\eeq
For $\mu,\nu \not= i,j$, they are trivially satisfied. The 
$(i,j)$ components of the Ricci tensor, on the other hand, lead to the
higher-dimensional covariant d'Alembert
operator \cite{GG92a,Dem93,GG93} 
\beq
\da^{(1)}R_{i}\,^j=-{1\over 2}\left(\hpp_i\,^j
+d{\dot a\over a} \hp_i\,^j + n{\dot b\over b}\hp_i\,^j- {\nabla^2\over
a^2} h_i\,^j\right)\equiv  -{1\over 2} \Box h_i\,^j  =0, 
\label{414}
\eeq
where $\nabla^2 = \pa_i^2$ and the indices of $h$ are raised and
lowered with the unperturbed metric. Using the identities (see also
\cite{Gas97}) \bea
&&
g^{jk}\hp_{ik}=\hp_i\,^j +2 H h_i\,^j , \nonumber\\
&&
g^{jk}\hpp_{ik}=\hpp_i\,^j +2 \dot H h_i\,^j +4H\hp_i\,^j +4H^2h_i\,^j, 
\label{415}
\eea
we are finally led to the S-frame tensor perturbation equation
\cite{GG93} 
\beq
\Box h_i\,^j -\dot \phi \dot h_i\,^j =0.
\label{416}
\eeq
In terms of the conformal time coordinate $\eta$, such that 
$dt = a d\eta$, this wave equation  can  also be  rewritten (for each
polarization mode $h$) as follows, 
\beq
 h^\se +(d-1) { a' \over  a}h'+n{b'\over b}h'- \phi' h' - \nabla^2  h =0 .
\label{417}
\eeq

Let us now repeat the computation in the  E-frame, where the
background equations following from the action  (\ref{48}) 
can be written  as 
\beq
\ti R_{\mu\nu}= {1\over 2} \pa_\mu \ti \phi \pa_\nu \ti \phi, 
~~~~~~~~~~~{\ti g}^{\mu\nu} \ti \nabla_\mu \ti \nabla_\nu \ti \phi=0. 
\label{418}
\eeq
Perturbing to first order, and using Eqs. (\ref{410}):
\beq
\da^{(1)}\ti R_\mu\,^\nu =0  
\label{419}
\eeq
(the perturbation of the scalar Klein--Gordon equation is trivially
satisfied). In the synchronous gauge, we can still apply Eq. (\ref{414})
(for the ``tilded" variables), and we are  finally led to the E-frame
perturbation equation
\beq
{\ti \Box} \ti h_i\,^j =0.
\label{420}
\eeq
In conformal time, and for each polarization component $\ti h$,
\beq
\ti h^\se + \left[(d-1) {\ti a' \over \ti a}+n{\ti b'\over \ti b} \right] \ti h'
-  \nabla^2 \ti h =0 . 
\label{421}
\eeq

This last equation seems to be different from the S-frame equation
(\ref{417}). We have to recall, however, that the conformal time is the
same in both frames, $d\eta= dt/a=d \ti t/ \ti a=d \ti \eta$ (see Section
\ref{Sec2.4}), and that, according to Eq. (\ref{45}), 
\beq
\ti a= a e^{-\phi/(d+n-1)}, ~~~\ti b= b e^{-\phi/(d+n-1)}, ~~~~
(d-1) {\ti a' \over \ti a}+ n{\ti b' \over \ti b} = 
(d-1) { a' \over  a}+ n{b' \over  b} -\phi'.
\label {422}
\eeq
We thus have  the same equation for $h$ and $\ti h$, the same solution
and, as a consequence, the same spectrum when the solution is
expanded in Fourier modes.  

In other words, the {\em observable} results of the perturbative analysis
are ``frame-independent" and, when computing the spectrum, we are
allowed to use the frame that is  most convenient  for practical
purposes. 

\subsection{Choice of the ``gauge" for scalar perturbations}
\label{Sec4.2}

Within each frame, we have a possible additional ambiguity due to the 
choice of the gauge, i.e. the choice of the parametrization
of the given metric manifold. The step corresponding to this choice  is
more delicate than the choice of the frame: in spite of the fact that the
final observable spectrum  has to be certainly  independent of the given
coordinate system, the perturbative analysis {\em  is}, on the contrary,
``gauge-dependent". It is possible,  for instance,  that a linearized
description of the perturbations  is valid in a given  system of
coordinates, and yet that the linear  approximation is broken in a
different system.  This problem obviously disappears in the context of a
fully covariant and gauge-invariant (to all orders) description of
perturbations \cite{ElBru89,BruEl92}. For the perturbative approach
usually adopted in a cosmological context \cite{MFB92}, however, gauge
invariance holds in the linear approximation only, and breaks down at
higher orders. 

An important example of this effect is provided,  in string cosmology,
by the time evolution of the scalar (metric and dilaton) perturbations. 
We will discuss this effect in the E-frame  (dropping the ``tilde", for
simplicity), starting from the unperturbed equations (\ref{418}), and
using, as in the previous subsection, a homogeneous, Bianchi-I-type,
higher-dimensional background, with $d+n$ factorized structure:  
\beq
g_{\mu\nu}= {\rm diag} (a^2, -a^2 \da_{ij}, -b^2\da_{mn}),~~~~~~~
\phi =\phi (\eta). 
\label{423}
\eeq
We shall consider, in particular, the exact background solution
parametrized by
\bea
&&
a=(|\eta|)^\a,~~~~~~b=(|\eta|)^\b,~~~~~~~
\phi=\sqrt{2\over d+n-1} ~ 
\frac{n-d-\sqrt{d+n}}{1+\sqrt{d+n}}\ln(|\eta|), 
\nonumber\\
&&
\a=\frac{\sqrt{d+n}+1-2n}{(1+\sqrt{d+n})(d+n-1)}, ~~~~~~~~~~~
\b=\frac{\sqrt{d+n}-1+2d}{(1+\sqrt{d+n})(d+n-1)},
\label{424}
\eea
and corresponding to the conformal-time parametrization of the vacuum
solution already introduced in Eq. (\ref{2113}). 
For $\eta<0$, $\eta \ra 0_-$, this unperturbed solution describes a
phase of accelerated evolution, growing curvature and growing dilaton,
in which the horizon expands more slowly (to be more precise, shrinks
faster) than the scale factor. It is thus a good background candidate to
amplify metric fluctuations. In the S-frame it describes a phase of
superinflation and dynamical dimensional reduction, in which $d$
dimensions are expanding and $n$ dimensions are contracting, with
scale factors related by the duality transformation $b= a^{-1}$. 

Let us  perturb the metric and the dilaton around this solution, by 
using the isometries of the factorizable geometry (\ref{423}) and 
assuming that all dynamical variables depend only on the ``external"
coordinates $x_i$, $i= 1,...,d$ (so that the translations along the internal
dimensions are also isometries of the full, perturbed background)  
\cite{GG97}.  In this case, modes with
different rotational transformation properties are decoupled, and the
scalar component of the background perturbations can be written in
general as \cite{MFB92,GG97}:
\bea
&&
\da^{(1)} \phi =\chi ,  ~~~~~~~~~~
ds^2= \left( g_{\mu\nu}+ \da^{(1)} g_{\mu\nu}\right) dx^\mu dx^\nu=
\nonumber\\
&&
a^2\left(1+2\varphi\right) d\eta^2 -a^2\left[
\left(1-2\psi\right)dx_i^2+2 \pa_i\pa_j E dx^idx^j+ 2 \pa_iB dx^id\eta
\right]- b^2(1-2\xi)dx_m^2,\nonumber\\
&& 
 \label{425}
\eea
where all variables  depend only on $\eta$ and $x^i$. 
It should be noted that this approach is appropriate to the standard
Kaluza--Klein scenario,  but it cannot be applied to non-factorized
geometrical structures such as those appearing in the context of the
Randall--Sundrum scenario \cite{RS2}. In this last case, the study of
scalar metric and dilaton fluctuations (see for instance
\cite{BaGasVe01,Giov01}) requires a more general metric decomposition
\cite{Dorca00}. In this report, however, we will restrict our
discussion to the  perturbations  of higher-dimensional Kaluza--Klein
backgrounds of the standard, factorizable type. 

In our case, it is important to stress that the 
six functions $\varphi,~ \psi,~ E,~B,~\xi,~\chi$ are {\em not} invariant
under local infinitesimal transformations of coordinates,
\beq
x^\mu \ra x^\mu + \ep^\mu (x), ~~~~~~~~~
g_{\mu\nu} \ra g_{\mu\nu}  -\nabla_\mu \ep_\nu -\nabla_\nu 
\ep_\mu. \label{426}
\eeq
However, if we consider infinitesimal coordinate transformations
depending on two scalar parameters  $\ep^0$ abd $\ep$,  
\beq
\ep^\mu =(\ep^0, \pa^i\ep, 0), ~~~~~~~~~
\eta \ra \eta + \ep^0, ~~~~~~~~~~ x^i \ra x^i+\pa^i \ep, 
\label{427}
\eeq
and thus preserving the scalar character of the
perturbations (in other words, coordinate transformations that do not
add  vector or tensor components to the perturbed metric), then the
scalar variables transform as 
\bea
&&
\varphi \rightarrow   \varphi - {a'\over a} \epsilon^0 -
{\epsilon^0}' ,~~~~~~
\psi \rightarrow   \psi  + {a'\over a} \epsilon^0, ~~~~~~
\xi  \rightarrow  \xi + {b'\over b}  \epsilon^0
\nonumber\\
&&
E \rightarrow E- \epsilon, ~~~~~~~~~~
B \rightarrow  B +\epsilon^0 - \epsilon' , ~~~~~~~~~
\chi \rightarrow  \chi - \phi' \epsilon^0 ,
\label{428}
\eea
and it is always possible to define a set of variables that are
gauge-invariant, in the linear approximation. For instance \cite{GG97}:
\begin{eqnarray}
&&\Phi = \varphi +\frac{1}{a}[(B-E')a]' , 
~~~~~~~~~~
\Psi =\psi -{a'\over a}(B-E') ,
\nonumber\\
&&\Xi =\xi - {b'\over b}(B-E') , 
~~~~~~~~~~~~~~
X = \chi +\phi' (B-E') .
\label{429}
\end{eqnarray}
The first two variables (also called Bardeen potentials
\cite{Bar80,Sas83}) are phenomenologically important, since their
spectral amplitudes  $\Phi_k, \Psi_k$, at the  recombination era are
directly related (via the Sachs--Wolfe effect) to the large-scale CMB
anisotropy $(\Delta T /T)_k$,  currently observed  by astrophysical
experiments (see Section \ref{Sec7}). 

A natural and convenient choice of
the gauge is thus the so-called longitudinal (or conformally
Newtonian) gauge, in which $E=0=B$, and the perturbed metric
depends directly on the Bardeen potentials, according to Eqs. 
(\ref{429}). In this gauge the coordinates are totally fixed because,
according to the variation of the metric under the transformation
(\ref{428}), the choice $E=0=B$ completely fixes $\ep^0$ and $\ep^i$,
and  leaves no residual degrees of freedom. 

In this gauge we can now write explicitly the set of perturbed field
equations 
\bea
&&
\da^{(1)} R_{\mu\nu} =\pa_\mu\phi \pa_\nu \chi ,\nonumber\\
&&
\da^{(1)}g^{\mu\nu}\nabla_\mu\nabla_\nu \phi + \nabla^2 \chi
-g^{\mu\nu}\da^{(1)} \Ga_{\mu\nu}\,^\a \pa_\a \phi=0,
\label{430}
\eea
obtained from the unperturbed Einstein equations (\ref{418}).
From the ($i,j\not=i$) component of the perturbation equations we
obtain a useful relation between three perturbation variables, 
\beq
\varphi=(d-2)\psi+n\xi.
\label{431}
\eeq
The ($0,i$) components give a constraint, while the  ($i,i$) and ($m,m$)
components of the perturbed Einstein equations, combined with the
($0,0$) component, provide the following interesting system of coupled
equations for the ``external" and ``internal"
perturbations
$\psi$ and $\xi$ \cite{GG97,Giov95}:
\bea
&& 
(d-1)\left\{\Box\psi +\psi'\left[3(d-1) {a'\over a}+ 3 n {b'\over b}
\right]\right\} =
-n\left\{\Box \xi +\xi'\left[3(d-1) {a'\over a} +3 n  {b'\over
b}\right]\right\}, \nonumber \\ 
&& 
d \left\{\Box \psi
+\psi'\left[3(d-1){a'\over a} +
\frac{b'}{bd}\left( 2(d-1)(n-1)+nd\right) \right]\right\} = \nonumber \\
&&
=-(n-1)\left\{\Box \xi +\xi'\left[ \frac{a'}{a(n-1)}\left(3 d
n - d -n +1\right)+3 n {b'\over b}\right]\right\}, 
\label{432}
\eea
where $\Box\equiv \left({\partial}^2/{\partial{\eta}^2}\right)
-{\nabla}^2$ denotes here the usual (flat-space) d'Alembert
operator. 

This system can easily be  diagonalized to find the (time-dependent)
linear
combination of $\psi$ and $\xi$ that represents the true ``propagation
eigenstates". For our purpose, however, the asymptotic behaviour of
the
modes
$\psi_{k}$, $\xi_{k}$ can be simply obtained (modulo logarithmic
corrections) by inserting into  the previous system the power-law
ansatz
\beq
\psi_k= A (-\eta)^x,~~~~~~~~~~~~~~~\xi_k=B(-\eta)^x. 
\label{433}
\eeq
For the background solution (\ref{424}) one then finds  that there are, in
the $|k\eta|\rightarrow 0$ limit,  non-trivial solutions for the
coefficients $A$ and $B$ only if
$x=0$ or $x=-2$. This means that, asymptotically, 
\beq
\psi_k= A_1+\frac{A_2}{ \eta^2},~~~~~~~~~~~~~~~~
\xi_k=B_1+\frac{B_2}{ \eta^2},
\label{solanis}
\eeq
and the amplitude is not frozen, unlike in the
standard inflationary scenario. Because of the growing mode, 
$\psi_k \sim \eta^{-2}$, the scalar perturbation amplitude 
blows up in the limit $\eta \ra
0_-$. Otherwise stated, at any given time $\eta$ there is  always a
low enough frequency band for which the typical fluctuations have a
dimensionless amplitude much larger than $1$ \cite{BruMu95} (see Eq. 
(\ref{475}) in the next subsection).  This breaks the validity of the linear
approximation and is not consistent, in general,  with the perturbative
expansion around a homogeneous background. 

This conclusion cannot be avoided, be it  by moving to a different frame 
(as discussed in the previous subsection), or by changing the number of
dimensions, since the previous result is dimensionality-independent. For
a $d=3$ isotropic background, in particular, one finds in the longitudinal
gauge $\varphi=\psi$, and the Fourier modes  $\nabla^2\psi_k= -k^2
\psi_k$ satisfy  the Bessel equation
\beq
\psi_k^\se +{3\over \eta}\psi_k'+k^2\psi_k=0
\label{435}
\eeq
(from Eq. (\ref{432})). 
The exact solution, in the limit $|k\eta|\ra 0$, has the asymptotic
expansion \cite{AbSte}
 \beq
\varphi_k = c_1 \ln |k\eta| + {c_2 \over \eta^2}
\label{436}
\eeq
($c_1, c_2$ are integration constants), with the same growing mode
$\sim \eta^{-2}$ as in higher dimensions. 

It is important to point out, however, that such a growing mode is absent
for tensor metric perturbations. If we consider the tensor
perturbation equation (\ref{421}), in the same background (\ref{424}),
we obtain indeed for each  Fourier mode the Bessel equation
\beq
h_k^\se +{1\over \eta}h_k'+k^2h_k=0,
\label{437}
\eeq
with the asymptotic solution
\beq
h_k = c_1+c_2 \ln |k\eta| , ~~~~~~~ |k\eta| \ra 0_- .
\label{438}
\eeq
All modes tend to stay constant, asymptotically, modulo a logarithmic
growth that can be easily kept under control if the
phase of accelerated evolution is not infinitely extended in the
high-curvature regime  but is bounded, for instance,  by the Planck  (or
string) scale (as expected). 

This is a first signal that the growing mode of scalar
perturbations, appearing in the longitudinal gauge, might be a pure
gauge effect, i.e. an artefact of the particular choice of  coordinate
system. This suspicion is confirmed by the application  of the ``fluid flow"
approach \cite{Haw66,Lid93} to the perturbations of a scalar-tensor
background. In this approach, the evolution of density and
curvature inhomogeneities can be described in terms of 
covariant scalar variables, which are gauge-invariant to all orders
\cite{BruEl92}. 

Let us consider, for simplicity, the isotropic case $d=3$,
$n=0$. There are two covariant variables, $\Da$ and $C$, 
defined in terms of the momentum density of the scalar field $\nabla
\phi$, of the spatial curvature $^{(3)}R$, and of their derivatives. By
expanding around the dilaton-driven background
(\ref{424}), which in our case reduces to  
\beq
a=(-\eta)^{1/2} , ~~~~~~~~~~~~~~~~ \phi=-\sqrt 3 \ln (-\eta) ,
~~~~~~~~~~~~~~  -\infty < \eta <0 ,
\label{439}
\eeq
one finds  for such variables, in
the linear  approximation,  the asymptotic solution ($|k\eta| \ll1$)
\cite{BruMu95}:  
\beq
\Da_k = {\rm const} ,~~~~~~~~~~~~
c_k= {\rm const} +A_k \ln |k\eta| . 
\eeq
Such variables tend to stay constant outside the horizon,
with at most a logarithmic variation (as in the tensor case),
which is not dangerous. 
Indeed,  in terms of $\Da$ and $C$,  the amplitude of density and
curvature fluctuations can be consistently computed using the linear
approximation (for all modes), and
their spectral distribution (normalized to an initial vacuum
spectrum, see the next subsection) turns out to be exactly the same as
the tensor spectral distribution (\ref{438}), which is bounded.

The same result, on the other hand, can also be  directly obtained in the
longitudinal gauge simply by neglecting the growing mode of Eq. 
(\ref{436}), as noted in \cite{GasVe94a}. This suggests that it should be
possible to move to a more appropriate gauge, in which the growing
mode is suppressed, to restore  the validity of the linear
approximation. 

This is exactly what happens  in the
off-diagonal gauge (also called ``uniform-curvature" gauge
\cite{Hwang91}), defined by  $\psi=0=E$,  
\beq
ds^2= 
a^2\left[\left(1+2\varphi\right) d\eta^2 -
dx_i^2- 2\pa_iB dx^i d\eta \right],  
\label{441}
\eeq
which represents another complete choice of coordinates, with no
residual degrees of freedom (just as the longitudinal gauge).
In this gauge there are two variables for scalar perturbations,
$\varphi$ and $B$, 
and the perturbation equations are solved, asymptotically, by
\beq
\varphi_k = c_1+c_2 \ln |k\eta| , ~~~~~~~ 
B_k = c_1 \eta \ln |k\eta| + {c_2\over \eta}.
\label{442}
\eeq
The growing mode is thus completely absent for homogeneous 
perturbations (since it is $\pa_i B$ that contributes to the perturbed
metric). The growing mode is still present in the off-diagonal part of the
metric for non-homogeneous perturbations,  but it is suppressed with
respect to the longitudinal gauge, since $\da^{(1)}g \sim$ $kB_k \sim
k\eta \psi_k \sim \eta^{-1}$, instead of $\psi_k \sim \eta^{-2}$. The 
lower is $k$ the stronger is the suppression, and this is enough for the
 linear approximation to be valid, as explicitly checked with a
computation up to second-order  corrections \cite{BruMu95}. This
off-diagonal gauge is in general useful for the study of scalar
perturbations also in the context of generalized scalar-tensor models
of gravity and of the higher-loop string effective action, as discussed
for instance in \cite{HwaNo00}. 

Even better, one may choose a gauge in which $\psi=0=B$. In this
gauge,  $E$ is related to the Bardeen potential $\Psi$ by $E \sim \eta^2
\Psi$, according to Eq. (\ref{429}). On the other hand, the gauge-invariant
Bardeen potential evolves asymptotically (in all gauges) like the
longitudinal variable $\psi$, i.e. like $\eta^{-2}$. Since $E$ enters the
metric perturbations with two spatial derivatives, it follows that, in this
gauge, $\da^{(1)}g_{ij} \sim (k\eta)^2 \Psi_k$, which is asymptotically
frozen, and sufficiently small (at small $|k\eta|$) for the linear
approximation to be valid. 

To conclude this subsection, let us stress that the asymptotic growth of
perturbations is a typical problem of string cosmology, with no
counterpart in the standard inflationary scenario. Consider, for instance, 
a $d=3$, conformally flat and accelerated background, which can be
conveniently parametrized in the negative range of the conformal time
coordinate by a generic power-law scale factor,  as follows:
\beq
a= (-\eta)^\a, ~~~~~~~~~~~~  \eta <0 .
\label{443}
\eeq
For $\a$ ranging over the whole real axis we can represent in this way
all classes of inflationary backgrounds introduced in Section
\ref{Sec1.3}, and reported in Table 1. Consider then the E-frame
perturbation equation (\ref{421}), which in this case reduces to 
the Bessel equation
\beq
h_k^\se +{2 \a\over \eta}h_k'+k^2h_k=0,
\label{444}
\eeq
with asymptotic solution (for $|k\eta| \ra 0$):
\beq
h_k =A+B\int^\eta {d\eta'\over a^2(\eta')} = A
+B\left|\eta\right|^{1-2\a}.
\label{445}
\eeq
The amplitude tends to stay constant for $\a<1/2$, while it tends to
grow for $\a>1/2$ (for $\a=1/2$ the growth is simply logarithmic). In
the E-frame the growing mode thus appears only for a phase of 
accelerated contraction (see Table 1), and it is  absent in the
standard inflationary context where the background is always
expanding (even in the E-frame).  

\begin{table}
\tabcolsep .4cm
\renewcommand{\arraystretch}{2.0}
\begin{center}
\begin{tabular}{|c||c||c|}
\hline
$\a=-1$ & \rm de~ Sitter & $\dot a >0, \ddot a >0, \dot H=0$ \\
\hline
$\a<-1$ & \rm power-inflation & $\dot a >0, \ddot a >0, \dot H<0$\\
\hline
$-1<\a<0$ & \rm superinflation & $\dot a >0, \ddot a >0, \dot H >0$ \\
\hline
$\a>0$ & \rm accelerated~contraction & $\dot a <0, \ddot a <0, \dot H<0$
\\ \hline 
\end{tabular}
\bigskip
\caption{Four classes of accelerated backgrounds,
parametrized by $a=(-\eta)^\a$.}
\end{center}
\end{table}

In the pre-big bang scenario, on the contrary, the comoving amplitude
of metric fluctuations may be growing, in the standard gauges, even in
the case of tensor perturbations. In some cases the validity of the
linear approximation may be restored in an appropriate gauge, as for
the scalar case discussed in this subsection. If this is impossible, the
growth of perturbations is a physical effect, signalling a (quantum)
instability of the given background \cite{KSS98,KS99}. 
In any case, however, the energy
spectrum of the perturbations can always be consistently estimated  by
truncating the comoving amplitude to the frozen part of the
asymptotic solution \cite{BruGas98}. This is a consequence of a duality
property of the perturbation equations that will be discussed in
Subsection \ref{Sec4.5}.

\subsection{Canonical variables and normalization}
\label{Sec4.3}

The linearized equations for the classical evolution
of the perturbations have been obtained, in the previous
subsection,   by perturbing directly (to first order) the equations of
motion of the background fields. 
The same equations, however,  can also be obtained with a different
approach,  by expanding the action up to terms quadratic in the
first-order fluctuations: 
\beq
g \ra g +\da^{(1)} g, ~~~~~~~~~~~~
\da^{(2)} S \equiv S\left[ \left( \da^{(1)} g\right)^2\right] .
\label {446}
\eeq
The variation of the perturbed action with respect to $\da^{(1)} g$ gives
in fact a set of  linearized equations, identical to Eqs.  (\ref{43}). 

This second
method is longer, in general, but has an advantage:  the diagonalization
of the quadratic perturbed action $\da^{(2)} S$ defines the
``normal modes" of oscillation for the total system formed by the metric
plus the matter sources. The so-called normal modes are the variables
that diagonalize the kinetic part of the action and that, once quantized,
satisfy canonical commutation relations \cite{MFB92}. They are
required to normalize the initial amplitude of the
perturbations to a quantum spectrum of zero-point
(vacuum) fluctuations, and  to study their amplification in the course of
the cosmological evolution.  

Let us  illustrate this point starting with the case of tensor metric
perturbations. The second-order action for tensor perturbations in a
cosmological background was first written in \cite{Ford77}, in the
E-frame and in $d=3$ dimensions (see also \cite{MFB92,Al94}).
Such a result can be easily generalized to the S-frame \cite{Gas97}
and to the ($d+1$)-dimensional, Bianchi-I-type metric background used in
the previous subsections, starting from the unperturbed, low-energy
gravidilaton string effective action (in units $2 \la_{\rm s}^{d-1}=1$): 
\beq
S=-\int d^{d+n+1}x \sqrt{-g} e^{-\phi} \left[ R+ \left(\pa_\mu
\phi\right)^2\right], 
\label{447}
\eeq 
and considering  the transverse, trace-free variable $h_{\mu\nu}$,
defined by the first-order perturbation (\ref{410}). As already discussed, 
the fluctuations of the $d$-dimensional external space are conveniently
described in the synchronous gauge of Eq. (\ref{411}). In this gauge, 
we expand to order $h^2$ the controvariant components of the metric,
\beq
\da^{(1)} g^{\mu\nu}=-h^{\mu\nu}, ~~~~~~~~~~ \da^{(2)}
g^{\mu\nu}=h^{\mu\a}h_\a\,^\nu , 
\label{448}
\eeq
the volume density, 
\beq
\da^{(1)} \sqrt{-g}=0, ~~~~~~ \da^{(2)} \sqrt{-g}=-{1\over
4}\sqrt{-g}h_{\mu\nu}h^{\mu\nu},  
\label{449}
\eeq
and so on for $\da^{(1)}R_{\mu\nu}$,  $\da^{(2)}R_{\mu\nu}$ (see
\cite{Gas97} for the details of the explicit computation). The
second-order  perturbed  action has then contributions from $\sqrt{-g}$,
$R$ and $g^{\mu\nu}$: 
\beq
\da^{(2)} S=
-\int d^{d+n+1}x e^{-\phi} \Bigg[
\da^{(2)}\left(\sqrt{-g} R\right)+
\da^{(2)}\left(\sqrt{-g}g^{\mu\nu}
\pa_\mu\phi\pa_\nu\phi\right)\Bigg].
\label{450}
\eeq
Integrating by parts, 
using the unperturbed equations of motion, and neglecting total
derivative terms, we finally arrive at the quadratic action:
\beq
\da^{(2)} S=
{1\over 4}\int d^{d+n+1}xa^d b^n e^{-\phi} \Bigg(\dot h_i^j\dot h_j^i + 
h_i^j{\nabla\over a^2} h_j^i \Bigg).
\label{451}
\eeq
Decomposing the fluctuations of the $d$-dimensional spatial metric into
the two physical polarization modes  $h_+$, $h_\times$, 
\beq
 h_i^j h_j^i = 2 \left( h_+^2 + h_\times^2\right),
\label{452}
\eeq
and introducing the conformal time coordinate $d\eta= adt$, 
we  eventually obtain,for each mode, the effective scalar action
\beq
\da^{(2)} S_h=
{1\over 2}\int d\eta a^{d-1}b^n e^{-\phi} \Bigg( h^{\prime 2}+ 
h{\nabla}^2 h \Bigg),
\label{453}
\eeq
where $h$ is now a scalar variable standing for either  of the 
polarization amplitudes $h_+$, $h_\times$.  The variation with respect to
$h$  now gives exactly Eqs. (\ref{417}), i.e. the S-frame tensor
perturbation equation (the sameas was obtained  by direct perturbation
of  the equations of motion). 

The above action describes a scalar field non-minimally coupled to a
time-dependent external field (also called ``pump field"), represented
in this case by the dilaton and by the internal and external scale factors
(the so-called ``moduli"). In order to impose the correct normalization
to a quantum spectrum of vacuum fluctuations  we need, however, the
canonical variable  that diagonalizes the kinetic part of the action and 
describes asymptotically a freely oscillating field
\cite{Deruelle92,Gris92a,MFB92}. In our case such a variable $u$ is
defined by \cite{GG93} 
\beq  u= z h , ~~~~~~~~~~~~~~~
z =a^{(d-1)/ 2}b^{n/ 2}  e^{-\phi/2}. 
\label{454}
\eeq
In fact, for each Fourier mode $u_k$,  we get   
from Eq. (\ref{453}), after integration by parts,  an effective action
in which the kinetic term appears in the standard canonical form: 
\beq
\da^{(2)} S_u(k)=
{1\over 2}\int d\eta  \Bigg(|u_k^{\prime}|^2- 
k^2|u_k|^2+ {z^\se\over z}|u_k|^2 \Bigg),
\label{455}
\eeq
and which leads to the canonical evolution equation: 
\beq
u_k^\se +\left[k^2-V(\eta)\right]u_k =0, ~~~~~~~~~~~
V(\eta)={z^\se\over z},
\label{456}
\eeq
with an effective potential $V(\eta)$ depending only on the external
pump field. An accelerated background, in particular, has an effective
potential that goes to zero as $\eta \ra -\infty$. We thus obtain,  
asymptotically, the free-field equation
\beq
u_k^\se +k^2u_k =0, ~~~~~~~~~~~
\eta \ra -\infty,
\label{457}
\eeq
and the variable $u_k$ can be normalized to an initial-vacuum
fluctuation spectrum,  
\beq
u_k={1\over \sqrt{2k}} e^{- ik\eta}, ~~~~~~~~~~~
\eta \ra -\infty,
\label{458}
\eeq
so as to satisfy canonical commutation relations  
$[u_{k_1}, u^{*'}_{k_2}]=i\da_{k_1k_2}$. The normalization of the
canonical variable  $u_k$ automatically fixes the normalization of the
metric  fluctuations,  $h_k=u_k/z$. 

It is important to stress  that the  procedure used to introduce  the
diagonalized  action (\ref{455}) can be exactly repeated in the E-frame.
The only difference in this new frame is that there is no dilaton coupling
$e^{-\phi}$ in front of the Lagrangian density, and the pump field is
simply determined by the E-frame scale factors, $\ti z= \ti\psi/\ti h$ 
$= \ti a^{(d-1)/ 2} \ti b^{n/ 2}$. 
But, according to the transformation rule
(\ref{422}), $\ti z =z$. The effective potential $z^\se/z$ is thus the
same in the two frames, the time evolution of 
$\psi$ and $\ti \psi$ is the same and, as a
consequence, we find the same spectrum in both frames,  as
anticipated in Subsection \ref{Sec4.1}. 

In the case of scalar metric perturbations, the computations are more
complicated, in practice, but the procedure is exactly the same in
principle. The scalar canonical variable has been computed for a $d=3$
isotropic background coupled to various types of sources: a perfect fluid 
\cite{Luk80,Chib82},  one scalar field \cite{Sas86,Mu88,Stew93}, two
scalar fields \cite{Deruelle92a}, $N$ scalar fields \cite{Ander94}.
Working in the E-frame (and omitting the ``tilde", for simplicity) the
scalar perturbation of the gravidilaton action (\ref{48}) can be easily
performed in the longitudinal gauge, where $\da^{(1)} \phi=\chi$, 
$\da^{(1)}g_{00}=2 a^2 \varphi$, $\da^{(1)}g_{ij}=2a^2\psi \da_{ij}$. One
then obtains the diagonalized second-order action \cite{MFB92}: 
\beq
\da^{(2)} S_v(k)=
{1\over 2}\int d\eta  \Bigg(|v_k^{\prime}|^2- 
k^2|v_k|^2+ {z^\se\over z}|v_k|^2 \Bigg),
\label{459}
\eeq
where the canonical variable $v$ is defined by
\beq
v= a \chi + z \psi, ~~~~~~~~~~~~~
z = {a^2 \phi'\over a'},
\label{460}
\eeq
and is gauge-invariant, as can be explicitly checked by  using the
infinitesimal transformations (\ref{428}). 

This result can be easily generalized to the scalar fluctuations of the
higher-dimensional, anisotropic background considered in the previous
subsections, and parametrized as in Eq. (\ref{425}). After the
diagonalization, the higher-dimensional action for the Fourier
modes of scalar perturbations can be written as \cite{GG97}
\beq
\delta^{(2)}S_{v,w}(k)= \frac{1}{2}\int d\eta\Bigg(  
|v_k^{\prime}|^2- 
k^2|v_k|^2+ {z^\se\over z}|v_k|^2+
|w_k^{\prime}|^2- 
k^2|w_k|^2+ {z^\se\over z}|w_k|^2\Bigg)
\eeq
(modulo a total derivative that does not contribute to the equations of
motion). There are two gauge-invariant canonical variables, $v$ and $w$,
which in the longitudinal gauge ($E=0=B$) are defined by 
\bea
&&
v = a^{(d-1)/2}b^{n/2}\chi +z\left(\psi +{n\over d-1}
\xi\right) \nonumber\\
&&
w = z\left[\frac{n(n+d-1)}{(d-1)}\right]^{1/2} \left(
\frac{{a'}}{a\phi'} \xi - \frac{{b'}}{b\phi'}\psi\right),
\eea
with the generalized pump field
\begin{equation}
z=\frac{a^{(d-1)/2}b^{n/2}\phi' }{{a'\over a}+\frac{nb'}{(d-1)b}}.
\end{equation}

The definition of the canonical variable and the correct normalization of
the perturbations are necessary ingredients to study the growth of
the quantum fluctuations, and to determine whether they backreaction
may become eventually so large as to destroy the initial
homogeneity of a pre-big bang solution (see also Section
\ref{Sec3}). It is the canonical normalization, in particular, that leads to
the standard relation between the amplitude of quantum fluctuations and
the curvature scale of the background geometry. 

In order to illustate this important point let us  consider tensor
perturbations in a $d=3$, isotropic, E-frame metric, 
evolving in time aymptotically according to Eq. (\ref{445}). The classical
perturbation equation (\ref{444}) only determines the time dependence
of the fluctuations. The canonical action, however,  determines not only
the $k$-dependence through the normalization of $u_k$, but
also the correct canonical dimensions of the fluctuations, $[u]= M$. By
inserting in fact the required dimensional factors in the (E-frame)
perturbed action,
\beq
\da^{(2)} S_h=
{1\over 32 \pi G}\int d\eta a^2 \Bigg( h^{\prime 2}+ 
h{\nabla}^2 h \Bigg),
\label{464}
\eeq
we obtain the scalar field action (\ref{455}) with canonical dimensions
by setting, for each Fourier mode, 
\beq
h_k(\eta) ={\sqrt 2 \over M_{\rm P}} {u_k(\eta) \over a(\eta) }
\label{465}
\eeq
(recall that $8 \pi G = M_{\rm P}^{-2}$ with our conventions).

The typical fluctuation amplitude over a comoving length scale $r$, on
the other hand, is determined by the two-point correlation function
\beq
\xi(\vec r)= \langle h(\vec x) h(\vec x+\vec r) \rangle,
\eeq
where the brackets denote either a quantum expectation value, if we
work in the formalism in which the perturbations are quantized and
expanded into annihilation and creation operators (see Subsection
\ref{Sec4.4}), or a macroscopic ensemble averaged over a spatial
volume, if  we work in the classical limit. In any case we consider real
perturbations ($h_k = h_{-k}^*$), satisfying the isotropy condition (i.e.
$|h_k|$ is a function of $k=|\vec k|$ only) and the stochastic
condition  
\beq
\langle h(\vec k) h(-\vec k')\rangle={(2\pi)^3} \da^3(k-k')  
|h( k)|^2  
\label{467}
\eeq
(which is a consequence of their quantum origin). An
explicit computation then gives \cite{MFB92,Gas98b}
\beq
\xi(r) \sim \int {dk\over k} {\sin kr\over kr} |\da_h(k)|^2,
\eeq
where $|\da_h(k)|$ represents the typical, dimensionless amplitude of
tensor fluctuations over a comoving length scale $r= k^{-1}$:
\beq
|\da_h(k)|\equiv k^{3/2} |h_k| \sim \left[\xi^{1/2}(r)\right]_{r=k^{-1}}.
\label{469}
\eeq

The amplitude of the quantum fluctuations of the metric is now fixed by
Eq. (\ref{465}) and by the normalization of the canonical variable,
Eq. (\ref{458}). Initially, for $\eta \ra -\infty$, all fluctuations  
oscillate inside the horizon, and 
\beq
|\da_h(k)| = {k\over aM_{\rm P}} ={\om \over M_{\rm P}},
\eeq
i.e. the amplitude of a mode is proportional to its proper frequency
$\om= k/a$ (and then  it is adiabatically decreasing, in an
expanding background). In the opposite regime $\eta \ra
0_-$, where the fluctuations are well outside the horizon of an
inflationary background with $a \sim |\eta|^\a$, we have to use the
asymptotic solution (\ref{445}), and  we may distinguish three
possibilities. 

\begin{itemize}
\item{}
If $\a <1/2$, then $h$ is frozen, while the canonical variable grows
adiabatically, following the growth of the pump field $z =a$ (see Eq. 
(\ref{445})): $u_k \sim a (u_k/a)_{\rm hc}$, where ``hc"  denotes the
time of horizon crossing, $|k\eta|\simeq 1$). The final amplitude can thus
be expressed in terms of the background curvature scale at the crossing
time $|\eta| \simeq 1/k$:
\beq
|\da_h(k)| \simeq {k\over M_{\rm P} a_{\rm hc}}  
\simeq {1\over M_{\rm P} (a\eta)_{\rm hc}}\simeq \left(H\over
M_{\rm P}\right)_{\rm hc} =   
\left(H_1\over M_{\rm P}\right)|k\eta_1|^{1+\a},
\label{471}
\eeq
where $H_1$ is some reference scale (a convenient choice is the final
curvature scale at the end of the inflationary phase), and $\a$ is the
conformal power of the scale factor. The spectral amplitude $\da_h$ is
frozen in time, and it is then clear from the above equation why pre-big
bang models, characterized by a growing (in time) curvature scale
($\a+1>0$, see Table 1), are associated to a growing (in frequency) tensor 
perturbation spectrum. 

\item{} If  $\a >1/2$, then $h$ is growing according to
Eq. (\ref{445}) and, as a  consequence, the spectral amplitude $\da_h$
grows not only in frequency but also in time:
\beq
|\da_h(k)| \simeq \left(H\over M_{\rm P}\right)_{\rm hc}|k\eta|^{1-2\a}.
\eeq

\item{}
The low-energy, dilaton-driven pre-big bang solution of Eq. (\ref{439})
corresponds to the limiting case $\a=1/2$. In that case the amplitude of
tensor perturbations only acquires a mild, logarithmic dependence
\cite{BruMu95,BruGas95}:
\beq
|\da_h(k)| \simeq \left(H\over M_{\rm P}\right)_{\rm hc}|\ln |k\eta|| = 
\left(H_1\over M_{\rm P}\right)
|k\eta_1|^{3/2}|\ln |k\eta||,
\eeq
which is under control for  inflation scales smaller than Planckian,
$H_1<M_{\rm P}$. 
\end{itemize}

It is important to stress that the behaviour of scalar fluctuations, in the
background with $\a=1/2$, is apparently very different from the tensor
one, as already discussed in the previous subsection. By eliminating the
dilaton fluctuation $\chi$, and using the scalar  perturbation equations,
one finds indeed that the scalar longitudinal variable $\psi$ is related to
the canonical variable $v$ of Eq. (\ref{460}) by \cite{MFB92}:
\beq
\psi_k= -{\phi'\over 4M_{\rm P} k^2}\left( v_k\over a \right)'.
\eeq
The normalized amplitude of scalar quantum fluctuations,
\bea
&&
|\da_\psi(k)| \equiv k^{3/2}|\psi_k|\simeq 
\left(H\over M_{\rm P}\right)_{\rm hc}|k\eta|^{-2} = 
\left(H_1\over M_{\rm P}\right)
|k\eta_1|^{3/2}||k\eta|^{-2} \nonumber\\
&&
= \left(H_1\over M_{\rm P}\right)
\left(\eta_1\over \eta\right)^2 |k\eta_1|^{-1/2}, 
\label{475}
\eea
is thus unbounded since, for any $|\eta|>|\eta_1|$, we can always find a
frequency band $|k\eta_1|\ll1$ for which $|\da_\psi|$ is arbitrarily
large.

By using the correct quantum normalization, however, it is now easy to
check that this difficulty disappears in the off-diagonal gauge
\cite{BruMu95}, where the off-diagonal perturbation variable $B$
satisfies $B_k \sim k\eta \psi_k$ and the amplitude of the 
corresponding quantum fluctuations is controlled by
\beq
|\da_B(k)|  \sim |k\eta||\da_\psi(k)| \sim 
\left(H_1\over M_{\rm P}\right)
\left|\eta_1\over \eta\right| |k\eta_1|^{1/2}.
\eeq
For the typical inflation scale of string cosmology, $H_1 \simeq 
M_{\rm s}<M_{\rm P}$, this amplitude stays smaller than $1$ for the
whole duration of the pre-big bang phase, $|\eta|>|\eta_1|$, and for all
modes with $k<k_1= |1/\eta_1|$ (higher-frequency modes are not
amplified, see the next subsection). It should be mentioned, however,
that this conclusion does not apply to {\em classical} fluctuations, whose
initial amplitude is not constrained by the quantum normalization
(\ref{458}), and can be fixed to much higher values (see a recent
discussion in \cite{BuoDa01}). 

To conclude this subsection we note that the above procedure used to
define the canonical variables can also be applied to more
complicated models, and in particular to the actions 
including the  higher-derivative corrections typical of the ``stringy", 
high-curvature regime \cite{MeTsey,Cal88}, appearing towards the end
of the phase of pre-big bang evolution. To first-order in $\ap$, and in the 
S-frame, such an action can be parametrized as \cite{GasMaVe97} 
\beq
S=\int {d^{4}x\over 2 \la_{\rm s}^2} \sqrt{-g}e^{-\phi} 
\left\{- R-
\pa_\mu \phi\pa^\mu \phi+{\ap \over 4} \left[R^2_{GB} - 
(\pa_\mu\phi\pa^\mu\phi)^2\right]\right\}, 
\label{477}
\eeq
where  $R^2_{GB} 
\equiv R_{\mu\nu\a\b}^2-4  R_{\mu\nu}^2+
R^2$ is the Gauss--Bonnet invariant (see Section \ref{Sec8.2} for further
details). Starting with Eq. (\ref{477}), the second-order perturbed action
for each polarization mode of tensor fluctuations becomes \cite{Gas97},
in conformal time,  
\beq 
\da^{(2)} S_h= {1\over 4\la_{\rm s}^2} 
\int d^3x d\eta \left[z^2(\eta)h^{\prime 2}+ y^2(\eta)h\nabla^2
h\right],
\label{478}
\eeq
where
\bea
&&
z^2(\eta)=e^{-\phi}\left(a^2-\ap {a'\over a}\phi'\right),
\nonumber\\
&&
y^2(\eta)=e^{-\phi}\left[a^2+\ap 
\left(\phi^{\prime 2}-\phi^{''}+
 {a'\over a}\phi'\right)\right]
\label{479}
\eea
(the inequality of the two pumping fields $z$ and $y$, which prevents
factorization, is a consequence of the breaking of the scale-factor
duality symmetry, to this order in $\ap$; see \cite{Meis97} for an
action with $\ap$ corrections, which preserves scale-factor duality).  By
setting $u=zh/\sqrt 2 \la_{\rm s}$ we get the action in canonical form,  
\beq
\da^{(2)} S_h= {1\over 2} 
\int d\eta \left(u^{\prime 2}+ {z^{''}\over z} u^2+
{y^2\over z^2}u\nabla^2 u \right),  
\label{480}
\eeq
and the corresponding perturbation equation
\beq
u_k^{\prime\prime}+\left[k^2-V_k(\eta)\right]u_k=0,~~~~~~
V_k(\eta)={z^{''}\over z}-{k^2\over z^2}(y^2-z^2), 
\label{481}
\eeq 
looks the same  as Eq. (\ref{456}), 
with the only difference that the effective potential  is now, in general,
$k$-dependent. The above equation takes into account (to first order)
higher-curvatures effects on the perturbation spectrum \cite{Gas97},
and  provides the starting point for studying the evolution of tensor
perturbations in the high-curvature regime typical of the pre-big bang
scenario \cite{CaCoGas}. A similar approach has been developed for
the study of scalar and vector fluctuations including $\ap$
corrections \cite{CaHwang01}. 

\subsection{Spectral distribution of the energy density} 
\label{Sec4.4}

For all types of perturbations the evolution in time of the 
canonical modes $u_k$ is described by the  Schr\"odinger-like
equation (\ref{456}), with an effective potential that depends on the
pump field, and thus on the  peculiar kinematics of the given 
background.  For all inflationary backgrounds, however, such a potential
tends to vanish, asymptotically, at large positive and negative values of
the conformal time. 

Consider, for instance, the evolution  of tensor 
perturbations in the E-frame: the pump field is the scale factor,
$z=a$, and $V=a^\se/a$. For a typical inflationary scenario the 
background is initially accelerated,  parametrized
by a power-law scale factor,  
$a \sim |\eta|^\a$, so that $|V|\sim \eta^{-2}$ for $\eta \ra -\infty$; at
large times the background eventually evolves into the
standard radiation-dominated era, $a \sim \eta$, so that $V=0$ for $\eta
\ra +\infty$. The analysis of perturbations thus reduces to a problem of
scattering of the ``wave function" $u_k$, induced by a  
 potential $V(\eta)$, which is asymptotically vanishing (see Fig.
\ref{f41}). 

The important difference between the present problem and 
an ordinary scattering problem, however, is that
the differential variable of Eq. (\ref{456}) is {\em time}, not space.
As a consequence, the oscillation frequencies represent {\em energies},
not momenta. With an initial normalization to a state of positive energy, 
\beq
u_{\rm in} \simeq {e^{-ik\eta}\over \sqrt 2},
~~~~~~~~~~~~~~~~~~~~~~~~
\eta \ra -\infty ,
\label{482}
\eeq
the final state is thus in general a superposition of positive and
negative energy modes,
\beq
u_{\rm out} \simeq c_+e^{-ik\eta} + c_-e^{+ik\eta}, 
~~~~~~~~~~~~~~~
\eta \ra +\infty .
\label{483}
\eeq
In the context of quantum field theory, it is well known that such
a mixing represents a process of pair creation from the vacuum
\cite{Bir82}. The so-called  Bogoliubov coefficients $c_\pm$ parametrize
the unitary  transformation connecting the set of $|{\rm
in}\rangle$ and $|{\rm out}\rangle$ annihilation and creation
operators:  
\bea
&&
\{u_{\rm in}, b_k, b_k^\dagger\} ~~ \Longrightarrow ~~
\{u_{\rm out}, a_k, a_k^\dagger\},\nonumber\\
&&
a_k=c_+b_k+c_-^\ast b_{-k}^\dagger, ~~~~~~
a_{-k}^\dagger=c_-b_k+c_+^\ast b_{-k}^\dagger ,
\label{484}
\eea
and define the 
expectation number of pairs produced from the vacuum in the mode
$k$:
\beq
\overline n_{\rm in}(k)=\langle0|b_k^\dagger b_k|0\rangle=0 , ~~~~~~~
\overline n_{\rm out}(k)=\langle0|a_k^\dagger a_k|0\rangle =|c_-(k)|^2
\not= 0 . 
\label{485}
\eeq

In a second quantization language, the amplification of perturbations
can thus be described as a process of particle production or, 
equivalently, as the  evolution of the initial vacuum into a
final ``squeezed" vacuum state \cite{Gris92,Gris92a,GriSid89,GriSid90}, 
\beq
|0\rangle  \Longrightarrow |S_k\rangle =\Sigma_k |0\rangle, 
\label{486}
\eeq
generated by the two-mode squeezing operator \cite{Schu86}
\beq
\Sigma_k =\exp \left(s_k^*b_kb_{-k} -s_k b_k^\dagger
b_{-k}^\dagger\right).
\label{487}
\eeq
Here $s_k=r_ke^{2i\theta_k}$ is the so-called squeezing parameter,
whose modulus and phase are related to the Bogoliubov coefficients
$c_{\pm}(k)$ by \cite{Schu86}
\beq
 c_+(k)= \cosh r_k,~~~~~~~~~~~~~
c_-(k)= e^{2i\theta_k} \sinh r_k . 
\label{488}
\eeq

Itshould be noted that the squeezed-state formalism is particularly
convenient for the analysis of the statistical properties of the produced
radiation,  and for the study of the entropy growth associated to particle
production. Representing the effective decoherence of the squeezed
density matrix through a suitable ``coarse-graining" procedure, which
accounts for the loss of information associated to (averaged)
macroscopic observations, one finds indeed that the entropy growth
associated to particle production, for each mode $k$, is given by 
$\Da S_k = 2 r_k$ \cite{GG93a,GG93b}. In the large squeezing limit,
summing over all modes, the total entropy inside a proper volume $V=(a
L)^3$ can then be estimated as
\beq
S \simeq V\int d \om \om^2 \ln |c_-(\om)|^2 =
V\int d \om \om^2 \ln \overline n (\om),
\label{488a}
\eeq
in agreement with different approaches to the same problem 
\cite{Bran92,Bran93a}, and with different definitions of
quantum (von Neumann) and informatic (Shannon--Wehrl) entropy 
\cite{GG98}. For a growing, power-law spectral distribution $n(\om)$,
typical of pre-big bang models, one obtains a constant entropy per
comoving volume, $S \sim (a \om_1)^3$, where $\om_1$ is the maximal
amplified frequency inside the given volume. This result has useful
applications to the problem of the graceful exit, as will be discussed in
Section \ref{Sec8.4}. 

\begin{figure}[t]
\centerline{\epsfig{file=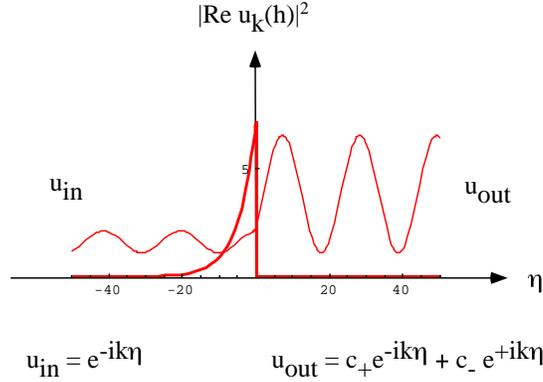,width=72mm}}
\vskip 5mm
\caption{\sl Parametric amplification of the  vacuum
fluctuations, illustrated by the qualitative evolution of the canonical
variable. The plot of the real part of $u_k$ (thin curve) has  been
obtained  through a numerical integration of the canonical
perturbation equation for a step-like effective potential (bold curve),
exponentially damped  in  the inflationary region.} 
\label{f41}
\end{figure}

In a semiclassical language,  the amplification of
perturbations can be represented instead as a parametric
amplification \cite{Gris75,Star79} of the wave function $u_k$ , i.e. as
an  ``antitunnelling" process \cite{Gas98b} in which the positive
frequency part of  the $|{\rm out}\rangle$ mode (\ref{483}) 
plays the role of the wave incident (from the
right) on the barrier and its negative part the role of the reflected
wave, while the $|{\rm in}\rangle$ mode (\ref{482}) plays the role of the
transmitted wave  (see Fig. \ref{f41}). The Bogoliubov coefficient $c_-$,
which controls the number of produced particles, can then be written as
the ratio of the  reflection and
transmission coefficients, $|c_-|^2=R/T$. In the parametric
amplification regime,  characterized by a large number of produced
particles  ($\overline n_k \gg 1$), one finds $|c_-| \simeq |c_+|$, so that 
$R\simeq 1$, and then 
\beq
|c_-|^2\simeq  T^{-1}.
\label{489}
\eeq
Hence the term  antitunnelling.

Quite independently of the  adopted language, the energy density of
the produced particles, for each mode $k$ (summing over two
polarization states), is given by 
\beq
d \r_k = 2 k\overline n_k {d^3 k \over (2\pi)^3}, ~~~~~~~~~~
\overline n_k = |c_-(k)|^2.
\label{490}
\eeq
The computation of the spectral energy distribution per logarithmic
interval of frequency,
\beq
{d \r_k\over d \ln k}\equiv k {d \r_k\over d  k}=
{k^4\over \pi^2}|c_-(k)|^2, 
\label{491}
\eeq
thus requires the computation of the Bogoliubov coefficient 
$c_-(k)$. By recalling Eqs. (\ref{482}), (\ref{483}), this amounts to
solving the perturbation equation in the limit of large positive times,
with the  initial normalization imposed at $\eta= -\infty$. 

We will consider, as a
simple example, a model of background characterized by two
phases: an initial accelerated evolution up to the time $\eta_1$, and a
subsequent radiation-dominated evolution for $\eta >\eta_1$:
\bea
&&
a \sim (-\eta)^\a,  ~~~~~~~~~~~~~~~~\eta<\eta_1, \nonumber\\
&&
a \sim \eta,  ~~~~~~~~~~~~~~~~~~~~~~~\eta>\eta_1.
\label{492}
\eea
In the first phase the canonical equation (\ref{456}) for tensor
perturbations reduces to a Bessel equation,
\beq
u_k^\se +\left[k^2-{\a(\a-1)\over \eta^2}\right]u_k=0,
\label{493}
\eeq
with general solution \cite{AbSte}
\beq
u_k= |\eta|^{1/2}
\left[AH_\nu^{(2)}(|k\eta|)+BH_\nu^{(1)}(|k\eta|)\right], ~~~~~~
\nu=|\a-1/2|, 
\label{494}
\eeq
where $H_\nu^{(1,2)}$ are the first- and second-kind Hankel functions,
of index $\nu=|\a-1/2|$ determined by the kinematics of
the background (in this case, by the time evolution of the scale factor;
more generally, however, $\a$ parametrizes the evolution of the pump
field in conformal time). 
By using the large argument limit \cite{AbSte}  for $\eta \ra
-\infty$,
\beq
H_\nu^{(2)}(k\eta)\sim {\sqrt{ 2\over \pi |k\eta|}}e^{-ik\eta},
~~~~~~~~~~~~ H_\nu^{(1)}(k\eta)\sim {\sqrt{ 2\over \pi |k\eta|}}
e^{+ik\eta}, \label{495}
\eeq
we normalize the solution to a vacuum fluctuation spectrum,
\beq
A=\sqrt{\pi / 4}, ~~~~~~~~~~~B=0.
\label{496}
\eeq
In the second phase $V=0$, and we have the  simple oscillating
solution, 
\beq
u_k= {1\over \sqrt{2k}}\left(c_+ e^{-ik\eta}+c_- e^{+ik\eta}\right).
\label{497}
\eeq

The coefficients $c_\pm$ are now determined by the continuity of
$u$ and $u'$ at $\eta=\eta_1$. More precisely, assuming a non-singular
and continuous background,  the matching would  require the continuity
of the perturbed metric projected on a space-like hypersurface
containing $\eta_1$, and the continuity of the extrinsic curvature of that
hypersurface \cite{HwaVi91,DeMu95};  but in many cases these
conditions are equivalent to the continuity of the canonical variable $u$,
and of its first time derivative (see however \cite{BMUV98a}). Such a
matching prescription might require modifications, however, in the case
of bouncing backgrounds \cite{PePi02} evolving from contraction to
expansion). 

When matching the solutions (\ref{494}), (\ref{497}), it is convenient
to distinguish the two regimes in which the comoving  frequency $k$
is much higher or much lower than the frequency associated to the
top of the effective potential
barrier, $|V(\eta_1)|^{1/2} \simeq |\eta_1|^{-1}$. 
In the first case, $k\gg {1/|\eta_1|}\equiv k_1$, we can use the
large-argument limit of the Hankel functions to find that there is no
significant particle production, i.e. 
 \beq
|c_+|\simeq 1, ~~~~~~~~~~
 |c_-|\simeq 0 . 
\label{498}
\eeq
Actually, $c_-$ is not exactly zero, but  it is  exponentially
suppressed \cite{Al88,Gar89} as a function of frequency, just like the
quantum-mechanical reflection probability for a wave function with a
total energy well above the top of a potential barrier.  Such an effect
is,  however, negligible for the purpose of a first-approximation 
estimate in a  cosmological context, and in any case  it only
affects the shape of the very high-frequency tail of the spectrum. A 
sharp cut-off, in that region, is a good enough approximation. 

In the second case,  $k\ll {1/|\eta_1|}\equiv k_1$, we can use the
small-argument limit of the Hankel functions \cite{AbSte}, 
\beq
H_\nu^{(2)} \sim a (k\eta_1)^\nu -i b  (k\eta_1)^{-\nu}, ~~~~~~~~~~
H_\nu^{(1)} \sim a^* (k\eta_1)^\nu +i b  (k\eta_1)^{-\nu}, 
\label{499}
\eeq
where $a$ and $b$ are dimensionless numbers of order $1$, 
and we find
\beq
|c_+|\simeq 
 |c_-|\simeq |k\eta_1|^{-\nu-1/2},
\label{4100}
\eeq
corresponding to a spectral distribution:
\beq
{d \r_k\over d \ln k}=
{k^4\over \pi^2}|c_-(k)|^2\simeq {k_1^4\over \pi^2}
\left(k\over k_1\right)^{3-2\nu}, ~~~~~~~~~~~~k<k_1.
\label{4101}
\eeq

For  a comparison with present observations it is convenient to
use proper frequencies, $\om(t)=k/a(t)$, and to express the spectrum
in units of critical energy density, $\r_c(t)=3M_{\rm P}^2H^2(t)$. We
thus obtain the dimensionless spectral distribution
\beq
\Om(\om, t)={\om \over \r_c(t)}{d\r (\om)\over d\om} \simeq
{\om_1^4\over 3 \pi^2 M_{\rm P}^2H^2} \left(\om\over
\om_1\right)^{3-2\nu}, ~~~~~~~~\om<\om_1,  \label{402}
\eeq
where
\beq
\om_1={k_1\over a} \simeq {1\over a\eta_1}\simeq {H_1a_1\over a}
\label{4103}
\eeq
is the maximal amplified frequency (approximately, the frequency
``hitting" the top of the potential barrier $|V(\eta)|$). Such a distribution
can also be rewritten (modulo model-dependent numerical factors) 
in the final useful form 
\beq
\Om(\om, t)\simeq g_1^2 \Om_\ga(t)
\left(\om\over \om_1\right)^{3-2\nu}, 
\label{4104}
\eeq
where $g_1=H_1/M_{\rm P}$ is the curvature scale at the transition
epoch $t_1$ (a fundamental parameter of the given cosmological model),
and  
\beq
\Om_\ga(t)={\r_\ga\over \r_c}=
\left(H_1\over H\right)^2\left(a_1\over a\right)^4
\label{4105}
\eeq
is the energy density (in critical units) of the radiation which becomes
dominant at $t=t_1$, rescaled down to a generic time $t$. 

It is important to stress that the spectral slope $3-2\nu$  is directly
related to the kinematic behaviour of the background during the phase 
of accelerated evolution, as illustrated in Table 2. In particular, the
spectrum tends to follow the behaviour of the curvature scale. In the 
standard inflationary scenario, represented by a phase of de
Sitter-like evolution, $\Om$ does not depend on $\om$, and one
obtains the so-called scale-invariant Harrison--Zeldovich spectrum
\cite{Har70,Zel72}. For the low-energy, dilaton-driven phase
 typical of string cosmology (the solution (\ref{439})) one has  $\a=1/2$,
and the corresponding  spectrum  $\Om\sim \om^3$ simulates the 
low-energy (Rayleigh--Jeans) tail of a  thermal black-body distribution
(see Subsection \ref{Sec5.2}).  Note, however, that a flat, scale-invariant
spectrum with $\nu =3/2$ can also obtained for $\a=2$, corresponding
to a phase of contraction dominated by a cold ``dust" source with
effective equation of state $p=0$. This possibility, first pointed out in
\cite{Gas94a}, was recently discussed in \cite{FiBra01} for the
pre-big bang and ekpyrotic \cite{KOST1,KOSST} scenarios. 

\begin{table}
\tabcolsep .4cm
\renewcommand{\arraystretch}{2.0}
\begin{center}
\begin{tabular}{|c||c||c||c|}
\hline
{\bf Curvature scale}   & {\bf Metric} &{\bf  Bessel index} & 
{\bf Spectrum}  \\ 
$H^2$ & $a =|\eta|^\a$ & $\nu= |\a-1/2|$& $ \Om \sim
\om^{3-2\nu}$ \\ 
\hline
de Sitter, constant  &$\a=-1 $   &   $3-2\nu =0$   & flat \\ 
\hline
power-inflation, decreasing  
& $\a<-1  $ &  $3-2\nu<0$  & decreasing\\ 
\hline
pre-big bang inflation, growing  
& $\a>-1 $    &  $3-2\nu >0$  & increasing\\ 
\hline 
\end{tabular}
\bigskip
\caption{Accelerated backgrounds and the slope of the spectrum.}
\end{center}
\end{table}

It is also to be remarked that the above results for the spectrum have
been obtained by matching, as usual, the (perturbed) induced metric and
the extrinsic curvature across the hypersurface of constant total 
energy density, $\r + \da \r=$ const. In the case of scalar
metric perturbations,  this implies the continuity of the Bardeen potential
$\Phi$ and of the variable $\zeta=v/z$ (representing curvature
perturbations), even for bouncing backgrounds \cite{HwaNo01}.
Performing the matching across different hypersurfaces may lead, 
however, to a different spectrum of scalar perturbations (without
changing the tensor spectrum), especially in the case of transitions from
a collapsing to an expanding phase, and in the case of non-continuous
backgrounds, as shown for a special case in \cite{PePi02} and discussed
in general in \cite{DuVe02} (see Section \ref{Sec7} for possible
applications of this effect to the scalar perturbation spectrum of the
pre-big bang scenario). 

It should be noted, finally, that the above procedure for  an
approximate determination of the spectrum can be  applied to any
type of perturbation described by the canonical equation (\ref{456}).
Also, it can be easily generalized to
the case of $n$ transition scales $\eta_1, \eta_2$, ...$\eta_n$, for a
background characterized by $n$ phases, each of them approximated
by a power-law evolution  with parameters 
$\a_1, \a_2$, ... , $\a_n$. In each of these phases we have a general
solution with Bessel index $\nu_i=|\a_i-1/2|$, $i=1,2,...,n$.  By
matching $u$ and $u'$, using the large-argument limit of the
Hankel functions for  modes above the potential barrier, and the
small-argument limit for modes below the barrier, we obtain a 
spectrum with $n$ different frequency bands. In each band, the slope of
the spectrum is only determined by the background kinematics of the
two phases in which a given mode ``hits" the barrier and ``re-enters"
 the horizon, respectively.  The amplitude of the spectrum, on the
contrary, keeps track of the relative duration and kinematics of all
phases during which the mode stays ``under the barrier" \cite{Gas98b}. 

By expanding the coefficients of the exact Bessel solutions (determined
by the matching conditions) at each transition scale $|\eta_i|=k_i^{-1}$,
and including the next-to-leading terms when the leading ones are zero
\cite{BMUV98a}, we can also formulate a set of synthetic prescriptions
for a diagrammatic computation of the spectrum \cite{Tesi00}. The
method requires the drawing of a simple plot in which we insert a
vertical line corresponding to  each transition,  the height of the
line being proportional to the associated transition frequency, 
$k_i=|\eta_i|^{-1}$. It is thus possible to identify at a glance the
various frequency bands of the spectrum, and the knowledge of the
various Bessel indices $\nu_i$  turns out to be sufficient to write down
a quick estimate of the various amplitudes and spectral distributions
\cite{Tesi00}. Such a computation can also be made exact by including
the  full numerical coefficients in the asymptotic expansion of the Hankel
functions.

\subsection{Duality of the perturbation equations}
\label{Sec4.5}

In the previous subsections we have applied the usual Lagrangian
formalism to describe the evolution in time of the perturbations, and 
their cosmological amplification. The Hamiltonian formalism, however, is
more appropriate for the study of an interesting duality symmetry which
connects the amplitude of quantum fluctuations to their conjugate
momentum, and which may be useful for estimating their 
energy--density spectrum \cite{BruGas98} (see also \cite{Wands99}). 

We shall assume, as discussed in Subsection \ref{Sec4.3}, that the 
perturbations of the effective action  leads to a quadratic Lagrangian
density which, in the conformal-time gauge, can be parametrized as 
\beq
{\cal L} = {1\over 2} z^2 \left(\psi'^2+ \psi \nabla \psi\right),
\label{4106}
\eeq
where $z(\eta)$ is the pump field and $\psi$ the comoving amplitude of
the given fluctuations (see for instance Eq. (\ref{453})). This Lagrangian
may describe not only the (scalar and tensor) metric perturbations, but
also the quantum fluctuations of gauge fields and antisymmetric tensors
(see  Section \ref{Sec7}). By introducing the conjugate
momentum
\beq
\pi= {\da {\cal L}\over \da \psi'}= z^2 \psi',
\label{4107}
\eeq
we can define the corresponding Hamiltonian density, written 
in Fourier space as
\beq
{\cal H}= {1\over 2} \sum_k \left(z^{-2} |\pi_k|^2 + z^2k^2
|\psi_k|^2\right),
\label{4108}
\eeq
where $\psi_{-{k}} = \psi_{{k}}^*$ and $\pi_{-{k}} =
\pi_{{k}}^*$. The first-order perturbation equations then read, for each
mode, 
\beq
\psi_k'=z^{-2} \pi_{-k}, ~~~~~~~~~~~~~~
\pi_k'=-z^{2}k^2 \psi_{-k},
\label{4109}
\eeq
and lead to the decoupled equations
\begin{equation}
\psi_k''+2\frac{z'}{z} \psi_k'+k^2\psi_k = 0 \, , ~~~~~~~~~~~
\pi_k''-2\frac{z'}{z} \pi'_k+k^2\pi_k = 0 .
\label{4110}
\end{equation}

It is now easy to check that the transformation
\begin{equation}
\pi_{{k}} \rightarrow \widetilde \pi_{{k}}=k \psi_{{k}} \, ,
 ~~~~~ \psi_{{k}} \rightarrow \widetilde{\psi}_{{k}}=- k  
^{-1} \pi_{{k}}\, , ~~~~~
z\rightarrow \widetilde z=z^{-1}
\label{4111}
\end{equation}
leaves the Hamiltonian, Poisson brackets, and equations of motion
unchanged. It is a duality transformation that reduces to the usual
strong--weak coupling duality \cite{Sen94,Pol96,Schwarz97} in the
special case in which $z=e^{-\phi/2} \equiv g_{\rm s}^{-1}$ (as for the
perturbations of heterotic, four-dimensional gauge bosons
\cite{GGV95,GGV95a,Lem95}, see Section \ref{Sec7}), and to 
scale-factor duality  \cite{GV91,Tse91,Sen91} in the
special case in which $z=a$ (as for tensor metric perturbations). In
addition, in the case of electromagnetic perturbations, the contribution
of $k\psi_k$ to the Hamiltonian can be identified with that of the
magnetic field, whereas $\pi_k$ is proportional to the electric field:  
hence, in that case, the transformation (\ref{4111}) exactly represents
the well-known electric--magnetic duality transformation.

As a consequence of this duality property, the energy-density spectrum
of the quantum fluctuations can be consistently estimated by truncating
the solution (outside the horizon) to the frozen part of the fluctuations,
even in backgrounds where the comoving amplitude $\psi$ grows with 
time and the growing modes dominate the spectrum. Only in a
duality-invariant context is such a truncation
consistent, as the contribution to the Hamiltonian of
the growing mode of $\psi$ is simply replaced by the  contribution
of the frozen part of its duality-related conjugate momentum
\cite{BruGas98}.

In order to illustrate this important effect, it is convenient to introduce
the new variables
\beq
\widehat \psi = z \psi, ~~~~~~~~~~~~~
\widehat \pi = z^{-1} \pi,
\label{4112}
\eeq
which are canonically normalized and which transform as $\psi,\pi$
under the duality transformation (\ref{4111}). Our Hamiltonian density 
(\ref{4108}) can then be written, formally\footnote[1]{Equivalently, we
can introduce ${\widehat\psi}, {\widehat\pi}$ via a time-dependent
canonical transformation and work with a corrected Hamiltonian
${\widehat {\cal H}}$.}, in a
free canonical form,  \begin{equation}
{\cal H}=\frac{1}{2} \sum_{\vec{k}}
 \left( |{\widehat\pi}_k|^2 + k^2 |{\widehat\psi}_k|^2\right) ,
\label{4113}
\end{equation}
where the variables ${\widehat\psi}, {\widehat\pi}$ satisfy  the
Schr\"odinger-like equations of motion: 
\begin{equation}
{\widehat\psi}_k{''}+\left[k^2-(z){''}
z^{-1}\right]{\widehat\psi}_k
= 0 \, ,  ~~~~~~~
{\widehat\pi}_k{''}+ \left[k^2-(z^{-1}){''} z\right]
{\widehat\pi}_k = 0, 
\label{4114}
\end{equation}
characterized by  two duality-related ``effective potentials", 
$V_\psi(z)=z''/z$ and $V_\pi(z)=z(z^{-1})''= V_\psi (\ti z)$.  

The initial evolution of perturbations, for all modes with
$k^2> \{|V_{\Psi}|$, $|V_{\Pi}|\}$, can now be described by the
WKB-like approximate solutions of Eqs. (\ref{4114}): 
\begin{eqnarray}
&&
{\widehat\psi}_k(\eta)=\left( k^2-V_{\psi}\right)^{-1/4}\exp 
\left[ -i\int\limits_{\eta_0}^{\eta} d\eta' \left( k^2-V_{\psi}\right)^
{1/2} \right],  \nonumber \\
&&
{\widehat\pi}_k(\eta)=k \left( k^2-V_{\pi}\right)^{-1/4}\exp
\left[ -i\int\limits_{\eta_0}^{\eta} d\eta' \left( k^2-V_{\pi}\right)
^{1/2} \right],
\label{4115}
\end{eqnarray}
which we have normalized to zero-point vacuum fluctuations, and
where the extra factor of $k$ in the solution for ${\widehat\pi}_k$
is inserted for consistency with the first-order  Hamiltonian equations.
We  have ignored a possible relative phase in the solutions.  
Asymptotically, at $\eta \rightarrow -\infty$, we have $V_{\psi}$,
$V_{\pi} \rightarrow 0$, and the solutions (\ref{4115}) reduce to
the canonically normalized vacuum fluctuations 
\begin{equation}
{\widehat\psi}_k(\eta) =  {
e^{\ -i k\eta+ i\varphi_{{k}}} \over \sqrt k}\, , ~~~~~~~~~~
{\widehat\pi}_k(\eta) =  \sqrt k
e^{\ -i k\eta+ i\varphi_{{k}}'},
\label{4116}
\end{equation}
where $\varphi_{{k}}$, $\varphi_{{k}}'$ are random phases,
originating from the random initial conditions. The initial state thus
satisfies
\begin{equation}
\langle z^{-2} \pi^2\rangle= \langle z^2 (\nabla\psi)^2\rangle,
\label{4117}
\end{equation}
where $\langle\cdots\rangle$ denotes ensemble average, or
expectation value if perturbations are quantized. 
Note that, because of the random phases, the initial state
is duality-invariant  only on the average, in the sense of the above
equation.

In the opposite regime,  $k^2< \{|V_{\psi}|$, $|V_{\pi}|\}$,  it is possible to
write ``exact" solutions to
Eq. (\ref{4114}) as follows \cite{BruGas98},
\begin{eqnarray}
&&
{\widehat\psi}_k(\eta)=  z  \Biggl[\widehat
A_k\ {\rm T}\!\cos(z^{-1},z)+  \widehat B_k\
{\rm T}\!\sin(z^{-1},z)\Biggr],\nonumber \\
&&
{\widehat\pi}_k(\eta)= k
z^{-1}\Biggl[\widehat B_k\ {\rm T}\!\cos(z,z^{-1})-\widehat A_k\
{\rm T}\!\sin(z,z^{-1})  \Biggr],
 \label{4118}
\end{eqnarray}
where $\widehat A_k, \widehat B_k$ are arbitrary integration
constants, and the functions
${\rm T}\!\cos (k, \eta)$,  ${\rm T}\!\sin(k, \eta)$
satisfy
\begin{equation}
 \left[{\rm T}\!\cos(z^{-1},z)\right]'= -\frac{k}{z^2} \ {\rm T}\!\sin(z,z^{-1}) \, ,~~~~~
 \left[{\rm T}\!\sin(z^{-1},z)\right]'= \frac{k}{z^2} \  {\rm
T}\!\cos(z,z^{-1}). 
\label{4119}
\end{equation}
They are defined through the following ``ordered" expansions (see also
\cite{MFB92}): 
\begin{eqnarray}
&&
{\rm T}\!\cos(z^{-1},z; k, \eta)=
1- k \int\limits_{\eta_{\rm ex}}^\eta d\eta_1 z^{-2}(\eta_1)
\ k \int\limits_{\eta_{\rm ex}}^{\eta_1} d\eta_2 z^2 (\eta_2)+
 \nonumber \\
&&
+~k \int\limits_{\eta_{\rm ex}}^\eta d\eta_1 z^{-2}(\eta_1)
\ k \int\limits_{\eta_{\rm ex}}^{\eta_1} d\eta_2 z^2 (\eta_2)
\ k \int\limits_{\eta_{\rm ex}}^{\eta_2} d\eta_3 z^{-2} (\eta_3)
\ k \int\limits_{\eta_{\rm ex}}^{\eta_3} d\eta_4 z^2 (\eta_4) +\cdots \;
;\nonumber \\
&&
{\rm T}\!\sin(z^{-1},z; k, \eta)
=k \int\limits_{\eta_{\rm ex}}^\eta d\eta_1 z^{-2}(\eta_1)-
\nonumber\\
&&
-~ k \int\limits_{\eta_{\rm ex}}^\eta d\eta_1 z^{-2}(\eta_1)
\ k \int\limits_{\eta_{\rm ex}}^{\eta_1} d\eta_2 z^2 (\eta_2)
 \ k \int\limits_{\eta_{\rm ex}}^{\eta_2} d\eta_3 z^{-2} (\eta_3)
+\cdots , 
\label{4120}
\end{eqnarray}
where, for matching purposes, a convenient choice of the time 
$\eta_{\rm ex}$, appearing as an arbitrary lower limit of integration, is
the horizon-exit time, such that $|k\eta_{\rm ex}|\simeq 1$.

Suppose now that the background, during the accelerated period, can be
parametrized by a power-law evolution of the pump field,  i.e. $z \sim
|\eta|^\a$ for $\eta \ra 0_-$.  From the general solution
(\ref{4120}) we obtain, in the limit $|k\eta| \ll 1$,
\beq
{\widehat\psi}_k= \widehat
A_k|\eta|^\alpha+ {k|\eta|\over 1-2\alpha } \widehat B_k
|\eta|^{-\alpha} , 
~~~~~~~
{\widehat\pi}_k= k
\left(\widehat B_k|\eta|^{-\alpha}-
{k|\eta|\over 1+ 2\alpha }\widehat A_k|\eta|^{\alpha}
\right),
\label{4121}
\eeq
with logarithmic corrections for $\alpha = \pm 1/2$. The
corresponding expansion for the  comoving amplitudes
$\psi_k$, $\pi_k$, is then
\beq
\psi_k = A_k + B_k |\eta|^{1-2\a}, 
~~~~~~~~~~~
\pi_k= k\left( B_k - A_k |\eta|^{1+2\a}\right).
\label{4122}
\eeq
They may contain a growing mode, which leads in general to a growing
Hamiltonian, and then to a growing energy--density spectrum. The
Hamiltonian, however, is always dominated by a joint contribution
of the {\em frozen} modes of $\psi_k$ and $\pi_k$. We may consider, in
fact, three possibilities. 
\begin{itemize}
\item{}$\a>1/2$. The growing mode dominates $\psi_k$, while $\pi_k$
is frozen, as the constant mode is dominant (see Eq. (\ref{4122})).
Asymptotically, however, the Hamiltonian (\ref{4113}) is
momentum-dominated,  $\widehat \pi^2 \gg k^2 \widehat \psi^2$,
according to Eq. (\ref{4121}). 

\item{}$\a<-1/2$. The  comoving amplitude $\psi_k$ is frozen, while the
growing mode dominates  $\pi_k$ (see Eq. (\ref{4122})). Asymptotically,
however, the Hamiltonian (\ref{4113}) is dominated by $\psi_k$,  i.e.
$\widehat \pi^2 \ll k^2 \widehat \psi^2$, according to Eq. (\ref{4121}). 

\item{}$-1/2<\a<1/2$. The comoving amplitude and the conjugate
momentum are both frozen, so that
\bea
&&
k \widehat \psi_k =k z \psi_k =k \left(\eta\over \eta_{\rm
hc}\right)^\a \psi_k\sim \sqrt k |k\eta|^\a, \nonumber\\
&&
\widehat \pi_k =k z ^{-1}\pi_k =\left(\eta\over \eta_{\rm
hc}\right)^{-\a }\pi_k\sim \sqrt k |k\eta|^{-\a}, 
\eea
and
\beq
{\cal H}_k = {1\over 2} {\rm Max} \left\{\widehat \pi_k^2, k^2 
\widehat \psi_k ^2\right\}.
\label{4124}
\eeq
\end{itemize}
Thus, for super-horizon wavelengths, the Hamiltonian density is always
growing but,  irrespectively of the value and sign of $\alpha$,
the joint contribution of
the constant modes $\psi_k \sim \widehat  A_k$ and
$\pi_k \sim k \widehat B_k$ always provides an
accurate estimate of the leading contribution to the canonical
energy density, which can be written as
\beq
{\cal H}\simeq \frac{1}{2} \sum_{{k}} k^2
 \Biggl[|\widehat  A_k|^2  |\eta|^{2\alpha} +
|\widehat  B_k|^2  |\eta|^{-2\alpha} \Biggr].
\label{4125}
\eeq

This estimate, based on the low-energy effective Lagrangian
(\ref{4106}), could no longer  be appropriate when including  $\ap$ and
loop corrections, typical of  the  high-curvature string phase of pre-big
bang models. Such corrections could indeed break  the low-energy
duality symmetry. In that case, we cannot exclude that other 
modes in $\psi$ and $\pi$ will dominate the Hamiltonian over the
constant modes. In that case, however, the above estimate can be
applied as a {\em lower bound} on the total energy spectrum, provided
the perturbation equations still contain a frozen part in their asymptotic
solutions, as shown, for instance, in the example discussed in detail in
\cite{Gas97}.

\section{Relic gravitons}
\label{Sec5}
\setcounter{equation}{0}
\setcounter{figure}{0}

One of the most firm predictions of
all inflationary models is the amplification of the traceless, transverse
part of the quantum fluctuations of the metric tensor, and the formation
of a stochastic background of relic gravitational waves
\cite{Gris75,Star79,Al97,Tur97}, distributed over  quite a large range of
frequencies (see \cite{Al97,AlRo99} for a detailed discussion of
the stochastic properties of such a background, and 
\cite{Gris98,AlPapa99} for a possible detection of the associated
``squeezing").   

In this section we will  apply the formalism and the results  established
in the  previous section to compute the possible  spectral distributions 
of relic gravitons  in the context of the pre-big bang
scenario. We will then estimate the maximal expected intensity
(allowed by phenomenological bounds and by model-dependent
constraints) of such backgrounds and discuss their possible
detection by present (or near-future) gravitational antennas. For
additional studies of this aspect of the pre-big bang scenario we also
refer the reader to  recent review papers, especially devoted to
the primordial graviton background  in string cosmology 
\cite{Ve97,Gas98b,Gas98c,Gas99a,Bru98,Mag00}. 

\subsection{Phenomenological bounds on the graviton spectrum}
\label{Sec5.1}

As discussed in the previous section, the spectrum of the tensor
perturbations of the metric, amplified by inflation, contains various
important parameters: the inflation scale $H_1$, the slope $3-2\nu$, the
end-point frequency $\om_1$. Such parameters are (partially) 
constrained by the direct (or indirect) observational bounds existing
at present on the  energy density of a primordial stochastic
background of gravitational waves. To discuss the effects of  such
bounds, let us recall, first of all, the gravitational spectrum expected in
the context of the standard inflationary scenario. 

In that context, the overall cosmological
evolution can be approximated by a three-phase model of 
background, in which the accelerated de Sitter (or quasi-de Sitter)
epoch is followed by the standard radiation-dominated and 
matter-dominated eras:  
\bea
&&
{\rm de~ Sitter~ inflation}, ~~~~~~~~~~~~~~~~~a\sim (-\eta)^{-1},
~~~~~~~~~~~~~~ \eta<\eta_1, \nonumber\\
&&
{\rm radiation~ domination}, ~~~~~~~~~~~~~a\sim \eta, ~~~~~~~~~~~
~~~\eta_1<\eta<\eta_{\rm eq}, \nonumber\\
&&
{\rm matter~ domination}, ~~~~~~~~~~~~~~~~a\sim \eta^2,
~~~~~~~~~~~~~ \eta_{\rm eq}<\eta<\eta_{0}. 
\label{51}
\eea
The effective potential appearing in the perturbation equation
(\ref{456}) is then  characterized by two transition scales, 
inflation $\ra$ radiation at $\eta\sim\eta_1$, and 
radiation $\ra$ matter at $\eta\sim\eta_{\rm eq}$. As a
consequence, there are two different branches of the spectrum,
corresponding to high-frequency modes,  $|k\eta_{\rm eq}|>1$, that are
affected by the first background transition only, and low-frequency
modes,  $|k\eta_{\rm eq}|<1$, that are affected by both transitions.
Following the procedure described in Section \ref{Sec4.4} we can easily
compute the corresponding Bogoliubov coefficients, and the resulting
spectral energy distribution (\ref{4104}), evaluated at the present
time $t_0$, is \cite{Rub82,Fab83,AbbWi84}:
 \bea
\Om_G(\om,t_o) &\simeq & \left(H_1\over M_{\rm P}\right)^2
\Om_\ga (t_0),
~~~~~~~~~~~~~~~~~~~~~~~~~~~~~~~~~\om_{\rm eq}<\om<\om_1,
\nonumber\\ 
& \simeq &\left(H_1\over M_{\rm P}\right)^2
\Om_\ga(t_0)
\left(\om\over \om_{\rm eq}\right)^{-2}, 
~~~~~~~~~~~~~~~~~~~~\om_{0}<\om<\om_{\rm eq}.
\label{52}
\eea

Here $H_1$ is the curvature scale corresponding to the 
inflation $\ra$ radiation transition, $\om_1$ is the maximal amplified
frequency, corresponding today to $\om_1 \simeq \sqrt{H_1/M_{\rm P}}
~10^{11}$ Hz;  $\om_0$  is the
minimal amplified frequency,  corresponding to a mode crossing today
the Hubble horizon, $\om_0 \simeq 10^{-18}$ Hz; $\om_{\rm
eq}\simeq 10^2 \om_0$ is the frequency of a mode  crossing the horizon
at the epoch of the radiation $\ra$ matter  transition (the maximal
frequency affected by the potential due to the matter era); 
finally, $\Om_\ga(t_0)\sim 10^{-4}$ is the present fraction of critical 
energy density in radiation. 

The expected distribution  decreases like $\om^{-2}$ between 
$\om_0$ and $\om_{\rm eq}$; it is flat (i.e. scale-invariant) between
$\om_{\rm eq}$ and $\om_1$ and then dies off exponentially for
$\om>\om_1$. The position of the spectrum in the $(\Om, \om)$ 
plane is completely controlled by $H_1$, which is the only unknown
parameter in this simple example of inflationary model. The existing 
phenomenological bounds able to constrain  the
position of the spectrum in the $(\Om, \om)$ plane can thus  
constrain, indirectly, the inflation scale $H_1$. At present,  there are
three main experimental bounds. 

A first bound follows from the analysis of the pulsar-timing data:
there is at present no detectable distortion of pulsar timing due to 
the influence of a cosmic graviton background, at the frequency scale
$\om_p \sim 10^{-8}$ Hz. The present level of experimental 
sensitivity provides the bound \cite{Kaspi94}
\beq
\Om_G(\om_p) \laq 10^{-8}, ~~~~~~~~~~~~~~~~~
\om_p \sim 10^{-8}~ {\rm Hz}.
\label{53}
\eeq

A second bound comes from the standard nucleosynthesis
analysis \cite{nucleo69}:  the total integrated energy density of the
graviton background, at  the epoch of nucleosynthesis, cannot exceed,
roughly, the energy density of one massless degree of freedom in thermal
equilibrium at that epoch. This gives a
bound for the integrated spectrum \cite{BruGas97}, 
\beq
h^2_{100}\int d \ln \om \Om_G (\om,t_0) ~\laq~ 0.5 \times 10^{-5}, 
~~~~~~~
h^2_{100}= H_0/(100 ~{\rm km~s}^{-1}{\rm Mpc}^{-1}), 
\label{53a}
\eeq
which also applies to a possible peak of the spectrum, at all scales. Note,
however, that the above bound can be relaxed by about an order of
magnitude if significant matter--antimatter domains are present at the
onset of nucleosynthesis, as recently pointed out in \cite{GioKu02}.

A third bound (the most stringent one for  flat or decreasing
spectra) comes from the COBE measurements \cite{Cobe92,Cobe94} of
the large-scale CMB anisotropy. Those observations imply that the
contribution of a graviton background to the CMB anisotropy, at the
horizon scale $\om_0$, cannot exceed the upper bound
\beq
\left(\Da T\over T\right)_{\om_0} \sim 
\left|\da_h(\om_0)\right|\laq 10^{-5}, 
\label{55}
\eeq
where $|\da_h(\om)|$ is the typical fluctuation amplitude (\ref{469}) of 
tensor perturbations  over a proper length scale $\om^{-1}$.
The spectral energy distribution, on the other hand,   can
be expressed in terms of the amplitude $\da_h$ by noting that, from
the E-frame action (\ref{464}) of tensor perturbations, the
average proper energy density  of the background (summing over
polarizations) is given by  
\beq
\r_G={d\langle E_g \rangle \over a^3 d^3x}={M_{\rm P}^2}
\langle |\dot h|^2\rangle .
\label{56}
\eeq
Using  the Fourier expansion, the isotropy and the stochastic condition
(\ref{467}) for $h_k$, we finally obtain the relation \cite{Gas98b}
\beq
\Om_G(\om, t)=
 {d(\langle E_g \rangle /\r_c)\over  a^3 d^3x ~d\ln \om}=
{1\over 3\pi ^2}
\left|\da_h(\om)\right|^2 \left(\om\over \om_0\right)^2 .
\label{57}
\eeq
The COBE bound can thus be rewritten as a condition on the graviton
spectrum, 
\beq
\Om_G(\om_0) \laq 10^{-10}, ~~~~~~~~~~~~~~~~~
\om_0 \sim 10^{-18}~ {\rm Hz}.
\label{58}
\eeq

A further, model-dependent bound can be obtained  by
considering the production of primordial black holes
\cite{Cop98,Cop98a}, that could be in principle closely related to the
process of  graviton production. The process of black-hole evaporation,
at the present epoch, is constrained in fact by a  number of
astrophysical observations: the absence of evaporation may thus 
impose an indirect upper limit on the graviton background.   In a string
cosmology context the best  upper limit obtained from primordial black
holes  turns out to be roughly of the  same order as the nucleosynthesis
bound (\ref{53a}), namely $\Om_G(\om_{\rm s}) \laq 0.5 \times 10^{-5}$
\cite{Cop98}, where $\om_{\rm s}$ is the maximal frequency amplified
at the end of the low-energy dilaton-driven phase (see the next
subsection). It should be mentioned, finally, that another bound
comparable with the nucleosynthesis one, $\Om_G \laq 10^{-5}$, and
only valid however in the frequency range $10^{-8}$ -- $10^{-16}$ Hz, is
also obtained from present gravitational lensing experiments
\cite{Barkana96}. 

Taking into account all significant bounds, the  maximal allowed 
spectrum for the standard inflationary scenario is represented  in Fig. 
\ref{f51} by the line labelled ``de Sitter".  The CMB isotropy bound
(\ref{511}) is the most significant one and,   when imposed on the
spectrum  (\ref{52}),  it implies \cite{Kra92,Star96} 
\beq
H_1 \laq 10^{-5} M_{\rm P} .  
\label{59}
\eeq
Also shown in Fig. \ref{f51} are two possible 
spectra for the case in which the accelerated de Sitter phase of Eq. 
(\ref{51}) is replaced by a phase of power-law inflation, with $a\sim
(-\eta)^\a$, $\a<-1$. In that case the spectrum is decreasing even
for  $\om >\om_{\rm eq}$ (see Table 2), and the isotropy bound on
$H_1$ becomes more and more stringent as the negative slope
becomes steeper and steeper \cite{GG92,GG93}. 

The bound (\ref{59}) on the final inflation scale thus applies to all
models characterized by a flat or decreasing energy spectrum. It may
be evaded, however, if the spectrum is growing, as in the context of
the pre-big bang models that will be considered in the next subsection. 

\begin{figure}[t]
\centerline{\epsfig{file=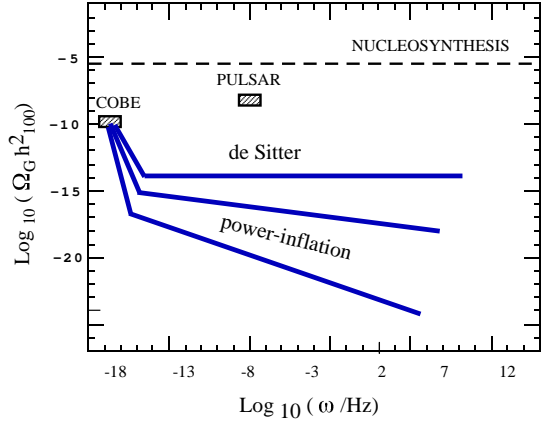,width=72mm}}
\vskip 5mm
\caption{{\sl Expected graviton spectra in the standard inflationary 
scenario.}}
\label{f51}
\end{figure}

\subsection{The graviton spectrum in minimal pre-big bang models}
\label{Sec5.2}

According to the pre-big bang scenario, the very early cosmological
evolution starts from a state approaching the string perturbative
vacuum, goes through an initial low-energy phase driven by the dilaton
kinetic energy,  and then necessarily reaches  a
purely ``stringy" and quantum regime of high curvatures and large
couplings. In that regime, the perturbative expansion of the string
effective action becomes inappropriate (see Section \ref{Sec8}), and
the details of the background kinematics become (with present
understanding) uncertain and strongly model-dependent;  for the 
illustrative purposes of this subsection we shall assume that, in spite of
these uncertainties, it is still possible to parametrize the kinematics of
the string phase by a power-law evolution of the metric and  of the
dilaton field. 

Neglecting the final matter-dominated era, which only affects the
low-energy tail of the spectrum (and which is not very relevant to 
spectra that are growing very rapidly with frequency), we are thus 
led to  consider a simple class of pre-big bang models in which
the phase of accelerated evolution preceding the standard radiation era
is divided into two physically distinct regimes: an initial,
low-energy, dilaton-dominated phase, and a late, high-energy, string
phase where $\ap$ and loop corrections become important (see
Section \ref{Sec8}). We shall denote by $\eta_{\rm s}$ the time scale 
marking the transition to the high-curvature string phase and by 
$\eta_1$ the time scale marking the transition to the
radiation-dominated cosmology (see Fig. \ref{f52}), associated 
to a final curvature of order $1$ in string units.  In the E-frame we may
thus parametrize the following class of  backgrounds:  
\bea
&&
{\rm dilaton~phase}, ~~~~~~~~~~~~~a\sim (-\eta)^{1/2},
~~~~~~~~~\phi\sim -\sqrt{3} \ln (-\eta),~~~~~~~~~~~~~ 
\eta<\eta_{\rm s},\nonumber\\ &&
{\rm string~phase}, ~~~~~~~~~~~~~~~a\sim (-\eta)^\a, 
~~~~~~~~~~~\phi\sim -2\b \ln (-\eta),
~~~~~~~~~~~~~\eta_{\rm s}<\eta<\eta_{1}, \nonumber\\
&&
{\rm radiation~era}, ~~~~~~~~~~~~~~a\sim \eta,
~~~~~~~~~~~~~~~~~\phi = {\rm const},~~~~~~~~~~~~~~~~~~~~~~
\eta>\eta_1. \nonumber\\
&& 
\label{510}
\eea
The model seems to contain four parameters, $\a,~\b,~\eta_1, ~
\eta_{\rm s}$ or, equivalently, the curvature scales and coupling
constants at the two transition epochs:
\beq
H_1, ~~~~~~H_{\rm s}, ~~~~~~g_1=e^{\phi_1/2}, 
~~~~~~g_{\rm s}=e^{\phi_{\rm s}/2}.
\label{511}
\eeq
We may expect, however, that during the string phase the curvature
scale stays controlled by the fundamental string mass  $M_{\rm s}$,
\beq
H_{\rm string~phase} \simeq M_{\rm s} = e^{\phi/2}M_{\rm P}, 
\label{512}
\eeq
so that $H$ keeps constant in the S-frame, while it grows (driven by
the dilaton) in the E-frame, where $M_{\rm P}$ is a constant.  This
condition has two important consequences.

The first is that the final curvature scale, $H_1 =g_1 M_{\rm P}$,  turns
out to be fixed  by the  present value of the fundamental ratio between
the string and the Planck mass \cite{Kap85}: 
\beq
{M_{\rm s}/ M_{\rm P}} \simeq 0.3 - 0.03.
\label{513}
\eeq
The second is that, in the E-frame, according to the parametrization
(\ref{510}),  we have $\left|H_{\rm s}/ H_1\right|={g_{\rm s}/ g_1}$, so that  
\beq
\left|\eta_1/\eta_{\rm s}\right|^{\b}=
\left|\eta_1/\eta_{\rm s}\right|^{1+\a} ,
\label{514}
\eeq
i.e.  $\b =1+\a$.

\begin{figure}[t]
\centerline{\epsfig{file=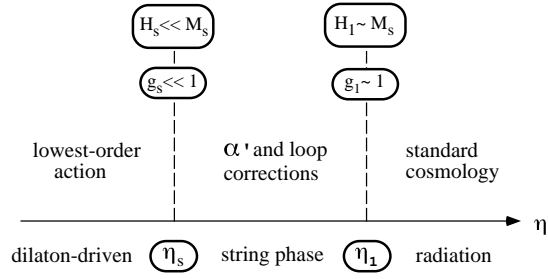,width=72mm}}
\vskip 5mm
\caption{{\sl A ``minimal" model of pre-big bang background.}}
\label{f52}
\end{figure}

Since $H_1$ is fixed, and $\a$ and $\b$ are related, we are thus left with
a ``minimal" class of pre-big bang models \cite{BruGas95,GGV95}, 
characterized by a two-dimensional parameter space. As a convenient
set of physical parameters we may choose,  for instance, the duration of
the  string phase in conformal time, $z_{\rm s}=\eta_{\rm s}/\eta_1$, and the
value of the string coupling $g_{\rm s}(\eta_{\rm s})$ at the beginning of the
high-curvature string phase. 

For such a background, the effective potential $V(\eta)$ of Eq. 
(\ref{456}) is characterized by two transition scales, and the
corresponding spectrum by two frequency bands, separated by the
limiting frequency $\om_{\rm s}$: we have in fact high-frequency modes
with $\om >\om_{\rm s}$ hitting the barrier (or crossing the horizon) 
during the string phase,  and low-frequency modes with $\om <\om_{\rm
s}$ hitting the barrier  during the dilaton  phase. By applying the general
procedure illustrated in the previous section one then finds the
following spectral distribution \cite{BruGas95}: 
 \bea
\Om_G(\om,t_o) &\simeq & g_1^2
\Om_\ga (t_0) \left(\om\over \om_{1}\right)^{3-2\nu},
~~~~~~~~~~~~~~~~~~~~~~~~~\om_{s}<\om<\om_1,
\nonumber\\ 
& \simeq & g_1^2
\Om_\ga(t_0) \left(\om_1\over \om_{s}\right)^{2\nu}
\left(\om\over \om_{1}\right)^{3}, 
~~~~~~~~~~~~~~~~~~~\om<\om_{s} 
\label{515}
\eea
(modulo logarithmic corrections, see Section \ref{Sec4.2}).  Such a 
spectrum has three important properties.

\begin{itemize}
\item{}The low-frequency band is characterized by a nearly thermal (i.e.,
Rayleigh--Jeans) behaviour, $\Om \sim \om^3$, that simulates the
low-energy part  of a black-body spectrum.

\item{}At high frequencies the slope is at most cubic, but in
general flatter, as $\Om \sim \om^{3-2\nu}$, with $\nu>0$. 

\item{}The end-point values of the spectrum are approximately known,
and fixed in terms of the fundamental ratio $g_1=H_1/M_{\rm P} \simeq
M_{\rm s}/M_{\rm P}$: 
\beq
\om_1 \simeq g_1^{1/2} 10^{11}~{\rm Hz}, ~~~~~~~~~~~~~~~
\Om(\om_1) \simeq 10^{-4}g_1^2 .
\label{516}
\eeq
\end{itemize}

The spectral behaviour of this class of
backgrounds, computed by perturbing the lowest-order string 
effective action,  is expected to remain valid even when the $\ap$
corrections are included  (to first order) in the perturbed action
\cite{Gas97}: the spectrum remains the same, only the amplitude is
slightly modified,  with no  appreciable consequence for an
order-of-magnitude  estimate (as confirmed also by recent numerical
simulations, \cite{CaCoGas}). What remains model-dependent, however,
is the position of the break in the spectrum, determined by the
coordinates $\om_{\rm s}$ and $\Om(\om_{\rm s})$. For comparison
with the  standard predictions of the de Sitter inflationary scenario, we
have thus plotted in Fig. \ref{f53} three possible spectra, for a graviton
background of pre-big bang origin, corresponding to different durations 
of the string phase, and to different slopes of the string branch of the
spectrum. 

It should be mentioned that the full spectrum of tensor metric
perturbations has recently been computed for a regular class of
minimal pre-big bang models, including in the linearized perturbation
equations also the contributions from the higher-derivative and loop
corrections needed for a smooth transition to the post-big bang regime 
\cite{CaCoGas}. It has been checked, in such a context, that the
spectrum of the low-frequency modes $\om <\om_{\rm s}$, crossing the
horizon in the low-curvature regime, is unaffected by the higher-order
corrections. Also, it has been explicitly confirmed that the ``string
branch" of the spectrum ($\om >\om_{\rm s}$) is flatter than at low
frequency, and that its amplitude can be reliably estimated also by
means of the low-energy perturbation equation. It has been found,
however, that if the transition to the post-big bang regime goes
through a phase described by the fixed-point solutions of the (first order
in $\ap$) truncated string effective action (see Subsection \ref{Sec8.3}),
then the slope of the string branch of the spectrum is at least
quadratic, i.e. still too steep for a possible detection by present
gravitational antennas (see Subsection \ref{Sec5.4}).

\begin{figure}[t]
\centerline{\epsfig{file=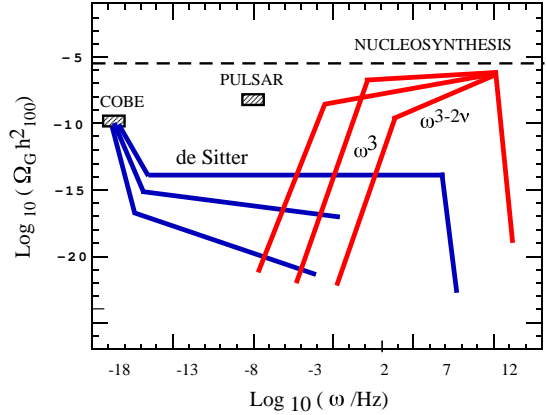,width=72mm}}
\vskip 5mm
\caption{{\sl Expected graviton spectra in minimal  
pre-big  bang models, compared with the flat (or decreasing) spectra
of the standard inflationary scenario.}} 
\label{f53}
\end{figure}

In any case, the graviton spectrum of the minimal pre-big bang scenario
is monotonically growing and, as a consequence, the most constraining
bound is  provided by nucleosynthesis, as is evident from Fig. \ref{f53}.
The string branch of the spectrum can be extended (at least in principle)
to arbitrarily low frequencies,  provided that the slope is not too 
flat, to avoid a conflict with the pulsar bound. It should be stressed,
however, that the nucleosynthesis bound  applies to the total integrated
energy density, and thus becomes  more and more stringent as the
spectrum is flatter and the string phase is longer. 

Given our present ignorance of the dynamical
details of the string phase, where high-curvature and quantum-loop
effects are expected to play  a significant role, the precise shape of
the spectrum turns out to be (at present) strongly model-dependent 
(as also stressed by the examples analysed in \cite{CaCoGas}).   It is
remarkable that, in spite of this uncertainty, it may be possible to
provide an accurate (and rather model-independent) estimate
of the theoretically  allowed region for the expected spectrum,  in the
plane of Fig. \ref{f53}. 

\subsection{Allowed region in the $\om$ -- $\Om$ plane}
\label{Sec5.3}

The determination of the allowed region requires a precise estimate 
of the coordinates of the end-point of the spectrum \cite{BruGas97},
$\om_1$ and $\Om(\om_1)$. The first one is the maximal amplified
frequency, corresponding approximately to the production of one graviton
per polarization mode and per unit of phase-space volume. The 
present value of this frequency is: 
\beq
\om_1(t_0)={H_1 a_1\over a(t_0)}. 
\label{517}
\eeq
The second  one is the associated peak value,
\beq
\Om(\om_1)= {\om_1^4\over \pi^2\r_c}
\label{518}
\eeq
(from Eq. (\ref{491})).  The present value of $\om_1$ 
contains two  elements of intrinsic uncertainty.

A first uncertainty corresponds to the identification $H_1=M_{\rm s}$. It
is true  that the final transition scale should be controlled by the string
mass scale,  but the exact value of $H_1$ might be slightly different from
$M_{\rm s}$.  This uncertainty is expected to be small, however, 
and can be neglected in a first approximation. 

A second (and just moreimportant) uncertainty is associated to the
rescaling of $\om_1$ from the transition time $t_1$ down to $t_0$.
There are in fact two possibilities. If the CMB radiation that we
observe today was all produced by the transition at
$t=t_1$,  together with the gravitons, then the ratio 
$\Om_G/\Om_\ga$ remains fixed throughout the subsequent
cosmological evolution from $t_1$ down to $t_0$, the redshift of
$\om_1$ is known and can be computed exactly. If, on the contrary,
there was some process of radiation production due to some  reheating
phase occurring at  time scales much later than $t_1$, then the present 
value of the ratio $\Om_G/\Om_\ga$  can be arbitrarily diluted with 
respect to its original value at $t_1$, corresponding to a redshift of 
$\om_1$ which is in principle unknown and arbitrarily large. 

To take  this last possibility into account, we may introduce the
phenomenological parameter $\da s$, representing the fraction of the 
present CMB entropy density due to all 
reheating processes following 
the end of the string phase at $t=t_1$. This parameter is defined by
\beq
\da s= {s_0-s_1\over s_0}, 
\label{519}
\eeq
where $s_0$ and $s_1$ are the thermal entropy density of the CMB
background at $t_0$  and $t_1$, defined in terms of the 
temperatures ($T_0, T_1$) and of the numbers ($n_0, n_1$) of particle
species contributing (with their own statistical weight) to the 
entropy: 
\beq
s_0= {2\pi^2\over45}n_0(a_0T_0)^3, ~~~~~~~~~~~~~~
s_1= {2\pi^2\over45}n_1(a_1T_1)^3. 
\label{520}
\eeq
 By using $n_1/n_0 \sim 10^3$, and assuming thermal equilibrium at
$t=t_1$, we then obtain \cite{BruGas97}: 
\bea
&&
\om_1(t_0) \simeq  T_0\left(M_{\rm s}\over
M_{\rm P}\right)^{1/2}\left(1-\da s\right)^{1/3} , \nonumber \\
&&
\Om_G(\om_1, t_0)\simeq 7\times 10^{-5}h_{100}^{-2}
\left(M_{\rm s}\over M_{\rm P}\right)^{2}
\left(1-\da s\right)^{4/3},
\label{521}
\eea
where $T_0 =2.7$ K $\simeq 3.6 \times 10^{11} ~{\rm Hz}$ is the
present CMB temperature. 

At fixed $\da s$, the main uncertainty on the end-point values is thus
controlled by the fundamental ratio $M_{\rm s}/M_{\rm P}$, as illustrated in  
Fig. \ref{f54}, where the boxes around the end point display an 
uncertainty corresponding to the  allowed range 
$0.01 <(M_{\rm s}/M_{\rm P})<0.1$. 
By varying $\da s$, the end point moves along the thin line with
slope $\om^4$. The present peak intensity is thus depressed for large
values of $\da s$;  however,  as we can see from the picture, even if $99
\%$ of the present entropy had been produced during the latest stage of
evolution, the intensity of the relic graviton background would stay well
above the full line labelled ``de Sitter", which represents the most
optimistic prediction of the standard inflationary
scenario in this frequency range. 

We note, finally, that  the
same numerical value of the ratio $M_{\rm s}/M_{\rm P}$ also controls the
amplitude of the bulk graviton spectrum possibly produced in the
context of a pre-big bang scenario with large extra-dimensions 
\cite{Arkani98,Anto98}. Indeed, even if the $D$-dimensional Planck scale
is smaller (for larger volumes of the internal space), the string scale is
correspondingly lowered, so as to keep the grand-unified gauge
coupling in the above allowed range. 

The above discussion refers to the minimal models illustrated in the
previous subsection, where the end point and the peak of the spectrum
are strongly correlated, according to Eq. (\ref{521}). Actually, what is
really fixed in the pre-big bang scenario is the maximal height of the
peak, but not necessarily the position in frequency, and it is not
impossible, in more complicated,  {\em non-minimal} models, to shift the
peak to lower frequencies. 

In the minimal models, in fact, the end of the string phase coincides with
the freezing of the coupling parameter $g=e^{\phi/2}$, and with the
beginning of the standard radiation era. In non-minimal models
\cite{Gas98b} (see also \cite{BabGio99}) the coupling is still very small at
the end of the string phase, $g(t_1) \ll1$, the dilaton keeps growing in a
decelerated way while the curvature starts decreasing, and  the
produced  radiation becomes dominant only much later (such models are
also important in the context of reheating, see \cite{Buo00}). 

The main difference between the two classes of models is that in  
the non-minimal case the effective potential $V(\eta)$ appearing in the
canonical perturbation equation is non-monotonic, so that the highest
frequency modes re-enter the horizon during the intermediate 
dilaton-driven phase that follows the string phase and preceds the
beginning of the radiation era.  This modifies the slope of the
spectrum in the high-frequency sector, with the possible
appearance of a negative power \cite{Gas98b} (see also \cite{Occhio97}).
The gravity-wave spectrum may thus become non-monotonic, and the 
peak may no longer coincide  with the end point, as illustrated
in Fig. \ref{f54}. A similar behaviour for the spectrum is also obtained in
the context of non-minimal pre-big bang models that include a phase
of ``thermal inflation", or a period of early matter domination 
\cite{MenLid98}. A phase of matter-dominated pre-big bang evolution
can also produce a flat graviton spectrum \cite{FiBra01}, but in that
case the intensity of the relic background is much smaller than the peak
values reported in Fig. \ref{f54}. 

\begin{figure}[t]
\centerline{\epsfig{file=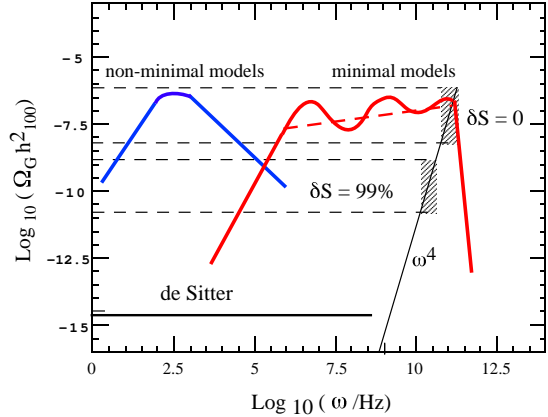,width=72mm}}
\vskip 5mm
\caption{{\sl Peak and end point of the graviton spectrum in minimal
and non-minimal models, as a  function  of $\delta s$. The boxes at the
end of each strip represent the uncertainty due to the unknown  value of
the ratio $M_{\rm s}/M_{\rm P}$.}} 
\label{f54}
\end{figure}

These are good news from an experimental point of view, because
they suggest the possibility of a large detectable signal, even at
frequencies much lower than the gigahertz band. However, they also
provide a warning against  too naive an interpretation and extrapolation
of possible  future experimental data, because of the complexity of the
parameter space of the string cosmology models. It should be noted,
for  this purpose, that  the spectrum associated to the string phase
could be monotonic only on the average, and could locally oscillate even
in minimal models  \cite{BMU97}, as illustrated by the wavy line
appearing in the high-frequency branch of the spectrum of  Fig. \ref{f54}.

Given the present theoretical uncertainties, it seems appropriate to
define the maximal allowed region for the expected graviton
background,  i.e. the region spanned by the spectrum in the $\{\om,
\Om \}$ plane when all its parameters are varied. Such a region is
illustrated in Fig. \ref{f55}, for the (phenomenologically interesting)
high-frequency range $\om > 1$ Hz. The figure emphasizes the possible,
large enhancement (of about eight orders of magnitude) of the  expected
graviton background in pre-big bang models, with respect to the
standard inflationary scenario.  The maximal allowed intensity of the
graviton background coincides with the upper border of the strip with
$\da s=0$ in Fig. \ref{f54}, i.e. from Eq. (\ref{521}):  
\beq
\Om_{G}^{\rm max} \simeq 10^{-6}h^{-2}_{100}.
\label{522}
\eeq
Note that this upper limit is automatically compatible with the
nucleosynthesis bound, Eq. (\ref{53a}), even if the spectrum is nearly 
flat from the end point down to the hertz scale. 

\begin{figure}[t]
\centerline{\epsfig{file=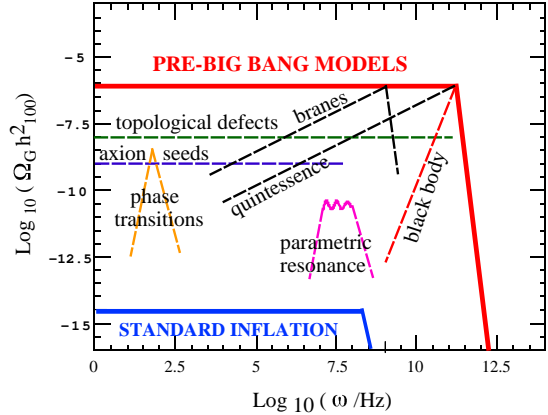,width=72mm}}
\vskip 5mm
\caption{{\sl Allowed region for the vacuum-fluctuation spectrum in
string cosmology and in the standard inflationary scenario, compared 
with other possible tensor perturbation spectra of 
primordial origin.}} 
\label{f55}
\end{figure}

This maximal amplitude can be conveniently expressed in terms of a
useful experimental parameter, the strain density $S_h$, defined by
\cite{Fla93} 
\beq
\langle h(\nu) h^*(\nu')\rangle = {1\over 2} \da(\nu-\nu') S_h(\nu),
\label{523}
\eeq
and related to the energy density (\ref{57}) (using the
stochastic condition (\ref{467})) by \cite{Gas98b,Mag00}:
\beq
S_h(\nu)={3H_0^2\over 4\pi^2\nu^3}~ \Om_G(\nu) ,~~~~~~~~~
\nu=\om/2\pi .
\label{524}
\eeq
The minimal experimental sensitivity required for a direct detection (by
a single gravitational antenna) of the maximal  background (\ref{522}) 
of pre-big bang gravitons is  thus given by \cite{BruGas97}
\beq
\left(S_h^{\rm min}\right)^{1/2}\simeq 3 \times 10^{-26}\left({\rm
kHz}\over \nu \right)^{3/2}~{\rm Hz}^{-1/2}. 
\label{525}
\eeq

This shows that the minimal sensitivity needed to reach the border of
the allowed region grows with the frequency band to which the
detector is tuned, even if such a border is the same at all frequency
scales (see Eq. (\ref{522}) and Fig. \ref{f55}). This effect may favour 
--from an experimental point of view-- detectors working at lower
frequencies. Unfortunately --from a theoretical point of view-- the
probability of a large background intensity seems to be larger at
higher frequencies, since the typical spectrum of pre-big bang models is
growing. 
It is important to stress, however, that the minimal required sensitivity
(\ref{525}) may be somewhat relaxed, for realistic detection strategies
in which the search for a stochastic background proceeds through the
cross-correlation of the outputs of two independent detectors (see the
next subsection). 

The relic background associated  with the parametric amplification of
the vacuum  fluctuations  is compared, in Fig. \ref{f55},  with other
possible backgrounds of primordial origin, obtained with different
mechanisms: graviton radiation from cosmic strings
\cite{Cald92,Shell96}  and other topological defects \cite{MarVi96},  from
axion seeds \cite{Vern00},  from bubble collision at the end of a
first-order phase transition \cite{TurWil90}, and  from parametric
resonance of the inflaton oscillations \cite{Khleb97,Bass97}. The
spectrum obtained from a phase transition, plotted in the figure, refers
to a typical reheating temperature $T_{\rm r} \sim 10^8$ -- $10^9$ GeV,
but higher peak values are possible for higher $T_{\rm r}$ \cite{TurWil90}.  

Also shown in the figure is the linearly growing spectrum obtained in
recently proposed models of ``quintessential" inflation
\cite{PeeVil99,Giov99}  and of ``brane-world" inflation \cite{SaSaso01},
and a thermal black-body spectrum corresponding to a present
temperature $T_0\sim 1$ K. In the standard inflationary scenario, a
thermal background might originate at the Planck scale, when the
temperature is high enough to maintain gravitons in  thermal
equilibrium. Such a background, however,  should be strongly diluted
(with respect to the present CMB radiation)  by the action of the
subsequent inflationary phase, occurring at curvature scales lower than
Planckian. As a consequence, the surviving  spectrum should correspond
today to an effective temperature so depressed as to be practically
invisible. In the  context of the pre-big bang models, on the contrary, a
thermal spectrum like that of Fig. \ref{f55} could be simulated by the
sudden transition of the low-energy dilaton-driven phase to the  
standard radiation-dominated era -- corresponding, for instance, to a
minimal model of Section \ref{Sec5.2} with $\eta_{\rm s} \sim \eta_1$. 

It may be noted  that all the primordial backgrounds of Fig. \ref{f55} are
higher (in intensity) than the background  obtained
from the amplification of the vacuum fluctuations in the standard
inflationary scenario,  but not higher than the analogous background
expected in the context of the pre-big bang scenario. In any case, given
the rich phenomenology of possible spectra, it seems appropriate to
report at this point the experimental sensitivities  of present
gravitational antennas, referred in particular to the  frequency range of
Fig. \ref{f55}. This will be the subject of the next subsection. 

\subsection{Experimental sensitivities}
\label{Sec5.4}

The limiting sensitivity (\ref{525}) (corresponding to the upper border of
the allowed region in Fig. \ref{f55}) is unfortunately too high for the
detectors that are at present in operation: the cryogenic,
resonant-mass,  gravitational antennas (see \cite{Pizz97} for a review).
In fact, if we take for instance the resonant bar NAUTILUS
\cite{CoFaMo92},  the resonance frequencies are at $907$ and $922$ Hz
and, at these frequencies, the strain sensitivity is approximately   
\beq
\sqrt{S_h} \simeq 5 \times 10^{-22}~ {\rm Hz}^{-1/2}. 
\label{526}
\eeq
With some improvements, the sensitivity is expected to reach in a few
years the target value $8\times 10^{-23}~{\rm Hz}^{-1/2}$.  Similar
resonance frequencies (around $900$ Hz) and similar sensitivities are
also typical of the other resonant detectors such as ALLEGRO
\cite{Ham98}, AURIGA \cite{Prodi98}, and EXPLORER \cite{Pall97}. The
resonant bar NIOBE \cite{Heng98}, however, is working  at a
smaller frequency ($\sim 700$ Hz) and could reach, in principle, a slightly
better sensitivity. 

Interferometric detectors, such as LIGO \cite{Abramo92}, VIRGO
\cite{Caron97}, GEO600 \cite{Luck97} and TAMA300 \cite{Kawa97}, are 
already basically built and will soon become
operative. The best planned sensitivity, for the first-generation 
instruments, is around  
\beq
\sqrt{S_h} \simeq  10^{-22}~ {\rm Hz}^{-1/2}, 
\label{527}
\eeq
very similar to the present bar sensitivities. This sensitivity, however, is
available at a smaller frequency band, $\nu= 100$ Hz, and thus
corresponds to a smaller limit in terms of the energy density of the
graviton background (see Eq. (\ref{524})). In practice,  the
interferometers of first generation will be directly sensitive to
$\Om_G h_{100}^2 \sim 10^{-1}$ -- $10^{-2}$, better than resonant bars,
which can at most reach the critical density, but still well above the
border of the allowed region. 

In order to approach the border sensitivity $\Om_G{\rm h}_{100}^2 \sim
10^{-6}$, we must wait for the second (and higher) generation of
interferometers. The Advanced LIGO project \cite{AdLigo98}, for
instance, might lead to a two orders of magnitude improvement of the
sensitivity (\ref{527}), at $\nu= 100$ Hz, thus reaching the borderline 
(\ref{522}). An even better result is expected for the space
interferometer LISA \cite{Hough98}, whose goal is to reach the strain
sensitivity \cite{Bender98}  $\sqrt{S_h} \simeq 4 \times 10^{-21}$
Hz$^{-1/2}$ in the frequency band from $3$ to $10$ mHz, corresponding
(in this band) to the remarkable limit  $\Om_G{\rm h}_{100}^2 \sim
10^{-11}$. 

Another interesting possibility concerns resonant detectors with
spherical (or truncated icosahedron) geometry, such as the TIGA
\cite{Tiga1,Tiga2} or SFERA \cite{Asto98a} projects. Hollow
spheres  \cite{Coccia98} are particularly promising: a large
copper--aluminium detector ($4$ -- $5$ m in diameter), pushed to
its  extreme quantum limit may reach a strain 
sensitivity\footnote{M. Cerdonio and L. Conti, private
communication.}  
$\sqrt{S_h} \simeq  3 \times 10^{-24}$ Hz$^{-1/2}$, with resonant 
frequencies of $600$ and $2500$ Hz (the resonance frequency
can be tuned, in principle, by varying the geometrical radius of the
sphere). Such a limiting sensitivity can be approached by the recently
proposed  wide-band ``dual-sphere" detector \cite{CCLOZ00},
which consists of a massive solid sphere suspended inside a larger
hollow one. In that  case, a sapphire detector of an overall size of $2.6$
m could reach a strain sensitivity $\sqrt{S_h} \simeq 10^{-23} {\rm
Hz}^{-1/2}$, for frequencies  between $1$ and $4$ kHz.

Concerning higher frequencies (well above the kilohertz), the only
possibility at present seems to be the use of resonant electromagnetic
cavities \cite{Peg78,Reece84} or waveguides \cite{Cruise00}. There are
also recent experimental studies for detecting small displacements
based on  two coupled microwave cavities \cite{Bernard98}: such a 
detector reaches the maximum sensitivity when the frequency of the
incident gravitational wave is equal to the frequency-difference of the
two cavity-modes. In this way, it seems possible at present to
reach the  strain sensitivity $\sqrt{S_h} \simeq 10^{-20}$ Hz$^{-1/2}$
for frequencies around the kilohertz range \cite{Chinca02}.
Unfortunately,  however, this value is well above the minimal required
sensitivity (\ref{525}).

As shown explicitly by the above examples, the limiting sensitivity
(\ref{525}) seems not to be  within easy reach of present
technology, for the case of a single detector. In addition, an
unambiguous detection of a stochastic gravity-wave background with a
single experimental apparatus would require a complete and exact
knowledge of all the intrinsic instrumental noises, and of all the
backgrounds of different origins that could interact with the 
gravitational antenna. Fortunately, an efficient answer to both
difficulties is known: it is provided by the cross-correlation of the
outputs of two (or more) detectors \cite{Chris92,Fla93,Al97,AlRo99}. 

Suppose that the outputs of two detectors, $s_i(t)$, $i=1,2$, are 
correlated over an integration time $T$, to define a signal: 
\beq
S = \int_{-T/2}^{T/2} dt~ dt' s_1(t)s_2(t') Q(t-t'). 
\label{528}
\eeq
Here $Q(t)$ is a real ``filter" function, determined so as to optimize the
signal-to-noise ratio (SNR), defined by an ensemble average as:
\beq
{\rm SNR}= \langle S \rangle /\Da S \equiv \langle S \rangle
\left(\langle S^2 \rangle- \langle S \rangle^2\right)^{-1/2}.
\label {529}
\eeq
The outputs $s_i(t)= h_i(t)+n_i(t)$ contain the physical strain induced by
the cosmic background, $h_i$, and the intrinsic instrumental noise,
$n_i$. The two noises are supposed to be uncorrelated (i.e. statistically
independent), $\langle n_1(t)n_2(t')\rangle=0$, and much larger  in
magnitude than the physical strains,  $|n_i | \gg |h_i|$. Also, the cosmic
background is assumed to be isotropic, stationary and Gaussian, with
$\langle h_i\rangle=0$. It follows that \cite{AlRo99}:
\beq
\langle S\rangle = \int_{-T/2}^{T/2} dt ~dt'\langle h_1(t)h_2(t')\rangle
Q(t-t').  
\label{530}
\eeq

The physical strain is given by the projection of the metric fluctuation
$h_{ab}(x_i,t)$ (computed at the detector position $x=x_i$) over the
detector response tensor $D_i^{ab}$, which characterizes the spatial
orientation of the arms of the $i$-th detector. If  $\hat n$ is a unit
vector specifying a direction on the two-sphere, and $e_{ab}(\hat n)$ is
the gravity-wave polarization along $\hat n$, we can expands the strain
in momentum space as 
\bea
&&
h_i(t)= \int dp \int d^2 \hat n ~h_A(p, \hat n) F_i^A(\hat n)  e^{2\pi
ip\left( \hat n \cdot {\vec x}_i -  t \right)}, \nonumber \\
&&
F_i^A(\hat n)=e_{ab}^A(\hat n)D_i^{ab},
\label{532}
\eea
where $d^2\hat n$ denotes the angular integral over the unit
two-sphere, and the index $A$ labels the two polarization modes of the
wave; $F_i^A(\hat n)$ is the so-called ``pattern function", depending on
the geometry of the detector. Note that in the above equation we have
used the proper momentum ${\vec p}= {\vec k}/a$, even if the
cosmological variation induced by the expansion of the scale factor is
negligible over a typical experimental time scale $T$. Note also that, for
a better comparison with experimental variables, we have used units in
which $h=1$, so that, for a gravity wave, $p=\nu=\om/2\pi$ (but not in
general for  massive waves, see Section \ref{Sec6.4}). 

For an isotropic, unpolarized and stationary background we can then use
the stochastic condition to define \cite{AlRo99,Mag00}
\beq
\langle h_A^* (p, \hat n), h_{A'}(p', \hat n') \rangle= {1\over 4\pi}
\da_{AA'}\da (p-p')  \da^2 (\hat n,\hat n') {1\over 2}S_h 
\label{533}
\eeq
(the integration over the angular variables $d^2\hat n, d^2\hat n'$ then
reproduces Eq. (\ref{523})). By inserting the momentum  expansion into
Eq. (\ref{530}), and assuming, as usual, that the observation time $T$ is
much larger than the typical time intervals $t$ -- $t'$ for which $Q\not=
0$, we finally obtain, using Eqs. (\ref{523}) and (\ref{524}): 
\beq
\langle S\rangle=N T {3 H_0^2\over 8 \pi^2} \int {dp\over p^3} \ga (p)
Q(p) \Om(p),  
\label{534}
\eeq
where $Q(p)$ is the filter function in momentum space, and $\ga(p)$ is
the so-called ``overlap-reduction function": 
\beq
\ga(p)= {1\over N}\int {d^2 \hat n \over 4 \pi } F_1(\hat n) F_2 (\hat n) 
 e^{2\pi i p \hat n \cdot ({\vec x}_2 - {\vec x}_1)}.  
\label{535}
\eeq
Here $N$ is an overall normalization factor, depending on the type of the
detectors (for two interferometers, a convenient choice is $N=2/5$, see
\cite{AlRo99}; see \cite{Mag00} for the normalization of
different antennas). 

We now need to compute the variance $\Da S^2$, which, for
uncorrelated noises much larger than the physical strains,  can be
expressed as \cite{AlRo99}:
\beq
\Da S^2 \simeq \langle S^2\rangle = \int _{-T/2}^{T/2} dt dt' d\tau
d\tau' \langle n_1(t) n_1(\tau)\rangle \langle n_2(t') n_2(\tau')\rangle
Q(t-t') Q(\tau-\tau'). 
\label{536}
\eeq
Introducing the (one-sided) noise
power spectrum in momentum space, $P_i(p)$, defined by 
\beq
\langle n_i(t) n_i(\tau)\rangle = {1\over 2} \int dp P_i(p) 
e^{-2\pi i p (t-\tau)},
 \label{537}
\eeq
and assuming as before that $T$ is much larger than the typical
correlation intervals $t$ -- $t'$, $\tau$ -- $\tau'$, we obtain 
\beq
\Da S^2 = {T\over 4} \int {dp} P_1(p) P_2(p) Q^2(p).
\label{538}
\eeq
The optimal filtering is now determined by the choice (see \cite{AlRo99}
for details)
\beq
Q(p)= \la {\ga(p) \Om(p)\over p^3 P_1(p) P_2 (p)},
\label{539}
\eeq
where $\la$ is an arbitrary normalization constant. With such a choice
one finally arrives, from Eq. (\ref{534}) and (\ref{539}), to the optimized
signal-to-noise ratio:
\beq
{\rm SNR} = {\langle S\rangle \over \Da S}=   N {3 H_0^2\over 4 \pi^2}
\left[T\int {dp\over p^6}{\ga^2(p) \Om^2(p)\over P_1(p)
P_2(p)} \right]^{1/2}. 
\label{540}
\eeq
The background can then be detected, with a detection
rate $\ga$, and a false-alarm rate $\a$, if \cite{AlRo99}: 
\beq
{\rm SNR} \geq \sqrt 2 \left({\rm erfc}^{-1} 2\a - {\rm erfc}^{-1} 2\ga\right).
\label{541}
\eeq

Armed with the above results, it is now possible to estimate the
minimum  detectable background at a given confidence level, for any
pair of gravitational antennas of given noise power spectrum $P_i$ and
overlap $\ga$. The correlation of all existing pairs of interferometers has
been studied  in \cite{BaFoLosurdo} and \cite{AlRo99}. With the
sensitivities of the first-generation interferometers, for an observation
time $T=4$ months, a detection rate of $0.95$, a false-alarme rate of 
$0.05$, and a flat graviton spectrum $\Om_G=$ const, the minimum
detectable value is  
\beq 
\Om_G{\rm h}_{100}^2 \simeq 5 \times 10^{-6}.  
\label{542}
\eeq
This value can be improved only slightly  by increasing the number of
correlated detectors. With the planned sensitivity of Advanced LIGO,
however, one obtains the much better limit \cite{Al97,AlRo99}: 
\beq
\Om_G{\rm h}_{100}^2 \simeq 5 \times 10^{-11}. 
\label{543}
\eeq

The correlation between two resonant bars, and between a bar and an
interferometer,  has been studied in
{\cite{Asto94,Asto97a,Vit97,BaFoLosurdo}. The minimum detectable
values for the VIRGO-AURIGA or VIRGO-NAUTILUS or AURIGA-NAUTILUS
pairs is at present \cite{Mag00} $\Om_G{\rm h}_{100}^2 \sim$ a few
$\times  10^{-4}$ (depending on the bar orientation) for a flat
spectrum, one year of observation, at the $90\%$ confidence level. 
The correlation of two spherical detectors seems to be more promising,
however. Two spheres (of $3$ m diameter) located at the same
site could reach a sensitivity \cite{Vit97} $\Om_G{\rm h}_{100}^2 \sim 4 
\times  10^{-7}$. Also, studies reported in  \cite{Coccia98}
suggest that the correlation of hollow spheres could reach $\Om_G{\rm
h}_{100}^2 \sim  
 10^{-9}$. 

All the above results refer to a flat graviton spectrum. The
signal-to-noise ratio, and therefore the sensitivity, depend, however, 
on $\Om_G(p)$, and then on the slope of the  spectrum (see Eq. 
(\ref{540})), as  also stressed in \cite{Mag00,BabGio99a}. The sensitivity
to a growing spectral distribution such as that predicted by pre-big bang
models has been computed in particular in \cite{AlBru97,UVe0} and
\cite{BabGio99,BabGio99b}, for a pair of LIGO and a (virtual) pair of VIRGO
interferometers, respectively. The minimum detectable intensity of the
background, in that case, does not show any improvement with respect
to a flat spectrum; on the contrary, the sensitivity could even
suffer from a slight decrease, if the maximum of the signal is near the
border of the accessible frequency band, where the noise of the
interferometer is higher. 

Finally, the cross-correlation in the case of space interferometers such
as LISA has recently been discussed in detail \cite{UVe1} by taking into 
account that, at lower frequencies, the sensitivity to a stochastic
graviton background of  primordial inflationary origin is fundamentally
limited by the possible presence of other astrophysical backgrounds,
generated at much later times, due for instance to binary systems of
compact objects that cannot be resolved as individual sources.  A
detailed analysis,  performed for the cross-correlation of two identical
LISAs, leads to a minimum detectable signal $\Om_G{\rm h}_{100}^2
\sim  10^{-12}$, in the mHz range, for a flat spectrum. The three arms of
LISA could also be combined in such a way as to simulate two
inteferometers located in the same place; unfortunately, they would
be rotated by an angle of $\pi/4$ with respect to one another, and the
corrresponding overlap function would be identically zero over the
whole frequency range. This is the reason why two space
interferometers are needed, even in that case. 

To conclude this subsection, let us recall  the best (direct) experimental
upper bound existing at present  on the energy density of a stochastic
graviton background, which has been obtained from the
cross-correlation of the two resonant bars NAUTILUS  and  EXPLORER
\cite{Asto98,Asto99}:  
\beq 
\Om_G {\rm}h_{100}^2
\laq 60, ~~~~~~~~~~~~~~~~~~~ \nu\simeq 907 ~{\rm Hz}.
\label{544}
\eeq
This results corresponds to a very small observation time ($T\sim 12$
hours), and it is thus far from the theoretical upper limit (\ref{525}), but
the future perspectives seem to be promising, as reported in the above
discussion. The cross-correlations of advanced (ground-based)
interferometers, space interforemeters, spheres, hollow spheres, are all
expected to cross the borderline sensitivity (\ref{525}), and then to
explore, in a very near future, the allowed region of the string 
cosmology spectrum. Thus, future data from gravitational antennas will
directly constrain the parameter space of the minimal and non-minimal
classes of pre-big bang models considered in this section.

We should add, finally, that useful information and constraints on the
relic graviton background will also be obtained from future
measurements of the CMB polarization \cite{SeZa96,KamJaf00,Les99}.
The presence of a gravitational-wave background should in fact induce a
characteristic ``curl component" in the large scale CMB polarization 
\cite{KamKosSteb97}, which might be detectable by dedicated
experiments (already at the sensitivity level of the Planck satellite 
\cite{Planck}) for the gravitational spectra predicted by the standard
inflationary models. No polarization signal is expected to be detectable,
on the contrary, for the graviton spectra of pre-big bang models, whose
amplitude is extremely suppressed at the large scales accessible to the
polarization experiments (see, however,  Ref. \cite{FiBra01} for possible
exceptions).

\section{Relic dilatons}
\label{Sec6}
\setcounter{equation}{0}
\setcounter{figure}{0}

The effective action of string cosmology always contains at least two
fundamental fields, the metric and the dilaton. The amplification of the
dilaton fluctuations leads to dilaton production, just like the
amplification of the transverse, traceless components of the metric
fluctuations  leads to graviton production. The possible formation of a
stochastic background of relic dilatons \cite{Gas94b,GasVe94a,Gas97a},
with statistical and squeezing properties similar to those of the graviton
background, is one of the most peculiar phenomenological aspects of
pre-big bang models. 

The main physical difference with  the graviton background, 
discussed in the  previous section, concerns the possible
contribution of the mass to the dilaton spectrum (see Subsection
\ref{Sec6.1}). The dilaton should indeed  be massive, according to
conventional models of supersymmetry breaking \cite{Banks93}. Also,
the dilaton {\em must} be massive if it is coupled non-universally to
macroscopic matter with gravitational strength \cite{Tay88,Ellis89},  in
order to avoid unacceptable violations of the equivalence principle. The
value of the dilaton mass, however, is largely unknown from a 
theoretical point of view. There are phenomenological constraints on the
 mass, but they depend on the effective coupling of dilatons to
macroscopic matter, which we shall discuss in Subsection \ref{Sec6.2}.
The value of the coupling, in  turn, is strongly model-dependent
\cite{DP94a,DP94b}, and phenomenologically constrained, at present,
only by tests of the equivalence principle and of Newtonian gravity
available on a macroscopic scale \cite{Fis92,Hoyle01}. 

As a consequence, the present amplitude of the relic dilaton 
background, controlled by the string and dilaton mass, is more
uncertain than in the graviton case. This is disappointing, 
on the one hand, as it prevents  the existence of an unambiguous link
between the spectral properties of the dilaton background and the
parameters of pre-big bang models. On the other hand, however, it
leaves open two interesting phenomenological possibilities: a relic
background of non-relativistic dilatons could represent a significant
fraction of the present dark matter density \cite{Gas94b} (see
Subsection \ref{Sec6.3}) and, if the mass is small enough, it could be
detectable by the cross-correlation of gravitational antennas
\cite{Gas00b} already at the  sensitivity  level of enhanced
interferometric detectors \cite{GasUn01}, as will be discussed in
Subsection \ref{Sec6.4}.

\subsection{Dilaton production}
\label{Sec6.1}

In this subsection we will estimate the spectral distribution of the
dilatons obtained from the amplification of the vacuum fluctuations, in
the context of pre-big bang models. There are, of course, also other 
cosmological mechanisms of production, such as particle collisions at
high (Planckian) temperature, coherent oscillations around the minimum
of the scalar potential (and more exotic mechanisms, such as dilaton
radiation from cosmic strings; see for instance \cite{DaVi96a}). We will
concentrate here on the parametric amplification  because, even if
the  temperature remains too low to allow thermal production, and
oscillations are avoided through a symmetry that ensures the coincidence
of the minima of the potential at early and late times \cite{DaVi96},
quantum fluctuations cannot be eliminated; also, they may be expected to
represent the dominant source \cite{Lyt96} when the inflation scale is
not smaller than about $10^{16}$ GeV, as in the context of the pre-big
bang scenario.

The production of dilatons through the amplification of the quantum 
fluctuations of the scalar background is very similar to the amplification
of tensor perturbations, discussed in Section \ref{Sec5}.  Unlike  the
graviton case, however, the analysis of  dilaton perturbations is
complicated by their coupling to the scalar component of  metric
perturbations, and to the scalar perturbations of the matter sources. The
matter sources may even be absent from the pre-big bang phase, but
they are certainly not absent when  dilatons re-enter the horizon, in the
radiation- and matter-dominated epoch. 

In order to discuss this effect, we shall work in the E-frame, in the
longitudinal gauge, and we shall assume that the $d$-dimensional,  
homogeneous and isotropic, unperturbed background is driven by the
dilaton and by a perfect-fluid model of  matter sources. Our 
independent variables are  the scalar metric, dilaton and fluid
perturbation, respectively  $\varphi$, $\psi$, $\chi$, $\da \r$, $\da p$,
$\da u_i$, defined by: 
\bea
&&ds^2=a^2\left[d\eta^2\left(1+2\varphi\right)-
\left(1-2\psi\right)dx_i^2\right] , ~~~~~~~~~~ \da \phi =\chi, 
\nonumber \\
&& \da T_0^0=\da \r , ~~~~~~~~~
 \da T_i^j=-\da p \da_i^j , ~~~~~~~~~~~~
 \da T_i^0={p+\r \over a}\da u_i. 
\label{61}
\eea

In the E-frame the fluid sources are non-trivially coupled to the
dilaton, and the unperturbed equations take the form (in units $16\pi
G=1$):
\bea
&&
2R_\mu~^\nu-\da_\mu^\nu R = \pa_\mu \phi \pa^\nu \phi +
\da_\mu^\nu \left[V-{1\over 2}(\nabla \phi)^2\right] +T_\mu~^\nu 
\nonumber \\
&&
\nabla_\mu \nabla^\mu \phi + {\pa V\over \pa \phi} + c T =0 ,
\label{62}
\eea
where the numerical factor $c=\sqrt{2/(d-1)}$ arises
from the conformal transformation of the S-frame string cosmology
equations (see for instance Eq. (\ref{2112})). The perturbation of these
equations, for the $i\not=j$ components, gives $\varphi=\psi$, while the
combination of the perturbed $(0,0), (i,i)$ and dilaton equation leads to
the vector-like  equation for the doublet $Z_k$, representing the Fourier
components of the metric and of the dilaton fluctuations \cite{GasVe94a}: 
\beq
Z_k''+2{a'\over a}{\cal A}Z_k' +\left(k^2{\cal B} + {\cal C} \right) Z_k 
=0, ~~~~~~~~~~~~~Z_k^{\dagger}=(\psi_k, \chi_k).
\label{63}
\eeq

Here ${\cal A, B, C}$ are $2\times 2$, time-dependent mixing matrices,
whose explicit form depends on the unperturbed solutions, 
on the equation of state $\ga(t)=p/\r$, on the model of fluid
perturbations $\ep (t)= \da p/\da \r$ and, finally, 
on the possible presence of a dilaton potential $V(\phi)$. In particular,
for an unperturbed background in which the dilaton can be parametrized
by $\phi= \b \ln a$, with $\b=$ const, the mixing matrices are given by 
\bea
&&
{\cal A}=\pmatrix{{1\over 2}(2d-3+d\ep), & -{1-\ep \over 4(d-1)}\b
\cr -(d-1)[{cd\over 2}(1-d\ep)+\b], & {1\over 2}(d-1)-{c\over
4}(1-d\ep)\b \cr},
\label{64}\\
&&
{\cal B}=\pmatrix{\ep, & 0\cr -c(1-d\ep)(d-1), & 1 \cr},
\label{65}\\
&&
{\cal C}=\pmatrix{2(d-2){a'' \over a} +(d-2)[d-4+d\ep + {1-\ep
\over 2(d-1)} \b^2]\left(a'\over a \right)^2, & 
{\ep+1\over 2(d-1)}{V'} a^2 \cr
cd(d-2)(\ep-\ga)[d(d-1)-{\b^2\over 2}]\left(a'\over a \right)^2 
+2(d-2){V'} a^2 , 
& [{V''} -{c\over 2}(1-d\ep)
{V'}] a^2  \cr}, \nonumber\\
&&
\label{66}
\eea
where the primes denote differentiation with respect to conformal
time, except for the potential, where $V'=\pa V/\pa \phi$.
Once $Z$ is
known, the density and velocity contrast are fixed in terms of $\psi$
and $\chi$ by two additional equations
$\da \r= \da \r (\psi, \chi)$, $\da u= \da u (\psi, \chi)$,
following, respectively, from the perturbation of the $(0i)$ and $(00)$
background equations:
\bea
&&
\pa_i\left[2(d-1)\left({a'\over a}(d-2)\psi
+\psi'\right)-\chi \phi' \right] =(\r+p)a\da u_i , 
\label{67}\\
&&
\nabla^2\psi -d{a'\over a}\psi' -\left[d(d-2)\left(a'\over a \right)^2
-{d-2\over 2(d-1)} \phi^{\prime 2} \right]\psi =\nonumber \\
&&
={1\over 2(d-1)}
\left(\phi ' \chi' +{\pa V\over \pa \phi} a^2\chi + a^2\da \r \right).
\label{68}
\eea

An exact computation of the dilaton fluctuations thus requires an 
 exact solution of the coupled equations (\ref{63}). Also, the
correct normalization of the dilaton spectrum  to 
the quantum fluctuations of the vacuum  requires in general the
diagonalization of the system {\sl metric + dilaton + fluid sources}, to
find the normal modes of oscillation  satisfying canonical
commutation relations, as discussed in Subsection \ref{Sec4.3}.

This program can be easily performed if we limit ourselves to a
dilaton-dominated pre-big bang phase with negligible matter sources
($T_{\mu\nu}=0=\da T_{\mu\nu}$), and we consider the high-frequency
part of the spectrum, i.e. those modes re-entering the horizon during the
radiation-dominated phase, characterized by adiabatic fluid
perturbations ($\ga=\ep=1/d$) and by the dilaton frozen ($\b=0$) at the
minimum of the non-perturbative potential ($\pa V /\pa \phi=0$).  In
that case the canonical variables are known, and one finds that the 
perturbation equations are decoupled in the radiation era, where the
canonical variable for the dilaton fluctuation is simply (in $d=3$) 
$\overline \chi = a \chi$, and satifies the (covariant) Klein--Gordon
equation (from the system (\ref{63})): 
\beq
\overline \chi_k^{\se}+ (k^2+m^2a^2)\overline \chi_k =0. 
\label{69}
\eeq

For relativistic modes, of proper momentum $p=k/a \gg m$, the solution
describes (in conformal time) free plane-wave oscillations, and we thus
recover for the dilaton the same spectrum as in the graviton case
\cite{GasVe94a}. Working in the framewok of minimal pre-big bang
models (see Subsection \ref{Sec5.2}) we find, in particular, that the
dilaton modes amplified during the low-energy phase are characterized
by the distribution $\Om_\chi(p) \sim p^3 \ln^2 p$. In order to include the
possible corrections due to the modified kinematics of the
high-curvature string phase, we may thus parametrize the
high-frequency, relativistic branch of the dilaton spectrum as  \beq
\Om_\chi(p, t)= g_1^2
\left(H_1\over H\right)^2 \left(a_1\over a\right)^4
\left(p\over p_1\right)^\da , ~~~~~~~~~~~ m< p<p_1. 
\label{610}
\eeq
Here $0 \leq\da \leq 3$ is the slope parameter corresponding to a 
growing, but unknown, spectral index; $g_1=H_1/M_{\rm P}$ and
$p_1=k_1/a\simeq H_1a_1/a$ are the parameters associated to the
transition scale $H_1 \simeq M_{\rm s}$ of  minimal pre-big bang models.
It should be noted that, unlike  the graviton case, we no longer 
identify  proper momentum $p$ and proper frequency $\om$,
as we are dealing, in principle, with massive particles for which 
$\om^2=p^2+m^2$. 

The above spectral distribution  cannot
be extrapolated down to momentum scales re-entering the
horizon after equilibrium, since in the matter era the
perturbation equations  (\ref{63}) are no longer decoupled, even if the
dilaton background is frozen at the minimum of the potential (they
remain coupled not only in the longitudinal gauge, but also in the
uniform-curvature gauge, defined in Eq. (\ref{441})). In addition, for
modes of low enough momentum $p \laq m$, we must take into account
the non-relativistic correction to the above spectrum (indeed, even if the
mass term is negligible at the inflation-radiation transition scale, i.e.
$k<ma_1$, the proper momentum is redshifted with respect to the mass
in the subsequent radiation era, and all modes tend to become
non-relativistic). 

When including the mass term, the general solution of Eq. (\ref{69}) in
the radiation era can be written in terms of the parabolic cylinder
functions \cite{AbSte}, and its matching to the (massless) pre-big bang
solution provides the correct inclusion of the mass contributions to the
spectrum. The details of the computation are the same as in the case of
massive axions \cite{DurGas99}, and will be reported in Section
\ref{Sec7}. Here we simply note that if a mode becomes
non-relativistic well inside the horizon, i.e. when $p(t)  >H$, then the
number of  produced dilatons is fixed after re-entry, when the mode
is still relativistic, and the effect of the mass  is a
simple rescaling of the energy density: $\Om_\chi \ra (m/p)
\Om_\chi$. If, on the contrary, a mode becomes non-relativistic
outside the horizon,  when $p(t) <H$, then the final energy
distribution turns out to be determined by the background
kinematics at the exit time, and the effective  number of
non-relativistic dilatons has to be adjusted by continuity, in such a way
that $\Om_\chi$ has the same spectral distribution as in the absence of
mass. The two regimes are separated by the limiting momentum scale 
$p_m$ of a mode that  becomes non-relativistic just at the time
it re-enters the horizon \cite{Gas94b,GasVe94a}:
\beq
p_m=p_1\left(m\over H_1\right)^{1/2}.
\label{611}
\eeq

Taking into account such mass corrections, we may thus add to the
relativistic dilaton spectrum (\ref{610}) two lower-frequency,
non-relativistic bands (see also \cite{Gas00e}):
\beq
\Om_\chi(p, t)= g_1^2{m\over H_1}
\left(H_1\over H\right)^2 \left(a_1\over a\right)^3
\left(p\over p_1\right)^{\da-1} , ~~~~~~~~~~~ p_m< p<m, 
\label{612}
\eeq
\beq
\Om_\chi(p, t)= g_1^2 \left(m\over H_1\right)^{1/2}
\left(H_1\over H\right)^2 \left(a_1\over a\right)^3
\left(p\over p_1\right)^\da , ~~~~~~~~~~~  p<p_m. 
\label{613}
\eeq
Note that these spectral distributions evolve in time like $a^{-3}$, and
thus keep constant during the matter-dominated era, for $t>t_{\rm eq}$.
Note also that, for simplicity, the non-relativistic corrections have been
referred  to a single relativistic branch, with $\da =3$
for the low-energy pre-big bang phase, and with $\da \leq 3$ for the
string phase. However, the mass corrections can  be extended
without  difficulty to a more general situation in which there are two
(or more)  branches in the relativistic spectrum, as in minimal pre-big
bang models characterized by the two frequency bands $p<p_{\rm s}$ 
and $p>p_{\rm s}$ (separated by the string scale $p_{\rm s}=H_{\rm s}$). 
In such a case, the non-relativistic corrections will affect only the lower,
or also the higher, frequency band of the spectrum,  depending on
whether  $m<p_{\rm s}$ or $m>p_{\rm s}$, respectively. 

If the dilaton mass is large enough, the present amplitude of a possible
relic dilaton background is thus controlled by the (unknown) value of its
mass. The cosmological bounds on the spectrum can thus be translated
into phenomelogical bounds on the mass, depending (but only weakly)
on the spectral index $\da$. Such bounds are to be combined with other,
model-dependent bounds on the mass related to the strength of the 
 effective dilaton coupling to macroscopic matter, which
will be discussed in the next subsection.

\subsection{Effective coupling to macroscopic matter}
\label{Sec6.2}

The motion of macroscopic test bodies in an external, non-trivial, 
gravidilaton background is in general non-geodesic, if the body
posseses a non-negligible dilatonic charge. Suppose in fact that the
interaction of macroscopic matter with the external scalar-tensor
background is described by the effective action of Subsection
\ref{Sec4.1}:  
\beq
S=\int d^4x \sqrt{-g} e^{-\phi} \left[ -R +\om \left(\nabla
\phi\right)^2\right] + S_M(g,\phi)
\label{614}
\eeq
(in units $ 2\la_{\rm s}^2=1$), 
where the Brans--Dicke parameter $\om$ is equal to $-1$ for the
lowest-order action of string theory, and we have added the action
$S_M$ for the matter fields describing a test body non-trivially 
coupled to the metric and to the dilaton. The variation with respect to
$g_{\mu\nu}$ and $\phi$ provides respectively the field equations 
(with $16\pi G=1$): 
\bea
&& 
G_{\mu\nu} + \nabla_\mu\nabla_\nu\phi -(\om+1) 
\nabla_\mu\phi\nabla_\nu\phi -g_{\mu\nu} \nabla^2 \phi
+\left({\om\over 2} +1\right)g_{\mu\nu} \left(\nabla \phi\right)^2
={1\over 2} e^\phi T_{\mu\nu}, 
\nonumber\\
&&
R +\om \left(\nabla \phi\right)^2 - 2 \om \nabla^2 \phi+e^\phi \sg =0,
\label{615}
\eea
containing two source terms: the
``gravitational charge density" $T_{\mu\nu}$ (the dynamical stress
tensor), and the dilatonic charge density $\sg$:
\beq
\sqrt{-g} ~T_{\mu\nu}= {2} {\da S_M \over \da
g^{\mu\nu}}, ~~~~~~~~~~~~~~~~
\sqrt{-g} ~\sg= {\da S_M\over \da\phi}. 
\label{616}
\eeq

The combination of the above equations, and the use of the contracted
Bianchi identity $\nabla_\nu G^{\mu\nu}=0$, then implies the covariant
conservation of the stress tensor, $\nabla_\nu T^{\mu\nu}=0$, {\em for
any} $\om$, {\em provided} $\sg=0$. If the dilatonic charge is non-zero,
however, the gravitational charge density is not separately conserved;
for $\om=-1$, in particular, we obtain from Eqs. (\ref{615}):
\beq
\nabla^\nu T_{\mu\nu}+\sg \nabla_\mu \phi=0.
\label{617}
\eeq

To obtain the equation of motion of the test body we now integrate this
equation over a space-like hypersurface, and perform a multipole
expansion of the external metric and dilaton field around the world line
$x^\mu(\tau)$ of the centre of mass of the body \cite{Pap51}. In
the point-particle approximation, the gravitational and dilatonic charge
densities are defined by 
\bea
&&
T^{\mu\nu}(x')= {p^\mu p^\nu \over \sqrt{-g}p^0}\da^{(3)}
\left(x'-x(\tau)\right),
\nonumber \\
&&
\sg(x')= q {m^2\over \sqrt{-g}p^0}\da^{(3)}
\left(x'-x(\tau)\right) 
\label{618}
\eea
(where $p^\mu= mu^\mu=m dx^\mu/d\tau$), and we obtain the
general equation of motion \cite{Gas99z}
\beq
{d u^\mu\over d\tau} + \Gamma_{\a\nu}\,^\mu u^\a u^\nu + q 
\nabla^\mu \phi=0,
\label{619}
\eeq
where $q$ is the
dilatonic charge per unit of gravitational mass (representing  the relative
strength of scalar to tensor forces). 

It is now evident that a non-zero $q$ induces deviations from pure
geodesic motion, and that the experimental tests on the gravitational
forces provide limits on the allowed values of the dilaton charge. Large
values of $q$ can only be allowed if the dilatonic interactions, on a
macroscopical scale, are suppressed by a sufficiently small range. We
may also note, for future reference, that the relative acceleration
between two world lines satisfying Eq. (\ref{619}), with infinitesimal
separation $\eta_\mu$, is given by a generalized equation of ``geodesic
deviation" \cite{Gas99z}: 
\beq
{D^2 \eta^\mu \over D\tau^2} + R_{\b\a\nu}\,^\mu u^\a u^\nu \eta^\b+
q\eta^\b \nabla_\b\nabla^\mu \phi =0.
\label{620}
\eeq
We shall come back to this point in Subsection \ref{Sec6.4}. 

A precise estimate of $q$, in a string theory context, should start from
the effective action describing the coupling of the dilaton to the
fundamental quark and lepton fields building up ordinary macroscopic
matter. Including all possible loop corrections, such an action can be
written in the general form
\beq
S= \int d^{d+1}x \sqrt{-g} \left[ -Z_R (\phi) R -Z_\phi (\phi)\left(\nabla
\phi\right)^2 -V(\phi) + {1\over 2} Z_k^i (\phi)\left(\nabla
\psi_i\right)^2 + Z_m^i (\phi)\psi_i^2 \right], 
\label{621}
\eeq
where $Z^i$ are the dilatonic ``form factors", to all orders in the loop
expansion, and where we have used, for simplicity, a scalar model of
fundamental matter fields $\psi_i$. The effective masses ($m_i$) and
dilaton couplings ($g_i$) of the fields $\psi_i$ can then be defined in
terms of the canonical rescaled variables $\widehat\psi_i$, which
diagonalize the kinetic term and have canonical dimensions. In the
low-energy limit we may expand the $\phi-\psi$ interaction Lagrangian 
around the value  $\phi_0$ of the dilaton, which extremizes the
effective  potential \cite{Tay88,GasVe94a}:
\beq
 L(\phi,  \psi_i)\equiv Z^i_m \psi_i^2 = {1\over 2} m_i^2 \widehat 
\psi_i^2 + {1\over 2}  g_i   \phi \widehat\psi_i^2+ ...
\label{622}
\eeq
Performing the computation in both the String and Einstein frames
\cite{Gas99z}, we find that the dilatonic charges per
unit of gravitational mass, respectively $q_i$ and $\ti q_i$   in the two
frames, are related by (in units $16\pi G=1$):
\beq
 q_i \simeq {g_i\over m_i^2} =\left[{\pa\over \pa \phi }\ln \left(
Z^i_m \over Z^i_k \right)\right]_{\phi=0}= \ti q_i-1. 
\label{623}
\eeq
Unless $\ti q_i$ is fine-tuned to 1 (corresponding to an exact
Brans--Dicke theory for the field $\psi_i$), the dilatonic charge is
non-vanishing in both frames and, in general, is non-universal, as the
form factors $Z_i(\phi)$ are different for different fields. 

As originally stressed in \cite{Tay88}, this suggests a large dilatonic
charge ($q_i \sim 40$ -- $50$) for the confinement-generated
components of hadronic masses, and a smaller charge, but of
gravitational intensity ($q_i \sim 1$), for leptonic masses (see also
\cite{Ellis89,Anto99}). If this is the case, the total charge of a
macroscopic body is large (in gravitational units), and
composition-dependent. 

Consider in fact a macroscopic body of mass $M$, composed of
$B$ baryons with mass and charge $m_b, q_b$, and $Z$ electrons 
with mass and charge $m_e, q_e$. The total dilatonic charge per unit
mass, $q=\sum_i m_i
q_i/\sum_i m_i$, for $Z\sim B$, $m_e \ll
m_b$, $q_e \ll q_b$ is then 
\beq
q \simeq {Bm_b q_b\over M}= \left(B\over \mu\right)q_b,
\label{624}
\eeq
where $\mu=M/m_b$ is the mass of the body in units of baryonic
masses. Since $B/\mu \sim 1$, the total dilaton charge of a
macroscopic body is controlled by the dilaton coupling to baryons,
$q\sim q_b \gg 1$, it is
of the same order of magnitude in the String and Einstein frames,
$q\sim \ti q$, and it is composition-dependent (as $B/\mu$ depends on
the internal nuclear structure), with variations, across different types of
ordinary matter, which are typically of order \cite{Gas99z}
\beq
{\Da q\over q} \simeq \Da \left(B\over \mu\right) \sim 10^{-3}.
\label{625}
\eeq

The only way to escape the conclusion that the dilaton charge of
macroscopic bodies is composition-dependent and large is to assume
that the coupling of the dilaton to the fundamental matter fields is
universal (the same for all fields, to all orders in the loop expansion), 
and to choose appropriate values of the parameters in the resulting
scalar-tensor effective action. This is not impossible, as illustrated by
the model discussed in \cite{DP94a,DP94b}. In such a context, if the
effective coupling is appropriately suppressed ($q \ll 1$), the dilaton
mass can be very small, or even vanishing. The precise measurements
of the gravitational  force over  quite a wide range of
distances provide in fact exclusion plots in the plane ($q^2,
m^2$), which constrain the dilaton charge as a function of its mass (see
for instance \cite{Fis92} for a comprehensive compilation of experimental
bounds on possible ``Yukawa" deviations from Newtonian gravity, both
for universal and composition-dependent interactions). 

For later use, let us recall here the phenomenological bound on the
dilaton mass for a large, composition-dependent dilaton charge
\cite{Fis92}:
\beq
m \gaq 10^{-4}~ {\rm eV}.
\label{626}
\eeq
In the opposite, small-coupling regime, and in the mass range
 appropriate to gravitational antennas, i.e. from $10^{-14}$ to 
$10^{-12}$ eV, corresponding to the frequency range from $10$ to $1000$
Hz, the phenomenological  bounds \cite{Fis92} can be parametrized as 
\bea
&&
\log q^2 \laq -7,~~~~~ ~~~~~ ~~~~~ ~~~~~~~~~~~~~~~
1~ {\rm Hz} \laq m \laq 10~{\rm Hz}, 
\nonumber\\
&&
\log q^2 \laq -7 +\log (m/10~{\rm Hz}), ~~~~~ ~~~
10 ~{\rm Hz} \laq m \laq 1 ~{\rm kHz},
\label{627}
\eea
for universal dilaton interactions, and
\bea
&&
\log q^2 \laq -8,~~~~~ ~~~~~ ~~~~~ ~~~~~~~~~~~~~~~
1 ~{\rm Hz} \laq m \laq 10 ~{\rm Hz}, 
\nonumber\\
&&
\log q^2 \laq -8 +\log (m/10~{\rm Hz}), ~~~~~ ~~~
10 ~{\rm Hz} \laq m \laq 1 ~{\rm kHz},
\label{628}
\eea
for composition-dependent interactions.  

We can now discuss the intensity of the relic-dilaton background, taking
into account two distinct phenomenological possibilities: 1) 
(heavy) massive dilatons, gravitationally (or more strongly) coupled to
macroscopic matter, and 2) massless (or very light) dilatons, universally
(and weakly enough) coupled to matter. In the last case, the
conclusions about a possible detection of the dilaton background
produced by pre-big bang models are not completely negative, as we
shall illustrate in Subsection \ref{Sec6.4}.

\subsection{Bounds and allowed windows for heavy (strongly coupled) 
dilatons} 
\label{Sec6.3}

Let us first discuss the large-coupling case $q \gaq 1$ (the most
unfavourable, probably, for a direct detection). In such a case 
$m \gaq 10^{-4}$ {\rm eV}, so that today even the highest mode present
in the spectrum, $p_1(t_0)$, is non-relativistic (see Eq. (\ref{516})):
\beq
p_1(t_0)={k_1\over a(t_0)}\simeq \left(H_1\over M_{\rm P}\right)^{1/2}
10^{11} ~{\rm Hz}\simeq \left(M_{\rm s}\over M_p\right)^{1/2} 
10^{-4}~ {\rm eV}  < m  
\label{629} 
\eeq
(since $M_{\rm s}<M_{\rm P}$). In other words, the branch (\ref{612})  of the
dilaton spectrum extends from $p_m$ to $p_1$. We shall start assuming
that the slope is steep enough, $\da >1$, for the high-frequency branch
from $p_m$ to $p_1$ to represent the dominant part of the spectrum. In
that case all the cosmological bounds that apply to the total integrated
dilaton energy density will become $\da$-independent
\cite{Gas94b,GasVe94a} (the case $\da <1$ will be considered below). 

When the spectrum becomes non-relativistic, the dilaton energy density
$\rho_\chi \sim a^{-3}$ starts to grow  with respect to the
radiation energy density, $\rho_\ga \sim a^{-4}$. In particular, 
when the Universe enters the matter-dominated era, the
non-relativistic part of the dilaton spectrum  remains
frozen (in critical units) at the value reached at the time of the
matter-radiation equilibrium, $\Om_\chi(t_{\rm eq})$. The present value
of the dominant (non-relativistic) branch of the spectrum can thus be
expressed in terms of the equilibrium scale as: 
\beq
\Om_\chi (t_0,p)\equiv 
\Om_\chi (t_{\rm eq}, p)\simeq\left(H_1\over M_{\rm P}\right)^2
\left(m^2\over H_1H_{\rm eq}\right)^{1/2}\left(p\over
p_1\right)^{\da-1}, ~~~~~~~~ p_m<p<p_1, 
\label{630}
\eeq
where we have used Eq. (\ref{612}), and the fact that $(H_1/H_{\rm
eq})^2 (a_1/a_{\rm eq})^3$ = $(H_1/H_{\rm eq})^{1/2}$, assuming
radiation-dominance from $t_1$ to $t_{\rm eq}$. We recall that, for the
equilibrium curvature scale,  $H_{\rm eq}\sim 10^6 H_0\sim
10^{-55}M_{\rm P}$. 

A first important constraint \cite{ElNa86,ElTsa86,DeCa93} is now 
obtained  by imposing that the energy density stored in the coherent
oscillations of non-relativistic dilatons, integrated over all modes,
remains smaller than critical, at all times:
\beq
\int^{p_1} d \ln p ~\Om_\chi(t_0,p) \laq 1.
\label{631}
\eeq
For $\da >1$ this gives the stringent constraint
\beq
m\laq \left(H_{\rm eq}M_{\rm P}^4\over H_1^3\right)^{1/2} , 
\label{632}
\eeq
which represents, in this context, the most restrictive upper bound on
the mass of (not yet decayed) dilatons. For $100$ keV $\laq m  
\laq 100$ MeV a more restrictive constraint on $\Om_\chi$ is provided
in principle by the observations of the astrophysical background of 
diffuse $\ga$-rays \cite{DaVi96a};  such a range of masses, however,
tends to be  excluded, as will be shown below,  in the context of pre-big
bang models where the inflation scale is controlled by the string mass, 
$H_1\simeq M_{\rm s}$. 

The critical density bound (\ref{632}) can be evaded if the dilatons are
heavy enough to decay, so that the energy stored in their coherent
oscillations was dissipated into radiation before the present epoch.  The
decay scale $H_d$ is fixed by the decay rate (for instance into
two photons) as 
\beq
H_d\simeq \Ga_d \simeq {m^3/M_{\rm P}^2}
\label{633}
\eeq
(which implies, in particular, that the produced dilatons have not yet
decayed, today, only if  $m \laq 100$ MeV). The
decay, however, generates radiation, reheats the Universe, and is
associated in general to a possible entropy increase 
\beq
\Da S \simeq
\left(T_r/T_d\right)^3,
\label{634}
\eeq
 where $T_r$ and $T_d$ are   the final reheating
temperature and the  radiation temperature immediately
before dilaton decay, respectively. The decay process is thus  the source
of additional phenomenological constraints on the dilaton spectrum,
which are complementary to the critical bound (\ref{632}), as they imply
in general a lower bound on the dilaton mass. 

In order to discuss the consequences of dilaton decay let us note, first
of all, that the induced reheating is significant ($T_r>T_d$, $\Da s > 1$)
provided dilatons decay when they are dominant with respect to
radiation, i.e. for $H_d <H_i$, where $H_i$ is the curvature scale marking
the beginning of dilaton dominance. Such a scale is defined by the
condition $\Om_\chi(t_i)= \Om_\ga(t_i)$, where $\Om_\chi(t_i)$ is
obtained by integrating over all modes the dominant branch (\ref{612})
of the non-relativistic dilaton spectrum. For $\da >1$ this leads to 
\beq
{m H_1/ M_{\rm P}^2} \simeq {a_1/ a_i},
\label{635}
\eeq
from which
\beq
H_i \simeq 
m^2 {H_1^3/M_{\rm P}^4}
\label{636}
\eeq
(we have assumed that such a scale belongs to the radiation era, $H_i >
H_{\rm eq}$; otherwise, the bound we obtain is less stringent). 

At the given scale $H_i$, the radiation temperature is fixed by the
Einstein equations (and by the Sefan law $\rho \sim T^4$) as
\beq
T_i= k (M_{\rm P} H_i)^{1/2},
\label{637}
\eeq
where $k$ is a dimensionless numerical factor of order unity, which 
disappears from the final estimate. The radiation temperature at the
epoch of dilaton decay, $T_d$, is then obtained by the adiabatic rescaling
\beq
T_d =\left(a_i\over a_d\right) T_i = \left(H_d\over H_i\right)^{2/3} T_i
\simeq k \left(m^{10}\over M_{\rm P} H_1^3\right)^{1/6},
\label{638}
\eeq
since the scale factor evolves like $t^{2/3}$ during the phase dominated
by the non-relativistic dilatons. After the decay, on the other hand, the
Universe becomes radiation-dominated, with a reheating temperature
fixed by the decay scale as
\beq
T_r= k (M_{\rm P} H_d)^{1/2}= k\left(m^{3}\over M_{\rm P}\right)^{1/2} ,
\label{639}
\eeq
and with a final, total entropy increase \cite{Gas94b,GasVe94a}:
\beq
\Da S \simeq
\left(H_1^3\over m M_{\rm P}^2\right)^{1/2}.
\label{640}
\eeq
This entropy injection can in principle disturb nucleosynthesis and/or
baryogenesis \cite{ElNa86,ElTsa86,DeCa93}, and we should consider two
possibilities. 

If the reheating temperature $T_r$ is too low to allow
nucleosynthesis, $T_r <1$ MeV, i.e. $m\laq 10$ TeV, we must assume
that nucleosynthesis occurred before, and we must impose 
\beq
\Da S \laq10, ~~~~~~~~~~~~~~~~~~~m\laq 10~ {\rm TeV},
\label{641}
\eeq
 to avoid destroying the light nuclei already formed (more precise
bounds can also be determined through a detailed analysis of
photodissociation \cite{ElNaSa85} and hadroproduction \cite{Dimo88}). 

If the reheating temperature is  large enough to allow
nucleosynthesis, $T_r>1$ MeV, i.e. $m\gaq  10$ TeV, the only possible
constraint comes from primordial baryogenesis. The bound is
model-dependent, but the constraint
\beq
\Da S \laq10^5, ~~~~~~~~~~~~~~~~~~~m\gaq 10~ {\rm TeV},
\label{642}
\eeq
seems to be sufficient \cite{ElNa86,ElTsa86,DeCa93} 
not to wash out any pre-existing baryon--antibaryon asymmetry
(this bound could be evaded in the case of low-energy
baryogenesis occurring at a scale $H<H_d$).

The above constraints can be easily extended to the case $\da \leq 1$, 
see \cite{Gas94b,GasVe94a}. Consider, for instance, the critical density
bound. For $0 \leq \da \leq 1$ the non-relativistic spectrum (\ref{630})
has a peak at $p=p_m=p_1(m/H_1)^{1/2}$, which dominates the integral
(\ref{631}), and which leads to a total integrated energy density:
\beq
\Om_\chi(t_0)\simeq \left(H_1\over M_{\rm P}\right)^{2}
\left(m^2\over H_{\rm eq}H_1\right)^{1/2} 
\left(m\over H_1\right)^{\da-1\over 2}.
\label{643}
\eeq
The bound $\Om_\chi <1$ now imposes \cite{Gas94b,GasVe94a}
\beq
m \laq M_{\rm P}^{4\over \da +1}H_{\rm eq}^{1\over \da +1}
H_1^{\da-4\over \da+1}, ~~~~~~~~\da \leq 1,
\label{644}
\eeq
which for $\da=1$ exactly reduces to the condition (\ref{632}). 

For the
dilaton decay the situation is similar: if  $\da\leq1$ the total energy
density is dominated by the contribution of $p=p_m$, and  all bounds
become  $\da$-dependent. The scale $H_i$ of dilaton dominance,
obtained by equating $\Om_\chi (p_m)$ to $\Om_\ga$ in the
radiation era,  becomes $ H_i \simeq m^{\da+1}H_1^{4-\da}M_{\rm
P}^{-4}$, and the corresponding increase of entropy is
\beq
\Da S = \left(T_r \over T_d \right)^3= \left(H_i \over H_d \right)^{1/2}
\simeq \left(m^{\da-2} M_{\rm P}^{-2}H_1^{4-\da}\right)^{1/2}.
\label{645}
\eeq

Taking into account all bounds, we obtain the allowed region in the plane
$(m,H_1)$ illustrated in Fig. \ref{f61}. The allowed values are below the
upper (bold) solid lines for a growing ($\da \geq 1$) spectrum
\cite{Gas94b}, and below the lower (thin) solid lines for a flat ($\da=0$)
spectrum \cite{Goncha84,ElNa86,ElTsa86}. For $0<\da<1$ the allowed
region interpolates between the two limiting cases illustrated in the
figure. 

To the left, the region is bounded by the tests of the equivalence
principle, Eq. (\ref{626}), as we are considering dilatons with large
couplings to macroscopic matter, $q\gaq 1$. To the right, the mass is
bounded by the condition $m<M_{\rm P}$ (not shown in the picture),
which has to be imposed to avoid overcritical density in case 
$m>H_1$, when  the produced dilatons are already non-relativistic from
the beginning \cite{GasVe94a} (in that case there are no additional
bounds, as dilatons decay before becoming dominant). Finally, the upper
limit on the inflation scale, $H_1<M_{\rm P}$, is required to avoid
overcritical production of massless particles since, for a growing
spectrum, $\int d \ln p \Om (p) \sim g_1^2 = (H_1/M_{\rm P})^2$ (see for
instance Eq. (\ref{610})). 

\begin{figure}[t]
\centerline{\epsfig{file=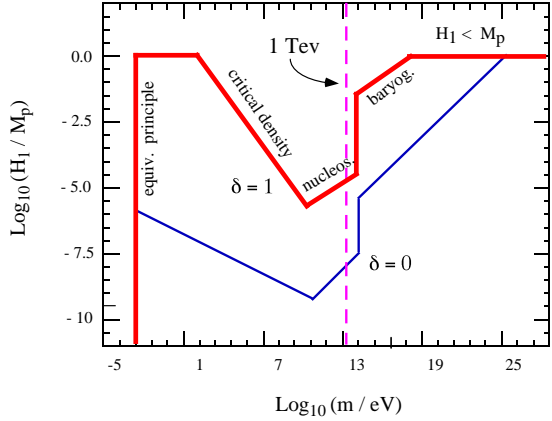,width=72mm}}
\vskip 5mm
\caption{{\sl Allowed region 
for a growing (below the bold lines) and flat (below the thin lines) 
dilaton spectrum, for strongly coupled dilatons.}}
\label{f61}
\end{figure}

In the standard inflationary scenario the spectrum  tends to be flat,
and the ``natural" value for the dilaton mass required by models of
supersymmetry breaking \cite{Banks93}, $m \sim 1$ TeV, is not
consistent with the preferred value of the inflation scale, $H_1\sim
10^{-5} M_{\rm P}$, typically determined by the grand-unification phase
transition (this is the so-called problem
of the dilaton mass).  In string  cosmology the spectrum tends to be
growing, the allowed region ``inflates" in parameter space, and the TeV
mass scale becomes marginally compatible with the wanted inflation
scale. Unfortunately, however, in a string cosmology context, $H_1$ is
expected to be determined by the string scale, and thus fixed
around $(0.1$ -- $0.01) M_{\rm P}$, so that the
problem remains. 

A possible solution of this problem can be obtained in the presence of a
late, post-inflationary reheating phase,  preceding nucleosynthesis and
dilaton decay,  and  producing a suitable dilution of the original dilaton
density by the factor \cite{Gas97a} 
\beq
\Om_\chi \ra \Om_\chi(1-\da s)^{4/3}\left(n_f\over n_b\right)^{4/3}, 
~~~~~~~~~ \da s ={(s_f-s_b)\over s_f} . 
\label{646}
\eeq
Here $s_b, s_f, n_b, n_f$ are,  respectively, the thermal entropy
density of the CMB radiation and the number of particles species in
thermal equilibrium, at the beginning ($t_b$) and at the end ($t_f$) of
this  reheating process. An efficient reheating, $s_f \gg s_b$, $\da
s\ra 1$, reduces in a significant way the dilaton fraction of critical
density $\Om_\chi$: as a consequence, the scale (\ref{636}) of dilaton
dominance is lowered, the decay temperature (\ref{638}) is raised, and
the bounds on $m$ following from the entropy constraint ($\Da S <10$)
and the critical bound ($\Om_\chi<1$) are relaxed. Such a reheating is 
possible, for instance, in the context of models of 
``intermediate scale" inflation \cite{Ran95}, or ``thermal" inflation
\cite{LySte96}. 

Even without such a relaxation, the minimal pre-big bang scenario 
leaves open two interesting mass windows,  obtained by intersecting
the allowed region of  Fig. \ref{f61}, for $\da \geq 1$, with the allowed
value of the string mass scale \cite{Kap85}, 
\beq
0.01 \laq{M_{\rm s}/ M_{\rm P}} ~\laq~ 0.1 
\label{647}
\eeq
(under the assumption $H_1 \simeq M_{\rm s}$). It follows that there are two
possible ranges for the dilaton mass \cite{Gas97a}:
\beq
10^{-4}~ {\rm eV} \laq m  \laq  10~ {\rm keV}, ~~~~~~~~~~~~~
m \gaq 10~ {\rm TeV}
\label{648}
\eeq
illustrated in Fig. \ref{f62}. The vertical solid line at the $100$ MeV mass
scale, shown in Fig. \ref{f62}, corresponds to a dilaton decay rate of
the same order as the present Hubble scale,  
$H_d \simeq m^3/M_{\rm P}^2 
\simeq H_0$. To the right of this line all produced dilatons have already
decayed, while to the left, and in particular inside the window of Eq.  
(\ref{648}), dilatons are still around us, and could in principle be
observed. 

\begin{figure}[t]
\centerline{\epsfig{file=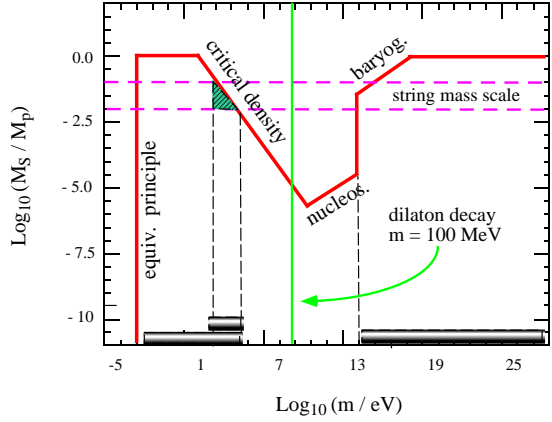,width=72mm}}
\vskip 5mm
\caption{\sl Allowed mass windows for heavy (strongly coupled)
dilatons in  minimal pre-big bang models 
with $\da \geq 1$. The region within the shaded triangle corresponds to
a significant dilaton contribution to the present cold dark matter
density.}  
\label{f62}
\end{figure}

If, in addition, the mass is within the restricted range
\beq
100~ {\rm eV} \laq m \laq  10~{\rm keV},
\label{649}
\eeq
obtained by intersecting the string mass scale with the critical 
density bound (\ref{632}), then 
the produced dilatons could even saturate the critical density
value, or at least could provide a significant fraction of the
present cold dark-matter density \cite{Gas94b}, with
\beq
0.01 \laq  \Om_\chi \laq 1,
\label{650}
\eeq
according to Eq. (\ref{630}). 
It is important to recall, in such a context, that the possibility  of a 
small dilaton mass is not theoretically excluded (in spite of the
``standard" TeV mass range), as shown for instance by models of
supersymmetry breaking with light dilatons \cite{FKZ94}.   
Also, it is important to stress that the range of masses saturating the
critical density bound  moves to lower and lower values as the spectrum
becomes flatter and flatter (see Eq. (\ref{644}) with $H_1=M_{\rm s}$). 
For $\da \ra 0$, in particular, the critical bound is saturated by 
\beq
m=H_{\rm eq} \left(M_{\rm P}/M_{\rm s}\right)^4,
\label{651}
\eeq
corresponding to masses as small as $10^{-19}$ eV (or  lower). 

Such values of mass are  excluded  neither theoretically nor
experimentally, as discussed in Subsection \ref{Sec6.2}, provided the
dilaton coupling to macroscopic matter is sufficiently small, $q \ll 1$. It
is then possible to imagine a cosmological scenario in which a relic
background of  non-relativistic dilatons, produced as a consequence of
the pre-big bang evolution, could provide today   a significant
contribution to the critical energy density and, simultaneously, 
have a  mass  in the appropriate range to stimulate a resonant
response of present gravitational antennas \cite{Gas00b} (i.e. $m \sim
10^{-12}$ -- $10^{-14}$ eV). In that case, the high intensity of the
background (not constrained by nucleosynthesis, as in the graviton
case), could possibly compensate the weakness of the dilatonic charge
of the antennas, and could determine in principle a detectable signal.
This interesting possibility will be discussed in the next subsection.

\subsection{Detection of ultra-light dilatons}
\label{Sec6.4}

A stochastic background of relic dilatons, produced by a phase of
pre-big bang evolution, and light enough to have a lifetime larger than
the present Hubble scale, could interact today in two ways with the
existing gravity wave detectors: either indirectly, through a 
geodesic coupling of the detector to the scalar part of the induced 
metric fluctuations contained in the Riemann tensor 
\cite{BiaCo96,BiaCo98,BruCo98,Shiba94,MaNi99,Nakao00},
or even directly through the non-geodesic coupling of the dilaton to the
scalar charge  of the detector \cite{Gas99z,Gas00b} (see Eqs. 
(\ref{619}), (\ref{620})). 

In the first case the coupling has a gravitational strength, but the
amplitude of scalar metric perturbations, at the typical resonant scales
of present detectors, could be much lower than the amplitude of the
original dilaton fluctuations. In the second case the detector could be
coupled to a background of very high intensity (of order $1$ in critical
units), but with a coupling constant strongly suppressed for
phenomenological reasons, as discussed in Subsection \ref{Sec6.2}. 

In order to take into account both possibilities, we will estimate the
scalar-to-noise ratio associated to a stochastic dilaton background by
assuming that the physical strains $h_i(t)$ ($i=1,2$), induced on a pair of
gravitational antennas, vary in time like the scalar fluctuations
$\phi(x_i,t)$ that  perturb the detectors, and are proportional to the
scalar pattern functions $F_i(\hat n)$ through the scalar charges $q_i$
of the detectors \cite{Gas00b,GasUn01}:
\beq
h_i(t)= \phi (x_i,t) F_i(\hat n), ~~~~~~~~~
F_i(\hat n)=q_i e_{ab}(\hat n)D_i^{ab}. 
\label{652}
\eeq
Here $e_{ab}(\hat n)$ is the polarization tensor
of the scalar wave propagating along the $\hat n$ direction, and 
$D_i^{ab}$ is the detector response tensor (see Subsection \ref{Sec5.4});
finally, $q_i=1$ for the coupling to scalar metric fluctuations, while  $q_i
\ll 1$ for the direct coupling of long-range scalar fields. 

We expand the strain in momentum space, taking into account the mass
of the scalar fluctuations ($\vec p = p \hat n$), 
\bea
&&
h_i(t)= \int_0^\infty dp \int d^2 \hat n ~ F_i(\hat n) 
\left[\phi (p, \hat n) e^{2\pi i\left[ p \hat n \cdot {\vec x}_i 
- E(p) t \right]}+ {\rm h.c.}\right],
\nonumber\\ &&
E(p)=(\ti m^2+p^2)^{1/2}, ~~~~~~~~~\ti m =(m/2\pi),
\label{653}
\eea
and we use the stochastic condition which, for a scalar
spectrum $\Om(p)$,
 \beq
\r= \r_c \int d\ln p ~\Om (p) = {M_{\rm P}^2\over 16 \pi }\langle |\dot
\phi|^2\rangle, 
\label{654}
\eeq
can be written as
\beq
\langle \phi^\star (p, \hat n), \phi(p', \hat n') \rangle= 
{3 H_0^2 \Om (p) \over 8 \pi^3 p E^2(p)}
\da (p-p') 
\da^2 (\hat n-\hat n').
\label{655}
\eeq
By imposing optimal filtering \cite{AlRo99}, and following the same
steps as in Subsection \ref{Sec5.4} for the graviton case, we finally
arrive at the signal-to-noise ratio \cite{GasUn01}:
\beq
{\rm SNR}= \left(\frac{H^2_0}{5\pi^2}\right)\,
\Bigg[2\,T \,\int_{0}^{\infty}{dp\over  p^3\,(p^2+\tilde{m}^2)^{3/2}}
\frac{\Omega^2(p)\,\gamma^2(p)}
{P_1(\sqrt{p^2+\tilde{m}^2})\,P_2(\sqrt{p^2+\tilde{m}^2})}\Bigg]^{1/2},
\label{656}
\eeq
where $T$ is the total integration time, $P_1(|\nu|), P_2(|\nu|)$ 
are the one-sided noise spectral density of the detectors, 
defined as Fourier transforms of the frequency $\nu=E(p)=
(\ti m^2+p^2)^{1/2}$, and $\ga (p)$ is the overlap  reduction function,
defined in momentum space by: 
\beq
\ga(p)= {15\over 4 \pi}\int {d^2 \hat n}
F_1(\hat n) F_2 (\hat n) 
 e^{2\pi i p \hat n \cdot ({\vec x}_2 - {\vec x}_1)}
\label{657}
\eeq
(normalized so that $\ga=1$ for coincident and coaligned
interferometers). 
For $m \ra 0$ and $q \ra 1$ we recover the signal-to-noise
ratio (\ref{540}), modulo a different normalization of $\ga$. 

The result (\ref{656}) determines the cross-correlation and the
detectability of a signal induced by a massive scalar spectrum $\Om(p)$.
We may distinguish three phenomenological possibilities.

If $m$ is much larger than the typical sensitivity band $\nu_0$ of the
antenna, then the noises $P_i$, estimated at the frequency $\nu=
(\ti m^2+p^2)^{1/2}\gg \nu_0$, tend to infinity, and we cannot expect to
detect a signal. If, on the contrary,  $m \ll \nu_0$ (which includes the
massless case, discussed in \cite{BaBa01}), then the detectors in their
sensitivity band will respond to the relativistic branch of the spectrum,
whose integrated amplitude is constrained by the nucleosynthesis 
bound, $\Om \laq 10^{-6}$ (see Subsection \ref{Sec5.1}). In addition, the
response is possibly suppressed by the factor $q^2 \ll1$, because  the
dilaton field has a  long range. As a consequence, we may expect in such
a case a signal not larger  (and possibly much smaller) than the signal
produced by a relic graviton background. 

The interesting possibility corresponds to the dilaton mass in the
sensitivity band of the detectors ($m \sim \nu_0$), where the overlap
is large and the noises as small as possible. In that case we have a
resonant response also to the non-relativistic branch ($p<m$) of the
spectrum, which is not constrained by nucleosynthesis and can approach
the critical density bound, as emphasized in the previous subsection. In
order to discuss the possibility of detection let us then consider the
most favourable case, in which the non-relativistic branch dominates
the spectrum and saturates the critical density, 
$h^2_{100}\int^m d \ln p \Om \sim 1$ (more precisely, we should impose
the saturation of $\Om$ at $\simeq 0.35$, to be consistent with present 
dark-matter estimates \cite{Turner00}). Let us suppose, also, that the
spectrum is peaked around $p \sim m$, and that the contribution of the
relativistic branch $p>m$, if present, is negligible. 

The integral (\ref{656}), in such a case, can be estimated by integrating
over the non-relativistic modes only, and approximating $P_i$ with their
constant values at $ \nu=m$. We consider the ideal case of two identical
detectors ($P_1=P_2=P$, $q_1=q_2=q$), with maximal overlap ($ \ga
=1$) at $\nu=m$ and integration time $T=10^8$ s. The dilaton
background is then detectable, SNR $ >1$, provided \cite{Gas00b}:
\beq
m^{5/2} P(m) \laq {q^2}\times 10^{-32} ~{\rm Hz}^{3/2}. 
\label{658}
\eeq

The intersection of this condition with the noise power spectrum
$P(\nu)$ of present gravitational antennas provides a rough estimate
of  the allowed values of the dilaton mass,  possibly compatible with a
{\em direct} detection of a stochastic dilaton background. The mass
windows depend, of course, on the values of the dilaton coupling. By
using a simple, power-law parametrization \cite{Cuoco98} of the VIRGO
noise curve, for instance, we find \cite{Gas00b} that 
the
detectable mass windows extends, for $q^2=1$,  over the full
sensitivity band of the detector ($1$ Hz to $10$ kHz). However, the
window decreases as $q$ decreases, and closes completely for $q^2 \laq
10^{-8}$. 

As already stressed at the beginning of this section (and in Subsection
\ref{Sec6.2}), a large value of $q^2$ can only refer, in this context,  to
the indirect coupling of the gravitational antennas to the (scalar) metric
perturbation spectrum generated by the dilatons. It seems unlikely, 
however, that such a spectrum, even including the non-relativistic
corrections due to the dilaton mass, can be high enough (at the kHz
scale) to saturate the critical density bound, as assumed to obtain the
condition (\ref{658}). 

For the direct coupling to the dilaton background, on the other hand,  we
have to recall that the maximum allowed value of $q^2$, for
composition-dependent and -independent interactions, are summarized
in Eqs. (\ref{627}), (\ref{628})). Taking into account such bounds, we are
led to the situation illustrated in Fig. \ref{f63}, where the noise power
spectrum of VIRGO \cite{Cuoco98} (bold curve) is compared with the
condition (\ref{658}) at fixed $q$ (thin dashed lines) and with the
maximal allowed values of $q^2$, for universal (bold upper lines) and
composition-dependent (thin lower lines) scalar interactions. A scalar
background of nearly critical density, non-universally coupled to 
macroscopic matter, turns out to be only marginally compatible with
detection (at least, at the level of sensitivity plotted in the picture). If
the coupling is instead universal (for instance, as in the dilaton model
discussed in \cite{DP94a}) but the scalar is not exactly massless, then
there is a mass window open to detection:   
\beq 
10^{-14} ~{\rm eV} \laq 
m \laq 10^{-12} ~{\rm eV}. 
\label{659} 
\eeq

\begin{figure}[t]
\centerline{\epsfig{file=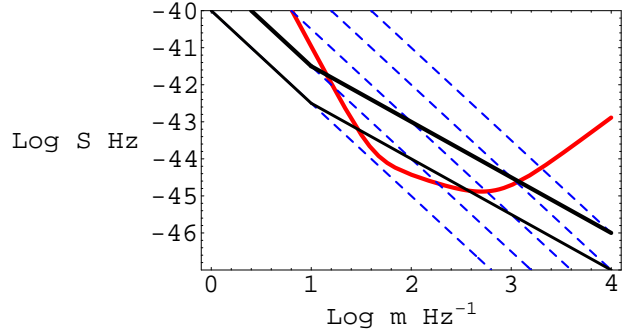,width=82mm}}
\vskip 5mm
\caption{\sl The noise spectrum of VIRGO (bold curve) 
is compared with the maximal values of $q^2$ allowed
by gravitational phenomenology, in two cases: universal (bold upper 
lines) and composition-dependent (thin lower  lines) scalar
interactions. The thin dashed lines correspond, from left to right, 
to $q^2=10^{-8}$, $10^{-7}$, $10^{-6}$, $10^{-5}$, $10^{-4}$.  The region
compatible with a detectable signal is above the noise spectrum and
below the upper bounds on $q^2$.} 
\label{f63} 
\end{figure}

The above results, to be interpreted only as a first qualitative
indication, are confirmed by a more realistic analysis of the cross
correlation of the two LIGO interferometers \cite{GasUn01}, performed
by applying Eq. (\ref{656}) to the phenomenological power spectrum: 
\beq
\Om(p)=\da ~\Om_0\left(p/ \ti m\right)^\da, ~~~~~~~~~~~~~~
p \laq \ti m,~~~~~~~~~~~~~~ 0< \da \leq  3. 
\label{660}
\eeq
In spite of the reduced value of the overlap factor we find (for any
$\da$, but with $\Om_{\rm s} \sim 1$) a significant  SNR in the relevant
mass range, already at the level of the Enhanced LIGO configuration. 
No possibility of detection, even with the Advanced LIGO configuration,
is found instead for a non-relativistic spectrum  peaked outside the
detector sensitivity band, such as the ``minimal" spectral distribution 
(\ref{612}), (\ref{613}). It should be stressed, however, that
interferometric detectors are particularly unfavoured for the detection
of a non-relativistic scalar background, because their response tensor
$D_{ab}$ is traceless: spherical resonant detectors represent 
more promising devices, and it seems appropriate to conclude this
section by reporting the possible enhanced level of SNR when a
massive scalar background interacts with the monopole mode of a
resonant sphere \cite{CoGas02}. 

For a massive scalar wave there are in fact two
possible pattern functions, one associated to the geodesic coupling
of the detector to the scalar component of metric fluctuations
\cite{MaNi99} (the Riemann term in Eq. (\ref{620})), the other related to
the direct, non-geodesic coupling of the  charge $q_i$ of the
detector to the spatial gradients of the scalar background
\cite{Gas99z}. In terms of the transverse and longitudinal
decomposition of the polarization tensor of the  wave,
with respect to the propagation direction $\hat n$, 
\beq
T_{ab}=(\delta_{ab}-\hat{n}_a\,\hat{n}_b), ~~~~~~~~~~~~
L_{ab}=\hat{n}_a\,\hat{n}_b \,,
\label{661}
\eeq
the two pattern functions can be written, respectively, as \cite{GasUn01}
\beq
F(\hat n)=D^{ab}\left(T_{ab}+
\frac{\ti{m}^2}{E^2}\,L_{ab}\right), ~~~~~~~~~~
F_q(\hat n)= q \frac{p^2}{E^2} D^{ab}
\,L_{ab}.
\label{662}
\eeq

For the differential mode of an interferometer  \cite{Fla93}
$D^{ab}\da_{ab}=0$, so that, in both cases, the pattern function of
massive waves  turns out to be proportional to the massless pattern
function and, for non-relativistic modes, it is highly suppressed by
the factor $(p/E)^2$ \cite{GasUn01}. The monopole mode of a resonant
sphere, on the contrary, is characterized by the response
tensor $D^{ab}=\da^{ab}$ \cite{BiaCo98}. The non-geodesic 
pattern function $F_q(\hat n)$ is still suppressed, but the geodesic part
is not, and the overlap function (\ref{657}) takes the simple form 
\beq
\ga(p)={15\over 2\pi}\left(3 \ti m^2 +2 p^2\over \ti m^2 +p^2\right)^2
{\sin (2\pi p d)\over pd},
\label{663}
\eeq
where $d$ is the spatial separation of the detectors. The response to
scalar spectra peaked at $p \ll m$ is thus strongly enhaced with
respect to interferometric detectors (at least in the case of geodesic
coupling). 

Beside this effect there is a second possible enhancement, for flat
enough spectra, associated to a faster growth of SNR with the
observation time $T$. Consider, for instance, the cross-correlation of
two spherical detectors according to Eq. (\ref{656}), computed for the
spectrum (\ref{660}) with the overlap (\ref{663}). Assuming that
the instrumental noises are constant (and finite) for $p$ ranging from
$0$ to $\ti m$,  it follows that for $\da <1$ the momentum integral is
dominated by its lower limit, i.e. by the infrared cut-off scale $p_{\rm
min}$ determined, for $p\ra 0$, by the ``minimum'' observable frequency
interval $\Da \nu$ associated to the ``maximal" observation time $T$
\cite{CoGas02}: 
\beq
\Da \nu =(p^2+\ti m^2)^{1/2}-\ti m > T^{-1},
~~~~~~~~~~~~ p_{\rm min}= \left(2 \ti m /T\right)^{1/2}.
\label{664}
\eeq

This effect modifies, for $\da <1$, the final $T$-dependence of SNR,
and enhances the response of the detector with the growth of the
observation time. Performing the integral for $\da <1$ one finds in fact 
(modulo numerical factors)
\beq
{\rm SNR} \simeq  
{\Om_{\rm s}\over \ti m~ P(\ti m)}\left(H_0\over \ti m\right)^2
\left(\ti m T\right)^{1-{\da/ 2}}, 
~~~~~~~~~~~~~
\da<1, 
\label{665}
\eeq
to be compared with the more standard result
\beq
{\rm SNR} \simeq  
{\Om_{\rm s}\over \ti m~ P(\ti m)}\left(H_0\over \ti m\right)^2
\left(\ti m T\right)^{1/2},  
~~~~~~~~~~~~~~~~
\da \geq 1.  
\label{666}
\eeq
For $T=10^7$ s, and  a mass in the kilohertz range, the sensitivity of
spheres to a flat enough and massive spectra is thus increased by a 
factor  $10^{10(1-\da)/2}$ with respect to interferometric detectors (see
\cite{CoGas02} for a more detailed discussion). 

We may thus conclude, summarizing the results of this subsection, that
both interferometric and resonant gravitational antennas seem to be
able to explore the possible presence of a light (but non-relativistic) 
dilatonic component of dark matter, and thus to constrain the
parameters of the pre-big bang models, in a mass range that  overlaps
with their sensitivity  band. This is possible in spite of the fact that the
relic dilaton background could be directly coupled to the total mass of
the detector, with a charge that is much weaker than gravitational.
Spherical detectors are more efficiently coupled to massive scalar
waves, but enhanced and advanced interferometers are already
sufficient to provide significant constraints on a possible background of
relic dilatons.

\section{Relic photons, axions and CMB anisotropy} 
\label{Sec7}
\setcounter{equation}{0}
\setcounter{figure}{0}

As discussed in the previous sections, the primordial spectrum of (scalar
and tensor) metric fluctuations, amplified by a phase of pre-big bang
inflation, is characterized in general by a positive and rather steep
slope. Because of the high-frequency normalization of the peak of the
spectrum, controlled by the string scale, the amplitude of metric
fluctuations is strongly suppressed at the large-distance scales
relevant to  the anisotropy of the CMB radiation observed by COBE 
\cite{Cobe92,Cobe94}, or to the density fluctuations  spectrum (see for
instance \cite{Pea94}) required by the standard mechanism of structure
formation. It  thus seems difficult, in such a context, to generate a flat
spectrum of scalar curvature perturbations directly responsible for the
above effects, unless we accept rather drastic modifications of the
pre-big bang kinematics, as recently suggested in \cite{FiBra01,DuVe02}
(and briefly discussed in Subsection \ref{Sec7.2}). 

The slope of the spectrum, however, depends on the external ``pump"
field, which amplifies the quantum fluctuations (see Section \ref{Sec4}) 
and which contains, in general, the contribution of all the components of
the background (metric, dilaton, moduli fields, ...). Different fluctuations
are coupled to different pump fields and may have very different
spectra, even in the same background. 

It turns out, in particular, that the primordial spectrum of
electromagnetic \cite{GGV95,GGV95a} and of ``axionic" 
\cite{Cop97,Cop97a}  fluctuations (i.e. the fluctuations of the
four-dimensional NS-NS two-form and of other,
string- and M-theory-motivated, antisymmetric tensors) may be
characterized by a slope much flatter than the slope of the metric
perturbation spectrum (in some cases, even decreasing
\cite{BMUV98a,Cop98b,BH98,Giov99a,DurSak00}, depending on the
coupling to the moduli fields, on the anisotropy of space-time, and  on
the number of extra dimensions). A flat and primordial distribution of
(isocurvature) axionic fluctuations, in turn, can directly generate a
scale-invariant spectrum of CMB anisotropies through the ``seed"
mechanism \cite{DurGas98,DurGas99,GasVe99}, which will be reported in
Subsections \ref{Sec7.3}, \ref{Sec7.4}, or even generate an adiabatic,
scale-invariant spectrum of density  and curvature perturbations
through the ``curvaton" mechanism recently discussed in 
\cite{EnSlo01,LyWa02,Moroi,BaLi02,NoRio,Mor02,BGGV02}, and illustrated
in Subsection \ref{Sec7.5}. 

In addition, the amplification of electromagnetic fluctuations, with a
flat enough spectrum, seems to naturally provide the seed fields
required to explain the origin of the magnetic fields present on a
galactic and intergalactic scale (see \cite{Grasso00} for a recent
review). We will first illustrate this interesting possibility in the
next subsection. 

\subsection{Large-scale magnetic fields and photon production}
\label{Sec7.1}

In the context of the pre-big bang scenario, the background evolution
that amplifies  metric perturbations also 
amplifies the quantum fluctuation of the electromagnetic field,
because of their direct coupling to the dilaton and to the moduli fields. 
For the heterotic string model, such a coupling is described by the
effective action  \cite{GGV95}
\beq
S=-{1\over 4}\int d^4x \sqrt{-g} e^{-\phi}F_{\mu\nu}F^{\mu\nu}, 
\label{71}
\eeq
where  $\phi = \Phi_{10} - \ln{V_6} \equiv \ln (g^2)$ controls the
tree-level four-dimensional gauge coupling ($\Phi_{10}$ being the
ten-dimensional dilaton field, and $V_6$ the volume of the
six-dimensional compact internal space). The effective coupling to the
external geometry may be different, however, if the electromagnetic
$U(1)$ symmetry is a component of the Kaluza--Klein gauge group
produced in the compactification down to $D=4$ dimensions
\cite{BMUV98a} (see also \cite{Giov00}). 

The electromagnetic field is also coupled to the four-dimensional
geometry, of course.   If such a coupling is minimal, however,  it is also
conformally invariant, and therefore does not contributes to photon
production in a conformally flat metric such as that of a typical
inflationary background, unless conformal invariance is broken at the
classical \cite{TurWi88} or quantum \cite{Dol93} level. Even in that case,
however,  photon production turns out to be rather suppressed (with
respect to graviton production, for instance). 

In order to discuss the amplification induced by the dilaton, in a
conformally flat metric   $g_{\mu\nu}= a^2\eta_{\mu\nu}$,  
it is convenient to use  the radiation gauge $A_0=0$, $\pa_i
A^i=0$ explicitly. 
Because of  conformal invariance, the coupling to the metric
disappears from the action (\ref{71}); after  partial integration
the action  becomes, in this gauge, 
\beq
S={1\over 2}\int d \eta e^{-\phi}\left( A_i' A^{i\prime} +
A_i\nabla^2A^i\right), 
\label{72}
\eeq
and can be easily diagonalized by setting
\beq
\psi_i= z A_i, ~~~~~~~~~~~~~~z= e^{-\phi/2}.
\label{73}
\eeq
For each polarization mode we then obtain the effective action
\beq
 S_\psi={1\over 2}
\int d\eta  \Bigg( \psi^{\prime 2}+ \psi \nabla^2\psi 
+ {z^\se\over z}\psi^2 \Bigg),
\label{74}
\eeq
and the canonical evolution equation
\beq
\psi_k^\se +\left[k^2-V(\eta)\right]\psi_k =0, ~~~~~~~~~~~
V(\eta)={z^\se \over z}= e^{\phi/2}\left( e^{-\phi/2}\right)^\se, 
\label{75}
\eeq
which is exactly the same as the graviton equation (\ref{456}), with
the only difference that now the pump field is fully determined by
the dilaton. If we use, in particular, the minimal model of background 
introduced in Subsection \ref{Sec5.2}, the spectrum of the 
amplified  electromagnetic fluctuations will be characterized by the
same parameters as the graviton spectrum. In such a case,  the
constraints on the electromagnetic spectrum will provide constraints on
the parameters of the model and then, indirectly,  on the associated
graviton spectrum. 

An important application of the ``dilatonic" amplification of  the
electromagnetic fluctuations is the possible production of  ``seed" for
the galactic magnetic fields \cite{GGV95,Lem95}. The origin of the
large-scale cosmic magnetic fields (with coherence scale $\gaq 10$ kpc,
and typical amplitude $\sim 10^{-6}$ G) is in fact, to a large extent,  still
an open problem \cite{Grasso00}. Almost all mechanisms invoked to
generate the large-scale fields, e.g. the galactic ``dynamo"
\cite{Parker79,Zeldo83}, require the presence of small, primordial seeds 
 to trigger the subsequent magnetic amplification.  Many
mechanisms of seed production have been suggested 
\cite{Harri73,TurWi88,Vacha91,Ratra93,Garre92,Dol93,Cheng94,Kibble95}
\cite{Berto99,BrusOa99,GiovShap00,CopSaf00} 
\cite{Bassett00,Finelli00,Davis00,Mar00,Gas00c}, trying to bypass the
difficulty --due to conformal invariance-- that forbids a standard
inflationary amplification of the electromagnetic fluctuations. 
In the context of the pre-big bang models, on the contrary, the
seeds can be produced directly from the electromagnetic fluctuations of
the vacuum, thanks to the accelerated growth of the dilaton during the
pre-big bang phase (with possible additional contributions from dilaton
oscillations, during preheating \cite{Giov97a}). 

Consider in fact the two-parameter model of Eqs. (\ref{510}):
 \bea
&&
{\rm dilaton~phase}, ~~~~~~~~~~~~~~~e^{-\phi/2}\sim (-\eta)^{\sqrt
3/2}, ~~~~~~~~~~
\eta<\eta_{\rm s}, \nonumber\\ &&
{\rm string~phase}, ~~~~~~~~~~~~~~~~e^{-\phi/2}\sim (-\eta)^{\b}, 
~~~~~~~~~~~~~~~\eta_{\rm s}<\eta<\eta_{1}, \nonumber\\
&&
{\rm radiation~era}, ~~~~~~~~~~~~~~~e^{-\phi/2}\sim  {\rm const},
~~~~~~~~~~~~~~~~
\eta>\eta_1.  
\label{76}
\eea
The  effective dilatonic potential of Eq. (\ref{75}) grows during the
dilaton phase, keeps growing during the string phase, and goes to 
zero in the radiation era. Because of the two background transitions, 
there are two bands of the spectrum, corresponding to modes hitting
the barrier in the dilaton phase, $\om <\om_{\rm s}$, and in the string
phase,  $\om>\om_{\rm s}$. By applying the standard procedure of Subsection 
\ref{Sec4.4} we obtain:
\bea
\Om_{\rm e m}(\om,t_o) &\simeq & g_1^2
\Om_\ga (t_0) \left(\om\over \om_{1}\right)^{3-2\mu},
~~~~~~~~~~~~~~~~~~~~~~~~~~~~~~~~~~\om_{s}<\om<\om_1,
\nonumber\\ 
& \simeq & g_1^2
\Om_\ga(t_0) \left(\om_1\over \om_{s}\right)^{1+2\mu-\sqrt 3}
\left(\om\over \om_{1}\right)^{4-\sqrt 3}, 
~~~~~~~~~~~~~~~\om<\om_{s} 
\label{77}
\eea
where $\mu=|\b-1/2|$. By recalling that $\b=1+\a$ (see Eq. (\ref{514})),
we find that the electromagnetic spectrum  thus has the same
parameters as the graviton  spectrum  (\ref{515}), but the slope is in
general flatter. Even flatter spectra are possible for  vector fields of
Kaluza--Klein origin \cite{BMUV98a}. 

The generation of fluctuations of a strength large enough to seed the
galactic dynamo constrain the electromagnetic spectrum
at the megaparsec scale, and imposes in particular a bound that, in the
most conservative form, can be written as follows \cite{TurWi88}:
\beq 
\Om_{\rm e m}(\om_G,t_{\rm gal}) \gaq 10^{-34}\Om_\ga(t_{\rm gal}),
~~~~~~~~~~~~~ \om_G \sim (1~{\rm Mpc})^{-1} \sim 10^{-14} ~
{\rm Hz}, 
\label{78}
\eeq
where $t_{\rm gal}$ is the epoch of galaxy formation, i.e. $a_{\rm gal} 
\sim 10^{-2} a_0$. 
Remarkably, this condition is compatible with
the condition of negligible backreaction,
\beq 
\Om_{\rm em}(\om) <\Om_\ga
\label{79}
\eeq
(at all frequency scales), required by the consistency of the
perturbative approach around a homogeneous unperturbed 
background. 

There is indeed a wide region of parameter space in which the two
conditions are consistent without fine-tuning \cite{GGV95}, as illustrated
by the shaded and hatched areas of Fig. \ref{f71} (the hatched  area
refers to the dilaton branch of the spectrum, $\om <\om_{\rm s}$; the
shaded area to the string branch,  $\om >\om_{\rm s}$). As shown in
the picture, the production of seeds is strong enough to satisfy the
needs of the dynamo mechanism (Eq. (\ref{78})), provided the coupling is
sufficiently small at the beginning of the string phase:
\beq
\log_{10} \exp (\phi_{\rm s}/2) \laq -20,
\label{710}
\eeq
and the duration in time of the string phase is sufficiently long:
\beq
\left|\eta_{\rm s}/\eta_1\right| \gaq 10^{10}\sim e^{23},
\label{711}
\eeq
(in cosmic time, $\Da t \gaq~23 \la_{\rm s}$). 
For illustrative purposes, the plots of Fig. \ref{f71} have been
constructed by taking $g_1=1$ as a reference value. 
We also note that, besides the condition (\ref{79}), there are additional
bounds on the energy spectrum of the magnetic fields because of a
possible (classical) production of gravitational waves induced by the
magnetic stress tensor \cite{CaDu02}. Flat enough  spectra are not
strongly constrained, however, and we are assuming that this is indeed
the case for the electromagnetic distribution $\Om_{\rm em}(\om)$ that
we are considering. Possible additional bounds on the primodial
magnetic amplitude, due to Faraday rotation effects, may follow in
principle from future measurements of the CMB polarization, as
discussed in \cite{Giov97b}. 

\begin{figure}[t]
\centerline{\epsfig{file=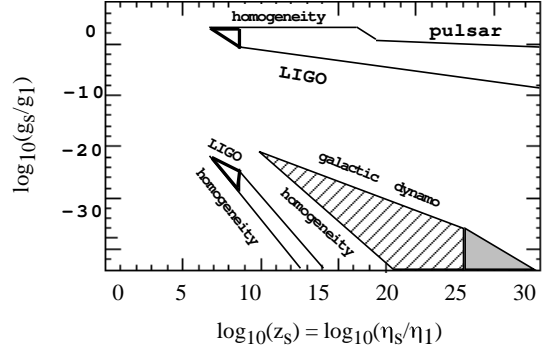,width=72mm}}
\vskip 5mm
\caption{{\sl Allowed region in the parameter space of minimal
pre-big bang models for the production of  seed  fields for the
galactic dynamo (shaded and hatched area). The other  two regions
correspond to the production of a graviton background  detectable by
advanced interferometric antennas.}} 
\label{f71}
\end{figure}

Also shown in Fig. \ref{f71} are the allowed regions of parameter space
compatible with the detection of the relic graviton background
produced in the context of minimal pre-big bang models. 
We have used the typical sensitivity expected for the
cross-correlation of two second-generation interferometers, with the
characteristic of  Advanced LIGO (see Eq. (\ref{543})). The allowed region
is then defined by the condition  $\Om_G(10^2~{\rm Hz}) \geq 10^{-10}$,
imposed on the spectrum (\ref{515}), together with the pulsar bound
(\ref{53}), and the condition of negligible backreaction ($\Om_G<1$ at all
$\om$ and $t$), required for the validity of our perturbative computation
of the  spectrum around a homogeneous background (for a detailed
implementation of all bounds see \cite{Gas96}). It turns out that there
are two allowed regions, corresponding to tensor perturbations whose
comoving amplitude stays constant outside the horizon during the string
phase (upper region) or grows (lower region). At the left-end of each
region, the marked triangular area corresponds to the possible detection
of low-energy modes, crossing the horizon  during the
dilaton-dominated phase; the rest of the allowed region refers to modes
crossing the horizon during the string phase. 

It should be noted that the gravitationally allowed regions have 
no possible overlap with the electromagnetic ones, 
implying that the interesting phenomenological possibilities of seed
production and of graviton  detection tend to exclude each other
\cite{Gas96}. Indeed, if we restrict to the class of backgrounds
consistent with the production of magnetic seeds (i.e. inside the shaded
and hatched regions of Fig. \ref{f71}), then the allowed  slope of the
high-frequency (string) branch of the graviton spectrum is constrained
to vary between  $2$ and $3$ \cite{Gas98b},  with a consequent drastic
reduction of the allowed region for $\Om_G$, as illustrated in Fig. 
\ref{f72}. In that case, the graviton background is certainly outside the
range of currently available  detectors (resonant bars and
ground-based  interferometers), which reach their maximal
sensitivity in the frequency band  $10^2$ -- $10^3$ Hz. 

\begin{figure}[t]
\centerline{\epsfig{file=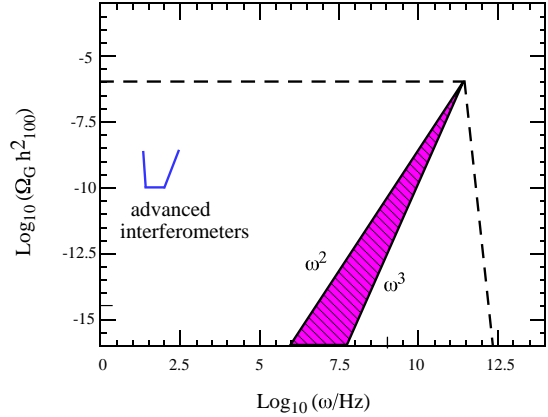,width=72mm}}
\vskip 5mm
\caption{{\sl Allowed region for the graviton spectrum in minimal
pre-big bang models (shaded area), assuming that the  amplification of
the electromagnetic fluctuations  is strong enough to seed the galactic
dynamo.}} 
\label{f72}
\end{figure}

According to the above analysis, a future detection of
cosmic gravitons in the hertz to kilohertz  range  would  seem to
exclude   a dilatonic origin of the seed magnetic fields, directly
amplified from the fluctuations of the vacuum. Conversely, an absence
of signals from the relic graviton background would seem to support,
indirectly, a primordial dilatonic origin of the cosmic magnetic fields. It
must be stressed, however, that such  conclusions, based on the plots of
Fig. \ref{f71}, cannot be applied in general to the pre-big bang scenario:
they are only pertinent to the class of  minimal models introduced in 
Subsection \ref{Sec5.2}, and could be evaded in the context of more
complicated models of background. 

A final remark concerns the fact that the photon
spectrum (\ref{77}) is in general flatter than the graviton spectrum
(\ref{515}), suggesting the possibility that the
electromagnetic fluctuations may  also act as seeds for the
(approximately) scale-invariant spectrum of temperature anisotropies
of the CMB radiation \cite{GGV95a,Gas95} and for the density
perturbations $\da \r/\r$ required for the structure formation. 
 According to a detailed analysis \cite{DurGas99}, however, such a
possibility has to be discarded in favour of an analogous possibility
based on the fluctuations of the fundamental antisymmetric tensors
appearing in the bosonic sector of string theory, which can be amplified
with a perfectly flat spectrum \cite{Cop97,Cop97a}, as will be
illustrated in the next subsection.

\subsection{Large-scale CMB anisotropy and axion production}
\label{Sec7.2}

In the context of the standard inflationary scenario, the observed CMB
anisotropy, at very large angular scales, can be consistently generated
by the primordial spectrum of scalar metric fluctuations $\Phi_k$ 
through the Sachs--Wolfe (SW) effect \cite{SW},  
$\left(\Da T/ T\right)_k \sim \Phi_k$. Indeed, when the metric  
fluctuations are directly amplified by the accelerated evolution of the 
background, their spectral  distribution is controlled by the value of
the   Hubble scale at the time of horizon crossing, 
$\Phi_k \sim \left(H/ M_{\rm P}\right)_k$.
For a standard de Sitter (or quasi-de 
Sitter) inflationary solution, $H$ is constant in time, so that the 
spectrum is scale-invariant. A typical normalization of the 
spectrum, corresponding to inflation occurring roughly at the GUT
scale \cite{Kra92},  
\beq 
{H\over M_{\rm P}} 
\sim {{\rm GUT~curvature~scale \over PLANCK~scale}}
\sim 10^{-5}, 
\label{712}
\eeq
is  thus perfectly consistent with the anisotropy observed 
at the present horizon scale \cite{Cobe92,Cobe94}, $\Da T/T \sim
10^{-5}$, and  with the fact  that the spectrum is scale-invariant. 

In the context of the pre-big bang scenario, on the contrary, the
curvature scale  grows with time, so that  the spectrum of the metric
fluctuations tends to grow with frequency, as illustrated in 
Section \ref{Sec4}.  In addition, the natural inflation scale corresponds to
the string   scale, so that the normalization of the spectrum, at the 
end-point frequency $k_1$, is controlled by the ratio 
\beq 
\left(H\over M_{\rm P}\right)_{k_1} 
\sim {{\rm STRING~curvature~scale \over PLANCK~scale}} \sim
10^{-1} - 10^{-2}.
\label{713}
\eeq
As a consequence, for a typical pre-big bang model,  the slope of the
spectrum is too steep,  and the  normalization too high, to be compatible
with the COBE observations.  The slope  is so steep, however, that the
contribution of the metric  fluctuations to $\Da T/T$ is certainly
negligible at the COBE scale. So,  on the one hand there is no contradiction
with observations, namely the COBE  data cannot be used to rule out
pre-big bang models \cite{Gas99b}. On the other hand,  the problem
remains: How to explain the observed CMB anisotropy? 

There are two possible answers to this question. The first one 
relies on somewhat drastic modifications of the kinematics of the
pre-big bang phase illustrated in the previous sections, in such a way
as to generate a primordial scale-invariant spectrum {\em directly} 
from the amplification of the vacuum fluctuations of the metric, as in
the standard inflationary scenario. The second possibility relies on the
{\em indirect} generation of scale-invariant metric perturbations
through the isocurvature fluctuations of another field, which may have
a flat spectrum even in the context of the pre-big bang 
solutions illustrated up to now in this paper. We will discuss in detail
the second possibility in the following subsections. Here we will briefly
report on the first possibility. 

There are at present two possible examples of pre-big bang models
characterized by a flat primordial spectrum of scalar perturbations
\cite{FiBra01,DuVe02}. They are both based on the introduction of a
negative, exponential dilaton potential $V(\phi)$, which can be
parametrized in the E-frame as follows (in units $M_{\rm P}^2=2$): 
\beq
V(\phi)=-V_0 e^{-\phi/\sqrt \ep}, ~~~~~~~~~~~~~~~
0<\ep<1, 
\label{714}
\eeq
and which is similar to the attractive potential existing between the
colliding branes in the context of the ekpyrotic scenario
\cite{KOST1,KOSST}. 

With such a potential, the corresponding (E-frame) pre-big bang
solution describes an accelerated contraction, controlled by the
dimensionless parameter $\ep$, 
\beq
a(t) \sim (-t)^\ep \sim (-\eta)^{\ep/(1-\ep)},
\label{715}
\eeq
which has to  match  the subsequent phase of decelerated,
radiation-dominated expansion. By assuming the existence of a
non-singular bounce for the background solution, and assuming the
continuity of the Bardeen potential $\Phi$ and of the
curvature perturbation $\zeta=v/z$ across the bounce  (their continuity
is guaranteed by performing the matching on the constant density
hypersurface), one then finds that after the bounce the spectrum of
$\Phi$ is determined by the pump field (\ref{715}), according to the
rules derived in Section \ref{Sec4}, namely
\beq
k^3 \left|\Phi_k\right|^2 \sim k^{3-2\nu}, ~~~~~~~~~~~
\nu= \left|{\ep\over 1-\ep}-{1\over 2}\right|.
\label{716}
\eeq
A flat spectrum of scalar perturbations can then be obtained for
$\ep=2/3$, as noted in \cite{FiBra01}, corresponding to a dilaton
potential that  simulates a ``dust" matter source with vanishing
effective pressure $p=0$. 

Alternatively, as recently discussed in \cite{DuVe02}, a flat spectrum of
scalar metric perturbations can be obtained from the  background
solution (\ref{715}) by assuming a very slow contraction, i.e.  $0<\ep\ll
1$ as in the ekpyrotic scenario \cite{KOST1,KOSST}, and assuming that the
matching to the expanding phase is performed across a hypersurface
different from that of constant energy density (for instance, across the
zero-shear hypersurface \cite{DuVe02}). In that case the scalar
spectrum after the bounce turns out to be different from Eq. 
(\ref{716}), and is given by 
\beq
k^3 \left|\Phi_k\right|^2 \sim k^{-2\ep/(1-\ep)}.
\label{717}
\eeq
The case $\ep \ll 1$ thus reproduces a (nearly) flat spectrum, for both 
the pre-big bang and ekpyrotic scenarios. 

This result can be easily explained by noting that, in the transition from
a contracting to an expanding phase, the asymptotic solutions of the
scalar perturbation equations are a linear combination of a constant
and of a growing mode before the transition, and of a constant and a
decaying mode after it. Consider, for instance, a bouncing
background evolving from the contracting solution (\ref{715}) valid for
$\eta< -\eta_1$, to the standard, radiation-dominated expansion $a
\sim \eta$, valid for $\eta > \eta_1$. The asymptotic solutions for the
super-horizon modes of the Bardeen potential, in the longitudinal
gauge, can be written as
\bea
&&
\Phi^1_k(\eta) = {A_1(k)\over \eta a_1^2(\eta)} +B_1(k), ~~~~~~~~~~~~~
\eta <-\eta_1, 
\nonumber \\
&&
\Phi^2_k(\eta) = {A_2(k)\over \eta a_2^2(\eta)} +B_2(k),
~~~~~~~~~~~~~~ \eta >\eta_1, 
\label{718}
\eea
where $a_1\sim (-\eta)^{\ep/(1-\ep)}$, and $a_2\sim \eta$. Here $A_1,
B_1$ are determined by the initial normalization to a vacuum
fluctuation spectrum, while $A_2,B_2$ are to be determined by the 
matching conditions. It is then clear that different matching
prescriptions may lead to different solutions for $B_2(k)$, the dominant
mode of the post-big bang expanding phase, and then to different final
spectra. 

Given a space-like hypersurface, with normal vector $n_\mu (\eta, \vec
x)$, the matching of two background manifolds \cite{Israel66} has to
be performed on the induced metric $q_{\mu\nu}$ and on the extrinsic
curvature $K_{\mu\nu}$ of the given hypersurface, where 
\beq
q_{\mu\nu}= g_{\mu\nu}+n_\mu n_\nu, ~~~~~~~~~~~~
K_{\mu\nu}={1\over 2}\left(\nabla_\mu n_\nu+\nabla_\nu
n_\mu\right). 
\label{719}
\eeq
For the given example of bouncing background the metric will be
continuous, i.e. $\left[g_{\mu\nu}\right]_\pm =0$, where
\beq
\left[g\right]_\pm= \lim_{\eta \ra \eta_1^+} g(\eta) - 
\lim_{\eta \ra -\eta_1^-} g(-\eta); 
\label{720}
\eeq
the curvature, however, will be characterized by a ``jump" from
negative to positive values, $\left[K_{\mu\nu}\right]_\pm
=S_{\mu\nu}$ (if we neglect the high-energy corrections, which are
expected to smooth out the singularity, see Section \ref{Sec8}). This
introduces in general a surface stress tensor, which in our case
corresponds to a tension (i.e. to a negative pressure), needed to avoid
the singularity and to implement the transition from contraction to
expansion. 

The matching of the metric fluctuations can now be performed by
perturbing the Israel junction conditions on the chosen hypersurface,
and leads to 
\beq
\left[\da g_{\mu\nu}\right]_\pm =0, ~~~~~~~~~~~~~~
\left[\da K_{\mu}^{\nu}\right]_\pm =\da S_\mu^\nu. 
\label{721}
\eeq
On a general hypersurface one then obtains  that the dominant
coefficient $B_2$, determining the post-big bang asymptotic expansion
of the Bardeen potential, is in general related to the pre-big bang
coefficients $A_1,B_1$ by \cite{DuVe02}
\beq
B_2(k) =f_A A_1(k)+f_B B_1(k)+ f_{\rm s} \da p_{\rm s},
\label{722}
\eeq
where $f_A, f_b, f_{\rm s}$ are (possibly $k$-dependent) coefficients, 
and $\da p_{\rm s}$ represents the perturbation of the surface tension. 
The final spectrum $k^3 |B_2(k)|^2$ thus depends, in general, not only on
the choice of the matching hypersurface, but also on its perturbed
tension (i.e. on the details of the regularizing mechanism that is
responsible for the bouncing transition). 

The matching across the hypersurface of total constant energy density 
($\r+\da\r=$ const) gives in particular $f_A=0$ \cite{DuVe02} so that,
assuming a negligible $\da p_{\rm s}$, $B_2$ simply inherits the spectrum of
the constant mode $B_1$, and one is led to the ``standard" result
(\ref{716}) for the scalar perturbation spectrum.  The matching across
a different hypersurface leaves, however, $f_A \not= 0$, so that the
final spectrum is dominated by the growing mode of the pre-big bang
solution and (again assuming a negligible $\da p_{\rm s}$) one is led to
the ``new" result reported in Eq. (\ref{717}), which leads to a flat
spectrum for all backgrounds with $\ep \ll1$. 

In the rest of this section we will adhere, however,  to the standard
matching prescription at constant energy density, and we will
concentrate on the alternative possibility of an indirect generation of
scale-invariant metric perturbations through the fluctuations of
another background field. Such an auxiliary field, in a string cosmology
context, can be naturally identified with the axion, whose quantum
fluctuations, unlike the metric fluctuations, can be directly amplified
with a flat spectrum, even in the context of the minimal pre-big bang
models (without the contribution of a dilaton potential). 

To explain the reason of such a difference, first pointed out in
\cite{Cop97,Cop97a}, let us consider the  four-dimensional reduced
action for the antisymmetric  tensor $H_{\mu\nu\a}$, 
\beq
S={1\over 12}\int d^4x \sqrt{-g} e^{-\phi}H_{\mu\nu\a}H^{\mu\nu\a}, 
\label{723}
\eeq
where
\beq
H_{\mu\nu\a}= \pa_\mu B_{\nu\a}+ \pa_\nu B_{\a\mu}+\pa_\a
B_{\mu\nu}, ~~~~~~ 
B_{\mu\nu}=-B_{\nu\mu}.
\label{724}
\eeq
The equation of motion,
\beq
\pa_\nu \left(\sqrt{-g} e^{-\phi}H^{\nu\mu\a}\right)=0, 
\label{725}
\eeq
can be automatically satisfied by introducing the ``dual" axion field
$\sg$: 
\beq
H^{\mu\nu\a}={ e^{\phi}\over \sqrt{-g}} \ep^{\mu\nu\a\b} \pa_\b \sg 
\equiv  e^{\phi}\eta^{\mu\nu\a\b}\pa_\b \sg 
\label{726}
\eeq
($\eta^{\mu\nu\a\b}$ is the totally antisymmetric, covariant tensor),
and the effective action (\ref{723}) may be rewritten
\beq
S={1\over 2}\int d^4x \sqrt{-g} e^{\phi}\pa_{\mu}\sg\pa^\mu \sg
\label{727}
\eeq
(note the coupling of $\sg$ to $e^\phi$ instead of $e^{-\phi}$). In
conformal time, and for a conformally flat metric 
\beq
S={1\over 2}\int d \eta a^2 e^{\phi}\left( \sg^{\prime 2} +
\sg\nabla^2\sg\right), 
\label{728}
\eeq
the action can be diagonalized by setting 
\beq
\psi= z \sg, ~~~~~~~~~~~~~~z= ae^{\phi/2},
\label{729}
\eeq
so that the evolution equation for the axion canonical variable $\psi$
is formally the same, 
\beq
\psi_k^\se +\left[k^2-V(\eta)\right]\psi_k =0, ~~~~~~~~~~~
V(\eta)={z^\se / z}, 
\label{730}
\eeq
but is characterized by a pump field $z$ different from that of gravitons
or photons.  Assuming a smooth evolution from  a power-law pre-big
bang background, with $z \sim |\eta|^\a$, to the standard radiation
era, the axion spectrum $\Om_\sg$ is then determined by  the
background kinematics according to the general prescriptions of
Subsection \ref{Sec4.4}, i.e.  
\beq 
\Om_\sg (\om)\sim \om^{3-2\nu},
~~~~~~~~~~~~~~ \nu=|\a-1/2|, ~~~~~~~~~~~ z\sim |\eta|^\a. 
\label{731}
\eeq

Consider now the low-energy branch of the spectrum, for modes  hitting
the effective potential  barrier during a higher-dimensional 
dilaton-driven phase, which we assume to be characterized by three
isotropically expanding dimensions, with scale factor $a(\eta)$, and $n$
``internal" dimensions, with scale factor $b_i(\eta)$. The background
evolution is known, determined by the $d=3+n$, anisotropic, S-frame
solutions of the string effective action, Eqs. (\ref{287}) and (\ref{288}).
The scale factors, and the dimensionally reduced dilaton $\phi$, can then
be expressed in conformal time as
\bea
&&
a \sim |\eta|^{\b_0/(1-\b_0)}, ~~~~~~~~~
b_i\sim |\eta|^{\b_i/(1-\b_0)}, ~~~~~~~~~
\Phi_d= {\sum_i \b_i +3 \b_0 -1 \over 1 -\b_0} \ln |\eta|,
\nonumber\\
&&
\phi=\Phi_d -\ln V_{n}=\Phi_d - \sum_i \ln b_i  \sim
{3\b_0-1\over 1-\b_0} \ln |\eta|, 
\label{732}
\eea
where $\b_0, \b_i$ are $n+1$ parameters related by the ``Kasner-like"
condition 
\beq
3 \b_0^2+\sum_i\b_i^2 =1
\label{732a}
\eeq
($\b_0<0$ for inflationary expansion), and $\Phi_d$  
is the dilaton field appearing in the $d$-dimensional,  non-reduced
action, 
\beq
\int d^{d+1} x\sqrt{-g_{d+1}}e^{-\Phi_d}= V_{n} \int d^{4}x
\sqrt{-g_{4}}e^{-\Phi_d}= \int d^{4}x
\sqrt{-g_{4}}e^{-\phi}.
\label{733}
\eeq 
Combining the behaviour of $a$ and $\phi$ we obtain  the power 
$\a$ of the pump field, and  the corresponding spectral index, which are
in general given  by : 
\beq
\a= -{5 \b_0-1\over 2(1-\b_0)}, ~~~~~~~~~~~  
3-2\nu= 3-|2\a-1|={3 \b_0+1\over 1-\b_0}. 
\label{734}
\eeq

In the particular case of an isotropic $n$-dimensional subspace,
$\b_i=\b$, $3\b_0^2 +n\b^2 =1$, one can also re-express the spectral
index (\ref{734}) in terms of the conformal power $s=\b/(1-\b_0)$ of
the modulus, as originally done in \cite{Cop97}, or even in terms
of a parameter $r= (\dot V_n/2V_n)(\dot V_3/V_3)^{-1}$, measuring
the relative evolution in time of the external and internal volumes
\cite{Mel99}. What is important, in our context, is that  a flat
spectrum (i.e. $\Om_\sg=$ const, $3-2\nu =0$) is now perfectly allowed
for $\b_0=-1/3$. For an isotropic $d$-dimensional background, in
particular, $\b_0 =-1/\sqrt d$, so that a flat spectrum exactly
corresponds to $d=9$, i.e. just  the number of dimensions of critical
superstring theory! (see also \cite{BMUV98a}). Flat (or nearly flat)
spectra are also possible for different exact solutions of the low-energy
string effective action \cite{Cop94,Cop95,GasVe99}, even in four
dimensions if the background is anisotropic \cite{Giov99a}, or if the
dilaton has an appropriate exponential potential and it is non-minimally
coupled to the axion \cite{FiBra01}.   

It should be stressed that for gravitons, on the contrary, a flat
spectrum cannot be obtained in the above background,
quite irrespective of the number and dynamics of the internal
dimensions. By using Eq. (\ref{732}) we find indeed that,
independently of $\b_0,\b_i$, the graviton pump field  $a
e^{-\phi}$  always has the same  power,  $\a=1/2$, 
\beq
z=ae^{-\phi/2} \sim |\eta|^{1/2}, 
\label{735}
\eeq
so that the spectrum (modulo logarithmic corrections) is always
characterized by the known cubic slope 
$3-2\nu= 3-|2\a-1|=3$. A flat spectrum, in the  background
(\ref{732}), is also forbidden for heterotic photons (coupled to the
dilaton as in the example discussed in the previous subsection). The
situation is different, however, for Kaluza--Klein photons, whose
spectrum is more sensitive to the dynamics of the internal dimensions
\cite{BMUV98a}. 

It is important to note, finally, that in the different backgrounds
obtained through $SL(2,R)$ transformations mixing the axion and dilaton
fields,  different evolution equations  are generally obtained for the
metric, the dilaton, and the axion fluctuations. After a diagonalization,
and a redefinition of new canonical variables, one finds however that the
evolution equations for the two fundamental scalar degrees of freedom,
i.e. the gauge-invariant (Bardeen) potential and the pseudoscalar axion
field, are unaltered by the $SL(2,R)$ transformation. Thus, the blue 
spectral tilt of metric perturbations and a possibly flat slope of the 
axion spectrum remain the same as in the standard, dilaton-dominated
background solutions \cite{Cop97,Cop97a} (see also \cite{Lidsey00} for
more details on the axion spectra). 

A flat, primordial spectrum of  axion fluctuations can be used for an
indirect generation of the observed CMB anisotropies in two ways.  A
first possibility is the ``seed" mechanism \cite{Dur90,Dur94}: indeed,
even if the axion 
fluctuations are  negligible as  sources of the metric 
background, $\r_\sg \ll \r_c$, their inhomogeneous stress tensor 
generates metric fluctuations according to the standard gravitational 
equations, and the metric, in its turn, is source of the 
temperature anisotropies  through the usual Sachs--Wolfe effect:
\beq
{\r_\sg / \r_c} \sim \Phi \sim {\Da T/ T}. 
\label{736}
\eeq

Such an indirect production of anisotropies could work, in principle,
because the contribution (\ref{736}) to $\Da T/T$ is quadratic in
the axion field, and not linear as in the case of metric perturbations. So,
even if  the amplitude of the axion fluctuations is still normalized at the
string curvature scale, the square of the amplitude is
not very far  from the expected value $10^{-5}$:
\beq 
{\Da T \over T} \sim \Phi \sim \sg^2 
\sim \left({\rm STRING~curvature~scale \over PLANCK~scale}\right)^2.
\label{737}
\eeq
The string normalization (\ref{713}), on the other hand,  is      
imposed at  the end point of the spectrum (typically, at the
gigahertz scale), while  COBE observations constrain the spectrum at the
present horizon scale  ($\sim 10^{-18}$ Hz): it is then evident that  a
very small, ``blue" tilt in the axion spectrum is  sufficient 
to make compatible the string and COBE normalizations, Eqs. (\ref{737})
and  (\ref{712}). 

The seed mechanism, however, seems to have difficulties in
reproducing the  peak structure of the observed anisotropies at
smaller angular scales, as will be discussed in Subsection \ref{Sec7.3}. A
more realistic scenario is probably offered by the ``curvaton" mechanism
\cite{LyWa02}: if the axion is heavy enough, the initial and flat
primordial spectrum of isocurvature axion fluctuations can be eventually
converted, after the axion decay, into a final scale-invariant spectrum
of curvature fluctuations. The standard results for the temperature
anisotropies then apply, with the only difference that the metric
perturbations are not primordial, but indirectly generated by the axion
decay. This possibility will be discussed in Subsection \ref{Sec7.5}, after
a brief report on the seed mechanism, which will be presented in the
next  two  subsections. 

\subsection{Massless axions as seeds}
\label{Sec7.3}

The quantum fluctuations of the so-called ``universal" (or Kalb-Ramond)
axion $\sg$ (i.e. the dual of the field strength  of the NS-NS two-form 
appearing in the string effective action)    
can be amplified with a  scale-invariant distribution of the spectral
energy density,   $\Om_\sg(k,\eta) = \r_c^{-1}(d\r_\sg(k,\eta)/\r_c d \ln
k)$ (as shown in the previous subsection); they are thus a possible
candidate for seeding the large-scale anisotropy  in the context of the
pre-big bang scenario. To illustrate this possibility we will proceed in two
steps. 

We will first show that the scalar metric fluctuation on a given scale
$k$, at the  time the scale re-enters the horizon, is precisely
determined by the axion  energy distribution evaluated at the
conformal time of re-entry,  $\eta_{\rm re} \simeq k^{-1}$:
\beq
\Phi_k(\eta_{\rm re}) \sim \Om_\sg (k, \eta_{\rm re}). 
\label{738}
\eeq
Secondly,  we will show that the dominant contribution  of a given scale
to  the Sachs--Wolfe effect comes  from  the time that scale re-enters
the horizon, in such a way that the final  temperature spectrum exactly
reproduces the primordial seed spectrum: 
\beq
\left(\Da T/ T\right)_k \sim \Phi_k(\eta_{\rm re}) \sim \Om_\sg (k, 
\eta_{\rm re}). 
\label{739}
\eeq 
These two results are far from being trivial, as they are consequences 
of the particular time dependence of the Bardeen spectrum  induced
by  axion fluctuations (the same  results do not apply, for instance, to
the case of electromagnetic  fluctuations). We will give here only a
sketch of the  arguments leading to the above results, which are
interesting in themselves, apart from their possible string cosmology
applications. For a detailed derivation we refer the reader to the
original papers \cite{DurGas98,DurGas99}. 

Let us first compute the spectrum of metric perturbations  seeded
by a  flat, primordial distribution of axion fluctuations. We define, as
usual,  the power spectrum of the Bardeen potential, $P_\Phi(k)$, in
terms of  the Fourier transform of the two-point correlation function,
\beq
\int {d^3 k\over (2 \pi k)^3} e^{i {\bf k}\cdot({\bf x}-{\bf x'})}
P_\Phi(k) =\langle \Phi(x) \Phi(x')\rangle
\label{740}
\eeq
(the brackets denote spatial average, or expectation value if 
perturbations are quantized), and we recall that the square root of this 
function,  evaluated at a comoving distance $k^{-1}$, represents the
typical  amplitude of fluctuations on a scale $k$: \beq
\left(\langle \Phi(x) \Phi(x')\rangle\right)^{1/2}_{|x-x'|=k^{-1}} 
\sim k^{3/2} |\Phi_k| 
\label{741}
\eeq
(see also Eq. (\ref{469})). 
We also define the power spectrum of the seed stress tensor, in  the
same  way (no sum over $\mu$, $\nu$):
\beq
\int {d^3 k\over (2 \pi k)^3} e^{i {\bf k}\cdot({\bf x}-{\bf x'})}
P_\mu^\nu(k) =\langle T_\mu^\nu(x) T_\mu^\nu(x')\rangle
-\langle T_\mu^\nu (x)\rangle ^2 .
\label{742}
\eeq

Metric fluctuations and seed fluctuations, on the other hand, are related
by the  cosmological perturbation equations \cite{Dur94}. By taking into
account the  important contribution of the off-diagonal components of
the axion  stress tensor one finds, typically, that the  Bardeen spectrum
$P_\Phi$ and axion  energy--density spectrum $P_\r$ are related by:
\beq
P_\Phi^{1/2}(k) \sim G \left(a \over k\right)^2 P_\r^{1/2}(k),
\label{743}
\eeq
where $\rho=T_0^0$, and 
\bea
\int {d^3 k\over (2 \pi k)^3} e^{i {\bf k}\cdot({\bf x}-{\bf x'})}
P_\r(k) &=&\langle \r_\sg(x) \r_\sg(x')\rangle
-\langle \r_\sg(x) \rangle^2\nonumber \\
&\sim & \langle \sg^{\prime 2}(x) \sg^{\prime 2}(x')\rangle
-\langle \sg^{\prime 2}(x)\rangle^2 + ...
\label{744}
\eea
Note that the two-point correlation  function
of  the energy density becomes a four-point function of the seed
field,  since the energy is quadratic in the axion field. 

Using now the stochastic average condition for the axion field,
\beq
\langle \sg'({\bf k}, \eta)\sg^{\prime \ast}
({\bf k'}, \eta)\rangle= (2 \pi)^3 \da^3 (k-k') \Sigma(
{\bf k}, \eta), 
\label{745}
\eeq
it turns out that the energy--density spectrum reduces to  a convolution
of  Fourier transforms,
\beq
P_\r(k) \sim {k^3\over a^4} \int d^3p~ \Sigma(p) 
\Sigma(|k-p|)+ ... ~~.
\label{746}
\eeq
Also, for a flat  enough axion  spectrum, one finds that the above
integral is dominated by the region $p\eta \sim 1$ \cite{DurGas99}. By
expressing the convolution through the spectral  energy density
$\Om_\sg$, evaluating the Bardeen potential at the  time of
re-entry $\eta_{\rm re} \sim k^{-1}$, and using Eq. (\ref{743}), one is
finally led to relating  the Bardeen spectrum and the
axion spectrum as follows: 
\beq 
P_\Phi^{1/2} (k, \eta_{\rm re}) 
\sim k^{3/2} \left|\Phi_k(\eta_{\rm re})\right| 
\sim \Om_\sg(k, \eta_{\rm re}).
\label{747}
\eeq
We should now explain why we are interested in the metric
fluctuations evaluated at the time of  re-entry.

To this purpose, let us compute the seed contribution to $\Da T
/T$. In the multipole expansion of the temperature anisotropies (in
terms of the Legendre polynomials $P_\ell$),
 \beq
\left\langle{\delta T\over T}({\bf
n}){\delta T\over T}({\bf n}') \right\rangle_{{~}_{\!\!({\bf n\cdot
n}'\equiv \cos\vartheta)}} ~~=~~~~
  {1\over 4\pi}\sum_\ell(2\ell+1)C_\ell P_\ell(\cos\vartheta)~,
\label{748}
\eeq
the coefficients $C_\ell$, at very large angular scales 
($\ell \ll 100$), are determined by the Sachs--Wolfe effect as 
\cite{Dur94}:  
\beq
C_\ell^{\rm SW} ={2\over \pi}\int
d~(\ln k) \left \langle \left[\int^{k\eta_0}_{k\eta_{\rm dec}} d(k\eta)~
k^{3/2} (\Psi +\Phi)({\bf k},
\eta)j_{\ell}^{\prime}\left(k\eta_0-k\eta \right)\right]^2 
\right\rangle .
\label{749}
 \eeq
Here $\Phi$ and $\Psi$ are the two independent components of the 
gauge-invariant Bardeen potential (see Subsection \ref{Sec4.2}), 
$j_\ell$ are the spherical Bessel functions, and a prime denotes the
derivative with respect to their argument. Equation (\ref{749}) takes
into  account both the  ``ordinary" and the ``integrated" Sachs--Wolfe 
contribution, namely the  complete  distortion of the geodesics of the
CMB photons (due to shifts in the  gravitational potential), from the time
of decoupling $\eta_{\rm dec}$ down  to the present time $\eta_0$. By
inserting the Bardeen potential  determined by the axion field, we find 
that the time-integral is  dominated by the region $k\eta \sim
1$.  Using Eq. (\ref{747}) we  finally obtain  
\beq
C_\ell^{\rm SW} \sim  \int d (\ln k)~ k^{3} |\Phi_k(\eta_{\rm re})|^2 
|j_\ell (k\eta_0)|^2, 
\label{750}
\eeq
which gives the multipole coefficients 
in terms  of the axion  distribution  $\Om_\sg (k, \eta_{\rm re})$,
through Eq. (\ref{747}).   This explains why it was important to evaluate
the Bardeen spectrum  at the time of re-entry, $\eta_{\rm re} \sim
k^{-1}$. 

From this final expression  we can extract, in 
particular, the value of the quadrupole coefficient $C_2$ \cite{DurGas99}:
\beq
C_2 \simeq \Om_\sg ^2 (k_0, \eta_0) \simeq 
\left(M_{\rm s}\over M_{\rm P}\right)^4 \left(k_0\over k_1\right)^{n-1},
\label{751}
\eeq
where $n$ is the spectral index (sufficiently near to $1$) 
characterizing the primordial axion distribution, $k_0$ is  the
comoving  scale of the present horizon, and $k_1$ the end point of
the spectrum,  namely the maximal amplified comoving frequency. We
have assumed, in the context of a minimal model of pre-big bang, that
the peak amplitude of the axion spectrum, at the  end-point frequency,
is  controlled by the fundamental ratio between string  and Planck
masses,  as it was for the graviton and dilaton spectra computed in the
previous sections. Note, however, that in the low-frequency band
corresponding to the large-scale observations the axion spectrum
contains the enhancement factor $(k/k_{\rm eq})^{-2}$ (see Eq.
(\ref{52})), which for $k=k_0$  almost exactly cancels the present
radiation energy density, $\Om_\ga(t_0) \sim10^{-4}$. 

The quadrupole coefficient, on the other hand, is at 
present  determined by COBE as  \cite{Banday97}
\beq C_2 = (1.9 \pm 0.23) \times
10^{-10}. 
\label{752}
\eeq
This experimental value, inserted into Eq. (\ref{751}),  provides an
interesting, direct  relation between the string mass and the spectral
index of the  temperature anisotropy. Using $k_1/k_0 \simeq
(M_{\rm s}/M_{\rm P})^{1/2} 10^{29}$ (see Eq. (\ref{516})), and using 
(as a reference value) the following  bounds on the spectral index
 \cite{Ben96}: 
\beq
1\leq n \leq 1.4
\label{753}
\eeq
(we have excluded  here the allowed values $0.9 \leq n\leq 1$, to avoid
introducing an infrared cut-off in the massless axion spectrum), we
obtain in fact an allowed range for the string-mass scale, 
 $0.0037 \laq M_{\rm s}/M_{\rm P} \laq 3.2$,
which is  compatible with the theoretical expectations 
\cite{Kap85}, $ 0.01\laq M_{\rm s}/M_{\rm P} \laq 0.1$. Conversely, the
theoretically expected range for $M_{\rm s}$  implies a  spectral index
$1.06 \laq n \laq 1.2$, 
which is also in agreement with the observational results. 

This promising result, however, is only a first step towards a more
complete comparison  of the seed model with
observations, comparison that must include also smaller
angular scales, where precise experimental data have recently been
obtained \cite{Mil99,Mau00,Debe00,Han00}, and will become available in
the next few years \cite{Map,Planck}. 

A numerical computation of the anisotropy power spectrum (\ref{749}), 
at intermediate angular scales (up to $\ell \sim 10^3$), has already
been  performed as reported in \cite{Mel99,Vern00}. The resulting
spectrum shows the familiar peak structure, typical of all inflationary
models. However, as was already shown very clearly  by a first analysis
\cite{Mel99}, the height and the position of the first acoustic peak can
be reproduced (at least approximately) only if the axion spectrum grows
fast enough  with frequency. In particular, using the notation of the
previous subsection, one must require at least $3-2\nu > 0.1$ (but
$3-2\nu \gaq 0.3$ seems to be favoured by the data). 

This requirement, however, is in general incompatible with the
high-frequency normalization typical of the pre-big bang scenario (if the
amplitude is large enough to match observations at the COBE scale and
the slope is too steep, then the energy density blows up and becomes
overcritical at the string scale). 

A possible solution, suggested in \cite{Mel99}, exists in the context of 
``non-minimal" models of background, leading to a slope that is steep
enough at low frequency, and flatter at high frequency, so as it limits 
the peak value. Since the axion spectrum depends both on the number
and on the kinematics of the internal dimensions (see Subsection
\ref{Sec7.2}), such a spectrum could be produced during a phase of
dynamical dimensional reduction,  in which the low-frequency modes are
frozen outside the horizon while the Universe is dimensionally different
from the subsequent phase of freezing of the high-frequency modes.  

A more detailed analysis \cite{Vern00}, based on the recent
BOOMERanG-98 \cite{Debe00} and MAXIMA-1 \cite{Han00} data, shows
indeed that it is not impossible to fit the height  of the first acoustic
peak, observed around $\ell \sim 200$, by using a (pre-big
bang-motivated) axion spectrum, which grows with a steep slope up to
a limiting scale $k_b$, and which is pretty flat above $k_b$:  
\beq
3-2\nu \simeq 0.33, ~~~~~~~~~ k<k_b, ~~~~~~~~
3-2\nu \simeq 0, ~~~~~~~~~ k>k_b.
\label{756}
\eeq
In order to be efficient, the break point $k_b$ has to be around the
equilibrium scale, $k_b \sim k_{\rm eq}$. However, the precise position
of the break is not strongly constrained, and one may exploit the
possibility of lowering the break scale, to lower the height of the
second (and subsequent) acoustic peaks, for a better comparison with
data (such a shift does not substantially affect the height of the first
peak).

A break in the axion spectrum is thus effective to match the {\em
height} of the peaks, but is not sufficient to fit the {\em position} of the
first acoustic peak, determined with small uncertainty by recent data 
\cite{Mil99,Mau00,Debe00,Han00} around $\ell \sim 200$. The position
depends on the full set of parameters of a given cosmological model
and, in the context of the seed mechanism based on massless axions, 
a peak at $\ell \sim 200$ seems to require a moderately closed
Universe , with $\Om_m+\Om_\La >1$. For instance \cite{Vern00},
\beq
\Om_m \simeq 0.4, ~~~~~~~~~~~~~
\Om_\La \simeq 0.85
\label{757}
\eeq
provides a good fit of the data, also compatible with current
supernovae results \cite{Perl99}. It should be noted, to confirm the
validity of a such model, that if the axion spectrum has a break it is also 
possible to fit  the dark-matter power spectrum, in reasonable
agreement with observations. 

However, even adjusting the break point $k_b$  and the value of
$\Om_m+\Om_\La$, the width of the peak is still not in very good
agreement with the present data, as the resulting normalized $\chi^2$
is about $1.8$, which confirms the model only at the $30$\% 
confidence level \cite{Vern00}. Nonetheless, it is
important to stress that the position of the second acoustic peak
predicted by the axion model is different from the one indicated by the
standard inflationary predictions (and by present data). In addition, the
axion model also predicts a small peak, or ``hump", around $\ell \sim
100$, due to the ``isocurvature" (rather than adiabatic, i.e. isoentropic)
nature of the induced metric perturbations. These two peculiar
predictions will be extremely important to confirm or disprove
definitely the axion model, when the future satellite  observations
\cite{Map,Planck} will become available.

Also, the axion fluctuations $\sg_k$ are described by a
Gaussian, stochastic variable, but the induced metric fluctuations
depend on the axion energy--momentum tensor: they are quadratic in
$\sg_k$ and therefore non-Gaussian, in general. This may change the
estimate of the confidence level (i.e.  of the probability) that the axion
model leads to the measured CMB anisotropy spectrum. 

It should be stressed, to conclude this subsection,  that the isocurvature
character of the primordial axion spectrum considered here is due to the
fact that the axions are treated as seeds, i.e. inhomogeneous
fluctuations of a background that {\em is not} axion-dominated. As a
consequence, the predicted anisotropies are different from previous
models of isocurvature axion perturbations \cite{Ax83,Sec85,Nambu90}.
So, even if this seed mechanism based on  isocurvature axion
perturbations were definitely ruled out by future observations, this
would not preclude the possibility of adiabatic axion perturbations (see
for instance \cite{Kawa96}), where the axions produced during the phase
of pre-big bang inflation represent  a significant fraction of the present
cold dark-matter density.

\subsection{Massive axion spectra}
\label{Sec7.4}

In the previous subsection we have discussed the CMB anisotropies
seeded by a stochastic background of massless axion fluctuations,
amplified by a phase of pre-big bang inflation. It is possible, or
even likely, however, that the axion fluctuations amplified by inflation
become massive in   the post-inflationary era: it is thus important, also
in view of future applications other than the seed mechanism, to discuss
whether,  even in the massive case, the  string normalization of the
primordial spectrum may be consistent with the large-scale
normalization determined by the CMB anisotropy, in
spite of some important differences that appear in the spectrum and in
the computation of the Sachs--Wolfe effect. 

In this subsection we shall keep assuming that the axion background is
negligible, $\langle \sg \rangle =0$, and that the whole contribution
to the axion energy density $\r_\sg$ comes from the vacuum
fluctuations, as in the previous subsection. There are, however,
various important differences.  A first difference between the massless
and massive axion cases concerns the relation between  $\r_\sg$ and the
Bardeen potential  $\Phi$. In the massless case the perturbation 
equations, taking into account the important contribution of all the 
off-diagonal terms of the axion stress tensor, lead to 
\beq
\Phi_k \simeq G \left(a\over k\right)^2 \r_\sg (k)
\label{758}
\eeq
(see Eq. (\ref{743})). 
In the massive case, on the contrary, the axion stress tensor can be 
approximated as a diagonal, perfect-fluid stress tensor, and we  
obtain \cite{GasVe99}
\beq
\Phi_k \sim G a^2 \eta^2 \r_\sg (k).
\label{759}
\eeq
Also,  the convolution (\ref{746}) 
for the axion energy density 
is dominated by the region $p \sim \eta^{-1}$ in the massless case, 
and by $p\sim k$ in the  massive case. In the massless case  the
integrated  Sachs--Wolfe effect is dominant, while it is the ordinary
one that dominates in the massive case \cite{GasVe99}.

In spite of these differences, the final result for the temperature
anisotropies is the same as before, with the quadrupole
coefficient determined by the axion  spectral energy density as
\cite{GasVe99}
 \beq
C_2 \simeq \Om_\sg^2 (k_0, \eta_0).
\label{760}
\eeq
In the massive case, however, the axion spectrum is affected by 
non-relativistic corrections, which are to be computed by including the
mass contribution to the canonical evolution equation (\ref{730}):
\beq
\psi_k^\se +\left[k^2-{z''\over z} + m^2 a^2\right]\psi_k =0,
~~~~~~~~~~~ \psi_k= z \sg_k.  
\label{761}
\eeq

To this purpose, let us consider a simple background transition at
$\eta=\eta_1$, from an initial pre-big bang phase in which the axion is
massless, to a final radiation-dominated phase in which the dilaton
freezes to its present value, $\phi=$ const, and the axion acquires a
mass, which is small in string units. Therefore, for $\eta <\eta_1$, we
 then have $m=0$, and the usual Hankel solution is normalized to a
vacuum fluctuation spectrum (see Section \ref{Sec4.4}),
\beq
\psi_k(\eta)={\sqrt \pi \over 2}|\eta|^{1/2}H_\nu^{(2)} (|k\eta|), 
\label{762} 
\eeq
where $\nu$ depends on the pre-big bang evolution of the  pump
field $z=a e^{\phi/2}$. In the subsequent radiation era, $\eta > \eta_1$,
the effective potential $z''/z$ is vanishing,  and the perturbation
equation reduces to
\beq
 \psi_k^\se +\left(k^2 +\a^2\eta^2\right)\psi_k=0,
\label{763}
\eeq
where we have put
\beq
m^2a^2 =\a^2\eta^2, ~~~~~~~~~~~~~~~~
\a= mH_1a_1^2,
\label{764}
\eeq
using the time behaviour of the scale factor $a \sim \eta$.

For the high-frequency modes that are relativistic at the transition scale,
$m \ll k/a_1$, the mass term can be neglected, and we can 
match the solution (\ref{762}) to the plane-wave
solution
\beq
\psi_k= {1\over \sqrt 2k}\left[c_+(k) e^{-ik\eta}+
c_-(k) e^{ik\eta}\right] ,
\label{765}
\eeq
to obtain:
\beq
c_\pm=\pm c(k) e^{\pm ik\eta}, ~~~~~~~~~
|c(k)|\simeq (k/k_1)^{-\nu-1/2} 
\label{766}
\eeq
(we neglect, for simplicity, numerical factors of order $1$,
which will be absorbed into the uncertainty relative to the inflation
scale, represented by the overall
numerical  coefficient $g_1$ in front of the final spectrum.) In the
relativistic regime the amplified axion perturbation then takes the
form:
\beq
\sg({\bf k}, \eta) = {c({\bf k})\over a \sqrt k}\sin (k\eta),
\label{767}
\eeq
and the corresponding energy density, $\rho_\sg^{\rm rel}(k) \sim
(\sg'/a)^2$, integrated over all modes, leads to the spectral distribution
(\ref{4101}), and to the usual relativistic spectrum (normalized to critical
units):
\beq
\Om_\sg(p,t)= g_1^2 \Om_\ga(t) 
 \left(p \over p_1\right)^{3-2\nu},  ~~~~~~~~~~
m<p<p_1, 
\label{768}
\eeq
where $p=k/a$ is the proper momentum (see Subsection \ref{Sec4.4}). 

In the radiation era the proper momentum is redshifted
with respect to the rest mass, and all axion modes tend to become
non-relativistic. When the mass term is no longer negligible, on the
other hand, the general solution of Eq. (\ref{763}) can be written in
terms of  parabolic cylinder  functions \cite{AbSte}. For an approximate
estimate of the axion field in the non-relativistic regime, however, it is
convenient to distinguish two cases, depending on whether a mode
$k$ becomes non-relativistic inside or outside the horizon.  Defining
as $k_m$ the limiting comoving frequency of a mode that becomes
non-relativistic ($k_m=ma_m$) at the time it re-enters the horizon
($k_m=H_ma_m$), we find, in the radiation era \cite{GasVe99},
\beq
k_m={H_ma_m\over H_1a_1} H_1a_1=
 k_1 \left(m\over H_1\right)^{1/2}.
\label{769}
\eeq
We will  consider the two cases $k \gg k_m$ and $k\ll k_m$ separately.

In the first case we can rewrite Eq.  (\ref{763})
as\beq {d^2\psi_k\over dx^2}+\left({x^2\over 4} -b\right)\psi_k=0,
~~~~~ x=\eta (2 \a)^{1/2}, ~~~~ -b= k^2/2\a , 
\label{770}
\eeq
and  the general solution we give has the form
\beq
	\psi=A W(b,x)+B W(b,-x),
\label{771}
\eeq
where $W(b,x)$ are the Weber parabolic cylinder functions
(see \cite{AbSte}, Chap. 19). In order to fix the integration constants
$A$ and $B$ we can match the solutions (\ref{771}) and (\ref{767})
in the relativistic limit where
\beq
{k^2 \over m^2 a^2}=
{k^2 \over \a^2 \eta^2}= {-4b\over x^2} \gg1.
\label{772}
\eeq
In this limit, since we are considering modes that become
non-relativistic when they are already inside the horizon,
\beq
\left(k\over k_m\right)^2 \sim {k^2\over \a} \sim (-b) \gg1,
\label{773}
\eeq
we can expand the $W$ functions for $b$ large with $x$ moderate
\cite{AbSte}. Matching to the plane-wave solution (\ref{765}), we obtain
$A=0$, and
\beq
\psi_k \simeq {c({\bf k})\over \a^{1/4}} W(b,-x) .
\label{774}
\eeq
In the opposite, non-relativistic limit $x^2 \gg |4b|$, the expansion
of the Weber functions gives \cite{AbSte}
\beq
\psi_k \simeq {c({\bf k})\over (\a \eta)^{1/2}} \sin\left(m\over
H\right)
\label{775}
\eeq
(we have used $x^2/4=ma\eta/2\sim m/H$). The corresponding 
axion field is then (inside the horizon)
\beq
\sg({\bf k}, \eta) = {c({\bf k})\over a \sqrt{ma}}\sin \left(m\over
H\right), ~~~~~~~~~~~~~k>k_m .
\label{776}
\eeq
The associated energy density, $\rho_\sg^{\rm non-rel}(k) \sim
(\sg'/a)^2\sim (ma/k) \rho_\sg^{\rm rel}(k)$, differs from the previous
one by the non-relativistic rescaling $m/p$, thus leading to the
spectrum  
\beq
\Om_\sg(p,t)= g_1^2 {m\over H_1} \left (H_1\over H\right)^2
\left (a_1\over a \right)^3
 \left(p \over p_1\right)^{2-2\nu}, ~~~~~~~
p_m<p<m, 
\label{777}
\eeq
where $p_m=k_m/a$ (see Eq. (\ref{768})).  

Finally, we consider  the case of a mode  that becomes non-relativistic
when it   is still
outside the horizon,  $k\ll k_m$. In this case, we cannot use the
large $|b|$ expansion as $|b| <1$, and it is convenient to express the
general solution of Eq. (\ref{770}) in the form 
\beq
	\psi=A y_1(b,x)+B y_2(b,x)~,\label{778}
\eeq
where $y_1$ and $y_2$ are the even and odd parts of the parabolic
cylinder functions \cite{AbSte}. The matching to (\ref{767}), in the
relativistic limit $x \ra 0$, gives $A=0$ and
\beq
\psi_k \simeq c({\bf k})\left(k\over 2\a \right)^{1/2} y_2(b,x) .
\label{779}
\eeq
In the non-relativistic limit $x^2 \gg |b|$ we use the relation
\cite{AbSte}
\beq
y_2 \sim \left[W(b,x)-W(b,-x)\right] \sim {1\over \sqrt x} \sin
{x^2\over 4} ,
\label{780}
\eeq
which leads to
\beq
\psi_k \simeq {c({\bf k})\over (\a \eta)^{1/2}} \left(k^2\over \a
\right)^{1/4} \sin\left(m\over H\right) ,
\label{781}
\eeq
from which
\beq
\sg({\bf k}, \eta) = {c({\bf k})\over a \sqrt{ma}}
\left(k\over k_1\right)^{1/2}\left(H_1\over m \right)^{1/4}
\sin \left(m\over H\right), ~~~~~~~~~~~~~k<k_m 
\label{782}
\eeq
with an associated spectrum 
\beq
\Om_\sg(p,t)= g_1^2 \sqrt{m\over H_1} \left(H_1\over H\right)^2
\left (a_1\over a \right)^3
 \left(p \over p_1\right)^{3-2\nu}, ~~~~~~~~~
p<p_m, 
\label{783}
\eeq
reproducing the low frequency band of the non-relativistic spectrum
already presented in the case of dilatons (see Eq. (\ref{613})). 

However, the above result is valid only if  the axion field becomes
massive {\em before} the limiting scale $p_m$ crosses the horizon. 
Let us call $T_m$ the temperature scale at which the mass turns on
(typically, $T_m\sim 100$ MeV if axions become massive at the epoch of
chiral symmetry breaking) and $p_T$ the proper frequency re-entering
the horizon precisely at the same epoch. The present value of $p_T$ is
then 
\beq
p_T(\eta_0) \simeq p_{\rm eq} \left(T_m\over {\rm eV}\right),
\label{784}
\eeq
 and the spectrum (\ref{783}) is
valid for $p_m <p_T$, namely for $m/H_{\rm eq} <(T_m/{\rm eV})^2$.
In the opposite case, $p_m >p_T$, the role of the transition
frequency, which separates modes that become non-relativistic inside
and outside the horizon, is played by $p_T$,
and the lowest-frequency
band of the non-relativistic spectrum (\ref{783}) has to be replaced by
\cite{GasVe99}:
\beq
\Om_\sg (p,t)= g_1^2 {m\over \sqrt{ H_1H_{\rm eq}}} \left({\rm
eV}\over T_m\right) \left (H_1\over H\right)^2 \left (a_1\over a
\right)^3
 \left(p \over p_1\right)^{3-2\mu}, ~
p<p_T, ~~\left(m\over H_{\rm eq}\right)^{1/2}\left( {\rm eV}\over
T_m\right) >1 
\label{785}
\eeq
(such a spectrum can be easily obtained by imposing the continuity with 
(\ref{777}) at $p=p_T$). 

In both cases, as  clearly illustrated in Fig. \ref{f73}, the effect of
non-relativistic corrections  is to enhance the amplitude of the 
spectrum at low frequency. The  enhancement is proportional to the
square root of the axion mass according to Eq. (\ref{783}), if the scale
$p_m$ is still outside  the horizon  at  the
time when axions become massive. In the opposite case,  the
enhancement  is linear in the axion  mass, according to  Eq. (\ref{785}). 
In both cases the slope is the  same as that of the massless spectrum,
$3-2\nu$. 

Since the  amplitude of the spectrum  now depends on the axion mass,  
the  condition imposed by the COBE normalization provides a
constraint on the allowed range of masses.  This might represent a
problem, in general: since the slope cannot be  too steep at low 
frequency (according to Eq. (\ref{753})), the allowed  mass has to be
small enough, typically in the range of $H_{\rm eq}$. An explicit analysis
\cite{DurGas99} then leads  to the window $10^{-27}$ eV $\laq m\laq
10^{-17}$ eV. 

It should be noted, at this point, that the mass of an axion field 
coupled to matter with a typical gravitational strength is not
significantly constrained (unlike the dilaton mass) by  the present
gravitational experiments, which are indeed to be performed with
polarized bodies \cite{Rit90} to be sensitive to pseudoscalar
interactions. Also, the Kalb-Ramond (KR) axion,  even if
(gravitationally) coupled to the QCD topological current, 
 is not to be necessarily identified  with
the ``invisible" axion \cite{Kim79} responsible for solving
 the strong CP problem. Other, more strongly coupled
 pseudoscalars, can play the traditional axion's role. 

In that case, the standard Weinberg--Wilczek formula 
\cite{Wein78,Wilc78} would give the mass of the appropriate
combination of pseudoscalars, which is coupled to the topological charge,
while the KR axion would mostly lie along the orthogonal combinations, 
which remain (almost) massless. In this context it is thus  possible that
KR axions are produced   neither  from an initial misalignment of the QCD
vacuum angle, nor from the decay of axionic cosmic strings, so that 
existing cosmological bounds on the axion mass \cite{Siki94} can be
evaded. In spite of all this, it is quite likely that the KR axion
 will be heavier than  $10^{-17}$ eV, in which
 case,  using the above spectra, it would be impossible even in principle 
to seed the observed CMB anisotropies. 

\begin{figure}[t]
\centerline{\epsfig{file=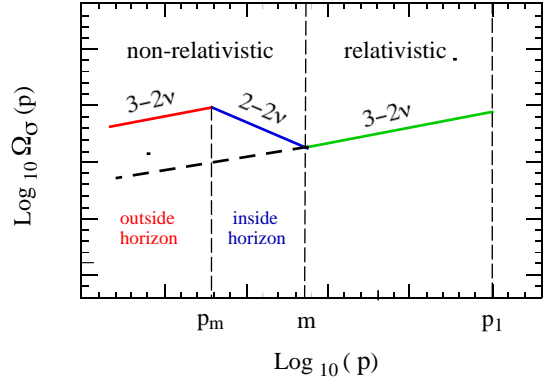,width=72mm}}
\vskip 5mm
\caption{{\sl Mass-dependent enhancement  of the spectrum at  low
frequency, due to  non-relativistic corrections.  If the slope is
sufficiently flat, $3-2\nu<1$, then the  non-relativistic
part of the spectrum may have a peak in
correspondence of the frequency  mode $p_m$ that becomes
non-relativistic at the time it re-enters the  horizon.}} 
\label{f73}
\end{figure}

This conclusion, based on the effect
illustrated in Fig. \ref{f73},  refers however to a relativistic spectrum
characterized by a constant  slope. It is quite easy to
imagine, however, and to implement  in practice, a model of background
in which the relativistic axion  spectrum is sufficiently flat  at low
frequency (as required by a fit of the  large-scale anisotropy), and much
steeper at high frequency. A simple  example is illustrated in Fig.
\ref{f74}, where we have compared two  spectra. The top one is flat
everywhere, except for modes becoming non-relativistic
outside the horizon.  The other one has a relativistic branch, which is
flat at low frequency $p< p_{\rm s}$,  and steeper at 
high frequency $p>p_{\rm s}$. It is evident that the steeper and the
longer the  high-frequency branch of the spectrum, the larger is the 
suppression of  the amplitude at low frequency, and the larger is the
axion mass allowed  by the COBE normalization at the
present Hubble scale $\om=\om_0$. 

\begin{figure}[t]
\centerline{\epsfig{file=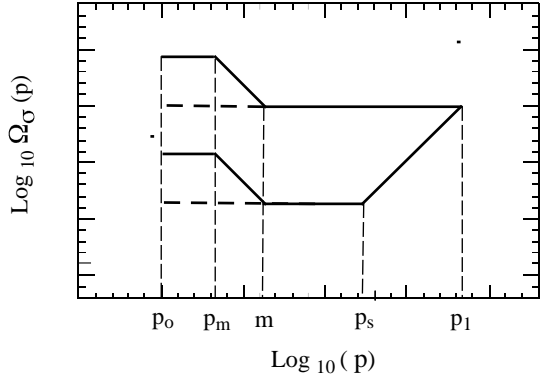,width=72mm}}
\vskip 5mm
\caption{{\sl Two examples of axion spectra with non-relativistic
corrections.  Note the common normalization at the end-point 
frequency $p_1$,  in spite of the different slopes in the different
frequency  regimes. }}
\label{f74}
\end{figure}

This possibility has been analysed through an explicit two-parameter
model of background \cite{GasVe99}, based on exact  solutions of the
low-energy string-cosmology equations with  classical string sources
\cite{GasVe94a,GGMV96,GGMV98}. By fixing the high-frequency
normalization at $Ms/ M_{\rm P} = 10^{-2}$, the low-frequency slope at
$n=1.2$, and the temperature of mass generation at $T=100$ MeV, we
can use as parameters the slope and the duration of the high-frequency
part of the spectrum, measured in particular by  $x= \log (p_1/p_{\rm
s})>0$ and by the corresponding variation of the pump field, $y=\log
(z_{\rm s}/z_1)<0$. By imposing the COBE normalization (\ref{760}),
$\Om_\sg(p_0)\simeq 10^{-5}$, the seed condition $\Om_\sg<0.1$, and
the condition of mass-dominated spectrum, $m>H_{\rm eq}$ (see
\cite{GasVe99}), the allowed region in parameter space
turns  out to be consistent with a very wide range of axion masses, 
from the  equilibrium scale $m\sim 10^{-27}$ eV up to $m \sim 100$ MeV
(higher  masses are not acceptable, because the axion  would decay into
photons before the present epoch, like dilatons, see Eq. (\ref{633})).  We
may thus conclude that there is no fundamental  incompatibility 
between the normalization imposed by the large-scale anisotropy   and
an axion mass in the  expected range of conventional axion models
\cite{Siki94}. The difficulties encountered to match the observed
spectrum at smaller angular scales, discussed in the previous
subsection, remains however . 

\subsection{Massive axions and adiabatic perturbations} 
\label{Sec7.5}

In the previous subsections we have analysed a possible scenario in
which no axion background is present in the post-big bang epoch
($\langle \sg \rangle =0$), and an {\em isocurvature}, {\em
non-Gaussian} spectrum of temperature anisotropies is eventually
induced by the quantum fluctuations of the (massless or massive) axion
field, amplified by inflation. This scenario, whose typical signatures
have  been discussed in Subsection \ref{Sec7.3},  seems to be strongly 
disfavoured with respect to  more conventional explanations of the CMB
anisotropy offered by the standard inflationary models.

In this subsection we will consider a possible alternative picture, 
based on the following assumptions: {\em i}) the constant value of the
axion background after the pre-big bang phase is displaced from the
minimum (conventionally defined as $\sg =0$) of the non-perturbative 
potential $V(\sg)$ generated in the post-big bang epoch; {\em ii})  the
axion potential  is strong enough to induce a phase of axion dominance
before its decay into radiation.  
Under these conditions one finds that the initial isocurvature axion
fluctuations, amplified by horizon-exit during inflation, can be
converted into {\em adiabatic} (and {\em Gaussian}) perturbations of
the spatial curvature, associated to super-horizon (scalar) metric
perturbations, which maintain the same spectrum as the original axion
fluctuations until the time of horizon re-entry:  these can then possibly
produce the observed CMB anisotropies.

This possibility, based on a mechanism originally  pointed out in
\cite{Mol90}, has recently been  discussed in detail 
(and not exclusively within a string cosmology framework) 
in \cite{EnSlo01,LyWa02,Moroi,BaLi02,NoRio,Mor02} for  the isocurvature
fluctuations of a generic scalar field (called ``curvaton" \cite{LyWa02}),
amplified with a flat spectrum. In the context of the pre-big bang
scenario, such a mechanism has been applied, in particular, to the
fluctuations of the KR axion \cite{EnSlo01}, as suggested in
\cite{Lidsey00}. Here, after a brief description of the conversion of
axion fluctuations into scalar curvature perturbations, we shall discuss
the constraints imposed by the CMB data, and its possible consistency
with the small-scale normalization and tilts typical of pre-big bang
models \cite{BGGV02}. 

The conversion of the axionic isocurvature modes into adiabatic
curvature inhomogeneities takes place  in the post-big-bang phase,
where we assume the dilaton $\phi$ to be frozen and the axion $\sg$ to
be displaced from the minimum of its potential. In
the E-frame the background evolution is then described by the Einstein
equations (\ref{62}), with $\phi$ simply replaced by $\sg$, and with
$c=0$ since the axion is minimally coupled to the metric. For a
conformally flat metric, $g_{\mu\nu}= a^2\eta_{\mu\nu}$, the time and
space components of such equations, together with the axion evolution
equation, can be written (in conformal time and in $d=3$) respectively as
\bea
&&
6 {\cal H}^2= a^2 \left(\r_r +\r_\sg \right), ~~~~~~~~~~~~
4 {\cal H}' + 2 {\cal H}^2 = -a^2 \left(p_r +p_\sg \right),
\label{786}\\
&&
\sg '' + 2  {\cal H} \sg ' + a^2 {\pa V \over \pa \sg}
=0,
\label{787}
\eea
where ${\cal H}=a'/a= d(\ln a)/d\eta$, $\r_r=3p_r$ is the energy density
of the radiation fluid, and
\beq
\r_\sg = {1\over 2 a^2} \sg'^2 + V(\sg), ~~~~~~~
\r_\sg = {1\over 2 a^2} \sg'^2 - V(\sg).
\label{788}
\eeq
The combination of Eqs. (\ref{786}) and (\ref{787})
leads to the conservation equation for the radiation
fluid, i.e.  $\r_r' + 4 {\cal H} \r_r=0$. In order to perform  analytical
estimates, we shall also assume here that the axion potential can be
approximated by the quadratic form  $V(\sg)=m^2 \sg^2/2$. This is
certainly true for $\sg_i \ll 1$, but it may be expected to be a realistic
approximation also for the range of values of $\sg_i$ not much larger
than $1$ (which, as we shall see, is the appropriate range for a
normalization of the spectrum compatible with present data). Actually,
for the periodic potential  expected for an axion, the  value of  $|\sg_i|$
is effectively bounded from above \cite{EnSlo01}.

Let us now consider the scalar perturbations of the Einstein equations
(\ref{62}). Working in the longitudinal gauge we set $\da \sg =\chi$,
and
we use the perturbation variables already defined in Eq. (\ref{61}) for
the metric and for the fluid sources. We impose, in addition, $\da p_r=
\da\r_r/3$ (adiabatic fluid perturbations), and we introduce the
convenient velocity potential $v_r$, such that $\da u_i= a \pa_i v_r$.

The perturbation of the $i \not=j$ component of the Einstein equations
then gives $\varphi= \psi = \Phi$, where $\Phi$ is the Bardeen
potential,
while the perturbation of the ($0,i$), ($0,0$), ($i,i$) components, and of
the axion (Klein--Gordon) equation, gives a set of four independent
equations for the four scalar variables $\Phi, ~\chi, ~\da \r_r, ~v_r$.
They can be written, respectively, in the following convenient form: 
\bea
&&
\Phi' + {\cal H} \Phi = {1\over 4} \chi \sg' +{1\over 3} a^2 \r_r v_r,
\label{789}\\
&&
\nabla^2 \Phi - 3 {\cal H} \left( \Phi' + {\cal H} \Phi\right)={1\over
4}a^2
\left(\r_\r \da_r + \r_\sg \da _\sg\right),
\label{790}\\
&&
\Phi'' +3 {\cal H} \Phi' +\left({2\cal H}' + {\cal H}^2\right) \Phi =
{1\over 4} a^2\left({1\over 3} \r_\r \da_r + \da p _\sg\right),
\label{791}\\
&&
\chi'' + 2 {\cal H} \chi' - \nabla^2\chi + a^2 {\pa^2 V \over \pa \sg^2}
\chi=4\sg' \Phi' -2 a^2 {\pa V \over \pa \sg}\Phi,
 \label{792}
\eea
where we have defined
\bea
&&
\da_r= \da \r_r /\r_r, ~~~~~~~~~~~
\da_\sg= \da \r_\sg/\r_\sg, \nonumber \\
&&
\da \r_\sg  = - \Phi \left(\r_\sg + p_\sg\right)+ {\sg ' \chi'\over a^2}
+{\pa V \over \pa \sg} \chi,
\nonumber\\
&&
\da p_\sg  = - \Phi \left(\r_\sg + p_\sg\right)+ {\sg ' \chi'\over a^2}
-{\pa V \over \pa \sg} \chi.
\label{793}
\eea
The above equations can also be obtained directly  from the
general system of scalar-perturbation equations (\ref{63})--(\ref{68})
(used to discuss dilaton production), for the particular case in which
$c=0,~ d=3, ~\b= \sg'/{\cal H}$, $ \ga=\ep=1/3,~ p=\r/3$,  and with
$\phi'$ replaced by $\sg'$. We note, for later use, that by  eliminating
$\da_r$ from Eqs. (\ref{790}) and (\ref{791}) we are led to
\beq
\Phi'' +4 {\cal H} \Phi' +\left({2\cal H}' + 2{\cal H}^2+{1\over 6}
\sg'^2 \right) \Phi -{1\over 3} \nabla^2 \Phi= {1\over 6} \sg' \chi' -
{1\over 3} a^2{\pa V\over \pa \sg}\chi,
 \label{11ax}
\eeq
which, together with Eq. (\ref{792}), provides a closed set of coupled
differential equations for the two variables $\Phi$ and $\chi$. By
using the above perturbation
equations, together with the background relations (\ref{786}) and 
(\ref{787}), two useful equations for the evolution of  $\da_r$ and
$v_r$  can be finally obtained:
\beq
\da_r' = 4 \Phi' +{4\over 3} \nabla^2 v_r, ~~~~~~~~~~~~
v_r' ={1\over 4} \da_r + \Phi.
\label{10ax}
\eeq

We now suppose that we start at $t=t_i$, with a radiation-dominated
configuration in which the axion background is initially constant and
non-vanishing, $\sg (t_i)=\sg_i \not=0$, $\sg'(t_i)=0$, providing a
subdominant (potential) energy density, $\r_\sg (t_i) =-p_\sg
(t_i)=V_i=m^2 \sg^2_i/2$ $ \ll H_i^2 \sim \r_r (t_i)$. We also give initial
conditions to our system of perturbation equations by assuming a given
spectrum of isocurvature axion fluctuations, $\chi_k(t_i)\not=0$, and
a total absence of perturbations for the metric and the radiation fluid,
$\Phi(t_i)= \da_r(t_i)= v_r(t_i)=0$. The initial values  of the first
derivatives of the perturbation variables  are then dedermined by the
momentum and Hamiltonian constraints (\ref{789}), (\ref{790}).

In order to discuss the possible generation of curvature fluctuations we
must now specify the details of the background evolution, taking into
account that the axion, initially constant and subdominant, starts
oscillating at a curvature scale $ H_{\rm osc} \sim m$ (see Eq.
(\ref{787})), and eventually decays (with gravitational strength) in
radiation,  at a scale $H_d \sim m^3/M_{\rm P}^2< H_{\rm osc}$. During
the oscillating phase the kinetic and potential energy densities are equal
on an average, so that $\langle p_\sg \rangle=0$ and $\langle\r_\sg
\rangle  \sim a^{-3}$ behaves like dust matter. Since $\r_r \sim
a^{-4}$, the radiation energy is always diluted faster, and the axion
background tends to become dominant at a scale  $H_\sg(t) \sim m
\sg(t)$.

For an efficient conversion of the initial  $\chi$ and
$\da_\sg$ fluctuations into $\Phi$ and $\da_r$ fluctuations it is further
required \cite{EnSlo01,LyWa02,Moroi}  that the decay occurs  after the
beginning of the axion-dominated phase, i.e. $H_\sg > H_d$.  Depending
upon the relative values of $H_{\sigma}$ and $H_{\rm osc}$ (i.e.
depending upon the value of $\sigma_{i}$, in Planck units) we have
thus different options which will  now be discussed separately. 

$(1)$ If $\sg_i <1$, and then $H_\sg < H_{\rm osc}$, the axion starts
oscillating  (at a scale $H \sim m$) when the Universe is still
radiation-dominated. During the oscillations the averaged potential
energy density decreases like $a^{-3}$, i.e. the typical amplitude of
the axion background decreases like $a^{-3/2}$, from its initial value
$\sg_i$ down to the value $\sg_{\rm dom}$ at which $H= H_\sg \sim m
\sg_{\rm dom}$. During this period $a \sim H^{-1/2}$ (as the background
is radiation-dominated), so that $\sg_{\rm dom} \sim \sg_i^4$, and
$H_\sg \sim m \sg_i^4$. Finally, the background remains 
axion-dominated until the decay scale $H_d \sim m^3/M_{\rm P}^2$. This
model of background is thus consistent for  $H_i > H_{\rm
osc}>H_\sg>H_d$, namely for 
\beq 
1>\sg_i > (m/M_{\rm P})^{1/2},
\label{13ax}
\eeq
which leaves a wide allowed range for $\sg_i$, if we recall the
cosmological bounds on the mass following from the decay of a
gravitationally coupled scalar \cite{ElNa86,ElTsa86}  (typically, $m>10$
TeV to  avoid disturbing standard nucleosynthesis). 

For this class of backgrounds  we can now solve the coupled
equations (\ref{792}), (\ref{11ax}), neglecting the $k^2$ terms  in  the
(Fourier transformed) perturbation equations, since we are interested
in modes well outside the horizon. We find (in cosmic time) that
the Bardeen potential $\Phi$ grows like $t^2$ during
the initial radiation phase in which the axion is slow rolling, and like
$t^{1/2}$ during the period of (radiation-dominated) axion oscillations:
\bea
\Phi_k(t) &=& A\sg_i\chi_k(t_i) \left[m(t-t_i)\right]^2,
~~~~~~~~~~~~~~~~~~~~~~~~ t_i<t<t_m, 
\label{14ax}\\
&= &\Phi_k(t_m) + B\sg_i\chi_k(t_i) \left[m(t-t_m)\right]^{1/2},
~~~~~~~ t_m<t<t_\sg
\label{15ax}
\eea
($A$ and $B$ are dimensionless numbers of order o$1$), 
where $t_m \sim H^{-1}_{\rm osc}\sim m^{-1}$ and $t_\sg \sim
H^{-1}_\sg \sim (m \sg_i^4)^{-1}$. In the subsequent axion-dominated
phase, preceeding the decay, the Bardeen potential finally oscillates
around the value $\Phi_k(t_\sg)$, with a resulting overall generation of
curvature perturbations from $\Phi_k=0$ to
\beq
\langle \Phi_k(t_\sg) \rangle \sim \sg_i\chi_k(t_i) + {\chi_k(t_i)\over
\sg_i} \sim  {\chi_k(t_i)\over \sg_i}, ~~~~~~~~~~~~\sg_i <1 .
\label{16ax}
\eeq
where $\langle \dots\rangle$ refers to averages over one oscillation
period.

$(2)$ If $\sg_i \sim 1$, and then $H_\sg \sim H_{\rm osc}\sim m$, the
beginning of the oscillating and axion-dominated phase are nearly
simultaneous, and the Bardeen potential only grows 
during the initial radiation-dominated phase, according to eq.
(\ref{14ax}). The final averaged value of the induced perturbations is
now  
\beq
\langle \Phi_k(t_\sg) \rangle \sim \sg_i\chi_k(t_i)  \sim 
{\chi_k(t_i)}, ~~~~~~~~~~~~~~~~~~~~~~~\sg_i \sim 1 . 
\label{17ax}
\eeq

$(3)$ Finally, if $\sg_i >$, and then $H_\sg >H_{\rm osc}$, the axion
starts dominating at the scale $H_\sg \sim m \sg_i$, which marks the
beginning of a phase of slow-roll inflation, lasting until the curvature
drops below the oscillation scale $H_{\rm osc} \sim m$. Such a model of
background is consistent for $H_\sg <H_1$, namely for 
\beq
{H_1\over m} >\sg_i >1,
\label{18ax}
\eeq
where $H_1$ (fixed around the string scale) corresponds to the
beginning of the radiation-dominated, post-big bang evolution. During
the inflationary phase the slow decrease of the Hubble scale can be
approssimated (according to the background equations (\ref{786}) and 
(\ref{787})) as  $H(t) =\a m\sg_i -\b m^2 (t-t_\sg)$, where $\a$ and $\b$
are dimensionless coefficients of order $1$. Inflation thus begins at the
epoch $t_\sg \sim 1/m\sg_i$, and lasts until the epoch $t_m \sim
(\sg_i-1)/m \sim \sg_i/m$. Let us estimate, for this background, the
evolution of the Bardeen potential generated by the primordial axion
fluctuations. 

For $t<t_\sg$ the solution is still given by eq. (\ref{14ax}), with $t_m$
replaced by $t_\sg =1/m\sg_i$. At the beginning of inflation the
averaged amplitude is then $\Phi_k(t_\sg) \sim \chi_k(t_i)/\sg_i$. 
During slow-roll inflation, on the other hand, it is known \cite{MFB92} 
that,  for superhorizon modes, $\Phi_k \sim \dot H/H^2$. The final
amplitude at the beginning of the oscillating phase is then  
\beq
\langle \Phi_k(t_m) \rangle =\Phi_k(t_\sg) \left(\dot H\over
H^2\right)_{t_m}  \left(H^2\over \dot H\right)_{t_\sg} \sim
\sg_i^2\Phi_k(t_\sg) \sim \sg_i \chi_k(t_i),
~~~~~~~~~~\sg_i > 1 ,   
\label{19ax}
\eeq
and no further amplification  is expected in the course of the
subsequent cosmological evolution. 

It is amusing to observe that the  results (\ref{16ax}), (\ref{17ax}),
(\ref{19ax}), which determine the amplitude of the Bardeen potential in
the oscillating (axion-dominated) phase preceding the axion decay, 
can be summarized by an equation that holds in all cases, namely
\beq
\langle \Phi_k(t_d) \rangle =-\chi_k(t_i)f(\sg_i), ~~~~~~~~~~~
f(\sg_i)= \left(c_1 \sg_i+{c_2\over
\sg_i}+c_3\right),
\label{20ax}
\eeq
where $c_1,c_2,c_3$ are numerical coefficients of the  order of unity. 
A preliminary fit \cite{BGGV02} based on numerical and analytical
integrations of the perturbation equations gives
$c_1=0.129,c_2=0.183,c_3=0.019$.  
The function $f(\sg_i)$ has the interesting feature that it is
approximately invariant under the transformation $\sg_i \ra \sg_i^{-1}$
and, as a consequence,  has a {\it minimal} value around $\sg_i =1 $.

Given $\Phi_k$,  we can easily compute the final value (before axion
decay) of the density constrasts $\da_\sg, \da_r$, and of the
variable $\zeta$  parametrizing in a gauge-invariant way the strength
of curvature perturbations. By using the
background equations   (\ref{786}) in the axion-dominated phase,
$\r_\sg \gg \r_r$, and taking the average of Eq. (\ref{790}), for
super-horizon modes, one finds indeed (averaging being understood) 
$\da_\sg (k) \simeq -2 \Phi_k$. Similarly, from Eqs. (\ref{10ax}),  and
for super-horizon modes,  $\da_r (k) \simeq 4\Phi_k$.

Consider then the variable $\zeta$ representing the perturbation of the
spatial curvature on uniform density hypersurfaces, which is conserved
(outside the horizon) for adiabatic perturbations, and which can be
written for a general background as \cite{MFB92}:
\beq
\zeta= -\Phi -{{\cal H} \Phi' + {\cal H}^2 \Phi
\over {\cal H}^2 - {\cal H}'}.
\label{799}
\eeq
Outside the horizon we can use $-\da_\sg/2$ for $\Phi$, the sum of
the
two background equations (\ref{786}) for the denominator
${\cal H}^2 - {\cal H}'$, and the Hamiltonian constraint
(\ref{790}) for the numerator ${\cal H} \Phi' + {\cal H}^2 \Phi$, in
order
to rewrite $\zeta$ in the form
\beq
\zeta_k= {\r_\sg \da_\sg(k) -(3/4) ( \r_\sg + p_\sg) \da_r
(k) \over 4 \r_r + 3 ( \r_\sg + p_\sg)}.
\label{7100}
\eeq
In the final phase dominated by an oscillating axion 
$\r_r$ is negligible, the (averaged) axion pressure is zero so that, using
$\da_\sg=- 2 \Phi=-\da_r/2$ we finally obtain
\beq
\zeta_k \simeq {5\over 6} \da_\sg (k) \simeq -{5\over 3} \Phi_k 
\label{7101}
\eeq
(as we have also  checked by an explicit numerical integration
\cite{BGGV02}). This result was derived and presented in
\cite{LyWa02}, using however different notations. 

The generated spectrum of super-horizon curvature perturbations is
thus directly determined by the primordial spectrum of isocurvature
axion fluctuations $\chi_k$, according to Eqs. (\ref{20ax}) and 
(\ref{7101}).
The axion fluctuations, on the other hand, are solutions (with pre-big
bang initial conditions) of the perturbation equation (\ref{792})  in the
radiation era (no additional amplification is expected, for super-horizon
modes, in the axion-dominated phase), computed for negligible
curvature perturbations ($\Phi=0=\Phi'$) in the massive, non-relativistic
limit (as we are already in the oscillating regime), and outside the
horizon.

In order  to exploit the results of the previous subsection, we can then
set $\chi_k= \psi_k/a$, where $\psi_k$ satisfies Eq. (\ref{770}). The
general solution for $\psi_k$, correctly normalized to a relativistic
spectrum of quantum fluctuations, has been  given in Eq. 
(\ref{779}) already, taking also into account  the details of the pre-big bang
amplification (encoded into the Bogoliubov coefficient $c(k)$). Outside
the horizon, $k |\eta| \ll 1$, and for non-relativistic modes, $k \ll
ma$,
we take (respectively) the limits $-b x^2 \ll 1$ and $-b \ll x^2$, in which
$y_2
\sim x = \eta \sqrt{2 \a}$. By inserting a generic, primordial power-law
spectrum with cut-off scale $k_1=H_1 a_1$, i.e. $c(k) =
(k/k_1)^{(n-5)/2}$, and using the identity $\eta/a= \eta_1/a_1\simeq
1/k_1a_1$, we finally obtain the generated spectrum of curvature
perturbations: 
\beq
k^3 \left| \Phi_k\right|^2 = f^2(\sg_i){k^3} \left|
\chi_k\right|^2
= f^2(\sg_i)\left(H_1\over  M_{\rm P}\right)^2
\left( k\over k_1\right)^{n-1}, ~~~~~~~~~~~~~~~k<k_1
\label{7102}
\eeq
We have re-inserted the appropriate Planck mass factors, keeping
however $\sg_i$ dimensionless in Planck units, and we have included
into $k_1$ possible numerical factors of order $1$, connecting the
cut-off scale to the string mass scale

This result, valid during the axion-dominated phase, has to be
transferred to the phase of standard evolution, by matching the (well
known \cite{MFB92}) solution of the Bardeen potential in the radiation
era (subsequent to axion decay) to the solution prior to decay, which is
in general oscillating. The matching of $\Phi$ and $\Phi'$,
conventionally
performed at the fixed scale $H=H_d$, shows that the constant
asymptotic value (\ref{20ax}) of super-horizon modes is preserved to
leading order by the decay process, modulo a random, mass-dependent
correction which typically takes the form 
$\left[1+ \ep \sin (m/H_d)\right]$, where $\ep$ is a numerical
coefficient of order $1$, and $m \gg H_d$. Such a
random factor, however,  is a consequence of the sudden
approximation adopted to describe the decay process, and disappears
in a more realistic treatment in which the evolution equations for
$\r_\sg$   and $\da\r_\sg$ are    
supplemented by the friction term $\Gamma {\sigma'}/{a}$  (with a
corresponding antifriction term $-\Gamma {{\sigma'}^2}/{a^2}$ in the
radiation equation). 

In such a way, as we have checked with an numerical integration, the
decay process preserves the  value of the Bardeen potential prior to
decay, damping the residual oscillations; $\zeta$ itself follows the
same behaviour and is finally exactly a constant. When the axion has
completely decayed, and the Universe is again dominated by radiation, 
we can properly match the standard evolution of $\Phi$ in the radiation
phase to the constant asymptotic value of Eq. (\ref{20ax}). The
expression we obtain for the Bardeen potential, valid until
the epoch of matter--radiation equality, can be written in the form 
\begin{equation}
\Phi_{k}(\eta) =-3\Phi_{k}(\eta_{d})  \biggl[ \frac{\cos{(
k c_s \eta)}}{(k c_s \eta)^2} - \frac{\sin{(k c_s\eta)}}{(k
c_{s}\eta)^3} \biggr] , ~~~~~~~~~\eta_d < \eta< \eta_{\rm eq}, 
\label{barfin}
\end{equation}
where $c_{s} = 1/\sqrt{3}$ and $\Phi_{k}(\eta_{d})$ is given in Eq.
(\ref{20ax}).

The above expression for the Bardeen potential  provides the initial
condition for the evolution of the CMB-temperature  fluctuations, 
and the formation of their oscillatory pattern. It is important to note
that, from the above equation,  $\Phi_{k}(\eta_i) = {\rm
constant}$ and $\Phi'_{k}(\eta_i) \simeq 0$, where
$\eta_d<\eta_i<\eta_{\rm eq}$, and $ k\eta_i \ll1$. This implies
\cite{BGGV02} that the induced temperature fluctuations  $(\Da T/T)_k$
will oscillate  in $k$ just as expected in the case of adiabatic
perturbations,  producing a peak structure clearly distinguishable 
(and, at present, observationally favoured) from that
produced by isocurvature fluctuations. 

Let us now discuss the large scale normalization of the induced
anisotropies, starting from the observation that the final amplitude of
the super-horizon perturbations  (\ref{7102}) seems to be unaffected
by the non-relativistic corrections to the axion spectrum. The
dependence on the axion mass  appears, however, when
computing the amplitude of the spectrum at the present horizon scale
$\om_0$, in order to impose the corrected normalization to the
quadrupole coefficient $C_2$, namely  \cite{Dur01}
\beq
C_2= 
\a^2_n f^2(\sg_i) 
\left(H_1\over M_{\rm P}\right)^2\left(\om_0\over
\om_1\right)^{n-1},  ~~~~~~~
\a_n^2 = {4^{n-3\over 2}\over 9}{\Ga (3-n) \Ga\left({3+n\over 2}\right)
\over \Ga^2\left({4-n\over 2}\right) \Ga\left({9-n\over 2}\right)},
\label{alpha}
\eeq
where the value of $C_2$ is experimentally determined by COBE
according to Eq. (\ref{752}). 

The present value of the cut-off
frequency, $\om_1(t_0)=H_1a_1/a_0$, depends in fact on the kinematics
as well as  on the duration of the axion-dominated phase (and thus on
the axion mass), as follows: 
\bea
\om_1(t_0)  &=& H_1
\left(a_1\over a_\sg\right)_{\rm rad} \left(a_\sg\over
a_d\right)_{\rm mat} \left(a_d\over a_{\rm eq}\right)_{\rm rad}
\left(a_{\rm eq}\over a_0\right)_{\rm
mat}, ~~~~~~~~~~~~~~~~~ \sg_i < 1,
\label{28ax}\\
&=&H_1\left(a_1\over a_\sg\right)_{\rm
rad} \left(a_\sg\over a_{\rm
osc}\right)_{\rm inf} \left(a_{\rm
osc}\over a_d\right)_{\rm
mat} \left(a_d\over a_{\rm eq}\right)_{\rm
rad} \left(a_{\rm eq}\over a_0\right)_{\rm
mat}, ~~~ \sg_i >1. 
\label{29ax}
\eea
Using $H_0 \simeq 10^{-6} H_{\rm eq} \simeq 10^{-61}
M_{\rm P}$, we find that  
the COBE normalization  imposes
\bea
&&
c_2^2\a^2_n \sg_i^{2{(n-4)\over 3}}
\left(H_1\over M_{\rm P}\right)^{(5-n)\over 2}
\left(m\over M_{\rm P}\right)^{-{(n-1)\over 3}}
10^{-29(n-1)} \simeq 10^{-10}, ~~~~~~~\sg_i < 1,
\label{32ax}\\
&&
c_1^2\a^2_n Z_\sg^{n-1} \sg_i^{(5-n)\over 2}
\left(H_1\over M_{\rm P}\right)^{(5-n)\over 2}
\left(m\over M_{\rm P}\right)^{-{(n-1)\over 3}} 
10^{-29(n-1)} \simeq 10^{-10}, ~~\sg_i >1,
\label{33ax}
\eea
where $Z_\sg= (a_{\rm osc}/a_\sg)>1$ denotes the
amplification of the scale factor during the phase of axion-dominated,
slow-roll inflation.

The condition (\ref{32ax}) is to be combined with the constraint
(\ref{13ax}), the condition (\ref{33ax}) with the constraint (\ref{18ax}),
which are required for the consistency of the corresponding classes of
background evolution. Also, both conditions are to be intersected
with the experimentally allowed range of the spectral index.
The allowed range of parameters compatible with all the constraints is
rather strongly sensitive to the values of the pre-big bang
inflation scale $H_1$. In the context of minimal models of pre-big
bang inflation  we have $H_1 \sim M_s$, and a flat spectrum
($n=1, \a_n^2=1/54 \pi$) is inconsistent with the normalization
(\ref{32ax}), (\ref{33ax}). A growing (``blue") spectrum is instead
allowed (see \cite{BGGV02} for a detailed discussion). It turns out, in
particular, that there is a wide range of allowed axion masses, but a
rather narrow range of allowed values for $\sg_i$, especially in the case
$\sg_i>1$. 

The allowed range of parameters is extended if the inflation scale $H_1$
is lowered, and a flat ($n=1$) spectrum may become possible if $H_1
\laq 10^{-5}(M_{\rm P} \sg_i/\a_1c_2)$, for $\sg_i<1$, and if $H_1 \laq
10^{-5}(M_{\rm P} /\sg_i\a_1c_1)$, for $\sg_i>1$ (see Eqs. (\ref{32ax})
and (\ref{33ax})). A lower value of the string scale could emerge indeed 
in a recently proposed framework \cite{Ven01} according to which, at
strong bare-coupling $e^{\phi}$,
 loop effects renormalize downwards the ratio $M_s/M_{\rm P}$
 and allow $M_s$ to approach the unification scale. 

In addition,  a flat
spectrum may be allowed even keeping pre-big bang inflation at a
high-curvature  scale, provided the relativistic
branch of the primordial axion fluctuations is characterized by a
frequency-dependent slope, which is flat enough at low frequency (to
agree with large-scale observations) and much steeper at high
frequencies (to match the string normalization at the end-point of the
spectrum). A typical example of such a spectrum, already introduced in
the previous subsection, and illustrated in Fig. \ref{f74},  can be
parametrized by a Bogoliubov coefficient with a break at the
intermediate scale $k_s$,  
\bea
|c_k|^2 &=& \left(k\over k_1\right)^{n- 5 + \delta},
~~~~~~~~~~~~~~~~~~~~k_s<k<k_1,
\nonumber\\
&=& \left(k_s\over k_1\right)^{n- 5 + \delta}
\left(k\over k_s\right)^{n-5},
~~~~~~~~~~~~ ~~k<k_s,
\label{35ax}
\eea
where $\delta > 0 $ parametrizes the slope of the break at high
frequency. Examples of realistic pre-big bang backgrounds producing
such a spectrum of axion fluctuations have been discussed in
\cite{GasVe99}. It is clear that the steeper (and the longer) the
high-frequency branch of the spectrum, the easier the matching of the
amplitude to the measured anisotropies (in spite of possible
$\sg_i$-dependent enhancements).

Using the generalized input (\ref{35ax}) for the spectrum of $\chi_k$,
the amplitude of the low-frequency ($k<k_s$) Bardeen spectrum
(\ref{7102}) is to be multiplied by the suppression factor
$\Da=(k_s/k_1)^{\delta} \ll 1$,
and the normalization condition at the COBE
scale becomes
\bea
&&
\a^2_n c_2^2 \left(H_1\over \sg_i M_{\rm P}\right)^{2}
\left(\om_0\over \om_1\right)^{n-1}
 \simeq C_2\Da^{-1},
~~~~~~~~~~\sg_i <1,
\label{36ax}\\
&&
\a^2_n c_1^2 \left(\sg_i H_1\over M_{\rm
P}\right)^{2} \left(\om_0\over
\om_1\right)^{n-1}
 \simeq C_2\Da^{-1},
~~~~~~~~~~~\sg_i >1.
\label{37ax}
\eea
A strictly flat spectrum is now possible, even for $\a_1H_1=\a_1 
M_s\simeq 10^{- 2}M_{\rm P}$, provided
\beq
\Da \left(c_1^2\sg_i^2
+ c_2^2\sg_i^{-2} \right) \laq 10^{-6}. 
\label{38ax}
\eeq
It thus becomes  possible, in this context, to satisfy the stringent
limits imposed by the most recent analyses of the peak and dip
structure of the spectrum at small scales \cite{Debe02}, which imply
$0.87 \leq n \leq 1.06$. 

We may thus conclude that,  in the context of the pre-big bang scenario,
a ``curvaton" model  based on the Kalb--Ramond axion is able to
produce the adiabatic curvature perturbation needed to explain the
observed large-scale anisotropies. The simplest, minimal models of
pre-big bang inflation seem to prefer blue spectra. A strictly
scale-invariant (or even slightly red, $n<1$) spectrum is not excluded 
but requires, for normalization purposes,
non-minimal models of pre-big bang evolution leading to axion
fluctuations with a sufficiently steep slope at high frequencies.

\section{Singularity and ``graceful exit"}
\label{Sec8}
\setcounter{equation}{0}
\setcounter{figure}{0}

As already stressed in Section \ref{Sec2}, the cosmological solutions of
the string effective action, thanks to their symmetries, are
characterized by four branches, which are inter-related by
time-reversal and  scale-factor duality transformations. Two branches
are of the pre-big bang type, and  evolve {\em towards} a singularity,
while the  other two branches are of the post-big bang type, and  
emerge {\it from} a singularity. A typical example is illustrated in Fig.
\ref{f81}, where we have plotted the asymptotic, $d$-dimensional
vacuum solutions of the lowest-order gravidilaton effective action, 
\beq
a(t)= (\mp t)^{\mp1/\sqrt d} , ~~~~~~~~~~~~~~
\fb (t)= -\ln (\mp t) , 
\label{81}
\eeq
which represent the bisecting lines of the ``phase space" spanned by
the variables $\{\dot{\fb},\sqrt{d} H\}$. The initial conditions of the
pre-big bang scenario and the present standard cosmological
configuration are thus disconnected by a singularity localized at the
future or past end, respectively, of the temporal half-line on which they
are defined. 

In the case of homogeneous and isotropic metric
backgrounds, with or without spatial curvature, there are ``no-go
theorems"   \cite{BruVe94,Kal95,Kal96,Eas96} that exclude the
possibility of a smooth transition from the pre- to the post-big bang
branch of the tree-level solutions, even including a ({\em local}) dilaton
potential  $V(\phi)$, and matter sources in the form of perfect fluids
and/or an homogeneous NS-NS two-form. In all these cases the
singularity cannot be removed. In addition, in the absence of sources and
of a  dilaton potential, it is known that the singularity also affects
anisotropic backgrounds, and can be shown to be generally present in all
homogeneous four-dimensional metrics of Bianchi type I, II, III, V,
${\rm VI}_0$ and ${\rm VI}_h$ \cite{GasRi95}. 

We are thus led to the problem of
explaining how the Universe may smoothly evolve from the string
perturbative vacuum, and from an initial accelerated growth of the
curvature and of the string coupling, to a final, decelerated, post-big
bang configuration, which is certainly included in the set of
string-cosmology solutions, but is possibly disconnected from the initial
pre-big bang branch by a curvature singularity. This is the so-called 
``graceful exit" problem. 

\begin{figure}[t]
\centerline{\epsfig{file=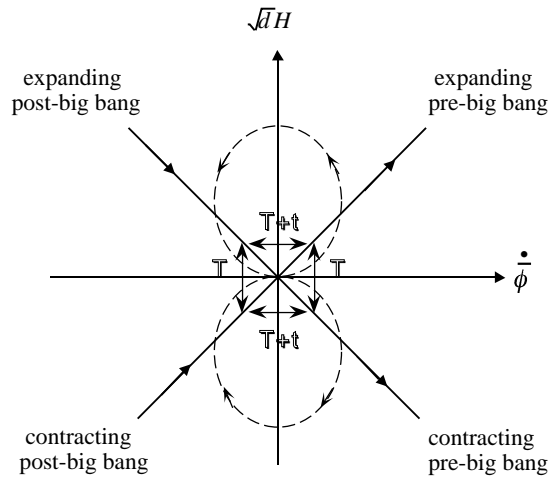,width=72mm}}
\vskip 5mm
\caption{\sl The four branches of the lowest-order string effective
action (solid lines), and a possible self-dual solution (dashed curve), 
characterized by $a(t)= a^{-1} (-t)$ and $\fb (t)= \fb (-t)$. The
symbols $\cal T$ and $t$ represent a T-duality and a
time-reversal transformation, respectively.} 
\label{f81}
\end{figure}

The exit from inflation, and a smooth description of the transition from
accelerated to decelerated expansion, is a typical problem of all
inflationary models (see for instance \cite{Bar86}). In the string
cosmology case, however,  such a problem is sharpened by the fact that
the transition from the pre- to the post-big bang phase is expected to
occur at high curvature and strong coupling,  possibly in the full 
quantum-gravity regime. This may suggest that a correct approach to 
the exit problem is to be developed in a quantum cosmology context, by
exploiting the Wheeler-De Witt equation in the string cosmology
minisuperspace, and the associated $O(d,d)$ symmetry, which helps
avoiding operator ordering ambiguities \cite{GasMaVe96,Ke96}. 

In this approach, interesting in itself,  the probability of
quantum transition from the pre- to the post-big bang branches of the
low-energy solutions can be finite and non-vanishing, even if the two
branches are classically disconnected by a singularity
\cite{GasMaVe96,GasVe96a}. This possibility will be presented in detail in
Section \ref{Sec9}. In this section we shall keep on a classical
(i.e. deterministic) description of the exit, which requires the smoothing
out of the background singularities. 

It should be stressed, in particular, that the phenomenological 
predictions made in the previous sections were based on the assumption
that the singularity can be regularized, and that {\em i}) a graceful exit  
does take place; {\em ii}) sufficiently large scales are only affected by it
kinematically,   i.e. through an overall redshift of all scales. 
Of course, one would  not only like to know that a graceful exit does  
take place: one would also like to describe the transition between the
two phases in a quantitative way. Achieving this goal would amount to
nothing less than a full description of what replaces the
big bang of standard cosmology in the pre-big bang scenario. As
mentioned in Section \ref{Sec1}, this problem of string cosmology is the
analogue, in some sense, of the (still not fully solved) confinement
problem of QCD. 

The problem of the background regularization is particularly hard
because, besides the smoothing out of the curvature, it also requires
the smoothing out of the dilaton kinetic energy, and a possible damping
of the dilaton (i.e. of the string coupling). We may consider, for a simple
illustrative example of this point, the lowest-order gravidilaton action, 
supplemented by a matter action $S_m$ that describes perfect-fluid
sources, with vanishing dilatonic charge, $\da S_m/\da \phi=0$: 
\beq
S = -\frac{1}{2\,\lambda_{\rm s}^{d-1}}\,\int\,d^{d+1}x\,\sqrt{|g|}\,
e^{-\phi}
\,\left[R+(\nabla_{\mu}\phi)^2+ V -{1\over 12} H_{\mu\nu \a}^2
\right]+S_m.
\label{82}
\eeq
By assuming an isotropic, spatially flat metric background, with spatial
sections of finite volume, we set
\bea
\sqrt{-g}~T_{\mu\nu} =2 {\da S_m\over \da g_{\mu\nu}}, ~~~~~~~~~~~~~~
&&T_\mu\,^\nu= {\rm diag} \left(\r(t), - p(t) \da_i^j\right),
\nonumber\\
g_{\mu\nu}={\rm diag} \left(1, -a^2(t) \da_{ij}\right), ~~~~~~~
&&\fb=\phi-{d}\ln a, ~~~~~~ \phi= \phi(t).
\label{83}
\eea
For $V=0$, $H_{\mu\nu\a}=0$, the field equations can be written in the
form 
\bea
&&
\fbp^2 -dH^2= e^{\fb} \rb, \label{84}\\
&&
\dot H-H \fbp={1\over 2} e^{\fb} \pb, \label{85}\\
&&
2 \ddot{\fb} -\fbp^2 -dH^2=0
\label{86}
\eea
where $\rb= \r a^d$, $\pb= p a^d$ (see Subsection \ref{Sec2.2}), and  we
have used units in which $2 \la_{\rm s}^{d-1}=1$.  Their combination
gives the conservation equation
\beq
\dot {\rb} +d H\pb=0. 
\label{87}
\eeq
The above equations can be satisfied by the following particular exact
solution 
\bea 
H= {a^{\sqrt d}\over (1+a^{\sqrt d})^2}, ~~~~
&&\rb= k H^2, ~~~~~~~~~~ \pb= -{2\over \sqrt d}{1-a^{sqrt d}\over 1+
a^{\sqrt d}} \rb , \nonumber \\
\fb =\ln {4 d\over k} {a^{\sqrt d}\over (1-a^{\sqrt d})^2}, ~~~~~~~~
&&\fbp = \sqrt d H {1+a^{\sqrt d}\over 1-a^{\sqrt d}} ,
\label{88}
\eea
where $k$ is an arbitrary  integration constant. 
The integration of $H=H(a)$ then leads  to the implicit expression for the
scale factor,
\beq
{t\over t_0}= {1\over \sqrt d}\left( a^{\sqrt d} - 
a^{-\sqrt d}\right) + 2 \ln a. 
\label{89}
\eeq
This solution is self-dual, in the sense that $a(t)= a^{-1}(-t)$, $\rb (a)=
\rb(a^{-1})$, $\fb (a)=\fb(a^{-1})$, and describes a smooth evolution of
the metric from an initial, accelerated phase of growing curvature and
negative pressure, to a final, decelerated phase of decreasing curvature
and positive pressure. The scale factor runs monotonically from zero to
infinity, without any curvature singularity. 

This example cannot be accepted as a successful model of exit,
however, because the dilaton kinetic energy ($\dot \phi^2$) blows up at
the transition point ($a=1$), together with the effective string coupling,
$\exp \phi$. This implies that both $\ap$ and loop corrections to the
effective action become non-negligible at the transition, and that the
above solution, obtained in the low-energy, tree-level approximation, 
cannot be trusted. In addition, the curvature is regular in the string
frame, but blows up in the Einstein frame defined by
\beq
\ti a= a e^{-\phi/(d-1)}, ~~~~~~~~~~~~
d \ti t = dt e^{-\phi/(d-1)}
\label{810}
\eeq
(see for instance Subsection \ref{Sec1.3}). Indeed, by setting  
$H = d(\ln a )/dt$ and  $\ti H = d (\ln \ti a )/d \ti t$, we obtain
\beq
\ti H= \left(H- {1\over d-1} \dot \phi \right) e^{\phi/(d-1)},
\label{811}
\eeq
so that, if the dilaton is unbounded, $\ti H$ may diverge even if  both $H$
and $\dot \phi$ are  regular. 

A successful exit thus requires a smooth dilaton evolution, with the
kinetic energy and $g_{\rm s}^2=\exp \phi$  bounded everywhere. In addition,
the dilaton has to evolve from the initial regime with $\fbp >0$ to a final
regime with $\fbp <0$ (see Fig. \ref{f81}); 
besides being bounded, at large positive times, its kinetic energy must
thus satisfy the condition  
\beq
\dot \phi <d H.
\label{812}
\eeq

This is not sufficient, however, for a complete exit. An even stronger
condition is imposed by requiring that the final, post-big bang
configuration may describe (even in the E-frame) a phase of
decelerated expansion, after the contraction representing pre-big bang
inflation (after all, the String and Einstein frames are doomed to
coincide, eventually, in the limit of dilaton stabilization).  This implies 
$\ti H>0$ at large positive times or, according to Eq. (\ref{811}),
\beq
\dot \phi <(d-1) H.
\label{813}
\eeq
The above constraints can be translated into energy conditions, to be
satisfied by the stress tensor of the effective sources driving the exit
transition \cite{BruMad97,BruMad98}. 

In spite of such constraints, there are simple examples of regular
backgrounds implementing a successful exit by means of appropriate 
sources, as we shall see in Subsection
\ref{Sec8.1}. The simplest examples, however, are not very realistic. The
most realistic physical mechanism of exit, in the perturbative regime, is
probably based on the backreaction of the quantum fluctuations, which
leads, however, to including loops (and possibly higher derivative)
corrections in the effective action, as we shall discuss in Subsections
\ref{Sec8.2} and  \ref{Sec8.3}. Such corrections suggest possible lines
along  which a quantitative description of the   exit might eventually 
emerge. 

The loop corrections, however, become important in the strong coupling
regime, i.e. in the realm of the still largely unknown M-theory
\cite{Wit95,Hor96}, as will be discussed in Subsection \ref{Sec8.5}. It is
thus possible that, in spite of the existing encouraging results, new
techniques and/or a deeper   understanding of string theory
in its non-perturbative regimes need  be developed (see 
\cite{BraLuOv01} for a recent attempt), before a fully satisfactory
description of the transition from pre- to post-big bang can be hoped for.

In the next subsection we will start presenting a few simple examples
of smooth solutions, without curvature singularities, obtained in the
context of the low-energy string effective action.

\subsection{Smoothing out the singularity at low energy}
\label{Sec8.1}

It is well known, from the singularity theorems of general relativity
(see for instance \cite{HP70,HawEl73,Wald}), that a smooth and complete
geometrical background, with all the curvature invariants bounded
everywhere, is only allowed if the sources of the gravitational field
satisfy appropriate energy conditions. 

In the context of the string cosmology equations, the situation is even
more constrained, because also the string coupling $g_{\rm s}$,  
parametrized by the dilaton, has to be bounded (and possibly smaller
than $1$) for a consistent and appropriate use of the string effective
action (and of the perturbative expansion). In addition, if the curvature
and the dilaton kinetic energy are regular in the S-frame, but the
dilaton is free to grow boundlessly, then  divergences may appear in
the E-frame according to Eq. (\ref{811}). This may occur, for instance, if
the regularization of the curvature in the S-frame leads to a phase of
constant $H$ and $\dot \phi$, as in the context of the higher-order
effective action \cite{GG92z,GasMaVe97}, which will be discussed in
Subsection \ref{Sec8.3}. The importance of a frame-independent
regularization is stressed by the fact that some particles, such as the
gravitons, follow the E-frame geodesics \cite{KaOl98}. 

In spite of such difficulties, examples of smooth, self-dual solutions
with no curvature singularities have been obtained, even in the context
of the lowest-order string effective action, and of homogeneous and
isotropic backgrounds, by introducing however {\em ad hoc}  a {\em
negative} and {\em non-local} dilaton potential
\cite{MeiVe91,GasVe93a,GasMaVe96}, $V(\fb)<0$. Such a potential, which
only makes sense for manifolds with spatial sections of finite volume, 
$\left(\int d^dx \sqrt{-g}\right)_{t={\rm const}}<\infty$ (for instance, a
torus), is non-local because $\fb$, to be a generally covariant scalar,
must contain the integral of the whole spatial volume (see Subsection 
\ref{Sec1.4}). 

Consider, as an example, the action (\ref{82}), with a dilaton potential
$V(\fb)$. The field equations, for an isotropic and spatially flat metric,
become ($2 \la_{\rm s}^{d-1}=1$)
\bea
&&
\fbp^2 -dH^2-V= \rb e^{\fb} , \label{814}\\
&&
\dot H-H \fbp={1\over 2}\pb e^{\fb} , \label{815}\\
&&
\fbp^2 -2 \ddot{\fb}+dH^2+{\pa V\over \pa \fb}-V=0, 
\label{816}
\eea
and the conservation equation (\ref{87}) is still valid. In the vacuum
case $\r=p=0$, the full set of equations can be integrated exactly
\cite{MeiVe91}, and one finds that non-singular solutions are in general
allowed. If we consider, for instance, the potential \cite{GasMaVe96}
\beq
V(\fb)= -V_0e^{4\fb},
\label{817}
\eeq
the above equations can be
exactly satisfied by the particular solution 
\beq
a=a_0 \left[{k^2t\over \sqrt{V_0}} +\left(1+{k^4t^2\over
V_0}\right)^{1/2}\right]^{1/\sqrt d}, ~~~~~~~~~~~
\fb= {1\over 2} \left({V_0\over k^2}+k^2t^2\right),
\label{818}
\eeq
where $k$ and $a_0$ are integration constants. 

This is a regular self-dual solution, $a(t)/a_0=a_0/a(-t)$, characterized
by a bounded, ``bell-like" shape not only of $H$ but also of the dilaton
kinetic energy (unlike the example given in Eq. (\ref{88})). For $t \ra
+\infty$ the solution satisfies the conditions (\ref{812}), ({\ref{813})
required for a successful exit \cite{BruMad97,BruMad98} and, together
with its time-reversed partner, it describes indeed a perfect
``$8$-shaped" curve in the plane of Fig. \ref{f81}. Such a solution
smoothly interpolates between the asymptotic branches (\ref{81}) of
the vacuum solutions, i.e. from the superinflationary expansion 
$a \sim (-t)^{-1/\sqrt{d}}$, ~$\fb=\sqrt{d}\ln a$ at $t\ra -\infty$, 
to the final state of decelerated expansion and decreasing curvature 
$a \sim t^{1/\sqrt{d}}$, ~$\fb=-\sqrt{d}\ln a$ at $t\ra +\infty$ 
(see Figs. \ref{f82} and  \ref{f83}). 

\begin{figure}[t]
\centerline{\epsfig{file=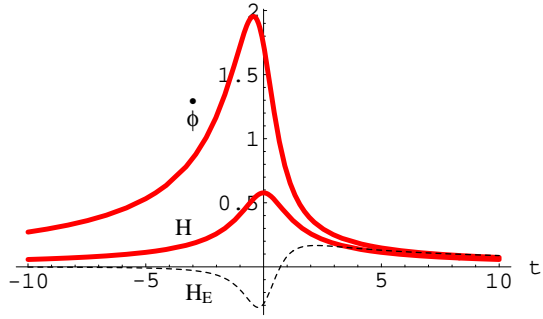,width=72mm}}
\vskip 5mm
\caption{{\sl Plot of the solution (\ref{818}) with
$V_0=1,~k=1,~d=3$. The bold curves show the time evolution of the
curvature and of the dilaton kinetic energy in the S-frame. The thin
dashed curve shows the Hubble parameter $H_E=\ti H$ of the E-frame.}} 
\label{f82}
\end{figure}

It is important to note that in this solution the dilaton is monotonically
growing (indeed, for $t \ra +\infty$, ~$\phi \sim (\sqrt{d}-1) \ln t$), but
the curvature is regular even in the E-frame, where $\ti H$ smoothly
evolves from the pre-big bang accelerated contraction at $\ti H <0$, to a
phase of decelerated expansion at $\ti H >0$. For $t\ra +\infty$, in fact, 
$\ti H \sim t^{(\sqrt{d}-3)/2}$, so that the E-frame curvature is bounded
for all $d \leq 9$. The fact that the dilaton keeps growing, however, 
is not consistent (asymptotically) with the perturbative approach, and
with the use of the lowest-order string effective action. In addition, the
curvature may blow up in some other frame, different from the String
and Einstein ones. 

A more satisfactory example of transition, in which the dilaton becomes
asymptotically frozen in the radiation-dominated regime, can be
implemented by using another form of non-local dilaton potential
\cite{GasVe93a},
\beq
V(\fb)=-V_0 e^{2 \fb}.
\label{819}
\eeq
In that case Eqs.  (\ref{814})--(\ref{816}) can be integrated
exactly, for a wide class of (possibly time-dependent) equations of
state. Let us report in some detail the integration procedure already
outlined in Subsection \ref{Sec2.3}, which is also useful for further
applications. 

We start by defining a new (dimensionless) time coordinate $x$, such
that  \beq
2dx= L \rb dt
\label{820}
\eeq
($L$ is a constant parameter with dimensions of a length). The
combination of Eqs. (\ref{814}), (\ref{816}), using the relation $2V=\pa
V/\pa \fb$, gives 
\beq
\rb\left(e^{-\fb}\right)' L^2= 2(x+x_0),
\label{821}
\eeq
where a prime denotes the differentiation with respect to $x$, and
$x_0$ is an integration constant. By assuming a time-dependent
equation of state, defined by the function $\Ga(x)$ such that 
\beq
\Ga'(x)=\pb/\rb,
\label{822}
\eeq
Eq. (\ref{815}) can be rewritten as
\beq
L^2~\rb\left(a'/a \right)e^{-\fb}= 2\Ga(x). 
\label{823}
\eeq
The matter-conservation equation (\ref{87}), on the other hand, gives
\beq
L^2~\rb^{~\prime}~ e^{-\fb} = -d \left(\Ga^2\right)'.
\label{824}
\eeq
By summing Eqs. (\ref{821}), (\ref{824}), and integrating, we are led to
defininig the function $D(x)$ such that
\beq
L^2~\rb ~e^{-\fb} =(x+x_0)^2-d \Ga^2(x) + \b \equiv D(x)
\label{825}
\eeq
($\b$ is an integration constant), and our system of equations for $a$
and $\fb$ can be reduced to quadratures:
\beq
{a'\over a}= 2{\Ga(x)\over D(x)}, ~~~~~~~~~~~~~~
\fb^{~\prime}= -2 {x+x_0\over D(x)}.
\label{826}
\eeq

We have to satisfy, finally, Eq. (\ref{814}), which determines the
constant $\b$ through the condition
\beq
{(x+x_0)^2\over D^2}-{d\Ga^2\over D^2}-{V\over L^2\rb^2}= {1\over D}.
\label{827}
\eeq
There are two possibilities. The first one is trivial: $V=0$ and $\b=0$. The
second solution, valid for the potential (\ref{819}), is
\beq
\b= V_0~L^2 >0.
\label{828}
\eeq
In such a case the integration of Eqs. (\ref{826}) may provide regular
solutions, with bounded curvature, kinetic terms and dilaton, describing
a smooth evolution from the pre- to the post-big bang regime
\cite{GasVe93a}. 

The simplest examples of regular solutions are obtained in the case of
perfect fluids, with equation of state $p/\r=\ga=$ const. From
Eq. (\ref{822}) one finds indeed $\Ga(x)=\ga(x+x_1)$, where $x_1$ is an
integration constant and, if $\b=V_0L^2 >0$, it is always possible to
choose $x_0,~x_1$ in such a way that $D(x)$ has no zeros, and the
solution is regular. See \cite{GasVe93a}, in particular, for the
radiation case with $\ga=1/d$. But, in general, we may also describe a
smooth evolution, which ends up with radiation, starting from its dual
equation of state, $\ga=-1/d$, corresponding to a gas of stretched
strings  (typical example of a pre-big bang phase dominated by string
sources, see Section \ref{Sec2}). If we take, for instance \cite{GasVe93a}, 
\beq
{p\over \r} (x)= {x\over d(x^2+x_1^2)^{1/2}}, ~~~~~~~~~~
\Ga(x)= {1\over d}(x^2+x_1^2)^{1/2},
\label{829}
\eeq
and we choose $x_0=0, ~x_1^2=L^2V_0$, the integration of Eqs.
(\ref{826}) gives in fact
\bea
&&
a=a_0\left[x+  (x^2+x_1^2)^{1/2}\right]^{2\over d-1}, ~~~~~~~
e^\phi=a_0^d e^{\phi_0}\left[1+ 
{x\over (x^2+x_1^2)^{1/2}}\right]^{2d\over d-1}, 
\label{830}\\ 
&&
\r e^\phi= {d-1\over dL^2} e^{2\phi_0}(x^2+x_1^2)^{-{d+1\over d-1}}, 
~~~
p e^\phi= {d-1\over dL^2} e^{2\phi_0}x~(x^2+x_1^2)^{-{3d+1\over
2(d-1)}}.
\label{831}
\eea

This solution is self-dual, $a(t)/a_0=a_0/a(-t)$, and describes a model
that  is always exapnding, $H>0$. The Universe, starting at $t\ra
-\infty$ from the flat space ($H \ra 0$) and weak coupling ($e^\phi \ra
0$) regime, evolves through a superinflationary phase ($a \sim
(-t)^{-2/(d+1)}$) dominated by unstable string matter ($p=-\r/d$),
towards the standard radiation-dominated phase ($a \sim t^{2/(d+1)}$, 
$p=\r/d$), with frozen dilaton ($\exp \phi \sim$ const). The curvature
parameters $H, ~\dot H$, and the coupling $\exp \phi$, are everywhere
bounded (see Fig. \ref{f83}).

\begin{figure}[t]
\centerline{
\includegraphics[width=6cm,height=3.7cm]{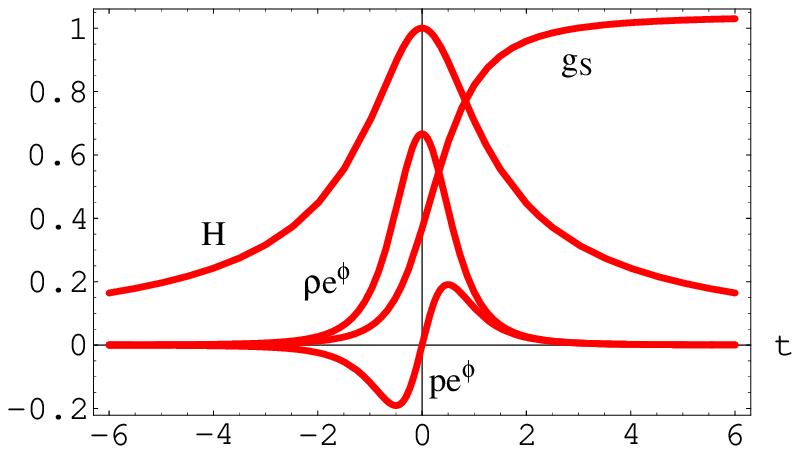}~~~~~
\includegraphics[width=6cm,height=3.7cm]{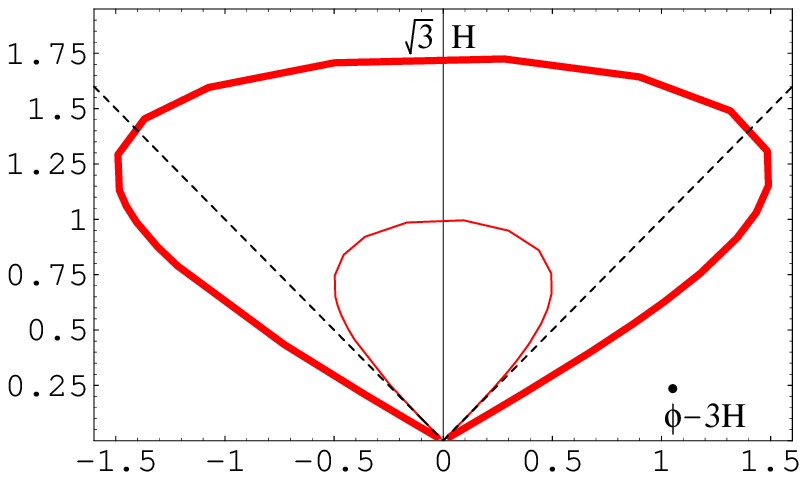}}
\vskip 5mm
\caption{\sl  Plot of the solution  (\ref{830}), (\ref{831})   
with $\phi_0=0,~L=1,~x_1=1,~ d=3,~ a_0=e^{-2/3}$. The left panel
shows the evolution of the curvature scale, of the coupling, of the
energy density and pressure. The right panel shows the trajectory of the
above solution (bold curve) in the plane parametrized by the shifted
dilaton and by the Hubble parameter. The dashed lines represent the
asymptotic vacuum solution (\ref{81}).  We have also reported  the
trajectory of the solution (\ref{818}) (thin curve).}  
\label{f83} 
\end{figure}

It should be noted that with different, more general choices of $x_0$
and $x_1$, the solution  contains, in general,  an intermediate stage of
contraction before the final radiation phase \cite{Angel95}, but even in
that case the curvature and the dilaton keep bounded. The above 
solution can also be generalized to anisotropic configurations,
describing  graceful exit from inflation and, simultaneously, dynamical
dimensional reduction \cite{GasVe93a}. In that case, besides the $d$
expanding dimensions, there are also $n$ shrinking internal dimensions
that evolve from accelerated contraction (driven by an internal positive
pressure)  to a final decelerated contraction (driven by negative
pressure). 

Without non-local potential and matter sources, smooth solutions can be
obtained by relaxing the isotropy and/or homogeneity assumptions.
There are indeed particular examples,  in the context of the lowest-order
string effective action, in which the curvature singularities can be
regularized by performing an appropriate $O(d,d)$ transformation, which
introduces, however, a non-trivial anisotropic axion background (it is
tempting to speculate that this softening of singularities, due  to a
non-trivial $B_{ij}$ field, could be related to recent developments in the
field of non-commutative  geometry \cite{noncomm}, also induced by a
$B_{ij}$ field). 

The simplest example \cite{GMV91} 
is the $d=2$ boost of the trivially flat, Milne-type
exact solution of the gravidilaton action (\ref{82}) (with $V=0=H$);
\beq
ds^2= dt^2-(bt)^2 dx^2-dy^2- (dz_i)^2, ~~~~~
\phi= {\rm const},
\label{832}
\eeq
or of its (singular) scale-factor duality partner:
\beq
ds^2= dt^2-(bt)^{-2} dx^2-dy^2- (dz_i)^2, ~~~~~
\phi= -2 \ln |bt|,
\label{833}
\eeq
($z_i$ are ``inert" coordinates spanning a $n$-dimensional Euclidean 
space). Consider in fact the $O(2,2)$ transformation, in the $(x,y)$ plane,
generated by the one-parameter matrix $\La(\ga)$ such that 
\bea
&&
M \ra \ti M = \La^T M \La ~, ~~~~~~~~~~~~~~~~~~~~~
M=\pmatrix{G^{-1} & -G^{-1}B \cr
BG^{-1} & G-BG^{-1}B \cr}, \nonumber \\
&&
\La(\ga)=\pmatrix{1+c & s & c-1 & -s \cr
-s & 1-c & -s & 1+c \cr
c-1 & s & 1+c & -s \cr
s & 1+c & s & 1-c \cr}, ~~~~
c= {\rm cosh} \ga, ~~ s={\rm sinh} \ga
\label{834}
\eea
(here $G$ and $B$ are the $2 \times 2$, $(x,y)$ part of the metric and of
the antisymmetric tensor). We obtain a new class of exact solutions of
the string effective action, with non-trivial $B_{\mu\nu}$, represented by
\cite{GMV91}
\bea
&&
\ti G(\ga) = \pmatrix{
{(c\mp 1)+ (c \pm 1) b^2t^2 \over (c\pm 1)+ (c \mp 1) b^2t^2} &
{s(1+b^2t^2)\over (c\pm 1)+ (c \mp 1) b^2t^2}\cr
{s(1+b^2t^2)\over (c\pm 1)+ (c \mp 1) b^2t^2} & 1\cr},
\nonumber\\
&&
\ti B(\ga)=\pmatrix{
0 &
{-s(1+b^2t^2)\over (c\pm 1)+ (c \mp 1) b^2t^2}\cr
{s(1+b^2t^2)\over (c\pm 1)+ (c \mp 1) b^2t^2} & 0\cr},
\nonumber\\
&&
\ti \phi =-\ln \left[(c\pm 1)+ (c \mp 1) b^2t^2\right],
\label{835}
\eea
where the upper and lower signs refer, respectively, to the boost of the
solution (\ref{832}) and (\ref{833}). 

For such solutions, all the curvature invariants are bounded
\cite{GMV91}, as well as the dilaton coupling $\exp \phi$, so that the
solutions are regular in all frames. They describe a smooth evolution
from (anisotropic) contraction to expansion, or vice versa, according to
the behaviour of the original metric (the Milne-type solution or its dual).
In any case, the boosted backgrounds (\ref{835}) evolve from growing to
decreasing curvature, and from growing to decreasing dilaton, as
appropriate to a transition from pre- to post-big bang configurations. 
The new solutions are highly anisotropic, but the associated shear
tensor rapidly decays  far from the high-curvature regime
\cite{GMV91}. Taking into account the backreaction of the produced
radiation it is also possible to obtain a more realistic scenario, in which
the background is eventually led to a phase of radiation-dominated,
isotropic expansion, at constant dilaton  \cite{Gas99}. 

More complicated, inhomogeneous examples of (non-trivial) regular
backgrounds at low energy are provided by the ``boosted" version of
the Nappi--Witten solution \cite{NW92}. We start with the following
(singular) exact solution of the action (\ref{82}) with
$V=0=H_{\mu\nu\a}$, 
\beq
ds^2=dt^2-dx^2-\tan^{-2}t~ dy^2 -\tan^2x~ dz^2, ~~~~~~
\fb= -\ln (\sin 2t) -\ln (\sin 2x) ,
\label{836}
\eeq
representing the direct product of a two-dimensional cosmological
metric in the $(t,y)$ plane and a two-dimensional Euclidean black
hole in the $(x,z)$ plane. Since the metric is independent of $y$ and $z$
we can apply to it any $O(2,2)$ transformation in the $(y,z)$ plane, to
obtain new classes of inequivalent solutions. In particular, the
transformation generated by the $ 4 \times 4$  $O(2,2)$ matrix 
\beq
\La_{\rm NW}(\a)=\pmatrix{\da & 0 & 0 &\da \cr
0 & 1& -1 & 0 \cr 
0 & \da^{-1} & \da^{-1} & 0 \cr
-1 & 0 & 0& 1 \cr}, ~~~~~~~~~
\da^2={1 -\sin \a \over 1+\sin \a},
\label{836a}
\eeq
generates the one-parameter family of inhomogeneous (and singular)
Nappi--Witten cosmologies \cite{NW92}. 

As in the previous example, the singularities of the solution (\ref{836})
can be removed by introducing an additional (originally flat) spatial
dimension, parametrized by the coordinate $w$. The model then
acquires a larger symmetry, isomorphic to $O(3,3)$, in the subspace
$(y,z,w)$. Performing two successive ``boost" transformations (with
the same parameter $\ga$ as in Eq. (\ref{834})) in the $(z,w)$ plane and
in the $(y,w)$ plane,  one can obtain a new class of
inhomogeneous exact solutions completely free from curvature and
dilaton singularities (see \cite{GMV92} for an explicit computation). As 
an example we report here the curvature scalar and the coupling of the
boosted solutions \cite{GMV92}: 
\bea
&&
e^{\ti \phi}= \left[ (c+1)\sin^2t + (c-1) \cos^2 t \right]^{-1}
\left[ (c+1)\cos^2x+ (c-1) \sin^2 x \right]^{-1},
\nonumber\\
&&
\ti R= {16 \cos^2t~ \sin^2 t- 20 s^2\over 
\left[ (c+1)\sin^2t + (c-1) \cos^2 t \right]^{2}}-
{16 \cos^2x~ \sin^2 x- 20 s^2\over
\left[ (c+1)\cos^2x+ (c-1) \sin^2 x \right]^{2}}.
\label{837}
\eea
The curvature and the dilaton are bounded, but the background is too
inhomogeneous to provide a realistic description of our present
Universe. Such solutions could be relevant, however, to descrive the
Universe near the transition regime, before the beginning of the
standard radiation era. 

The same comment  applies to another class of regular inhomogeneous
solutions of the lowest-order action (\ref{82}), which is obtained even 
in the absence of torsion ($B_{\mu\nu}=0$) and dilaton potential, and
which can be parametrized in diagonal form as \cite{Giov98}
\beq
ds^2=A\left(dt^2-dx^2\right)-B\left(Cdy^2+C^{-1} dz^2\right), ~~~~~~~
\phi= \a~ {\cal G}_d (\mu t) + {\rm const},
\label{838}
\eeq
where: 
\bea
&&
A(x,t)= e^{\a {\cal G}_d (\mu t)}\left(\cosh \mu t\right)^{2+\b}
\left(\cosh{\mu x\over 2}\right)^{2\b(\b+1)},
\nonumber\\
&&
B(x,t)= e^{\a {\cal G}_d (\mu t)}\left(\cosh \mu t\right)
\left(\sinh{\mu x}\right),
\nonumber\\
&&
C(x,t)=\left(\cosh \mu t\right)^{1+\b}
\left(\sinh{\mu x\over 2}\right)
\left(\cosh{\mu x\over 2}\right)^{1+2\b}.
\label{839}
\eea
Here $\mu$, $\a$ and $\b=\sqrt{\a^2+4}$ are constant parameters, and 
${\cal G}_d (\mu t)=\tan^{-1}\left(\sinh \mu t\right)$ is the so-called
``Gudermannian amplitude". As shown in \cite{Giov98,Giov99b}, there
are particular values of $\a$ for which the curvature invariants are
bounded, the solutions are geodesically complete and describe a phase
of growing dilaton coupling, evolving from a minimum up to a maximal
value. We note that there is also another family of inhomogeneous
solutions of the gravidilaton equations in which all curvature
invariants are bounded \cite{Pim99}, which is parametrized by a
non-diagonal metric and reduces, in the E-frame, to  known regular
solutions of the Einstein equations with stiff  sources
\cite{Mars95}. In that case, however, the dilaton grows linearly and
boundlessly. 

We may thus conclude that, as in general 
relativity \cite{Seno90,Mars95}, if the metric background is allowed to
be inhomogeneous there is no need of a non-local potential and/or
antisymmetric tensor densities to obtain regular cosmological solutions,
even in the contex of the lowest order gravidilaton action. 

To stress the difficulty in obtaining smooth solutions in the
homogeneous and isotropic case, on the contrary, we will conclude this
subsection with a simple example, based on a Bianchi-I-type metric, and
barotropic (but anisotropic) fluid sources with constant dilaton charge
per unit mass:
\bea
&&
g_{\mu\nu}={\rm diag }\left(1, -a_i^2 (t)\da _{ij}\right), ~~~~~~~~~
T_\mu\,^\nu=  {\rm diag} \left(\rho(t), -p_i(t) \da_i^j\right),
\nonumber\\
&&
p_i=\ga_i \r~, ~~~~~~ ~~~~~~~~~~~~~~~~~~~~~~~~~
\sqrt{-g} \sg= {\da S_m\over \da \phi}= \ga_0 \r~,
\label{840}
\eea
where $S_m$ is the matter action of Eq. (\ref{82}), and $\ga_i,\ga_0$
are $d+1$ constant parameters specifying the ``equation of state" of
the fluid sources. The corresponding cosmological equations can be
written in the form
\bea
&&
\fbp^2 -\sum_iH_i^2= \rb e^{\fb} , \label{841}\\
&&
\dot H_i -H_i \fbp={1\over 2} \rb (\ga_i + \ga_0) e^{\fb}, \label{842}\\
&&
\fbp^2 -2 \ddot{\fb}  +\sum_iH_i^2= \ga_0\rb  e^{\fb}.
\label{843}
\eea
Following the procedure already applied to the case of the non-local
potential (\ref{819}), the above equations can be integrated exactly in
terms of the time coordinate (\ref{820}), and we obtain \cite{DeRi01}
\bea
&&
\fb^{~\prime}  =-2(1+\ga_0) {(x+x_0)\over D(x)},
~~~~~~~~~~~~~~~~(1+\ga_0)\not=0, \label{844}\\
&&
{a_i'\over a_i}=2(\ga_i+\ga_0) {(x+x_i)\over D(x)},
~~~~~~~~~~~~~~~~~(\ga_i+\ga_0)\not=0, 
\label{845}
\eea
where
\beq
L^2 \rb e^{-\fb} = D(x) \equiv (1+\ga_0)^2 (x+x_0)^2 -
\sum_i (\ga_i+\ga_0)^2 (x+x_i)^2, 
\label{846}
\eeq
and $x_i,x_0$ are integration constants. In the case $1+\ga_0=0$
and/or $\ga_i+\ga_0=0$,  a slightly different definition of
the quadratic form $D(x)$ is obtained, as well as a different
expression for $\fb^{~\prime}$ and $a_i'$, but in that case no smooth
solution with bounded curvature and energy density is allowed, for any
value of $\ga_i, ~\ga_o,~x_i,~x_0$. We will thus concentrate on 
Eqs. (\ref{844})--(\ref{846}) (see \cite{DeRi01} for a discussion of the
other possibilities). 

By setting $D(x)= \a x^2+bx +c$ it can easily  be  checked that if the
background is isotropic (i.e. $\ga_i$ and $x_i$ have the same values
along all the $d$ spatial directions), then the discriminant of $D(x)$ is
always non-negative, i.e. $\Da = b^2-4\a c \geq 0$, so that 
$D(x)$ necessarily has zeros on the real axis, correponding to
singularities both in the curvature and in the dilaton kinetic energy. 
Such singularities can be avoided, however, for {\em anisotropic}
backgrounds,  provided we accept a model of sources with {\em
negative energy density}, $\r <0$. 

Consider for instance a model of background in which 
the spatial geometry is factorizable as the direct
product of two conformally flat manifolds with $d$ and $n$ dimensions,
respectively, so that we can set:
\bea
&&
a_i=a_1, ~~~~~~~~~ \ga_i=\ga_1, ~~~~~~~~~
x_i=x_1, ~~~~~~~~~ i=1, \dots d, 
\nonumber\\
&&
a_i=a_2, ~~~~~~~~~ \ga_i=\ga_2, ~~~~~~~~~
x_i=x_2, ~~~~~~~~~ i=d+1, \dots d+n, 
\label{847}
\eea
and choose a convenient set of integration constants, such
that the linear term in the quadratic form (\ref{846}) disappears. For
instance:
\beq
x_0=0, ~~~~~~~~~~~~
x_1= -x_2{n (\ga_2+\ga_0)^2\over d (\ga_1+\ga_0)^2}.
\label{848}
\eeq
It turns out that $c<0$, and that the absence of zeros in $D(x)$ can be
avoided, $\Da=-4 \a c <0$, provided 
\beq
\a = (1+\ga_0)^2-d(\ga_1+\ga_0)^2 -n (\ga_2+\ga_0)^2 <0.
\label{849}
\eeq

With the above assumptions, the integration of Eqs. (\ref{844}),
(\ref{845}) leads to the final solution 
\bea
&&
{a_i} = a_{i0} E_i (x)\left|D(x)\right|^{{\ga_i+\ga_0\over
\a}},  ~~~~~
E_i(x)= \exp\left[{2x_i (\ga_i+\ga_0)\over \sqrt{\a c}} \tan^{-1} \left(\a x
\over \sqrt{\a c}\right) \right], ~~~~~ i=1,2,
\nonumber\\
&&
e^{\phi}={e^{\phi_0}}a_{10}^d a_{20}^n
E_1^d(x) E_2^n(x) \left|D(x)\right|^{-[(1+\ga_0)-d(\ga_1+\ga_0)-n 
(\ga_2+\ga_0)]/\a}, \nonumber\\
&&
\r=- {e^{\phi_0}\over L^2}a_{10}^{-d} a_{20}^{-n}
E_1^{-d}(x) E_2^{-n}(x)
\left|D(x)\right|^{1-[(1+\ga_0)+d(\ga_1+\ga_0)+n  (\ga_2+\ga_0)]/\a}, 
\label{850}
\eea
where $a_{i0},\phi_0$ are integration constants. As shown in 
\cite{DeRi01}, there is a region of non-zero extension in the space of the
parameters $\ga_i,\ga_0$ where the condition (\ref{849}) is satisfied,
together with the other conditions needed to secure that the curvature,
the dilaton kinetic energy, the effective string coupling, the energy
density are everywhere bounded, that the density goes asymptotically
to zero at large times, and that the solution describes in its
final configuration $d$ expanding and $n$ contracting dimensions (as
appropriate to a smooth transition from a higher-dimensional pre-big
bang configuration to a post-big bang regime of dynamical dimensional
reduction). 

The condition (\ref{849}) implies, however, that $D(x)<0$ everywhere,
and thus $\r<0$, according to Eq. (\ref{846}) (the result $D<0$ in the
absence of zeros is independent from the particular choice $b=0$). Such
a result is not surprising, after all, because also the regular solutions
obtained in the case of negative non-local potentials correspond to a
negative energy density. See also \cite{ERSD99} for another class of
smooth,  self-dual solutions of the low-energy effective action in which
the energy density of the matter sources becomes negative, near the
transition, because of an ``exotic" equation of state. 

In the above example the negative energy density goes to zero,
asymptotically, far from the transition regime, and can thus be
interpreted as the backreaction of the quantum fluctuations outside the
horizon \cite{DeRi01}. During the initial pre-big bang phase of growing
curvature and shrinking horizons, in fact, the
quantum fluctuations are stretched outside the horizon, and in such
regime they may be characterized by a negative effective gravitational
energy density \cite{Nambu01,Nambu02},  which
may  favour the transition to the post-big bang branch of the classical
solution \cite{Gosh00}. Such a negative backreaction is eventually
damped to zero when the curvature starts decreasing, the horizon blows
up again, and all the fluctuations re-enter inside the horizon and in the
regime of positive energy density. 

The inclusion of the backreaction of the quantum fluctuation, on the
other hand, is also a semiclassical way of taking into account the
contribution of the quantum loop corrections to the effective action.
Thus, the examples of graceful exit reported in this subsection, based on
effective  sources (fluid or non-local potential) with $\r<0$, which
become dominant just around the epoch of transition,  suggest the
possible importance of the quantum backreaction --and of the loop
corrections to the string effective action-- for a successful mechanism
of curvature and dilaton regularization. This possibility will be
illustrated  in  Subsection \ref{Sec8.3}, after a preliminary discussion of
the effects of the higher-curvature corrections, which will be
presented in the next subsection.

\subsection{Growth of the curvature and $\ap$ corrections}
\label{Sec8.2}

The effective action of string theory is characterized by two
perturbative expansions: the higher-genus expansion in the world-sheet
topology, and the order-by-order condition of vanishing conformal
anomaly, also called $\ap$ expansion \cite{GSW87}. The first one is
equivalent to the quantum, field-theoretical loop expansion in powers of
the coupling; the second one is equivalent to a higher-curvature
expansion in powers of the field gradients (in units of $\la_{\rm s}=(2 \pi
\ap)^{-1/2}$). The second one is typical of strings, and disappears in the
point-like limit $\la_{\rm s} \ra 0$. 

These two expansions are indepedent in principle; it is possible,
depending on the initial conditions, that the Universe evolving from the
string perturbative vacuum reaches the strong-coupling regime when
the curvature is still small, and the $\ap$ corrections are negligible.
Alternatively, the Universe can reach the regime of high curvatures  (in
string units) when the  coupling is small: in that case, only the $\ap$
corrections are to be added to the tree-level action. 

What is important, in the context of the exit problem, is that such
higher-curvature corrections may help the transition to
the post-big bang regime, by damping the inflationary growth of the
dilaton and of the curvature, and driving the cosmological background
to a high-curvature ``string phase" of constant curvature and linearly
evolving dilaton \cite{GasMaVe97}, $H=$ const, $\dot \phi=$ const. 

Remarkably, such a regularized background can be a solution of the
tree-level effective action {\em to all orders in} $\ap$. Indeed, as
shown in \cite{GasMaVe97}, for a Bianchi-I-type configuration with
$a_i=\exp (H_i t)$, $\phi=ct$, with constant $c$ and $H_i$, the string
cosmology equations reduce (to all orders in $\ap$) to a system of $d+1$
algebraic equations for the $d+1$ unknowns $c,~H_i$. String-phase
solutions thus generally exist, provided the algebraic system admits
(non-trivial) real solutions. 

In addition, the (all orders) equations resemble those of a
renormalization group flow (in cosmic time), with the string-phase
solution like a ``fixed point" of the renormalization equations. It can
thus be shown \cite{GasMaVe97} that a string phase solution with $\fbp
<0$ can play the role of late-time attractor for an isotropic Universe
evolving from pre-big bang initial conditions, provided no other fixed
point, or singularity, separates the string phase from the string
perturbative vacuum (i.e. from the trivial fixed point with
$H=0=\dot\phi$). 

A simple example of this possibility, first presented in
\cite{GasMaVe97}, is based on the following S-frame effective
action, truncated to first order in $\ap$: 
\beq
S=-{1\over 2\la_{\rm s}^{d-1}}\int d^{d+1}x \sqrt{|g|}e^{-\phi} \left[ R+
(\nabla \phi)^2-{k\ap \over 4} \left(R^2_{GB} - 
(\nabla \phi)^4\right)\right] 
\label{859}
\eeq
($k$ is a number of order $1$ depending on the particular string model
adopted). Note that we have chosen a convenient 
field redefinition (preserving the conformal invariance of the
$\sg$-model  action \cite{MeTsey}) that eliminates higher
than second derivatives from the  equations of motion, but at the price
of introducing dilaton-dependent $\ap$ corrections. 

For a spatially flat anisotropic background, with $d$ external and $n$
internal dimensions, 
\beq
g_{00}=N^2(t), ~~~~~ g_{ij}=-\da_{ij} e^{\b(t)}, ~~~~~
g_{ab}=-\da_{ab} e^{\ga(t)}, 
\label{860}
\eeq
the action becomes 
\bea
 S =&&\int dt e^{d\b+n\ga-\phi} \left[{1\over
N}\left(-\dot\phi^2 - d(d-1)\bp^2-n(n-1)\dot \ga^2-2dn\bp \dot
\ga +2d\bp\dot\phi+2n\dot \ga\dot\phi\right)+\right.
\nonumber\\
&&
\left. {k\ap\over 4 N^3}\left(c_1\bp^4+c_2\dot \ga^4+
c_3\dot\phi\dot \b^3+c_4\dot\phi\dot\ga^3+c_5\dot\phi\dot \b
\dot\ga^2+c_6\dot\phi\dot \b^2\dot\ga+
 c_7\dot \b^2\dot\ga^2+\right. \right.
\nonumber\\
&&
\left. \left. 
c_8\dot \b\dot\ga^3+c_9\dot \b^3\dot\ga-\dot\phi^4\right)\right]\; ,
\label{861}
\eea
where $c_1, ~\dots , c_9$ are dimensionless numerical coefficients
depending on $d$ and $n$ (see \cite{GasMaVe97}). By varying the action
with respect to $\phi$ and $N$, and setting $n=0$, $\dot\phi=x=$ const,
$\dot \b = y =$ const (in the $N=1$ gauge ), we are led to the
algebraic equations for an isotropic fixed point: 
\bea
&&
x^2+d(d-1)y^2-2dxy-{k\ap\over 4} \left(c_1
y^4+c_3xy^3-x^4\right)-\nonumber \\
&&
(dy-x)\left[-2x+2dy+{k\ap\over 4}\left(c_3y^3-4x^3\right)
\right]=0,
\nonumber \\
&&
x^2+d(d-1)y^2-2dxy-{3\over 4}k\ap \left(c_1 y^4+c_3xy^3-x^4
\right)=0,
\nonumber \\
&&c_1= -{d\over 3}(d-1)(d-2)(d-3), ~~~~~
c_3= {4\over 3}d(d-1)(d-2), 
\label{862}
\eea
which have real solutions for any $d$ from $1$ to $9$. For instance 
(in units $k\ap=1$)
\beq
d=3, ~~~~~~~~~~\dot \phi=\pm1.40...,
~~~~~~\dot \b=\pm0.616...,
\label{863}
\eeq
where the same sign has to be taken for $\dot\phi$ and $\dot \b$. Note
that, with the addition of a non-zero $B_{\mu\nu}$ background,
anisotropic fixed-point configurations can also be obtained, as shown in
\cite{FoMa98}. 

A numerical integration of the isotropic cosmological equations, starting
from pre-big bang initial conditions ($H>0$, $\fbp >0$), shows that the
background, when the $\ap$ corrections become important, deviates
from the tracks of the low-energy, singular solutions,  and is eventually
attracted to the isotropic fixed point (\ref{862}), as illustrated in Fig.
\ref{f84}. In this sense, the curvature singularity is regularized by the
$\ap$ corrections, at least in the S-frame (but not in the E-frame, as
the linear growth of the dilaton is unbounded). Note that a similar
curvature regularization is also effective for a static, spherically
symmetric solution of the action (\ref{859}) \cite{BGU97}. 
It should be mentioned that a similar behaviour can also be obtained at
low energy (without $\ap$ corrections), including the
contribution of a supersymmetric gravitino--dilatino condensate, which
is able to damp the acceleration of the background and to regularize the
S-frame curvature, as shown in \cite{FoMaStu99}. 

\begin{figure}[t]
\centerline{
\includegraphics[width=6cm,height=3.7cm]{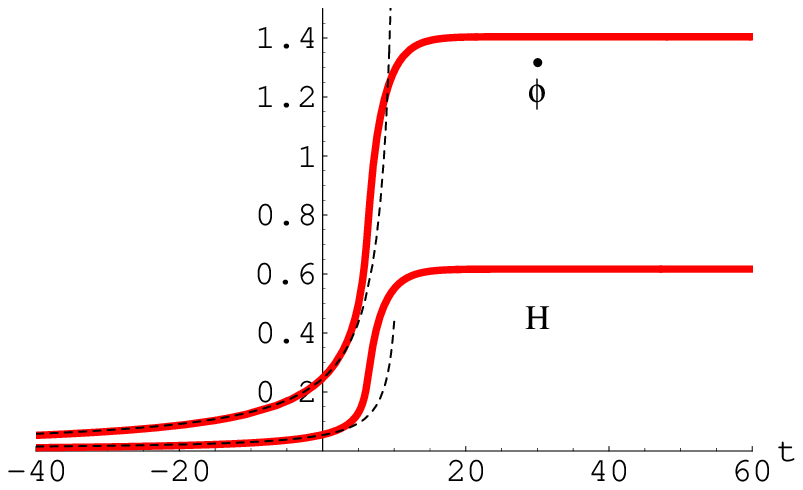}~~~~~
\includegraphics[width=6cm,height=3.7cm]{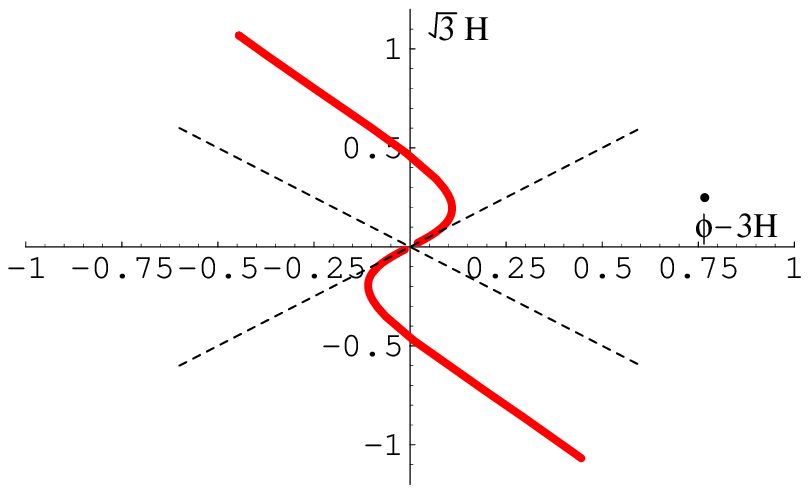}}
\vskip 5mm
\caption{\sl Numerical integration of the isotropic equations for the
action (\ref{861}) (solid curves) with $d=3$, $n=0$, and with initial 
conditions on the low-energy pre-big bang branch (dashed curves). 
The left panel shows the time evolution of $\dot \phi$ and $H=\dot
\b$ for an expanding background. The right panel shows the evolution
connecting the vacuum to the non-trivial fixed point, for an expanding
solution and its time-reversed counterpart.}
\label{f84}
\end{figure}

It may be noted, finally, that a solution with
$H=$ const, $\dot \phi=$ const  is in general allowed in many
higher-derivative models of  gravity.  Such  a fixed point of the
cosmological equations, however,  is in general  disconnected from the
trivial fixed point $H=0=\dot \phi$ (the  perturbative vacuum, in this case)
by a singularity, or by an unphysical  region in which the curvature
becomes imaginary. For the action  (\ref{859}), on the contrary, the
constant fixed point is a late-time attractor for  all isotropic
backgrounds emerging from the string perturbative  vacuum, and even
for some anisotropic backgrounds, provided sign\{$\dot \ga$\}=
sign\{$\dot \b$\} (the attraction basin of anisotropic configurations
covers regions of small but finite size in the space of initial conditions
\cite{GasMaVe97}). 

This attraction property, unfortunately, is not invariant under field 
redefinitions, as long as the action is truncated at a 
given finite order in $\ap$. Also, and most important, the growth of 
the curvature is stopped, but the transition is not completely
performed. Indeed, the final fixed point is in the post-big bang 
regime $\dot{\fb}<0$, as illustrated 
in Fig. \ref{f84}, but the transition cannot proceed further towards 
lower curvatures to complete the transition. 

The reason for such an incompleteness, which is manifest as an
asymmetry in the plane $\{\fbp, \sqrt 3 H\}$ of Fig. \ref{f84}, is that the
action (\ref{859}) is invariant under time reflections, but is not
duality-invariant. As a consequence, the regularization concerns the
expanding pre-big bang branch of the lowest-order solution, and its
time-symmetric counterpart, the  {\em contracting post-big bang}
branch. The expanding post-big bang branch
remains, however, singular, and thus cannot be smoothly connected to
the regularized pre-big bang branch. 

It is possible, of course, to modify the action with an additional field
redefinition truncated to order $\ap$, in such a way as to preserve
conformal invariance and to restore the $T$-duality symmetry (to the
same order in $\ap$) \cite{Meis97,KalMeis97}. The modified action has
then self-dual fixed points, but they are not smoothly connected to the
perturbative vacuum, as we have checked. See also
\cite{BruMad99,CarCopMad00} for a discussion of the existence and
position of the fixed points in relation to the scale-factor duality of the
higher-order action, and for  the difficulty of a higher-order, self-dual
model of exit. 

The evolution of the background away from the fixed points, towards
the low-energy, post-big bang regime, becomes however possible if the
action (\ref{859}) is further generalized by the inclusion of quantum
loop correction (after all, the dilaton keeps growing while the Universe
is trapped in the fixed point, so that higher-order effects in the string
coupling are doomed to become eventually important). This point will be
illustrated in the next subsection.

It should be noted, finally, that all examples of exit based on a truncated
$\ap$ expansion are characterized by a certain degree of ambiguity,
because the physical properties of such models are
field-redefinition-dependent \cite{GasMaVe97}, as already stressed. A
truly unambiguous example of exit should in principle correspond to the
solution of an exact conformal model, which automatically includes all
orders in $\ap$ (see  \cite{Mag98}, however, for a different
non-perturbative approach to higher-order  corrections, based on the
inclusion of the exponentially growing density of all massive
string-theory states, and on their possible role in regularizing  the
singularity). 

There are in fact higher-curvature, exact conformal-field theory
models that smoothly interpolate between two duality-related
low-energy solutions \cite{KK94,KK95}. In that cases, however, it turns
out impossible, in general, to describe the high-curvature regime in
terms of classical fields such as the metric and the dilaton, and we may
lose an intuitive and geometrical understanding of the transition (the
conventional space-time interpretation is also lost in other,  
higher-dimensional, string-theory resolutions of the cosmological
singularities \cite{LarWil96}). 

Exact conformal models, for $D=2$ and $D=3$ string-cosmology
backgrounds, have been recently studied \cite{VLS00} in the
context of gauged Wess--Zumino--Witten models \cite{Wit91}, for the
$SO(2,1)/SO(1,1)$ and $SO(2,2)/SO(2,1)$ cosets. It turns out that it is
possible, in some cases, to avoid the curvature singularities, but the
dilaton (and the string coupling) seems to remain unbounded in 
correspondence of inflationary backgrounds. 

To conclude this subsection, devoted to higher-derivative corrections,
we recall that a higher-curvature regularization of the cosmological
singularity can also be  obtained in the  context of models implementing
the so-called ``limiting-curvature hypothesis"
\cite{MuBra92,BraMuSor93,MoTro95}, applied to string cosmology in
both the Einstein \cite{BraEasMai98} and String \cite{EaBra99} frames.
They are based on a higher-derivative action which, in the S-frame,
takes the form
\beq
S=-{1\over 2}\int d^{4}x \sqrt{|g|}e^{-\phi} \left[ R+
(\nabla \phi)^2+ c \psi e^{\ga \psi} I_2 +V(\psi)\right],
\label{870}
\eeq
where $c$ and $\ga$ are constant parameters, $\psi$ is a scalar
Lagrangian multiplier, and 
\beq
I_2=\left(4 R_{\mu\nu}^2-R^2\right)^{1/2}.
\label{871}
\eeq
Finally, $V(\psi)$ is a potential that satisfies appropriate boundary
conditions, $V\ra \psi^2$ for $|\psi|\ra 0$, $V \ra $ const for 
$|\psi|\ra \infty$, and has non-trivial zeros. 

Numerical and analytical studies then show  that this action has
regular spatially flat and isotropic solutions, smoothly interpolating
from contraction to expansion in the E-frame \cite{BraEasMai98}, and
from superinflation to decelerated expansion in the S-frame
\cite{EaBra99}, even at constant dilaton. A dynamical dilaton may
increase the fraction of phase space for which the curvature is bounded
(in the absence of an appropriate potential, however, a dynamical
dilaton keeps growing). Unfortunately (and as in other models that will
be discussed in the next subsection),  the action (\ref{870}) is not derived
from a string-theory expansion (the higher-order  terms are artificially
constructed), so that the obtained results are only valid as
(interesting) phenomenological examples of possible approaches to the
exit  transition.

\subsection{Growth of the coupling and loop corrections}
\label{Sec8.3}

According to the pre-big bang scenario, the tree-level effective action
(\ref{82}) can provide a correct description of the Universe at low
energies,  during the initial evolution leading the Universe away from the
string perturbative vacuum, until the string coupling remains  small,
$\exp \phi \ll 1$. However, since $\phi$ is growing during pre-big bang
inflation, in the absence of a mechanism able to stop the growth of the
dilaton, the Universe will tend to evolve towards strong couplings, $\exp
\phi \gaq 1$.  In that regime all the higher-loop contributions will
become important, and since the fully corrected action is unknown, in
general, we have to restrict our studies to the simplest cases where we
know  the explicit form of the loop corrections. 

In all cases so far analysed in the literature, it was  found that the
inclusion of  appropriate higher-order  terms seems to
damp the growth of $H$ and $\dot \phi$, thus favouring the exit
transition. Indeed, in the presence of loop corrections,  the ``theorems"
on the impossibility  of branch changing have to be generalized
\cite{Sahar97}; it  becomes possible, in general, to evade the
obstructions present at the tree level if the loop functions satisfy
appropriate conditions.  For a complete and successful transition it
seems required, however, that the quantum corrections  be included at
least to second order \cite{BruMad98,CarCopMad00} 
(the two examples of smooth homogeneous isotropic solutions presented
in  Subsection \ref{Sec8.1} correspond indeed to a two-loop (\ref{819})
and to a four-loop (\ref{817}) (non-local) potential). 

We wish to recall, at this point, that the quantum loop corrections are
generated by the higher-genus expansion of the world-sheet topology
of the corresponding sigma-model action \cite{GSW87}, and can be
represented as an expansion in powers of $g_{\rm s}^2= e^\phi$. They
contain, already to first order, the backreaction of particle production,
which becomes important when $\ap H^2 e^\phi \sim 1$, and thus
naturally introduce higher-derivative terms in the effective action, as 
in general relativity (see for instance \cite{Vilko}), with the only
difference that the gravitational coupling is not fixed but  is controlled
by a dynamical dilaton. 

In string theory, as already mentioned, there are  additional
higher-derivative corrections (typical of the minimal finite size
$\la_{\rm s} =(2 \pi \ap)^{1/2}$ of fundamental strings), which are
generated by the $\ap$ expansion of the sigma model action
\cite{GSW87}, and which may become important even at small coupling,
if the curvature reaches the string scale. The possible effects of such
higher-curvature terms, without loop corrections, have been discussed in
the previous subsection. Here we shall start considering higher-genus
corrections only, by assuming that the strong-coupling regime is reached
when the $\ap$ corrections are still negligible (this is certainly possible
in some appropriate model of cosmological evolution, see
\cite{MagRio98}). As we have already anticipated, however,  both $\ap$
and loop corrections seem to be required \cite{Gas98d} for a complete
transition to the post-big bang regime. 

An unrealistic  (but instructive) example of curvature regularization
operated by the loop corrections is provided by (a generalization of) the
so-called CGHS model \cite{CGHS92} of dilaton gravity in $D=2$
dimensions, described by the action 
\bea
&&
S=-\int d^2x \sqrt{-g} e^{-\phi} \left[R+ (\nabla \phi)^2\right] +
S_{\rm one-loop}~~~,
\nonumber\\
&&
S_{\rm one-loop}= {k\over 2} \int d^2x \sqrt{-g} \left(R\nabla^{-2} R +
\ep \phi R\right).
\label{851}
\eea
Here $k=(N-24)/24$, where $N$ is the number of conformal scalar fields
possibly present in the model, and the trace-anomaly term of the loop
corrections has been supplemented by a local counterterm $\ep \phi R$,
needed to preserve other classical symmetries. The case $\ep=0$
reproduces the original CGHS model, the case $\ep=1$ preserves
conformal invariance \cite{DeAlvis92,Russo92}, and has been applied to
the string cosmology case in \cite{Rey96}. 

The cosmological equations for the action (\ref{851}) can be integrated
exactly for any constant value of $\ep$ \cite{GasVe96b}, and it has been
shown that for $\ep \geq 1$ there are regular solutions without
curvature singularities, describing a smooth evolution away from the
(two-dimensional version of the) pre-big bang branch of the vacuum
solution, $a \sim (-t)^{-1}$, $\phi \sim -2 \ln (-t)$. For $\ep =1$ the
curvature regularization requires $k<0$ \cite{Rey96}, which is known to
correspond to gravitational instabilities \cite{Russo92}. Regular
examples that do notspoil the physical requirement $k>0$ are possible,
however, provided $\ep >1$.
In that case the smooth solutions describe, for any $\ep$, the evolution
from a phase of pre-big bang superinflation to a final asymptotic
configuration with flat space, $a =$ const, and linearly growing dilaton,
$\phi = t$. The solutions can be expressed in parametric form as a
function of the monotonic coupling parameter $g_{\rm s}(t)= \exp [\phi(t)/2]$
as follows \cite{GasVe96b}:
\bea
&&
a(g_{\rm s})=e^\b=e^{\b_0}\left|{ g_{\rm s}^2\over \ep(r+g_{\rm s}^2) +\ep
-2 }\right|^{\ep/4}\left|2r+2 +(\ep-2)g_{\rm s}^2\over
g_{\rm s}^2\right|^{(\ep-2)/4} e^{r-1 \over 2g_{\rm s}^2}, 
\label{852}\\
&&
\dot\phi(g_{\rm s}) ={g_{\rm s}^2\over a t_0 r}, ~~~~~~~~
2\dot\b ={1\over a t_0 r}\left(1+{\ep\over 2}g_{\rm s}^2-r\right) , 
\nonumber\\
&&
r(g_{\rm s})=\sqrt{1+(\ep-2)g_{\rm s}^2+{\ep^2\over  4}g_{\rm s}^4},
\label{853}
\eea
where $t_0$ and $\b_0$ are integration constants, and 
$g_{\rm s}(t)$ is given
implicitly by \beq
{t\over t_0}= \int {dg_{\rm s}^2\over g_{\rm s}^4} 
r(g_{\rm s})~a(g_{\rm s}) . 
\label{854}
\eeq

In the above example the curvature and the dilaton kinetic enegy are
bounded everywhere. The dilaton, however, keeps growing as $t \ra
+\infty$. It is true that a monotonic evolution of the dilaton from weak
to strong coupling may be equivalent, via S-duality transformations 
\cite{Sen94,Pol96}, to a smooth interpolation between two different,
weak-coupling regimes. In a realistic scenario, however, the final value
of the dilaton should go to a finite constant, not to zero. To this aim, a
non-perturbative dilaton potential and/or the backreaction of the
produced radiation are probably to be  included. 

Unfortunately, however, the final state of the above regular
backgrounds is still characterized by $\fbp >0$. This means that the
conditions of ``branch changing" \cite{BruMad97} are not satisfied and
that the background, in spite of the bounce of the curvature, is still in
the pre-big bang regime. As a consequence, it is not ready to be attracted
to any stable minimum of the potential, so that dilaton stabilization is
impossible \cite{GasVe96b}, even asymptotically, in the context of the
above one-loop action. 

Similar conclusions apply to the regular solutions of another
(four-dimensional) model of one-loop string effective action
\cite{ART94} (which also provided, to the best of our knowledge, the
first example of loop regularization of the cosmological singularity in a
superstring-theory context). The action is based on moduli-dependent
loop corrections to the gravitational coupling for a heterotic string
compactified on a symmetric orbifold \cite{AnGav92,AnGav92a}. There
are no corrections to the Einstein term, only quadratic-curvature
corrections, and the action can be written (in the E-frame, and using the
notations of \cite{ART94}) as
\bea
&&
S = \frac{1}{2}\,\int\,d^{4}x\,\sqrt{|g|}\,
\left[R+\frac{1}{2}(\nabla_{\mu}\phi)^2+\frac{3}{2}(\nabla_{\mu}\sg)^2 
\right]+S_{\rm one-loop}~~,
\nonumber\\
&&
S_{\rm one-loop}= \int\,d^{4}x\,\sqrt{|g|}\,
\left[\la e^\phi- \da \xi(\sg)R_{GB}^2\right].
\label{855}
\eea
Here $\sg$ is the modulus field, $R_{GB}^2=R^2-4R_{\mu\nu}^2+
R_{\mu\nu\a\b}^2$ is the Gauss--Bonnet invariant, $\la$ and $\da$ are
constant parameters ($\la$ is fixed by $\ap$, while $\da$ depends on
the number of chiral, vector and gravitino massless supermultiplets
included in the model). Finally,
\beq
\xi(\sg)= \ln \left[ 2  e^\sg \eta^4(i e^\sg)\right],
\label{856}
\eeq
where $\eta$ is the Dedekind function \cite{Tricomi}. 

For an isotropic and spatially flat background, the equations have been
numerically integrated in \cite{ART94}, and it has been shown that, 
for $\da <0$,  the action (\ref{855}) admits regular solutions without
singularities, in which the curvature grows up to a maximum and then
decreases, with a behaviour that mimics that of the pre-big bang
scenario. However, the Hubble parameter turns out to be always positive
even in the E-frame, and the dilaton contribution is negligible, so that
such solutions do not describe a transition from the pre- to the post-big
bang regime. 

The same action also admits regular homogeneous and isotropic
solutions with positive spatial curvature, $k>0$ \cite{EasMae96}. For
such solutions the curvature invariants are bounded, the metric smoothly
evolves from contraction to expansion, and there is indeed a branch
changing. The dilaton has, however, a very small influence on the
background, which is dominated by the modulus and by the spatial
curvature. In addition, the dilaton is monotonically decreasing, so that
such solutions do not describe the evolution of the Universe away from
the string perturbative vacuum, as required by the pre-big bang
scenario. 

The regular, spatially flat solutions of the action (\ref{855}) can be
generalized to the anisotropic case, corresponding to a Bianchi-I-type
metric background \cite{KS98,YMO99}. The regular, spatially curved
($k>0$) solutions, however, cannot: for the corresponding anisotropic
curved metric, of Bianchi-IX-type, no regular solution of the action 
(\ref{855}) has been obtained \cite{YMO99}. This seems to be in
agreement with an instability of the regular solutions of the action 
(\ref{855}) against tensor perturbations \cite{KSS98,KS99}, since the 
Bianchi-IX-type metric may    indeed be regarded as a model of closed
Friedmann Universe with a non-linear tensor perturbation, represented by
a gravitational wave of given wave-number. In this sense, the
curvature regularization operated by the one-loop term (\ref{855}) is
generic for spatially flat backgrounds, but non-generic for curved ones 
\cite{YMO99}. 

It should be noted, finally, that the action (\ref{855}) has also motivated
the study of scalar-tensor models of gravity characterized by the higher
derivative, non-minimal coupling $\xi(\phi) R_{GB}^2$, with arbitrary
coupling functions $\xi(\phi)$. Even in this case it has been found that 
there are, for appropriate forms of $\xi(\phi)$, singularity-free,
homogeneous and isotropic solutions, without \cite{RizTam94} and with
\cite{KRT98} spatial curvature, and also (in very special cases) with a
scalar potential \cite{ATU99}. But, again, such solutions do not describe
the same initial configuration as in the pre-big bang scenario. In
addition, the higher-derivative, scalar-tensor coupling is postulated
{\em ad-hoc}, and is not the result of a reliable one-loop computation. 

An appropriate loop-corrected action for string cosmology has been
discussed instead in \cite{FofMagStu99}, considering the heterotic
string compactified to four dimensions on a $Z_N$ orbifold, and
restricted to the gravidilaton-moduli sector, as in \cite{ART94}, but
including in addition all string $\ap$ corrections (truncated to first
order). The action contains modular-invariant loop corrections to the
K\"ahler potential, known to all orders \cite{DFKZ91}, and one-loop
corrections to the gravitational couplings of the higher-derivative
terms. The final action, neglecting non-local terms, can be written in the
S-frame   (using the conventions of \cite{FofMagStu99}) as:
\bea
S &=&\frac{1}{2\ap}\,\int\,d^{4}x\,\sqrt{|g|}\,
e^{-\phi}\, 
\left[R +\left(1+e^\phi G(\phi)\right)
(\nabla_{\mu}\phi)^2-\frac{3}{2}(\nabla_{\mu}\sg)^2 \right.
\nonumber\\
&&
\left.
+{\ap\over4}\left(1+e^\phi \Da(\sg)\right)
\left( R_{GB}^2-(\nabla_{\mu}\phi)^4\right)\right],
\label{857}
\eea
where $\Da(\sg)$ is a modular function depending on the
compactification \cite{ART94}, and 
\beq
G(\phi)= {3 \a\over 2} {6+\a e^\phi\over (3+\a e^\phi)^2},
\label{858}
\eeq
where $\a>0$ is a model-dependent constant of order $1$. 

The cosmological equations of this action, for a homogeneous and
spatially flat background, have been numerically integrated  
\cite{FofMagStu99}, starting
from an initial pre-big bang regime with $H>0$ and $\dot \phi >0$. The
integration shows that, while the $\ap$ corrections can drive the
Universe to a fixed point with $\fbp <0$ (even in the absence of loop
corrections, see the next subsection), the loop corrections to the
K\"ahler potential are essential to drive the Universe to another fixed
point  with $\dot \phi <2H$, so that both conditions (\ref{812}) and
(\ref{813}) (necessary for a graceful exit \cite{BruMad97}) can be
satisfied. Such  conditions are not sufficient, since a regular solution 
requires that the curvature be bounded also in the E-frame. This result 
is achieved, however,  through the contribution of the
(moduli-dependent) higher-derivative terms --at least for the case of
small $\dot \sg$ and/or nearly constant (negative)  $\Da(\sg)$ 
discussed in \cite{FofMagStu99}. 

From the action (\ref{857}), numerical solutions can thus be obtained, 
which smoothly interpolate from the pre-big bang regime of the
S-frame, tree-level effective action, to a final decelerated expansion,
with decreasing curvature. The curvature and dilaton kinetic energy
are bounded everywhere in the S-frame and E-frame. However, the
post-big bang decelerated expansion becomes a phase of the de Sitter
inflationary expansion in the E-frame, because the loop corrections to
the dilaton kinetic term do not disappear, asymptotically, but approach a
constant value \cite{FofMagStu99}. Also, for all the regular solutions,
the dilaton remains growing, logarithmically, as $t \ra +\infty$.
The dilaton cannot be stabilized being trapped  in the minimum of a
potential, because it becomes unstable at strong couplings (its kinetic
term acquires the ``wrong sign" even in the E-frame).

In spite of this property, which means that the solutions do not describe 
a complete transition to the post-big bang regime, the existence of such
solutions represents a remarkable and encouraging result, especially in
view of the fact that the higher-order  terms of the action (\ref{857}) are
not {\em ad hoc}  (except for the truncation of the $\ap$ expansion), but are
the outcome of a well motivated string theory calculation.  A complete
example of transition (but with {\em ad hoc}  loop corrections) can  be
obtained instead, by considering the following higher-order  action
\cite{BruMad98} 
\beq
S=-{1\over 2\la_{\rm s}^{2}}\int d^{4}x \sqrt{|g|}e^{-\phi} \left[ R+
(\nabla \phi)^2\right] +S_{\ap}+S_q+S_m,
\label{864} 
\eeq
where $S_{\ap}$ contains the $\ap$ corrections of Eq.  (\ref{859}), $S_q$
provides the appropriate quantum loop corrections, and $S_m$ contains
radiation, or a dilaton potential, for the final dilaton stabilization. 

The loop corrections are required to give a negative contribution to the
energy density, are expected to become strong enough to drive the
Universe away from the fixed point, and to disappear fast enough at late
times, in order for the dilaton to  eventually be captured by a potential 
or be frozen by radiation production. In the numerical  examples
presented in \cite{BruMad98}, the one-loop corrections are suppressed
at late times through the introduction of a step function, or with the
addition of a two-loop term of opposite sign. A toy model is the
following  (in units $2 \la_{\rm s}^2=1$):
\beq
S_q=\int d^{4}x \sqrt{|g|} \left(c_1 +c_2 e^\phi\right) \left(\nabla
\phi\right)^4, ~~~~~
S_m=c_3\int d^{4}x \sqrt{|g|} \left(\phi-\phi_0 \right) ^2
e^{\phi-\phi_0},
\label{865}
\eeq
where $c_i$ are dimensionless coefficients. As shown in
\cite{BruMad98}, the numerical integration of the cosmological
equations with $c_1=-c_2=-10^3$, $c_3=-10^{-1}$, $\phi_0=1$, provides 
a complete model of graceful exit in which the background smoothly
evolves from the vacuum, pre-big bang configuration of the tree-level
action, to a final expanding, decelerated configuration, with the dilaton
performing damped oscillations around  $\phi_0=1$. The
model can be further improved with the introduction of radiation, which
 eventually becomes dominant and freezes out the dilaton. 

Other examples of a smooth and complete graceful exit, but again with
{\em ad hoc}  loop corrections, can be obtained with the following 
higher-order action \cite{CarCopMad00}: 
\beq
S=-{1\over 2\la_{\rm s}^{2}}\int d^{4}x \sqrt{|g|}e^{-\phi} \left[ R+
(\nabla \phi)^2+ {\cal L}_{\ap} + A e^\phi {\cal L}_{\ap}^{(1)}+
 Be^{2\phi }{\cal L}_{\ap}^{(2)} \right] ,
\label{866}
\eeq
where ${\cal L}_{\ap}$ is the most general form of the first-order $\ap$
corrections to the gravidilaton action \cite{MeTsey}, on which  the
condition has been imposed  that the equations of motion contain at
most second derivatives: 
\beq
{\cal L}_{\ap}= {\ap\over 4}\left[a R_{GB}^2+b G^{\mu\nu} \nabla_\mu
\phi \nabla_\nu \phi + c (\nabla \phi)^2 \nabla^2 \phi +
d (\nabla \phi)^4\right].
\label{867}
\eeq
The coefficients $a,b,c,d$, in order to reproduce string scattering
amplitudes, are constrained by
\beq
a=-1, ~~~~~~~ b+2(c+d)=-2a,
\label{868}
\eeq
but are otherwise arbitrary, because of possible further shifts due to
field redefinitions, truncated to first order in $\ap$ (note that our
notation is diferent from \cite{CarCopMad00}, where they use a
different normalization for the dilaton, i.e. $g_{\rm s}^2= e^{2\phi}$; see also
\cite{CaCoGas}). The one-loop and two-loop terms, ${\cal L}_{\ap}^{(1)}$
and ${\cal L}_{\ap}^{(2)}$, have the same functional form as  ${\cal
L}_{\ap}$, but the  coefficients $a,b,c,d$ are in principle
replaced by different numbers, $a_1,b_1,c_1,d_1$ and 
$a_2,b_2,c_2,d_2$ , respectively. Finally, the constant parameters $A$
and $B$ control the onset of loop corrections and their late-time
suppression for a successful exit. 

The study of the fixed points in the weak-coupling regime of the action
(\ref{866}), and of the subsequent transition to the decelerated
expansion of the post-big bang scenario, has shown that the case
$a=-d$, $b=c=0$ (discussed in \cite{GasMaVe97,BruMad98}) is not the
only one that is compatible with a graceful exit. There is indeed a wide
region in parameter space that allows fixed points with $\fbp<0$ and,
from the fixed points, the transition can further proceed successfully
even if $b$ and $c$ are non-zero \cite{CarCopMad00} and different from
$b_1,c_1$ and $b_2,c_2$ \cite{CaCoGas}. In all cases, the exit can be
completed by dilaton stabilization, obtained by introducing ``by hand"
some radiation density $\r_r$, coupled to the dilaton through the
conservation equation  
\beq
\dot \r_r +4 H \r_r -{1\over 2} \Ga \dot\phi^2 =0.
\label{869}
\eeq

The main problem with such models of exit is that the one-loop and
two-loop corrections are not the result of a reliable string theory
perturbative expansion. In particular, a successful exit seems to require
(as also stressed in \cite{BruMad98}) an appropriate sign of the
corrections, so that (in our case) sign$\{A\}= -$ sign$\{B\}$. 
It is unclear, however, whether the required sign may also 
be  the outcome of a correct loop computation (even if the results
of \cite{FofMagStu99} seem to provide a positive answer). A successful
example of exit, corresponding to a numerical integration of the
equations for the action (\ref{866}) with $A=1$, $B=-2\times 10^{-3}$,
is illustrated in Fig. \ref{f85}. 

\begin{figure}[t]
\centerline{
\includegraphics[width=6cm,height=3.7cm]{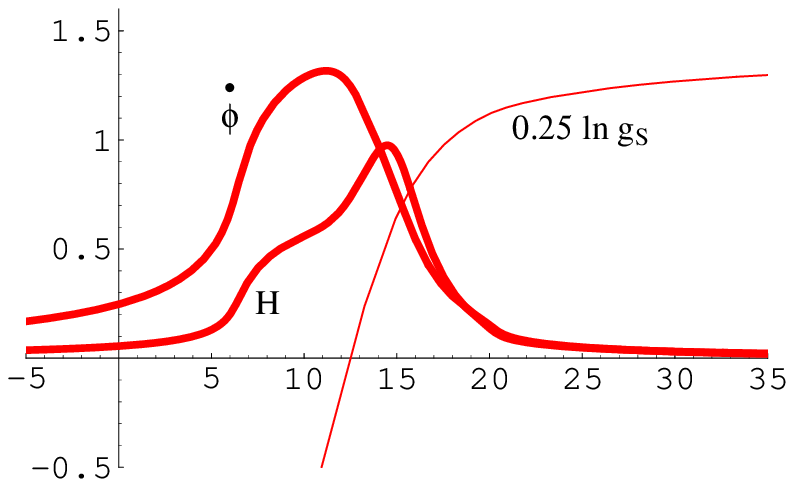}~~~~~
\includegraphics[width=6cm,height=3.7cm]{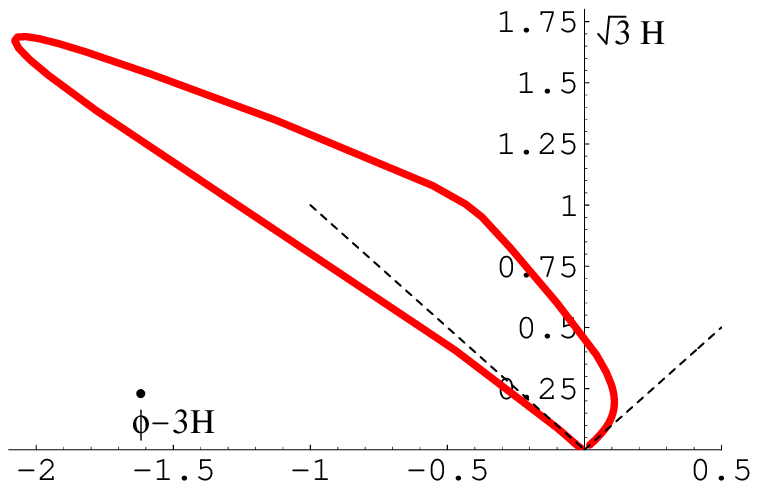}}
\vskip 5mm
\caption{\sl Numerical integration of the cosmological equations for the
action (\ref{866}) with $A=1$, $B=-2\times 10^{-3}$, 
$a=a_1=a_2=-d=-d_1=-d_2=-1$, $\Ga=5.63\times 10^{-4}$, and all
$b_i,c_i=0$. 
The left panel shows the time evolution of $\dot \phi$, $H$ and of the
log of the string coupling. The right panel shows the evolution from the
vacuum  to the fixed point of Fig. \ref{f84}, and the
subsequent exit induced by the loop corrections.}  
\label{f85}
\end{figure}

In addition, the perturbative approach at the two-loop level is
self-consistent only for a quite small value of the final coupling,
$\exp \phi \ll1$. A small value of the final dilaton, for the solutions of
the action (\ref{866}), seems to require, however, that $A$ and $B$ be 
both larger than $1$ and of the same order (in modulus) \cite{CaCoGas},
in contrast with the usual expectation for the coefficients of a
perturbative expansion. 

If the string coupling is not kept small enough, on the other hand, the
Universe necessarily enters the non-perturbative regime
where typical M-theory effects may become important, as will be
illustrated  in Subsection \ref{Sec8.5}. We shall first present, however,
some arguments stressing the importance of the notion of entropy for a
model-independent approach to the exit problem.

\subsection{Graceful exit and entropy considerations}
\label{Sec8.4}

After the various models of exit reported in the previous subsections, it
seems appropriate to recall, at this point, that entropy-related
considerations have recently led to {\em model-independent} arguments
in favour of the occurrence of a graceful exit in the pre-big bang
scenario. As we shall see, those are physically quite close to the
arguments based on backreaction and loop corrections, which we have
already discussed. 

Let us start by recalling that, as suggested by Bekenstein 
\cite{Bek1,Bek2}, the entropy of  a  limited gravity
system of energy $E$ and of size $R$ larger than its gravitational
radius, $R > R_g \equiv 2G_N E$, is limited by the upper bound 
\begin{equation}
 S_{\rm Bek} = ER \simeq R_g R ~\la_{\rm P}^{-2}.
\label{872}
\end{equation}
According to the holographic principle (see \cite{Thooft93,Sus95} and
references therein), on the other hand, the maximal entropy of a system
is bounded by $S_{\rm hol}$, 
\begin{equation}
S_{\rm hol}= A \la_{\rm P}^{-2}, 
\label{873} 
\end{equation}
where $A$ is the area of the space-like
surface enclosing the region of space whose entropy we wish to bound.
For systems of limited gravity, since $R > R_g$ and  $A=R^2$, the
Bekenstein bound  (\ref{872}) implies the holography bound (\ref{873}).

How can  these bounds be extended to be applied to the whole
Universe? At cosmological scales the Universe is {\it not} a system of
limited gravity, since its large-distance behaviour is determined by the  
gravitational
effect of its matter content through Friedmann's equations. 
Furthermore, the holography bound obviously fails for sufficiently large
regions of space since, for a given temperature, entropy grows like
$R^3$ while area grows like $R^2$. The generalization of entropy bounds
to cosmology turned out to be subtle.

A possible prescription  \cite{Bek3} for a cosmological extension
is to identify $R$ in Eq. (\ref{872}) with the particle horizon. In this way
one can then arrive at the conclusion that
the  bound is violated sufficiently near the big-bang singularity, implying
that the latter is fake (if the bound is always valid).
More recently, a similar extension of the
holographic bound to cosmology has been proposed \cite{FS98}, arguing
that the area of the particle horizon should bound  entropy on the
backward-looking light cone, according to (\ref{873}). It was soon
realized, however, that the such a proposal requires modifications, since
violations of it were found to occur in physically reasonable situations.
An improvement of the proposal  \cite{FS98}, applicable to light-like
hypersurfaces, was later made by Bousso \cite{Bousso99,Bousso99a} 
(see also \cite{Fla99}).

Of more interest, in our context,  are however the attempts  made at 
deriving cosmological entropy bounds on space-like hypersurfaces 
\cite{EL99,GV99,Rey99,BisMaPra98,KaLin99,BruVe99}.
These identify the maximal size of a spatial region for which holography
works:  the Hubble radius   \cite{EL99, GV99, KaLin99},
the so-called apparent horizon \cite{BruFoStu99}, or, finally, a 
causal-connection (Jeans) scale \cite{BruVe99}.

For our purpose there is no need here to enter into the relative merits
of these various proposals. Rather,
we will only outline the physical idea behind them. Consider, inside a
quasi-homogeneous Universe, a sphere of radius $ H^{-1}$. We may
consider ``isolated" bodies, in the sense of Ref. \cite{Bek1},   fully
contained in the sphere, i.e. with radius $R < H^{-1}$. For such systems,
 the bound (\ref{872}) holds, and  it is saturated by a black hole of  size
$R$. We may next consider several black holes inside our Hubble
volume,   each carrying an entropy proportional to the square of its
mass. If two, or more, of these   black holes merge,
their masses will add up, while the total entropy after the
merging,  being quadratic in the total mass, will exceed the sum of  
the initial entropies. In other words, in order to maximize entropy, it
pays to form  black holes  as large as possible.

Is there a limit to this process of entropy increase?
The suggestion made in \cite{EL99,GV99,Rey99,KaLin99,BruVe99},  which
finds support in old results by several groups \cite{Carr74,Carr75,
Nov80}, is that a critical length   of order $H^{-1}$
is the upper limit on how large a classically stable black hole can be.
If we accept this hypothesis, the upper bound on the entropy contained  
in a given region ${\cal R}$ of space will be given by the number of
Hubble volumes in ${\cal R}$, $n_H = V H^3$ times the
Bekenstein--Hawking entropy \cite{Bek73,Haw75} 
of a black hole of radius $H^{-1}$, $H^{-2}
\la_{\rm P}^{-2}$. The two factors can be combined in the suggestive
formula: 
\begin{equation} 
S({\cal R}) < \la_{\rm P}^{-2} \int_{{\cal R}}  d^3 x ~ \sqrt{h}
~\overline{H}   \equiv S_{HB} \;,
\label{874}
\end{equation}
where $\int_{{\cal R}}  d^3 x ~ \sqrt{h}$ is the volume of the  
space-like hypersurface whose entropy we wish to bound, and 
$\overline{H}$  differs from one proposal to another, but is,  
roughly, of the order of the Hubble parameter. Actually, since $H$ is
proportional to the trace of the second fundamental form on the
hypersurface, Eq.   (\ref{874}) reminds us of the boundary term that has
to be added to the gravitational action in order to correctly derive
Einstein's equations from the usual variational principle. This shows that
the bound (\ref{874}) is generally covariant for   $\overline{H}=H$.   It
can also be written covariantly for the   identification of $\overline{H}$
made in \cite{BruVe99}.

For the qualitative discussion that follows,  let us therefore take  
$\overline{H} = H$, and let us convert the bound to S-frame
quantities, taking into account   the relation between
$\la_{\rm P}$ and $\lambda_{\rm s}$, given in Eq. (\ref{125}). We obtain
\cite{GV99}: 
\begin{equation}
S({\cal R}) <  (VH^3) (H^{-2} \lambda_{\rm s}^{-2} e^{-\phi}) =  
e^{-\bar{\phi}} H \lambda_{\rm s}^{-2}  \;,
\label{875}
\end{equation}
where we have fixed an  arbitrary additive constant in the definition of
$\bar{\phi}$. Equation   (\ref{875}) thus connects very simply the
entropy bound of a region of fixed comoving volume to the most
important variables occurring in string cosmology (see, e.g., the phase
diagram  of Fig. \ref{f81}).

An immediate application of the bound (\ref{875}) to the exit problem 
was pointed out in  \cite{GV99}, noting that the bound is initially
saturated in the E-frame picture of pre-big bang inflation as a
collapse, with corresponding black-hole formation \cite{BDV99}. 
Since the entropy, and the bound itself, cannot decrease without a  
violation of the second law, one obtains: 
\begin{equation}
\fbp \le {\dot{H}}/{H}  \;.
\label{876}
\end{equation}
It is easy to check that this relation holds with the equality sign during
the initial, low-energy evolution from the string perturbative vacuum
(see Eq. (\ref{85})). In other words, the Hubble entropy bound  is
saturated  initially {\it and throughout} the low-energy, dilaton-driven
evolution described in \cite{BDV99}. But what happens if the curvature
stops its growth, $\dot H =0$ (for instance, during the high-curvature
string phase discussed in the previous subsections)?  It is quite clear
that Eq. (\ref{876}) does not allow $H$ to reach saturation ($\dot{H} =0$) 
in the upper-right quadrant of Fig. \ref{f81}, since  $\fbp>0$ there.
Instead, saturation of   $H$ in the upper-left quadrant (where $\fbp \le
0$) is perfectly all right. But this implies having attained the sought for
branch change!

Let us now look at the loop corrections. Physically, these  
correspond to taking into account the backreaction from particle
production, i.e. from the quantum fluctuations amplified by the
cosmological evolution. Let us check when
their entropy starts to threaten the bound. Using the results presented
in \cite{GG93a,GG93b,Bran92,Bran93a}, the entropy density carried by
quantum fluctuations inside a region of size $H^{-1}$ is given by 
\beq
\sigma_q \sim N_{\rm eff} H^3, 
\label{877}
\eeq
where $N_{\rm eff}$ is the effective number of species that are
amplified. This entropy equals the upper bound (\ref{875}) precisely
when
\beq
N_{\rm eff}~ H^2 ~\la_{\rm s}^2 ~e^\phi \sim N_{\rm eff}~ H^2 ~\la_{\rm P}^2 
\sim 1.
\label{878}
\eeq
But this is also the  line on which
the energy density in quantum fluctuations is expected to become 
critical, and the backreaction of the produced radiation is expected to
drive the Universe to the beginning of the standard radiation-dominated
phase \cite{Gas95,BMUV98a}. 

Indeed, one can easily estimate the total energy stored in the
quantum fluctuations amplified by the pre-big bang 
backgrounds  (for a discussion of generic perturbation spectra, see
\cite{BMUV98a,BH98}).  The result is, roughly,
\begin{equation}
\rho_{q} \sim N_{\rm eff} ~ H^4_{\rm max} \; ,
\label{879}
\end{equation}
where $H_{\rm max}$ is the maximal curvature scale reached around the
exit transition. The saturation of the entropy bound is thus equivalent
to the condition that the above energy density becomes critical (and
induces the exit) just around the transition scale $H_{\rm max}$. On the
other hand, we have already
argued that   $H_{\rm max} \sim M_{\rm s} = \lambda_{\rm s}^{-1}$,
and we know that, in heterotic string theory, $N_{\rm eff}$ is in the
hundreds. It follows, according to this picture, that the exit may occur
when the dilaton satisfies $\exp(\phi_{\rm exit}) \sim 
N_{\rm eff}^{-1}$  $\sim 10^{-2}$, i.e. at a value of the string coupling
very close to its present one. The saturation of the Hubble entropy bound
$S_{HB}$ thus supports a picture in which the exit transition occurs to
avoid violations of the (generalized) second law of thermodynamics, and
it is induced by the backreaction of quantum fluctuations  (see also
\cite{BruFoStu99} and the numerical examples of \cite{CarCopMad00}). 

Remarkably, this also explains why the radiation, which originates from
the quantum fluctuations and dominates our final Friedmann phase, has
an entropy that  roughly corresponds to the number of elementary
``Hubble spheres" (i.e. spatial regions of size $H^{-1}$) contained inside
our Universe just after the exit (namely, to the entropy of the quantum
fluctuations evaluated at the string scale): $S \sim n_H \sim 10^{90}$. A
very large number, on the one hand, but a very small entropy for the
total mass and size of our present observable Universe, on the other
hand, as often emphasized by Penrose \cite{Pen89}. The pre-big bang
scenario may thus neatly explain why the Universe, at the big bang,
{\it looks} so fine-tuned (without being so), and may provide a natural  
arrow of time in the direction of higher entropy \cite{GV99}.

In conclusion, the picture that finally emerges from all these
considerations is best  illustrated with reference to the diagram of  Fig.
\ref{f86}.  Two  lines are shown,  representing boundaries for the
possible evolution. The horizontal boundary is   forced upon by the
large-curvature corrections, while the tilted line in the first   quadrant
corresponds to the equation $\la_{\rm P}^2 H^2 N_{\rm eff} = 1$ that we
have just discussed. This line was also suggested  as a boundary   beyond
which copious production of $0$-branes would set in \cite{MagRio98} 
(see the next subsection). 
Thus, depending on initial conditions, the pre-big bang bubble
corresponding to   our Universe would hit first either the high-curvature
or the large-entropy boundary and initiate an exit phase. Hopefully, a
universal late-time attractor will emerge guiding the evolution into the
FRW phase of standard cosmology.

\begin{figure}[t]
\centerline{\epsfig{file=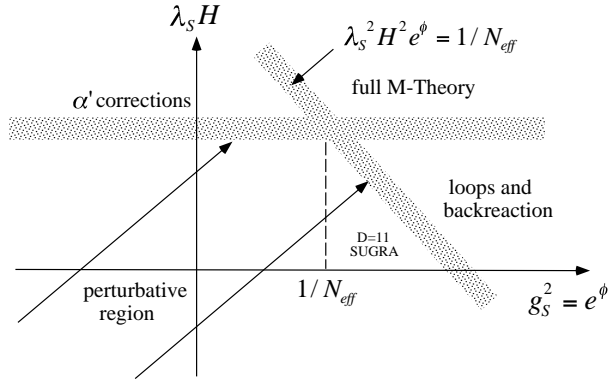,width=80mm}}
\vskip 5mm
\caption{\sl Logarithmic plot of the curvature scale versus the string
coupling $g_{\rm s}^2$. The two parallel diagonal lines emerging from the
bottom-left quadrant represent possible trajectories for the
(low-energy) pre-big bang evolution, corresponding to different initial
values of the dilaton.} 
\label{f86}
\end{figure}

Needless to say, all these arguments have to be considered, at best, as
having   heuristic value. If  we were to take them seriously, they would
suggest that the Universe will never enter the strong-coupling,
strong-curvature regime, where the largely unknown M-theory should
be used. The low-energy limit of the latter (the much better understood
$D=11$ supergravity) could suffice to deal with the fundamental exit
problem of  string cosmology. We refer to the next subsection for a short
discussion of this possibility.

\subsection{M-theory and brane cosmology}
\label{Sec8.5}

If the growth of the string coupling, which starts  during the phase
of accelerated pre-big bang evolution, is not stopped at some point by a
dilaton potential (or by some other mechanism), the Universe 
necessarily enters the strong-coupling regime described by the
so-called M-theory, which  is well approximated, at low-energy, by the
known $D=11$ supergravity theory \cite{Wit95}. In this subsection we
will briefly report various ideas on the possibility of smoothing out the
big bang singularity, and eventually implement the exit transition, in
the context of the low-energy (i.e. small-curvature)  limit of M-theory. 

We start be recalling the action for the bosonic sector of
eleven-dimensional supergravity, 
\bea
S_{11}=&&\int d^{11}x \sqrt{-g} \left[R-{1\over 48} F_{ABCD}^2~~-\right.
\nonumber\\
&&
\left.
{1\over (12)^4 \sqrt{-g}} ~\ep^{A_1A_2A_3B_1B_2B_3B_4
C_1C_2C_3C_4}A_{A_1A_2A_3}F_{B_1B_2B_3B_4}F_{C_1C_2C_3C_4}
\right],
\label{880}
\eea
where $F$ is the field strength of the antisymmetric three-form
potential $A_{BCD}$, and the Chern--Simons term arises as a
consequence of supersymmetry \cite{CJS78} (upper-case Latin indices
run from $0$ to $10$, Greek indices from $0$ to $9$, and we follow the
conventions of \cite{BiCoLid99}. The ten-dimensional effective action,
obtained by compactification of $S_{11}$ on a circle of radius $r_{11}$
(controlled by the dilaton \cite{Wit96}) can be written as 
\bea
S_{11}=&&\int d^{10}x \sqrt{-g} \left[~e^{-\phi}\left(
R+(\nabla \phi)^2-{1\over 12} H_{\mu\nu\a}^2\right)
-{1\over 48} F_{\mu\nu\a\b}^2
\right.
\nonumber\\
&&
\left.
-{1\over 384 \sqrt{-g}} ~\ep^{\a_1\a_2\mu_1\mu_2\mu_3\mu_4
\nu_1\nu_2\nu_3\nu_4}B_{\a_1\a_2}F_{\mu_1\mu_2\mu_3\mu_4}
F_{\nu_1\nu_2\nu_3\nu_4} \right],
\label{881}
\eea
where $H$ and $F$ are the field strengths of the antisymmetric
potentials $B_{\mu\nu}$ and $A_{\mu\nu\a}$, respectively, and we have
dropped the one-form potential arising from the dimensional reduction
of the metric. After the reduction, we have also rescaled the
ten-dimensional metric as $g_{\mu\nu} \ra r_{11~}g_{\mu\nu}$, and we
have defined the ten-dimensional dilaton as  $\phi=3 \ln r_{11}$. 

The above action contains not only the usual string theory action, but
also Ramond forms, i.e. higher-rank antisymmetric tensors uncoupled to
the dilaton (in the S-frame), and exactly reproducing the
Ramond--Ramond sector of the  low-energy Type II A superstring
action (see for instance \cite{Lidsey00}). This provides a new,  in
principle richer context to look for cosmological solutions
\cite{LuOWa97,Kalo97,LuMu97,CoLiWa98} (see \cite{BiCoLid99} for a
complete analysis of spatially flat models), suggesting also the
possibility of an M-theory version of the pre-big bang scenario
\cite{Cav01}. 

In such a context, the existence of black $p$-brane solutions, of the
type of those given in \cite{BeFor93,Poppe97}, provides a first indication
of a possible resolution of the problem of curvature singularities
\cite{LarWil96,LuOWa97a} (see also \cite{BeFoSchwa97} for examples
including $\ap$ corrections). The different branches of a cosmological
solution  can indeed be related by $U$-duality transformations
\cite{LuOv98}, but an explicit representation of a smooth exit transition,
in this context, is problematic \cite{BraLuOv01}. By $U$-duality it is
possible, however, to transform a singular solution into a non-singular
one (as in the case of $O(d,d)$-covariant backgrounds
\cite{GMV91,GMV92}). This possibility has been studied, in particular, in
the moduli space of $D=11$ supergravity compactified on a ten-torus
\cite{BaFiMo99}. 

In addition, the decompactification of the $11$-th dimension, produced
in the strong-coupling regime as a consequence of the identification of
the string coupling with the $11$-th dimensional scale factor
\cite{Wit96,Hor96,Hor96a}, is possibly associated with a ``softening" of
curvature singularities, as pointed out in \cite{KaKoOl98}. The study of
some explicit examples has shown \cite{FeVa00}, in particular, that the
curvature singularity of FWR models minimally coupled to a set of scalar
fields, in four dimensions, can be removed  by lifting the solutions to
higher dimensions, provided the four-dimensional background $M_4$ has
negative or vanishing spatial curvature ($ k \leq 0$). In that case, the
regular higher-dimensional backgrounds from which $M_4$ is obtained
by dimensional reduction has only one of the extra dimensions that  is
dynamical, and for $k=0$ it coincides with the trivial Minkowski
space-time parametrized by Milne coordinates. 

A genuinely new effect, in the context of dimensionally reduced
supergravity theories, is the presence of $p$-branes, and the possibility
of their production in the strong coupling regime where the brane
states become light and (possibly) unstable. This effect might lead to
regularizing the curvature singularities, as first proposed in some
pioneering papers on string \cite{LaMar95} and solitonic $p$-brane
\cite{Rama97} production in cosmological backgrounds. As suggested in
\cite{Riotto00}, brane production might lead for instance to a phase of
constant brane density, and then brane-driven (de Sitter) inflation
(reminescent of the old idea of string-driven inflation \cite{Tur88}), thus
freezing the growth of the background curvature. We should recall,
finally, that brane production in high-energy scattering processes is
also allowed for conventional Einstein gravity in models with large extra
dimensions \cite{Ahn1,Ahn2}. 

More generally, as suggested in \cite{MagRio98}, when approaching the
singularity one should describe the Universe not in terms of the metric
and of the  dilaton, but in terms of new low-energy modes, more
appropriate to the strong-coupling regime where D(irichelet) $p$-branes
are the fundamental players. Indeed, in the string frame, their tension
is controlled by $g_{\rm s}^{-2}=\exp (-\phi)$, so that they are very light at
strong coupling: for instance, a D$p$-brane of type II A has a mass
\beq
m \sim g_{\rm s}^{-1} (\ap)^{-(p+1)/2}.
\label{882}
\eeq

In addition, they become unstable at large enough values of the Hubble
parameter (an effect very similar to string instability in curved
backgrounds \cite{SV90,GSV91a}), so that they can be copiously produced
when approaching the singularity, and in particular around the epoch
where \cite{MagRio98} $g_{\rm s} H \la_{\rm s} \sim1$. As a consequence of this
huge production, the D$p$-brane energy density is expected to become
critical, inducing a transition towards the standard radiation-dominated
phase. If so, the curvature scale $H \sim \la_{\rm P}^{-1}$ would be the
largest value of curvature probed during the cosmological evolution
\cite{MagRio98}. 

More recently, the idea of a brane-dominated Universe has been further
developed in the context of the so-called ``brane gas cosmology"
approach \cite{ABE00,Ea01} to the M-theory phase of the very early
Universe (which generalised previous string-cosmology models
\cite{BraVa89,TseVa92}). The Universe, in the strong-coupling regime, is
assumed to have nine spatial dimensions in a toroidal topological state,
and to be filled with a hot gas of $p$-branes, with all modes in
(approximate) thermal equilibrium. This approach aims not only at the 
resolution of the singularity problem, but also at providing a possible
explanation of the number of spatial dimensions of our present Universe,
along the lines of  \cite{BraVa89}. 

The model contains all the branes that appear in the spectrum of the
(ten-dimensional) Type II A string theory. By recalling that M-theory
contains the graviton ($0$-brane), $2$-branes and $5$-branes 
as fundamental degress of freedom (see the action (\ref{880})), the
compactification on a circle $S^1$ leads to $0$-branes, strings
($1$-branes), $2$-branes, $4$-branes, $5$-branes, $6$-branes and
$8$-branes (see Eq. (\ref{881})) as the fundamental extended objects of
the $10$-dimensional theory. The total action for the brane gas model is
thus the sum of the low-energy effective action for the ``bulk"
(ten-dimensional) space-time manifold plus the ($p+1$)-dimensional
action of all the branes in the gas:
\beq
S= S_{\rm bulk} + \sum_i S_i ({\rm brane}),
\label{883}
\eeq
where $S= S_{\rm bulk}$ is the gravidilaton--axion sector of the action
(\ref{881}); and the Born--Infeld action of a $p$-brane, coupled to the
bulk via delta function sources with strength fixed by the tension
$T_p$, is given by
\beq
S_p= T_p \int d^{p+1} \xi~ e^{-\phi} \left[ - {\rm det} \left(
g_{mn} +b_{mn}+2 \pi \ap F_{mn}\right)\right]^{1/2}.
\label{884}
\eeq
Here $g_{mn}$ is the induced metric on the brane, $b_{mn}$ is the
induced antisymmetric tensor, and $F_{mn}$ is a gauge field possibly
living on the brane. 

All branes contribute to the energy of such a primordial gas with three
types of modes \cite{ABE00}. There are winding modes, corresponding 
to $p$-branes wrapping around $p$ cycles of the torus, momentum
modes, corresponding to the centre-of-mass motion of the brane, and
oscillatory modes, corresponding to fluctuations in the directions
transverse to the brane. The brane action (\ref{884}) can thus be
expanded in terms of these modes, and by taking the average
contribution of their stress tensor one obtains the effective ``equation
of state" for the gas of winding modes; 
\beq
\ti p /\r =- p/d
\label{885}
\eeq
($\ti p$ is the pressure, $p$ the rank of the brane). The momentum
modes have the ``dual" equation of state $\ti p /\r =p/d$ (for strings, 
$p=1$,  one recovers the results of \cite{GSV91a,GSV91}). Transverse
oscillations can be viewed as particles living on the brane, and their
equation of state is that of ``ordinary" matter, with $0 \leq \ti p \leq
1$. It follows, from the covariant conservation of the stress tensor,
that the energy $E_p$ of winding (w) modes increases with the scale
factor $a(t)$ as
\beq
E_p= \r a^d \sim T_p a^p
\label{886}
\eeq
(note, however,  that should be included at very high temperature 
thermal corrections , and that this can modify the effective equation of
state, as discussed in \cite{Gleiser85} for the case of strings near the
Hagedorn scale). 

The Universe, in the strong-coupling (M-theory) regime, is thus assumed
to behave as a hot soup of all modes of all $p$-branes. The curvature
singularity is expected to be avoided because of $T$-duality exchanging
w-modes and momentum modes, and then transforming a phase of
increasing temperature into decreasing temperature, with the fixed
point of the symmetry acting as a regularizing cut-off for all physical
observables \cite{ABE00,Ea01}. But the most interesting (and probably
unique, at present) virtue of this scenario seems to be the possibility
of explaining why our present Universe contains just three large spatial
dimensions. 

Indeed, as the primordial Universe expands, the w-modes tend to
become dominant as their energy increases, starting with the largest
value of $p$, according to Eq. (\ref{886}). The effect of the w-modes, on
the other hand, is to halt the expansion \cite{TseVa92}, so that the
spatial dimensions can decompactify (i.e. inflate away from the string
scale) only if the w-modes disappear.

They can disappear by annihilation with antiwinding ($\overline {\rm
w}$) modes (the initial number of w and $\overline {\rm w}$ is assumed
to be the same, by symmetry). On the other hand, assuming that the
space is periodic, it follows that the world-volume of two $p$-branes
will necessarily intersect in at most $2p+1$ spatial dimensions. In the
initial Universe $d=9$, so that winding $p$-branes with $p=8,6,5,4$ have
no problem to self-annihilate. We are thus left with $2$-branes and
strings. The annihilation of the (heavier) winding $2$-branes will first
allow five dimensions to become large. Within this distinguished torus
${\cal T}^5$, the $1$-branes (strings) w-modes will then only allow the
$d=3$ subspace ${\cal T}^3$ to expand. 

In this way, one can understand the origin of the  $d=3$ dimensionality
of our Universe and, interestingly enough, can predict a hierarchical
structure of the ten-dimensional compact manifold as follows
\cite{ABE00}:
\beq
{\cal M}=S^1 \times \left({\cal T}^4 \times {\cal T}^2 \times 
{\cal T}^3 \right),
\label{887}
\eeq
where $S^1$ comes from the original M-theory compactification on a
circle, and the other compact manifolds have been listed in order of
growing size, from left to right. 

In this context,  a possible  problem  is that, at the end of the
decompactification process, at least one w-mode is left per Hubble
volume, leading to the well known domain-wall problem. This problem
can be solved if the w-modes, before they  annihilate, drive the
background to a phase of slight contraction, also called ``loitering"
\cite{SaFeSte92}, during which the Hubble horizon becomes larger than
the spatial size of the Universe (as in the phase of pre-big bang
inflation, when it is seen as a contraction in the E-frame). 

By supplementing the string-cosmology equations by terms describing
the annihilation of w-modes into string loops \cite{BEK01}, it has been
shown that during the loitering phase all w-modes are eliminated and
that, from this point on, the Universe begins to expand again. The
w-modes annihilation leads to breaking $T$-duality, and it is tempting to
speculate that this effect could  be related to SUSY breaking, and to
the mechanism of dilaton mass generation needed to stabilize the
coupling after decompactification \cite{BEK01}.  

Outside the context of brane gas cosmology, but always in the context
of an M-theory approach to the strong coupling regime, a more drastic
resolution of the dimensionality problem is represented by the so-called
brane-world scenario (inspired by heterotic M-theory
\cite{Hor96,Hor96a}), in which our Universe  {\em is assumed} to
coincide with a $3$-brane embedded in a higher-dimensional bulk
manifold \cite{LOSW99,LOSW99a}. 

In a realistic picture, of course, our $3$-brane is not expected to be the
only extended object embedded in the bulk manifold, especially in the
regime of large bulk dilaton, and very strong coupling. In that case, the
transition to the post-big bang regime could be triggered by a 
head-on collision against another brane, just as happens in the context
of the so-called ``ekpyrotic" scenario \cite{KOST1,KOSST}, where the big
bang explosion, and the birth of a hot, radiation-dominated Universe, is
simulated by the collision of two $3$-branes along a (hidden)  fifth
spatial dimension (see also \cite{Gen01} for a similar but spherically
symmetric process of bubble collision). 

In the ekpyrotic scenario the $11$-th space-time is assumed to have
topology  $M^{10} \times S^1/Z_2$, with two branes at the orbifolds
fixed points,  where space-time has boundaries. Six dimensions are
compactified on a Calabi--Yau threefold, leaving the effective theory
five-dimensional. The boundary branes are called the visible and the
hidden brane, the visible one being identified with our Universe. 

The action for this model is the sum of three parts:
\beq
S=S_{\rm het}+S_I+S_M,
\label{888}
\eeq
where $S_I$ represents the brane interactions, $S_M$ describes the
matter on the brane created by the collision, and $S_{\rm het}$ is the
action of five-dimensional heterotic M-theory. With the conventions of
\cite{KOST1}:
\bea
&&
S_{\rm het}= {M_5^3\over 2} \int d^5 x \sqrt{-g} \left[R-{1\over 2}
(\nabla \phi)^2 -{1\over 5 !}e^{2\phi} F^2_{ABCDE}\right]
\nonumber\\
&&
-\sum_i 3 \a_i M_5^3\int d^4\xi_i \sqrt{-h_i}\left[~e^{-\phi}
-{1\over 4! \sqrt{-h_i}}~\ep^{\mu\nu\a\b}A_{ABCD} 
\pa_\mu X_i^A \pa_\nu X_i^B \pa_\a X_i^C \pa_\b X_i^D \right].
\nonumber\\
&&
\label{889}
\eea
Here $M_5$ is the five-dimensional Planck mass, $F$ is the field
strength of the four-form $A_{ABCD}$, $h_{\mu\nu}^i$ is the induced
metric on the $i$-th $3$-brane, whose embedding in five dimensions is
described by the five parametric equations $X_i^A=X_i^A(\xi_i^\mu)$. 
Finally, the tensions are given by  $T_i=\a_i M_5^3$. 

In the first version of the ekpyrotic scenario \cite{KOST1}, besides the
two boundary branes, there is  a third ``bulk" brane, possibly
originated from spontaneous bubble nucleation. In that case, the three
tensions satisfy the condition $\a_2=-\a_1-\a_3$, with $\a_3>0$ and
$\a_3<|\a_1|$, and the bulk brane is attracted towards the visible
brane,  until they collide. The kinetic energy of the bulk brane is then
converted into matter and radiation of the visible brane, with a
spectrum of density perturbations left impressed by the quantum
fluctuations of the bulk brane. 

In the second version of the scenario \cite{KOSST}, there are instead only
the two boundary branes, which collide and then bounce apart, in what
is hoped to be a non-singular process. This second possibility also
represents an important ingredient in a related, recently proposed
``cyclic" model of Universe \cite{SteiTu01}.  

In our context, this second possibility is particularly interesting
because the collision of the  two boundary branes is associated to
the collapse of the dimension transverse to the brane: namely, of the
$11$-th dimension, taking into account the Calabi-Yau
compactification. But the shrinking of the eleven-dimensional radius
$r_{11}$, in an M-theory context, is equivalent to a decreasing of the
string coupling (recall that $r_{11}=\exp (\phi/3)$). This means that
the collision, and the subsequent beginning of the radiation-dominated,
post-big bang evolution, occurs at very weak coupling, in the
perturbative regime. 

In conclusion, this model suggests that the space-time manifold, 
after having reached the strong-coupling, M-theory regime as a
consequence of the pre-big bang evolution, and after the associated 
production of branes, might be forced again to the perturbative regime
by the attraction (and by the subsequent collapse) of the branes at the
boundary of the $D=11$ space-time. The exit would thus proceed at weak
coupling (possibly with a mechanism of radiation production different
from the one discussed in the previous subsections). This may perhaps
justify the use of the tree-level effective action for the quantum
cosmology approach to the exit, presented in the next section.

\section{Quantum string cosmology}
\label{Sec9}
\setcounter{equation}{0}
\setcounter{figure}{0}

According to the pre-big bang scenario,
the present cosmological state of our Universe should emerge as 
the result of a long evolution, starting from the string perturbative
vacuum. Such an evolution necessarily includes the transition from a
phase of growing curvature and strong coupling to a phase of decreasing
curvature and (nearly) constant string coupling.  The full dynamical
description of such a process still contains various problematic aspects,
as discussed in the previous section. 

Since the
transition is expected to occur in the high-curvature (nearly Planckian) 
regime,  where quantum-gravity effects may become important, it is not
excluded that a quantum-cosmology approach may be appropriate to
describe the ``decay" of the string perturbative vacuum into our
present, post-big bang Universe. Also in the context of the standard
inflationary scenario, in fact, the quantum-cosmology approach is 
required to describe the ``birth" of our classical Universe out of the
Planckian regime \cite{Vil96}. With an important difference, however: in
the standard scenario the initial state of the Universe is unknown, and
has to be fixed through some {\em ad hoc} prescription. There are various
possible choices for the initial boundary conditions
\cite{Hartle83,Haw84,Vil84,Linde84,Zeldo84,Rub84,Vil88}, leading in
general to different quantum pictures of the early cosmological
evolution. In the pre-big bang scenario, on the contrary, the initial
state is fixed in such a way as to approach, asymptotically, the string
perturbative vacuum, and this unambiguously determines the initial
wave function. 

In a quantum cosmology context the Universe is described by a wave
function evolving in superspace, according to the Wheeler--De Witt
(WDW) equation \cite{Witt67,Wheeler68}, and it is always possible, in
principle,  to compute the transition probability between two different
geometrical configurations --in particular, from a pre- to a post-big
bang state.  We shall report here the results obtained in the context  of
what may be called a ``low-energy" approach to quantum-string
cosmology \cite{Gas98},  which is based on the lowest-order string
effective action, and in which no higher-order ($\ap$ and loop)
corrections are taken into account into the WDW equation, except those
possibly encoded into an effective, non-perturbative dilaton potential
(see \cite{Pol89,Pol92,SaMo01} for high-curvature contributions to the
WDW equation). 

Such a low-energy approach, first introduced in string cosmology in
\cite{CapDeRi93a,Bento95,Lid95}, is analogous (in the particle case) to
low-energy quantum mechanics, where one neglects relativistic and 
higher-order corrections. It is a first approximation, which is already
sufficient to take into account pure quantum-gravity effects, such as 
the possibility of transitions which are classically forbidden --in
particular, transitions between two low-energy geometric
configurations which are classically disconnected by a singularity
\cite{GasMaVe96,GasVe96a,CaUnga99}-- and thus to motivate a
``minisuperspace approach" to the exit problem \cite{Gas98d} (see also
\cite{MaMo00,PiCo01} for a recent quantum-cosmology approach to this
problem based on the Bohm--de Broglie ontological interpretation of
quantum mechanics). 

Quite irrespectively of its applications, quantum-string cosmology may
represent a rich and interesting field of research in itself;  it thus 
seems  appropriate to recall that it is affected by various
conceptual problems already present in the ``standard"
quantum-cosmology context: the meaning of the probabilistic
interpretation \cite{Vil86}, the existence and the meaning of a
semiclassical limit \cite{Bento95,Lid95}, the unambiguous identification
of the time-like coordinate (see however \cite{CaDe97}). Other problems
affecting the standard scenario, however, disappear. In particular, in the
context of the pre-big bang scenario, there is no problem of boundary
conditions, which are unambiguously prescribed by the choice of the
string perturbative vacuum (see Subsection \ref{Sec9.2});  no
problem either of operator ordering in the WDW equation (see for
instance \cite{Ashte74}), as the ordering is fixed by the duality
symmetry of the string effective action \cite{GasMaVe96,Ke96}. This
important property of quantum-string cosmology will be illustrated in
the next subsection. 

\subsection{The Wheeler--De Witt equation}
\label{Sec9.1}

The simplest example of quantum-string cosmology model is based on
the lowest-order, gravidilaton string effective action
\beq
S = -\frac{1}{2\,\lambda_{\rm s}^{d-1}}\,\int\,d^{d+1}x\,\sqrt{|g|}\,e^{-\phi}
\,\left[R+(\nabla_\mu\phi)^2 +V (\phi,
g_{\mu\nu})\right],
\label{91}
\eeq
where we have included a (possibly non-local
and non-perturbative) dilaton potential $V$ (see \cite{CaMo01} for more
complete models based on the M-theory action). 
If we are considering, in
particular,  an isotropic and 
spatially flat cosmological background,
\beq
\phi=\phi(t), ~~~~~~~~~~~~~
g_{\mu\nu} ={\rm diag} \left(N^2(t), -a^2(t) \da_{ij}\right), 
\label{92}
\eeq
the gravidilaton system has only two physical degrees of
freedom, the scale factor $a$ and the dilaton (the ``lapse" function
$N=\sqrt{g_{00}}$ can be arbitrarily fixed by a choice of gauge). The
quantum evolution of the system is thus associated to a
two-dimensional ``minisuperspace", which, assuming spatial sections
of finite volume, can be conveniently parametrized by
\beq
\b= \sqrt{d} \ln a, ~~~~~~~~~~~~~
\fb=\phi -\sqrt{d}\,\beta -\ln\,\int\,d^dx/\lambda_{\rm s}^d. 
\label{93}
\eeq 
Each ``point" $\{\b(t), \fb (t)\}$ of the minisuperspace will then
represent a classical solution of the action (\ref{91}). 

In terms of the coordinates of the minisuperspace, the action (after
integration by parts) can be explicitly written as 
\beq
S=\frac{\lambda_{\rm s}}{2}\,\int\,dt\,{e^{-\fb}\over N}\,
\left[\dot{\beta}^2-\dot{\fb}^2-
N^2\,V (\b, \fb)\right]. 
\label{94}
\eeq
Its variation with respect to $g_{00}$ defines the total energy density
of the gravidilaton system, and leads to the so-called Hamiltonian
constraint (in the cosmic time gauge, $N=1$):
\beq
{\cal H}=
\left({\da S\over \da N}\right)_{N=1}=
\dot{\fb}^2- \dot{\beta}^2-V=0.
\label{95}
\eeq
Introducing the canonical momenta 
\beq
\Pi_{\beta}=\left({\da S\over \da\dot{\beta}}\right)_{N=1}=
\lambda_{\rm s}\,\dot{\beta}\,e^{-\fb} , ~~~~~~~~~~~~
\Pi_{\fb}=\left({\da S\over \da\dot{\fb}}\right)_{N=1}=
-\lambda_{\rm s}\,\dot{\fb}\,e^{-\fb} ,
\label{96}
\eeq
the Hamiltonian constraint becomes 
\beq
\Pi^2_{\beta}-\Pi^2_{\fb}
+\lambda_{\rm s}^2\,V(\b,\fb)\,e^{-2\,\fb}=0~,
\label{97}
\eeq
and its differential implementation in the minisuperspace spanned
by $\b$ and $\fb$ finally leads  to the WDW equation for the
gravidilaton system:
\beq
\left [ \partial^2_{\fb} - 
\partial^2_{ \beta}
+\lambda_{\rm s}^2\,V(\b,\fb)\,e^{-2\fb}~ \right ]\, \Psi(\b,\fb)= 0 . 
\label{98}
\eeq

It is important to stress that such an equation is manifestly free from
operator-ordering problems, as the Hamiltonian (\ref{97}) has a flat
metric in momentum space. This is not a special feature of the case we
have considered because, thanks to the duality symmetry of the string
effective action, the associated WDW minisuperspace is globally flat,
and we can always choose a convenient parametrization leading to a
flat minisuperspace metric.  

For a more general discussion of this point we may add to the 
effective action (\ref{91}) a non-trivial antisymmetric tensor
background, $B_{\mu\nu}\not= 0$. The kinetic part of the action may
then be written in compact form as \cite{GasMaVe96,Ke96} (see
Subsection \ref{Sec2.3}): 
 \beq
S=-{\la_{\rm s}\over 2}\int dt e^{-\fb}\left[(\dot{\fb})^2+{1\over 8}{\rm Tr}
~\dot M(M^{-1})\dot{}\right] , 
\label{99}
\eeq
where  $N=1$, and $M$ is a symmetric $2d \times 2d$ matrix, including
the spatial part of the background fields, $G\equiv g_{ij}$,  $B\equiv
B_{ij}$, and already defined in Eq. (\ref{253}).  We recall that this action
is invariant under global $O(d,d)$ transformations  that  leave  the
shifted dilaton invariant,  
 \beq
\fb \ra \fb , ~~~~~~~~~~ M\ra \Om^T M \Om , 
\label{911}
\eeq
where
\beq
\Om^T\eta \Om =\eta, ~~~~~~~~~~ \eta =
\pmatrix{0 & I \cr I & 0 \cr} .
\label{912}
\eeq
The Hamiltonian associated to  torsion-graviton background,
\beq
{\cal H }={4 \over \la_{\rm s}} {\rm Tr}~(M~\Pi_M M~\Pi_M), 
~~~~~~~~~\Pi_M = \da
S/\da \dot M,  
\label{913}
\eeq
would seem to have ordering problems, because $[M, \Pi_M] \not= 0$.
However, thanks to the $O(d,d)$ properties of $M$, 
\beq
M\eta M =\eta , 
\label{914}
\eeq
we can always rewrite the kinetic part of the action in terms of  the
flat $O(d,d)$ metric $\eta$:
\beq
{\rm Tr}~ \dot M(M^{-1}) \dot{}=
{\rm Tr}~ (\dot M\eta)^2.
\label{915}
\eeq
The corresponding Hamiltonian
\beq
{\cal H}=-{4\over \la_{\rm s}}{\rm Tr}~(\eta\Pi_M \eta\Pi_M)
\label{916}
\eeq
has a flat metric in momentum space, with no ordering problems for
the corresponding WDW equation \cite{GasMaVe96,Ke96}: 
\beq
\left[ {\da^2 \over \da {\fb}^2} +8 {\rm Tr}~\left(\eta{\da \over \da M}
\eta{\da \over \da M}\right) + \la_{\rm s}^2 V e^{-2\fb}\right]
\Psi(M,\fb)=0.
\label{917}
\eeq

On the other hand, if we insist on adopting a curvilinear
parametrization  of the minisuperspace, the ordering fixed by
the duality symmetry is exactly the same as the ordering imposed by
the requirement of reparametrization invariance, as there are no
contributions to the ordered Hamiltonian from the scalar curvature of
minisuperspace \cite{Ashte74}, because minisuperspace is globally flat.  

To illustrate this point we may  
consider the pair of minisuperspace coordinates $\{a, \fb\}$,
different from the previous pair $\{\b, \fb\}$ used in Eq. (\ref{94}). The
kinetic part of the action (\ref{91})  then leads to the (kinetic
part of the) Hamiltonian
\beq
{\cal H}= {a^2\over d}\Pi^2_{a}-\Pi^2_{\fb}\equiv \ga^{AB}\Pi_A\Pi_B, 
\label{918}
\eeq
where
\beq
\Pi_{a}=\left({\da S\over \da a}\right)_{N=1}=d 
\lambda_{\rm s}\,{\dot a \over a^2}\,e^{-\fb} ,
\label{919}
\eeq 
corresponding to the non-trivial $2\times 2$ metric:
\beq
\ga_{AB}= {\rm diag} \left({d\over a^2}, -1\right) .
\label{920}
\eeq
The quantum operator associated to the Hamiltonian (\ref{918}) has
to be ordered,  because $[a,\Pi_a]\not=0$, 
and its differential representation can be written in
general as
\beq
{\cal H}= {\pa^2\over \pa \fb^2} -{1\over d} \left(a^2
{\pa^2\over \pa a^2}+ \ep a {\pa\over \pa a}\right) ,
\label{921}
\eeq
where $\ep$ is a c-number parameter depending on the ordering. Note
that there are no contributions to the ordered Hamiltonian from the 
minisuperspace scalar curvature \cite{Ashte74}, which is vanishing for
the metric (\ref{920}). 

Reparametrization invariance now imposes on the Hamiltonian the
covariant d'Alembert form 
\beq 
{\cal H} =-\nabla_A\nabla^A =
-{1\over \sqrt{-\ga}} \pa_A (\sqrt{-\ga} \ga^{AB} \pa_B), 
\label{922}
\eeq
and
consequently fixes $\ep=1$. The kinetic part of the action (\ref{94}), on
the other hand, is invariant under the $T$-duality transformation
\cite{GV91, Tse91} 
\beq
a \ra \ti a = a^{-1} , \,\,\,\,\,\,\,\,\,\,\,\,\,\,\,\,\,\,
\fb \ra \fb , 
\label{923}
\eeq
which implies, for the Hamiltonian (\ref{921}),
\beq
{\cal H} (a)= {\cal H}(\ti a) + {2\over d} (\ep -1) \ti a 
{\pa \over \pa \ti a }. 
\label{924}
\eeq
The invariance of the Hamiltonian requires $\ep= 1$, and thus fixes the
same quantum ordering as the condition of general covariance
in minisuperspace. 
A similar relation between quantum ordering and duality symmetry can
be easily established for more general effective actions, including  a
larger class of non-minimal gravidilaton couplings \cite{Lid95,Lid97}.

\subsection{Wave scattering in minisuperspace}
\label{Sec9.2}

In the absence of dilaton potential, the WDW equation (\ref{98})
reduces to the free d'Alembert equation and provides a plane-wave
representation in minisuperspace of the different asymptotic  branches
of the classical solutions. 

Let us recall, in fact, that the solutions of the low-energy effective
action (\ref{94}) always contain four branches. For $V=0$, in
particular, we have four (physically different) solutions, 
\beq
a(t)= (\mp t)^{\mp1/\sqrt d} , ~~~~~~~~~~~~~~
\fb (t)= -\ln (\mp t) , 
\label{925}
\eeq
related by the duality transformation (\ref{923}) and by time reversal,
$t \ra -t$. They satisfy the condition
\beq
\sqrt d H \equiv \dot \b = \pm (\mp t)^{-1} = \pm \fbp,
\label{926}
\eeq
and may thus be represented as the bisecting lines of the plane  
$\{\dot{\fb},\dot\b\}$, corresponding to
\begin{itemize}
\item{}expansion $\ra \bp >0$, 
\item{}contraction $\ra \bp <0$,
\item{}pre-big bang (growing dilaton) $\ra \fbp >0$,
\item{}post-big bang (decreasing dilaton) $\ra \fbp <0$,
\end{itemize}
as illustrated in Fig. \ref{f91}. We may note, for later use, that if we add
to the action a positive cosmological constant ($V=\La$), then the
classical solutions are represented in the plane $\{\dot{\fb},\dot\b\}$ by
the hyperbolas
\beq
{\dot{\fb}}~^2-(\sqrt d H )^2 =\La. 
\label{927}
\eeq
In that case, the initial configuration is shifted to a state with flat
metric and linearly evolving dilaton, but the solution is still
characterized by four branches coinciding, asymptotically, with the
free configurations of Eq. (\ref{925}). 

\begin{figure}[t]
\centerline{\epsfig{file=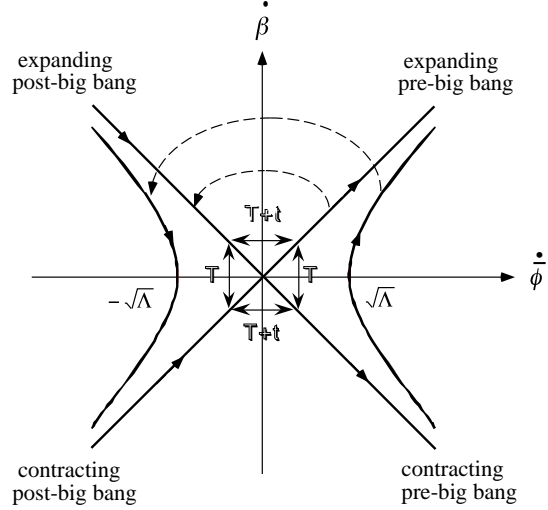,width=72mm}}
\vskip 5mm
\caption{{\sl The four branches of the lowest-order string-cosmology
solutions, with (hyperbolas) and without (bisecting lines) a positive
cosmological constant $\La$. The dashed curves represent the quantum 
transition from a pre- to a post-big bang configuration.}} 
\label{f91}
\end{figure}

The $V=0$ solutions of the WDW equations, on the other hand, 
can be factorized in the form of plane waves 
representing free energy and momentum eigenstates:
\beq
\Psi(\b,\fb)\sim \psi_{\b}^{(\pm)}\psi_{\fb}^{(\pm)} \sim e^{\pm ik \b
\pm ik \fb}, 
\label{928}
\eeq
where ($k>0$):
\beq
\Pi_{\b}\,\psi_{\b}^{(\pm)}= \pm k \,\psi_{\b}^{(\pm)} , ~~~~~~~~~~~
\Pi_{\fb}\,\psi_{\fb}^{(\pm)}= \pm k \,\psi_{\fb}^{(\pm)}.
\label{929}
\eeq
By recalling that $\Pi_\b \sim \bp$, $\Pi_{\fb} \sim - \fbp$ (see Eq.
(\ref{96})), the above plane waves represent the four branches of Eq. 
(\ref{925}), defined by $\Pi_\b = \pm \Pi_{\fb}$, with the following
correspondence:  
\begin{itemize}
\item{}expansion $\ra \psi_{\b}^{(+)}$, 
\item{}contraction $\ra \psi_{\b}^{(-)}$,
\item{}pre-big bang (growing dilaton) $\ra \psi_{\fb}^{(-)}$,
\item{}post-big bang (decreasing dilaton) $\ra \psi_{\fb}^{(+)}$. 
\end{itemize}

For an isotropic, low-energy solution \cite{GV91},  
the dilaton is growing ($\dot \phi>0$) only if the metric is
expanding ($\dot\b >0$), see Eq. (\ref{93}). If we impose, as our
physical boundary condition, that the Universe emerge from the
string perturbative vacuum (corresponding, asymptotically, to $\b \ra
-\infty$,  $\phi \ra -\infty$), then the initial state $\Psi_{\rm in}$ must
represent a configuration with positive eigenvalue of $\Pi_\b$ and
opposite eigenvalue of $\Pi_{\fb}$, i.e. $\Psi_{\rm in} \sim \psi_{\b}^{(+)}
\psi_{\fb}^{(-)}$ (see also \cite{DaKie97} for a detailed discussion of
boundary conditions in quantum-string cosmology). 

A quantum transition from pre- to post-big bang thus
becomes, in this representation, a transition from an initial state 
\beq
\Psi_{\rm in} \sim e^{ ik \fb - ik \b}, ~~~~~~~\Pi_\b >0,~~ ~~~~~~
\Pi_{\fb} <0,  \label{930}
\eeq
to a final state
\beq
\Psi_{\rm out} \sim e^{- ik \fb - ik \b}, ~~~~~~~\Pi_\b >0, ~~~~~~
~~\Pi_{\fb} >0   
\label{931}
\eeq
(see the dashed curves of Fig. \ref{f91}). The associated trajectory
describes a monotonic evolution along $\b$, and a reflection along
$\fb$. The above discussion suggests that we look at the quantum
evolution  (in minisuperspace) of the initial pre-big bang state  as at the
scattering, induced by an effective WDW potential, of an incoming wave
travelling from $-\infty$ along the positive direction of the axes $\b$
and $\fb$.

The effective potential, on the other hand,  is known to be strongly
suppressed as we approach the string perturbative
vacuum, $\b, \fb \ra -\infty$. We shall assume that a possible growth of
$V$ in the strong-coupling regime is not strong enough to prevent the
effective WDW potential from going to zero  also at large positive
values of $\beta$ and $\fb$, so that $V \exp(-2\fb) \ra 0$ for $\b,\fb
\ra \pm \infty$. It follows that also the final asymptotic configuration 
$\Psi_{\rm out}$ can be represented by the free eigenstates 
$\psi_{\b}^{(\pm)}, \psi_{\fb}^{(\pm)}$.
However, even if the initial state is fixed by the boundary conditions of
the pre-big bang scenario, the final state is not, and there are in
general four different types of
evolution \cite{Gas98}, depending  on whether the asymptotic outgoing
state $\Psi_{\rm out}$ is a superposition of waves with the same
$\Pi_\b$ and opposite $\Pi_{\fb}$, or with the same $\Pi_{\fb}$ and
opposite $\Pi_\b$, and also depending on the identification of the
time-like coordinate in minisuperspace \cite{CaDe97,CaUnga99}.  

These four
possibilities are illustrated in Fig. \ref{f92}, where  cases $(a)$
and $(b)$ correspond to $\Psi_{\rm out}^\pm \sim 
\psi_{\b}^{(+)}\psi_{\fb}^{(\pm)}$,
while cases  $(c)$ and $(d)$ correspond to $\Psi_{\rm out}^\pm \sim 
\psi_{\fb}^{(-)}\psi_{\b}^{(\pm)}$.  Also, in cases $(a)$ and $(d)$ the
time-like coordinate is identified with $\b$, in cases $(b)$ and $(c)$ with
$\fb$. It must be noted that cases $(c)$ and $(d)$ require a
duality-breaking dilaton potential, otherwise $[{\cal H}, \Pi_\b]=0$, and
then a reflection along the $\b$ axis is impossible, even at the quantum
level \cite{GasMaVe96}. 

\begin{figure}[t]
\centerline{\epsfig{file=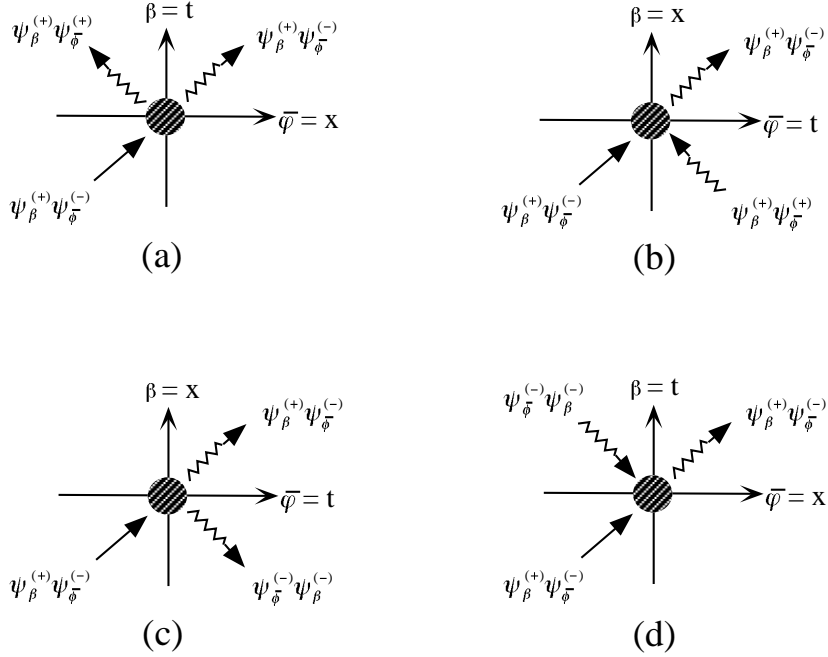,width=110mm}}
\vskip 5mm
\caption{\sl Four different classes of scattering processes for the
incoming wave function representing the string perturbative vacuum
(straight, solid line). The outgoing state is represented by a mixture of
positive and negative eigenfunctions of $\Pi_\b$ and $\Pi_{\fb}$.} 
\label{f92}
\end{figure}

The two cases $(a)$ and $(c)$ represent scattering and reflection along
the spacelike axes $\fb$ and $\b$, respectively. In case $(a)$ the
evolution along $\b$ is monotonic, so that the Universe always keeps 
expanding. The incident wave is partially transmitted towards the
pre-big bang singularity (unbounded growth of the curvature and of
the dilaton, $\b \ra +\infty$, $\fb \ra +\infty$), and partially reflected
back towards the low-energy, expanding, post-big bang regime  ($\b
\ra +\infty$, $\fb \ra -\infty$) \cite{GasMaVe96}. In case $(c)$ the
evolution is monotonic along the time axis $\fb$, but not along $\b$. 
The incident wave is therefore totally transmitted towards the
singularity ($\fb \ra +\infty$), but in part as an expanding configuration
and in part as a contracting one\footnote{R. Ricci, M. Gasperini
and G. Veneziano (1996),  unpublished.}.  

The other two cases, $(b)$ and $(d)$, are qualitatively different, as the
final state is a superposition of positive and negative energy
eigenstates, i.e. of modes of positive and negative frequency 
with respect to the time axes chosen in minisuperspace. 
In the language of third quantization
\cite{Rub88,Koz88,Guig88,Guig89,Guig90}  (i.e. second quantization of
the WDW wave function in superspace) they represent a ``Bogoliubov
mixing", describing the production of pairs of universes from the
vacuum. The mode moving backwards in time has to be
``re-interpreted", as in quantum field theory,  as an ``antiuniverse" of
positive energy and opposite momentum (in superspace). Since the
inversion of momentum, in superspace,  corresponds to a reflection of
$\dot \b$, the re-interpretation principle in this context changes
expansion into contraction, and vice versa. 

Case $(b)$, in particular, describes the production of
universe--antiuniverse pairs --one expanding, the other 
contracting-- from the string perturbative vacuum \cite{BuoGas97}.  The
pairs evolve towards the strong-coupling regime $\fb \ra +\infty$,   so
that both  members of the pair fall inside the pre-big bang singularity.
Case $(d)$ is more interesting, in our context, since there the
universe and the antiuniverse of the pair are both expanding: one falls
inside the pre-big bang singularity, the other expands towards the
low-energy, post-big bang regime, and may expand to infinity,
representing the birth of a Universe like ours in a standard
Friedmann-like configuration \cite{Gas00d}. 

Possible explicit examples of a quantum representation in
minisuperspace of the transition from pre- to post-big bang, based on
the above scattering processes, will be given in the next two
subsections. 

\subsection{Birth of the Universe as ``quantum reflection"}
\label{Sec9.3}

The simplest way to represent the transition from the string
perturbative vacuum to the present, post-big bang regime, is to identify
the time-like coordinate in the  minisuperspace with $\b$, and to
represent the process as a reflection of the incoming wave function
along the space-like coordinate $\fb$, according to Eqs. (\ref{930}),
(\ref{931}) (case $(a)$ of Fig. \ref{f92}).  The transition probability is
thus controlled by the reflection coefficient 
\beq
R_k=
{|\Psi_{-\infty}^{(-)}(\b,\fb)|^2\over|\Psi_{-\infty}^{(+)}(\b,\fb)|^2}, 
\label{932}
\eeq
where $\Psi_{-\infty}^{(\pm)}$ are the asymptotic components of the
WDW wave function at $\fb \ra -\infty$, containing the left-moving $(-)$
and right-moving $(+)$ part of the wave along $\fb$ (see
\cite{CaDe97} for a rigorous definition of scalar products in the
appropriate Hilbert space, and \cite{CaUnga99,CaUnga00,Giri00} for a
path-integral approach to the computation of the transition amplitudes). 

Without potential in the WDW equation there is no transition, of
course (in the absence of a dilaton potential, the possible decoherence
of the quantum-cosmological system, and its implications for the exit
problem, have been discussed in \cite{LuPo97}). 
With an appropriate potential, allowing a smooth classical
evolution from pre- to post-big bang, the transition probability $R$
tends to unity; $R$ may be non-zero, however, even if the two branches
of the classical solution are causally disconnected by a singularity. 

An instructive example, to this purpose, is provided by the
duality-symmetric (non-local) four-loop  dilaton potential
\beq
V_{\pm} (\b\fb) = \pm V_0 e^{4 \fb} , ~~~~~~~~~~~
V_0 = {\rm const}, ~~~~~~~~~~~V_0>0,
\label{933}
\eeq
already introduced in Subsection \ref{Sec8.1} 
(for generalized exponential potential, with the inclusion of the
antisymmetric tensor contribution to the WDW equation, see
\cite{MaMu97}).  The corresponding action (\ref{94}), in $d=3$, has the
following classical solution \cite{GasMaVe96} \beq
\fb= -{1\over 2}\ln \left ({k^2t^2\over
\la_{\rm s}^2} \mp {\la_{\rm s}^2 V_0\over k^2}\right) , ~~~~~~
a= a_0\left|{k^2t\over
\la_{\rm s}^2\sqrt{V_0}}+\left({k^4t^2\over
\la_{\rm s}^4V_0}\mp 1\right)^{1/2}
\right|^{1/\sqrt 3} , 
\label{934}
\eeq
where $a_0$ is an integration constant, and 
\beq
k=\la_{\rm s} \bp e^{-\fb}= 
{\sqrt3\over \la_{\rm s}^2}\int d^3x\sqrt{-g} e^{-\phi}\bp= {\rm const}
\label{935}
\eeq
is the conserved momentum along the $\b$ axis. 

For $V<0$, Eq. (\ref{934}) describes a regular ``self-dual" solution 
 --satisfying $a(t)/a_0=a_0/a(-t)$-- characterized by a ``bell-like"
shape of the curvature scale and of the coupling $e^{\fb}$ (see 
Subsection \ref{Sec8.1}). The solutions of the WDW equation (\ref{98}), in
the small coupling regime $ \fb \ra -\infty$, can be written as
\cite{GasMaVe96} \bea
\lim_{\fb \ra -\infty} \Psi_k(\fb,\b) &=& 
-{N \pi\over 2 \sin(ik\pi)}\left[
\left(\ls\sqrt{V_0}\over 2\right)^{ik}{e^{-ik(\b-\fb)}\over\Ga(1+ik)}-
\left(\ls\sqrt{V_0}\over 2\right)^{-ik}{e^{-ik(\b+\fb)}\over\Ga(1-ik)}
\right] \nonumber \\
&=&A_k \psi_{\b}^{(+)}\psi_{\fb}^{(-)}+B_k \psi_{\b}^{(+)}\psi_{\fb}^{(+)}
\label{936}
\eea
($N$ is a normalization coefficient), and contain the superposition of
the initial string perturbative vacuum and of the reflected component,
representing the post-big bang regime. Clearly, $R=1$ for all $k$, as the
two branches of the solution are smoothly connected already at the
classical level.

For $V>0$, on the contrary, the pre- and post-big bang branches are
disconnected by an unphysical region, of extension $|t| < \ls^2
\sqrt{V_0}/k^2$, where the expansion rate $\bp$ and the dilaton
coupling $e^{\fb}$ become imaginary. Such a region is bounded, on both
sides, by a curvature singularity. However, after fixing the boundary
conditions with the string perturbative vacuum and in the small coupling
regime, the solutions of the WDW
equation can be written  as \cite{GasMaVe96}: 
\bea
\lim_{\fb \ra -\infty} \Psi_k(\fb,\b) &=&
{iN  \csc(ik\pi)}\left[e^{k\pi}
\left(\ls \sqrt{V_0}\over 2\right)^{ik}{e^{-ik(\b- \fb)}\over\Ga(1+ik)}-
\left(\ls\sqrt{V_0}\over 2\right)^{-ik}{e^{-ik(\b+\fb)}\over\Ga(1-ik)}
\right] \nonumber \\
&=&A_k \psi_{\b}^{(+)}\psi_{\fb}^{(-)}+B_k
\psi_{\b}^{(+)}\psi_{\fb}^{(+)}. 
 \label{937}
\eea
The reflection coefficient 
\beq
R_k={|\Psi_{-\infty}^{(-)}|^2\over |\Psi_{-\infty}^{(+)}|^2}= e^{-2\pi k}
\label{938}
\eeq
is exponentially suppressed but non-zero, corresponding to a
non-vanishing transition probability from the pre- to the post-big bang
branches of the above solution. 

Exactly the same result (\ref{938}) is obtained for the simplest case of
dilaton potential, a positive cosmological constant $V(\b,\fb)=\La$. In
that case, by expressing the conserved momentum $k$ in terms of the
coupling $g_{\rm s}= e^{\phi_{\rm s}/2}$ at the string scale $H_{\rm s}
=\ls^{-1}$, the
transition probability between the left and right branches of the
hyperbolas of Fig. \ref{f91}, for a three-dimensional portion of space of
fixed proper volume $\Om_i$ at $t \ra -\infty$, can be written in the
form  \cite{GasVe96a}: 
\beq
R(\La, \Om_i, g_{\rm s})=\exp\left\{-{\sqrt{12}\pi \over
g_{\rm s}^2}{\Om_i \over  \ls^3}\left[{\sqrt3\over \ls \sqrt\La} +
\left(1+{3\over \ls^2\La}\right)^{1/2}\right]^{\sqrt3}\right\} ,
\label{939}
\eeq
Unfortunately, such a transition probability is exponentially suppressed,
unless the proper size of the transition volume is small, and the
cosmological constant is large, in string units. Note, however, that the
probability is peaked in the string coupling regime, with a typical
``instantonic" behaviour $R \sim \exp (-1/g_{\rm s}^2)$: this means that
the Universe tends to emerge from the transition in the coupling regime
appropriate to the present post-big bang phase. 

It is important to stress that the quantum-transition probabilities
(\ref{938}), (\ref{939}) have a  conceptual meaning different from 
similar results obtained in the context of the standard
inflationary scenario. In string cosmology, in fact, the quantum
(Planckian) regime is reached {\em at the end} of the pre-big bang phase,
when the Universe is expected {\em to exit} from (not to enter in) the
inflationary regime. Thus, quantum effects are not responsible for
inflationary initial conditions \cite{Gas00}, whereas they are in the
standard inflationary scenario \cite{Lin99}. In spite of this important
difference, the result (\ref{938}) is formally very similar to the
probability that our Universe may emerge from the Planckian regime
obtained in the context of the ``tunnelling from nothing" and other
similar inflationary scenarios
\cite{AkaTera83,Vil84,Linde84,Zeldo84,Rub84}, where the probability is
also exponentially suppressed, and the coefficient $k$ is inversely
proportional to the effective cosmological constant and to the
gravitational coupling, $P \sim \exp (-1/\la_{\rm P}^2 \La)$. 

The reason for this formal analogy is easy to  understand if we recall
that, by choosing the string perturbative vacuum as the initial state of
the pre-big bang scenario, it follows that in minisuperspace there are
only right-moving waves approaching the singularity at $\fb \ra
+\infty$. This is exactly equivalent  to imposing tunnelling boundary
conditions that select only outgoing modes at the singular space-time
boundary \cite{Vil86,Vil88}. 

In this sense, the quantum reflection illustrated in this subsection can
also be interpreted as a tunnelling process, not {\em ``from nothing"},
however, but {\em ``from the string perturbative vacuum"}. 
A different minisuperspace representation of the birth of the Universe
from the vacuum will be illustrated in the next subsection.

\subsection{Birth of the Universe as ``antitunnelling"}
\label{Sec9.4}

The process of quantum reflection (or tunnelling) is not the only
``channel" open to vacuum decay in the string cosmology
minisuperspace introduced in the previous subsections. There are also
other --in principle more efficient-- processes, such as the parametric
amplification of the WDW wave function, which could represent the
transition to our present cosmological state. We refer, in particular, to
the cases $(b)$ and $(d)$ illustrated in Fig. \ref{f92}. 

Case $(b)$, discussed in \cite{BuoGas97}, shows indeed that a
primordial conversion of expanding into contracting dimensions,
corresponding to the production of pairs of Universes from the vacuum,
can be efficiently described as the parametric amplification of the wave
function. In this subsection we shall report a particular, explicit
example of case $(d)$ \cite{Gas00d} to show that,  with
an appropriate model of dilaton potential,  the 
parametric amplification of the wave function can also efficiently 
represent  the transition from the pre- to the post-big
bang regime, via pair production from the vacuum. In that case, the
birth of the Universe can be described as a process of 
``antitunnelling from the string perturbative vacuum" (the term 
``antitunnelling", synonymous of parametric amplification, 
follows from the fact that the transition probability in that case is
controlled by the inverse of the quantum-mechanical transmission
coefficient, see Subsection \ref{Sec4.4}). 

First of all we note that, with a duality-invariant
dilaton potential, the string-cosmology Hamiltonian is translationally
invariant along the $\b$ axis: in that case,  the  initial
expanding pre-big bang configuration keeps expanding, and the  {\em
out} state cannot be a mixture of states with positive and negative
eigenvalues of $\Pi_\b$. To implement the process $(d)$ of Fig.
\ref{f92},  we thus need a non-local, duality-breaking potential, which 
contains both the metric and the dilaton, but {\em not} in the 
combination $\fb$.  We shall use, in particular, a two-loop dilaton
potential induced by an effective cosmological constant $\La$ 
(two-loop potentials are known to favour the transition to the
post-big bang regime already at the classical level
\cite{GasVe93a,BruMad98}, but only for appropriate repulsive
self-interactions, $\La <0$). We shall assume, in addition, that such a
potential is rapidly damped in the large-radius limit $\b \ra +\infty$, and
we shall approximate such a damping, for simplicity, by the Heaviside
step function $\theta$, by setting 
\beq
V =\La \theta (-\b) e^{2\phi}, ~~~~~~~~~~~~~~~\La >0.
\label{940}
\eeq 
With such a 
damping we represent the effective suppression of the cosmological
constant, required for the transition to a realistic post-big bang
configuration. 

Given the above potential,  the general solution of the WDW equation
can be factorized in terms of the 
eigenstates of the momentum $\Pi_{\fb}$, by setting
\beq
\Psi(\b,\fb)=\Psi_k(\b) e^{ik\fb}, ~~~~~~~~~~
\left [\partial^2_{\b}
+k^2-\lambda_{\rm s}^2\,\Lambda\,\theta(-\b)\,e^{2\sqrt{d}\b} \right ]\,
\Psi_k(\b)= 0.
\label{941}
\eeq
In the region $\b >0$ the potential is vanishing, and  the 
outgoing solution is a superposition of eigenstates of $\Pi_\b$
corresponding to  positive and negative frequency
modes $\psi_\b^\pm$, as in case $(d)$ of Fig. \ref{f92}. In the  region
$\b <0$ the solution is a combination of  Bessel functions
$J_\nu(z)$, of imaginary index $\nu= \pm ik/ \sqrt{d}$ and argument 
$z= i\la_{\rm s} \sqrt{\La/d}~e^{\sqrt{d}\b}$. If we 
fix the boundary conditions at $\b \ra -\infty$, by imposing that the
Universe starts expanding from the string perturbative vacuum, 
\beq
\Psi_{\rm in}= \lim_{\b \ra -\infty} \Psi(\b,\fb)\sim e^{ik(\fb - \b)},
\label{942} 
\eeq
then the
WDW wave function is uniquely determined as: 
\bea
\Psi(\b,\fb)=N_kJ_{-{ik\over\sqrt{d}}}\left(i\la_{\rm s}
\sqrt{\La\over d}~e^{\sqrt{d}\b}\right) \,e^{ ik\fb}\,,
~~~~~~~~~~~~~~\b&<&0 , \nonumber\\
=\left[A_+(k) e^{ -ik\b}+A_-(k) e^{ik\b}\right]\,e^{ ik\fb}\,,
~~~~~~~~~~~~~~\b&>&0 .
\label{943}
\eea
With the matching conditions at $\b=0$ we
can  finally compute the Bogoliubov coefficients $|c_\pm(k)|^2 = 
|A_\pm(k)|^2/ |N_k|^2$ determining, in the third quantization
formalism, the number $n_k$ of universes produced from the vacuum, 
for each mode $k$ (where $k$ represents a given configuration in the 
space of the initial parameters). 

As shown by an explicit computation \cite{BuoGas97,Gas00d}, the wave
function is parametrically amplified (i.e. $n_k \gg1$), for all 
$k < \la_{\rm s}
\sqrt{\La}$. In that case, the birth of our present post-big bang Universe
may proceed efficiently, and may be represented in minisuperspace as the
forced production of pairs of Universes from the quantum fluctuations
of the string perturbative vacuum. However, for a realistic process
occurring at the string scale, with $\bp \sim \ls$ and $e^{\phi/2} \sim
g_{\rm s}$, the condition of parametric amplification can be written as 
\beq
k \sim \left(\Om_3\over \la_{\rm s}^3\right)
{1\over g_{\rm s}^2} \laq \la_{\rm s} \sqrt
\La, 
\label{944}
\eeq
where $\Om_3 = a^3 \int d^3x$ is the
proper spatial volume emerging from the transition in the post-big
bang regime. This implies that the transition is strongly favoured for 
configurations of small enough spatial volume in string units, large
enough coupling $g_{\rm s}$, and/or large enough cosmological constant $\La$
(in string units), just as in the cases discussed in the previous
subsection. 

For $k \gg \la_{\rm s} \sqrt{\La}$ the wave function does not  ``hit" the
barrier,  there is no parametric amplification, and the inital state runs
almost undisturbed towards the singularity.  Only a small,
exponentially suppressed fraction is able to emerge in the post-big bang
regime, exactly as in the case of tunnelling (or quantum reflection). In
the context of third quantization, this process can still be described as
the production of pairs of universes, but the number of pairs is now
exponentially damped, $n_k \sim \exp(-k/\la_{\rm s} \sqrt\La)$, with a 
Boltzmann factor corresponding to a ``thermal bath" of universes,  at
the effective temperature $T \sim \sqrt \La$ in superspace. 

It seems possible to conclude, therefore, that the low-energy
quantum-cosmology processes considered in this subsection may allow a
quantum transition from pre- to post-big bang configurations, even if
such configurations are classically disconnected. With a ``realistic"
dilaton potential, the ``bubbles" of  post-big bang phase nucleated  out
of the string perturbative vacuum seem, however, to be too small to
reproduce our present Universe without a further long period of post-big
bang inflation. The quantum processes reported in this Section, however,
could explain the ``foam" of infinitely many ``baby Universes" required,
for instance, in the context of chaotic inflation \cite{Lin95} and of other,
standard, inflationary scenarios. 

\vskip 5mm

\section{Conclusion}
\label{Sec10}
\setcounter{equation}{0}
\setcounter{figure}{0}

In this last section we will conclude our report by discussing a possible
``late-time" consequence of the pre-big bang scenario --the
large-scale dominance of the dilatonic dark energy density-- and by 
presenting a list of various problems and aspects of such a scenario that
are not included in the previous sections. Finally, we will summarize our
personal outlook of string cosmology, together with some speculations
about its possible future perspectives. 

\subsection{Towards the future: a dilaton-dominated Universe?}
\label{Sec10.1}

The dilaton is, undoubtedly, one of the most important ingredients  of
the pre-big bang scenario illustrated in this paper. Indeed, the dilaton
implements the duality symmetry, controls the strength of all
interactions, sustains the initial inflationary evolution, contributes to
the amplification of the quantum fluctuations (and, in particular, to the
production of seeds for the magnetic fields), leading eventually to the
formation of a cosmic background of massive scalar particles. All such
effects are typical of string cosmology and represent the ``imprint"
of the pre-big bang scenario with respect to other, more conventional,
inflationary scenarios. 

Not satisfied with all such effects, however, the dilaton could still be
hale and hearty, and still in action even today, so that it would 
affect in a determinant way not only the very early cosmological past,
but also the present (and, possibly, future) state of our Universe. The
dilaton potential energy, or a mixture of kinetic and potential energy
density, could represent in fact the dark component responsible for the
cosmic acceleration observed very recently \cite{Riess98,Perlmutter99}.
In this sense, string theory can automatically provide, with the dilaton, a
``non-minimal" model of quintessence \cite{Gas01,GPV01} (i.e. a model
of quintessence based on a scalar field non-minimally coupled to gravity
and to elementary matter fields). 

There are, basically, two possible scenarios, depending on the
(currently uncertain) shape of the non-perturbative dilaton potential
$V(\phi)$. In the context of superstring models for grand-unified
theories (GUTs), the present value of the dilaton should determine in
fact the whole set of gravitational and gauge coupling parameters,
which are today constant (or, if they are time-dependent, are
nevertheless running very slowly on a cosmological time scale). This
constraint can be implemented in two ways. 

The first possibility is a dilaton almost frozen at a
minimum of the potential, in the moderate-coupling regime with
$-\phi$ of order $1$, in such a way that the GUT gauge coupling is fixed
to \cite{Kap85}
\beq
\a_{\rm GUT} \sim (M_{\rm s}/M_{\rm P})^2 \sim e^\phi \sim 0.1 - 0.01 .
\label{101}
\eeq
The second possibility is a dilaton monotonically running towards
$+\infty$, and the couplings saturated at small values in the
strong-coupling regime because of the large number $N$ of fields 
entering the loop corrections, or the large value of the quadratic
Casimir $C$ for unification gauge groups like $E_8$. Realistic
values of $\a_{\rm GUT}$ and $M_{\rm s}$ can be obtained at $\phi \gg
1$, for $N \sim C \sim 10^2$, typically as \cite{Ven98,Ven01}:
\beq
\a_{\rm GUT} \sim {e^\phi \over 1+ C e^\phi}, ~~~~~~~~~~~~~~~~~
{M_{\rm s}\over M_{\rm P}} \sim {e^\phi \over 1+ N e^\phi}. 
\label{102}
\eeq

In both cases, if the present value $V_0\equiv V(\phi_0)$ of the
potential is of the order of the present Hubble scale $H_0^2$ (and,
possibly, $\dot \phi^2 \sim V_0$), the dilaton can reproduce the
observed ``dark-energy" effects, playing the role of the so-called
``quintessence" \cite{CalDaSte98}. The required fine-tuning of the
amplitude of the potential (and the corresponding ultra-light value of
the dilaton mass,  $m \sim H_0$), seems to be unavoidable. In this
context, however, it seems possible to alleviate the problem of the
``cosmic coincidence" \cite{Stein97}, generated by the observed
(approximate) equality  $V_0 \sim \r_0$, where $\r_0$ is the present
energy density of the dominant dark-matter component. Let us explain
how this could happen, in both  the weak-coupling and
the strong-coupling scenarios of dilatonic quintessence. 

In the first case we note that, even assuming that the dilaton gets
frozen at a value $\phi_0$ after the transition to the post-big bang
regime, with a potential energy that is initially negligible ($V_0 \ll
H_{\rm eq}^2$), it may keep frozen for the whole duration of the
radiation epoch, but it tends to be shifted away from equilibrium as
soon as the Universe enters the matter-dominated regime. 

Suppose in fact that we  add a potential $V(\phi)$ to the low-energy
gravidilaton action: 
\beq
S= -{1\over 2\la_{\rm s}^{2}} \int d^{4}x \sqrt{|g|}~
e^{-\phi}\left[R+ \left(\nabla \phi\right)^2+V(\phi) \right] +S_m, 
\label{103}
\eeq
where $S_m$ describes, for simplicity, perfect-fluid sources minimally
coupled to the background. For a homogeneous and conformally flat
metric, the dilaton equation can then be written in the form
\beq
\fpp+3H\fpu -\fpu^2 +{1\over 2} e^\phi (\r-3p) +V'+V=0.
\label{104}
\eeq
Combining this equation with the standard conservation equation of
the matter sources, 
\beq
\dot \r +3H(\r +p)=0, 
\label{105}
\eeq
it follows that constant and stable solutions $\phi=\phi_0$ (with
$\fpu=0=\fpp$) are allowed in three cases only: 
1) vacuum, $\r=p=0$, $V+V'=0$;  2)  cosmological constant,
$\r=-p=\r_0=$ const, $V+V'=-2e^{\phi_0}\r_0=$ const;  3)  
radiation, $\r=3p$,  $V+V'=0$. 

So, even if the dilaton is ``sleeping" in the radiation era, it ``wakes up"
and starts rolling away from the freezing position determined by 
$V+V'=0$ just after the equilibrium epoch. We are thus led to the
following question: For which values of the potential $V_0$ may the
dilaton bounce back to the minimum, and the matter era  be
followed by a quintessential, potential-dominated epoch?

If the answer would point {\em only and precisely} at $V_0 \sim
H_0^2$, then the cosmic coincidence would be explained. If the answer
would indicate for $V_0$ a restricted range of values, including $H_0^2$,
the coincidence problem would remain, but it would be alleviated. 

The answer to the above question depends on the shape of the
potential, and on the effective strength of the dilaton coupling to
macroscopic matter. To illustrate this point we may consider the 
E-frame cosmological equations (\ref{62}), obtained from the action
(\ref{103}) through the conformal transformation 
\beq
\ti g_{\mu\nu} =g_{\mu\nu} e^{-\phi}, ~~~~~
\ti \phi= \phi, ~~~~~ \ti V = e^\phi V, ~~~~~
\ti \r = \r e^{2\phi}, ~~~~~ \ti p =p e^{2\phi}. 
\label{106}
\eeq
In units of $16 \pi G =1$ (and omitting the tilde, for simplicity), the
dilaton equation becomes
\beq
\ddot {\phi}+3H \dot {\phi} +{1\over 2} \alpha(\phi)(\r-3p) +{V}'=0 ,
\label{107}
\eeq
where we have taken into account, through the coupling function
$\a(\phi)$, a possible loop renormalization of the effective dilaton
coupling in the matter action. In the radiation era, $\r=3p$ and a stable,
frozen dilaton ($\fpu=0$) thus corresponds to an extremum of the
E-frame potential, $V'=0$. In the matter era, when $p=0$, there is a
dilaton acceleration away from the minimum, $\fpp =- \a \r/2$,
possibly contrasted by the restoring force $-V'$. 

Consider now a typical, supersymmetry-breaking dilaton potential,
instantonically suppressed \cite{BG86} at $\phi \ra -\infty$, with a
non-trivial structure developing a minimum in the region of moderate
coupling, and exponentially growing at $\phi \ra +\infty$ because of
the conformal transformation to the E-frame. A ``minimal" example of
such a potential can be simply parametrized (in the E-frame) as follows
\cite{KaOl93}
\beq
V= m^2 \left[e^{k_1(\phi_-\phi_1)}+ \beta
e^{-k_2(\phi_-\phi_1)}\right] e^{-\ep \exp\left[-\gamma
(\phi_-\phi_1)\right]}, 
\label{108}
\eeq 
where $k_1,k_2, \phi_1,\ep ,\b,\ga$ are dimensionless numbers of
order $1$ For an illustrative purpose we will choose here the particular
values $k_1=k_2=\b=\ga=1$, $ \ep=0.1$, $\phi_1=-3$, in such a way that
the minimum is at $\phi_0 =-3.112...$, and that $ g_{\rm s}^2 =
\exp (\phi_0)\simeq 0.045$, in agreement with Eq. (\ref{101}). 

With the above choice of parameters, the dilaton potential 
is plotted in Fig. \ref{f101} for different values of $m$. It is important
to note that the extremum $\phi_0$ {\em is not} separated from the
perturbative regime  ($\phi \ra -\infty$) by an infinite potential
barrier, and that the lower  $V_0 \sim m^2$, the lower  the barrier,
the weaker is the restoring force $-V'$ and the easier it is  for the
dilaton  to escape from the minimum and run to $-\infty$. 

\begin{figure}[t]
\centerline{\epsfig{file=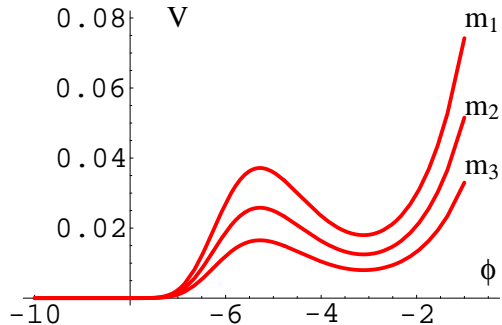,width=72mm}}
\vskip 5mm
\caption{\sl Plot of $V(\phi)$ from Eq. (\ref{108}), with 
 $k_1=k_2=\b=\ga=1$, $ \ep=0.1$, $\phi_1=-3$. The three curves, from
top to bottom, correspond respectively to $m=1/10, 1/12, 1/15$ in
units of $M_{\rm P}^2=2$.} 
\label{f101}
\end{figure}

To match present phenomenology,  the value of $V_0$, in such a context,
has to be chosen very small in string units, $V_0 \sim H_0^2 \sim
(10^{-33}~ {\rm eV})^2$ (see below for a possible justification). It
follows that the potential barrier is very low, and the dilaton would
certainly escape from the minimum at the beginning of the matter era,
{\em unless} the coupling to matter ($\a\r/2$) is also correspondingly
small. Such a coupling, on the other hand, {\em has} to be very small,
because the corresponding scalar force has a very long range,
$V''(\phi_0) \sim m^2\sim V_0\sim H_0^2$, and the dilaton must be
strongly decoupled ($\a \ll1$, at least today) to avoid unacceptable
violations of the equivalence principle \cite{DP94a,DP94b}. 

It can be shown, as a consequence, that the present values of $\a$
allowed by the gravitational phenomenology ($\a_0 \laq 10^{-3}$) are
compatible with a late cosmological phase dominated by the dilaton
potential only for a restricted range of $V_0$, which depends on 
the value of $\a$ at the equilibrium epoch, $\a_{\rm eq}$ \cite{Gas01}. 
Expanding $\a(\phi)$ around the minimum, and starting from    
$\a_{\rm eq}=0.1$, for instance, a numerical analysis shows that the
dilaton, after a small shift at the equilibrium epoch, bounces back to the
minimum provided $10^{-7} H_{\rm eq} \laq m \laq H_{\rm eq}$, which
includes the realistic case $m \sim V_0^{1/2} \sim H_0\sim 10^{-6}
H_{\rm eq}$. In such a context, therefore, the coincidence problem is
not  strictly solved, but possibly alleviated (see \cite{Gas01} for a more
detailed discussion). 

An alternative scenario for a dilatonic interpretation of the
observed ``quintessential" effects, is based on the assumption that the
dilaton never gets  trapped in a minimum, and evolves monotonically
(and boundlessly) from negative (pre-big bang) to positive (post-big
bang) values, with a potential smoothly approaching zero
as $\phi \ra +\infty$ (see Fig. \ref{f21}). The non-perturbative
potential, in this context, is typically characterized by a bell-like
shape,  which can be parametrized (in the S-frame) as \cite{GPV01}:
\beq
V=m^2 \left[ e^{-{1\over \b_1}\exp (-\phi)}-
e^{-{1\over \b_2}\exp (-\phi)}\right],
\label{109}
\eeq
where $\b_1>\b_2>0$ are dimensionless numbers of order unity and,
again, the mass scale $m$ satisfies $m \ll H_{\rm eq}$, to avoid that
the dilaton becomes dominant too early. 

As before, the potential  is instantonically suppressed at $\phi \ra
-\infty$, but now is also esponentially suppressed in the $\phi \ra
+\infty$ limit. The mass scale $m$, in the context of a non-perturbative
potential, is naturally related to the string scale by $m \sim
\exp\left[-\b \a^{-1}_{\rm GUT}\right] M_{\rm s}$, where $\b$ is a
model-dependent parameter of order $1$. This may explain the
smallness of the potential in string units, but it should not hide the fact
that the constant parameter $\b$ has to be precisely adjusted  {\em ad
hoc} if we want to start the accelerated dilaton-dominated phase not
much  earlier than at redshifts of order one \cite{Turner01}, and not later
than today. 

In the context of a running-dilaton scenario, the gauge couplings and
the ratio $M_{\rm s}/M_{\rm P}$ are to be stabilized, in the $\phi \ra +\infty$ 
limit, by the loop corrections. Such corrections, on the other hand, also
play a fundamental role in determining the effective dilatonic charge of
the elementary fields appearing in the matter action. Thanks to this
interplay, a dilatonic charge that switches on at late times, in the
(non-baryonic) dark-matter sector, seems able to provide a
possible dynamical explanation of the cosmic coincidence \cite{GPV01}
(see \cite{DaPiaVe02,DaPiaVe02a} for possible related violations of
free-fall universality, and time variation of the natural constants,
induced in this context). 

Such a dynamical approach to the coincidence problem is based, in
particular, on the following loop-corrected action
\beq
S = -{1\over 2\la_{\rm s}^2} \int d^{4}x \sqrt{-  g}~
\left[e^{-\psi(\phi)} R+ Z(\phi)
\left(\nabla \phi\right)^2 + {2\la_{\rm s}^2} V(\phi)\right] +S_m(\phi, {\rm
matter}),
\label{1010}
\eeq
where $V$ is the potential of Eq. (\ref{109}), and the loop form factors
$\psi$ and $Z$, in a minimal ``induced-gravity" scenario, are assumed to
be parametrized at large $\phi$ as follows \cite{Ven98,Ven01}:
\beq 
e^{-\psi(\phi)}\, = \,  e^{-\phi} + c_1^2 \, , ~~~~~~~~~~~
Z(\phi)\, = \, e^{-\phi} - c_2^2 ,  
\label{1011}
\eeq
where $c_1^2 \sim c_2^2 \sim 10^2$, in agreement with Eq. 
(\ref{102}). We also assume that the action $S_m$ contains
non-baryonic dark-matter whose dilatonic charge density per unit of
gravitational mass, $\sg_d/\r_d$, switches on as $\phi \ra +\infty$,
and can be parametrized as \cite{GPV01}: 
\beq
q(\phi)= {\sg_d\over \r_d}= -{2\over \sqrt{-g} \r_d} {\da S_m\over \da
\phi} =q_0 {e^{q_0 \phi}\over c^2+ e^{q_0\phi}},
\label{1012}
\eeq
where, again, $c^2 \sim 10^2$. Note that we are using here a definition
of  dilatonic charge density different from the one previously introduced
in Eqs. (\ref{616}) and (\ref{840}). 

The analytical and numerical study of such a model shows that, in a way
that is largely independent from the initial conditions, the Universe is
eventually driven to an asymptotic accelerated regime characterized
by a frozen ratio between the dark-matter and the dilatonic energy
density, $\r_d/\r_\phi=$ const. The present approximate equality of
$\r_d$ and $V_0$ is no longer a coincidence, in this context, but a
dynamical property of the asymptotic configuration, as in models of
``coupled quintessence" \cite{Wett00,AmenToc01}. 

For a better illustration of the ``late-time" post-big bang scenario, 
described by Eqs. (\ref{1010})--(\ref{1012}), it is convenient to
consider the E-frame cosmological equations for the tilted variables
defined through the conformal transformation 
\beq
 a = c_1 \ti a e^{\psi/2}, ~~~~~
d t = c_1 d \ti t e^{\psi/2}, ~~~~~
\ti \r = c_1^2  \r e^{2\psi} , ~~~~~
\ti p = c_1^2  p e^{2\psi} , ~~~~~
\ti \sg = c_1^2 \sg  e^{2\psi}
\label{1013}
\eeq
(we are considering, as before, perfect-fluid sources evolving in a
conformally flat metric background). Note that  $c_1^2$, according 
to Eq. (\ref{1011}), controls the asymptotic value of the ratio between
the string and the Planck scale, $c_1^2\la_{\rm s}^{-2}=\la_{\rm P}^{-2}$. The
E-frame equations thus become (omitting the tilde, and in units $2
\la_{\rm P}^2=16\pi G=1$) \cite{GPV01}
\bea
&&
6H^2 = \r +\r_\phi, \label{1014} \\
&&
4 \dot H + 6H^2 =-p -p_\phi, \label{1015} \\
&&
k^2(\phi) \left(\ddot{\phi}+3 H \dot{\phi}\right) +
k(\phi)\, k'(\phi)\, \dot{\phi}^2 
+ \hat{V}'(\phi) + \frac{1}{2}\left[{\psi'(\phi)} (\rho - 3 p) + q\r
\right] = 0, 
\label{1016}
\eea
where a dot denotes differentiation
with respect to the E-frame cosmic time, and we have defined: 
\bea
&&
k^2(\phi) = 3 \psi^{\prime 2} - 2 {e^\psi} Z , 
~~~~~~~~~~~~~~~~\hat V = c_1^4 e^{2\psi} V , \label{1017}\\
&&
\r_\phi= {1\over2} k^2(\phi) \dot \phi^2 +\hat V(\phi), ~~~~~~~~~~~~
p_\phi= {1\over2} k^2(\phi) \dot \phi^2 -\hat V(\phi). 
\label {1018}
\eea
The source terms $\r,p,\sg=q\r$ generically include the
contribution of radiation, baryonic and non-baryonic matter
components, but only for the non-baryonic dark  component $\r_d$
is the dilaton charge  assumed to be significantly different from zero,
and parametrized as in Eq. (\ref{1012}). By analysing the time evolution
of the (total) dilaton energy density, $\r_\phi$, we then find that the
post-big bang solutions are generally characterized by three dynamical
phases \cite{GPV01}. 

Suppose in fact that the radiation-dominated,
post-big bang regime starts with with a negligible dilaton potential,  a
negligible dark-matter energy density $\r_d \sim a^{-3}$, and  a fully
kinetic dilaton energy density ,which is rapidly diluted as $\r_\phi \sim
a^{-6}$. When $\r_\phi$ falls below $\r_d$ one finds that the Universe
enters a first ``focusing" phase with $\r_\phi \sim a^{-2}$, so that the
dilaton kinetic energy starts growing with respect to radiation, and
converges towards the other energy components. 

When the matter becomes dominant,  the Universe enters a subsequent
``dragging" phase, where $\r_\phi$ and $\r_d$ evolve in time with the
same behaviour, together with the baryonic matter density
(uncoupled to the dilaton), $\r_b \sim a^{-3}$.  Actually,  $\r_\phi$ and
$\r_d$ are diluted a little bit faster (like $ a^{-3 -\ep^2}$) than $\r_b$,
but the deviation is controlled by a parameter, which is constant and 
very small during the dragging phase, 
\beq
\left[\ep(\phi)\right]_{\rm drag} \equiv \left[\psi' + q\over k
\right]_{\rm drag} \sim {q_0 c_1\over c^2 c_2}\sim {1\over c^2} \ll1.
\label{1019}
\eeq

At late times (but not later than today) the potential $V(\phi)$ and the
dilaton charge $q(\phi)$ of dark-matter come eventually into play, and
the Universe enters a ``freezing" phase characterized by a constant
ratio $\r_\phi/\r_d$, and by accelerated evolution. In this asymptotic
regime, $\phi \ra + \infty$, the parameters of the model approach the
constant  values
\beq
k^2(\phi) = 2c_2^2/c_1^2 \equiv 2/\lambda^2,  \quad \sg  
= \sg_d,  \quad \r  = \r_d,  \quad q(\phi) = q_0, 
\label{1020}
\eeq
and the E-frame equations can be written as 
 \bea
&&
\label{dark1} \dot \r_d +3H\r_d -\frac{q_0}{2}\, \r_d \dot \phi=0,
\quad \quad \quad
\dot \r_\phi +6H\r_k +\frac{q_0}{2}\, \r_d\dot \phi=0, \label{1021}\\
&&
1=\Om_d+\Om_k+\Om_V, \quad \quad \quad \quad ~~~
1+{2\dot H\over 3 H^2}=\Om_V-\Om_k,
\label{1022}
\eea
where 
\bea
&&
\r_d= 6H^2 \Om_d,
~~~~~~~~~~~~~~~~~~~~~~\r_\phi=\r_k+\r_V,\label{1023}\\ 
&&
\r_k= 6H^2 \Om_k= \dot \phi^2/ \lambda^2, ~~~~~~~~~~
\r_V= 6H^2 \Om_V=\hat V. ~~~~~~~~
\label {1024}
\eea
They can be solved by an asymptotic configuration with constant
$\Om_k$, $\Om_V$ and $\Om_d$ (acting as late-time attractor
\cite{Amen00}), where the dilaton fraction of critical energy density,
$\Om_k+\Om_V$, the dilaton equation of state, 
$w_\phi= (\Om_k-\Om_V)/(\Om_k+\Om_V)$, and the asymptotic
acceleration parameter, are given by \cite{GPV01}
\bea
&& \label{1025}
\Omega_\phi\, =\, \frac{12
+  q_0(q_0+2)\lambda ^2}{ (q_0+2)^2\lambda ^2}, \quad \quad \quad 
w_\phi\, = \, - \frac{ q_0 (q_0+2)\lambda^2}{12
+  q_0(q_0+2)\lambda^2}, \\
&&
{\ddot a \over aH^2}= 1+{\dot H\over H^2}= {q_0-1\over q_0+2}.
\label{1026}
\eea

Such an asymptotic configuration is illustrated in Fig. \ref{f102}, where
we have plotted various curves at $\Om_\phi=$ const and $w_\phi=$
const in the $\{q_0,\la\}$ plane. As shown in the figure, a positive
acceleration ($q_0>1$) is perfectly compatible with the range of
$\Om_\phi$ and $w_\phi$ allowed by present
phenomenological constraints \cite{Wang00,Balbi01}, namely $0.6 \laq
\Om_\phi \laq 0.7$ and $-1 \leq w_\phi \laq -0.4$. 

\begin{figure}[t]
\centerline{\epsfig{file=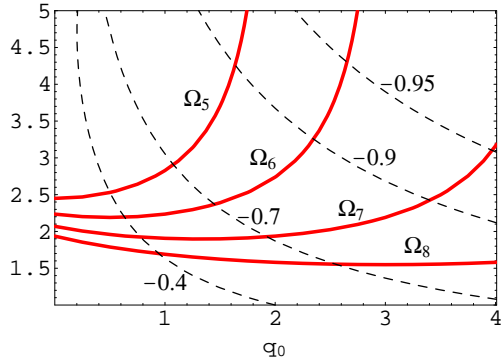,width=72mm}}
\vskip 5mm
\caption{\sl Asymptotic configurations in the plane  $\{q_0,\la\}$,
according to Eq. (\ref{1025}). The full curves correspond to the
constant values  $\Om_5=0.5$, $\Om_6=0.6$, $\Om_7=0.7$,
$\Om_8=0.8$, and the dashed curves to the constant values
of $w_\phi$ indicated in the figure.}   
\label{f102}
\end{figure}

We note, finally, that a simple integration of Eq. (\ref{1026}) gives the
late-time evolution of the dilaton and dark-matter energy density, 
\beq
\r_\phi \sim \r_d \sim H^2 \sim a^{-6/(2+q_0)}.
\label{1027}
\eeq
The baryon dark matter, uncoupled to the dilaton ($q=0, \r_b \sim
a^{-3}$), is rapidly diluted with respect to $\r_d$, and we thus obtain,
in the context of such a dilatonic scenario, also a possible explanation
of the present smallness of the ratio $\r_b/\r_d$ \cite{GPV01} (see also
\cite{AmenToc01a}). 

\subsection{Other open problems}
\label{Sec10.2}

In this report we have focalized our attention on many
important aspects of the pre-big bang scenario (in particular, on those
mainly studied in the past years), without pretending, of course, to
have presented an exhaustive discussion of all possible problems and
aspects of such scenario. There are important topics that we have left 
almost untouched, also because too few results ae available on these in
the present literature. Thus, it seems to us appropriate to conclude with
a list (and a brief discussion) of such topics, which we hope will be the
object of more intensive studies in the next years. 

In particular, we have given very little room to a discussion, in the
pre-big bang context, of a realistic model of (dynamical?) dimensional
reduction of the superstring manifolds, the (possible) compactification
and the associated stabilization of the internal dimensions. The best
approach to this problem, at present, is probably the model of
\cite{BraVa89}, generalized to a brane-gas in \cite{ABE00,BEK01}, which
assumes, however, a toroidal structure of the space-time manifold from
the very beginning (as in standard Kaluza--Klein compactification).
There are no results, at present, concerning the possible evolution of the
pre-big bang phase into a multi-dimensional non-compact structure,
eventually characterized by a warped geometry able to confine
long-range interactions on a four-dimensional brane \cite{RS2}. 

Also, we have not  discussed in detail the matching of the pre-big
bang regime to the subsequent Friedmann--Robertson--Walker (FRW)
phase, the possibility in this context of background oscillations (of the
metric and of the dilaton), with related ``preheating" and reheating
effects,  and the associated particle production (for instance gravitino
production, which is a potential problem  \cite{Wei82} for any inflationary
model based on a supersymmetric theory). All these problems, including
dimensional reduction, are (more or less directly) related to the fact
that there is at present no compelling model of a complete and realistic
exit transition (see Section \ref{Sec8}). 

It is worth mentioning, in this respect, that the problem of reheating
and of possible dangerous relics (moduli and gravitinos) in pre-big bang
cosmology has recently been faced in \cite{Buo00}. The main
difference from the reheating scenario of standard inflationary models
is that, for pre-big bang inflation, the matching to the FRW phase may
occur in principle at higher-curvature scales (typically, $H_1 
\sim M_{\rm s}
\sim 10^{17}$ GeV), where gravitinos and moduli fields should also be
 copiously produced in scattering processes. Another important difference
is that the particles present at the beginning of the FRW phase are
produced (with various spectra) directly from the amplification of the
quantum fluctuations of the background fields during inflation (as 
stressed in Subsection \ref{Sec8.4}), with typical densities $\r_i(t) \sim
N_i H_1^4 (a_1/a)^4$, where $N_i$ is the number of helicity states in
species $i$. 

In such a context, the thermalization scale of scalar and vector
particles, charged under the group of the observable sector, has been
estimated in \cite{Buo00} by assuming that the dilaton and the
internal dimensions are already frozen at the beginning of the FRW
phase, when $H=H_1$ and $g=g_1 \equiv M_{\rm s} /M_{\rm P}(t_1)$. The
charged particles, with typical energy $\om\sim H_1 a_1/a$ and number
density $n_r \sim \r_r/\om$, where
\beq
\r_r = \Om_r \r_c \sim {N_r\over N_{\rm tot}} 
{H_1^2 M_{\rm P}^2 a_1^4\over a^4} = 
{N_r\over N_{\rm tot}} \left(H_1M_{\rm s}\over g_1\right)^2
\left( a_1\over a\right)^4
 \label{1028}
\eeq
($ N_{\rm tot}$ is the total number of degrees of freedom present at
$t=t_1$), interact with cross section $\sg \sim \a^2 /\om^2$, where $\a
\sim g_1^2$. The thermalization scale $H_{\rm th} \sim n_r \sg$ is then
given by 
\beq
H_{\rm th} \sim g_1^4 \left(N_r \over  N_{\rm tot}\right)^2 {M_{\rm s}^4 \over
H_1^3}. 
\label{1029}
\eeq

If the charged particles are already dominat at $t=t_1$, i.e. $N_r \sim  
N_{\rm tot}$, $\Om_r \sim 1$, and if $H_1 \sim M_{\rm s}$ as in minimal
pre-big bang models, then the post-big bang Universe thermalizes at a
scale $H_{\rm th} \sim g_1^4 H_1\sim (10^{-4}$ -- $10^{-8})~ M_{\rm
s}$, with a corresponding reheating temperature $T_{\rm r} \sim (H_{\rm
th} M_{\rm P})^{1/2} \sim g_1^{5/2} M_{\rm P}\sim g_1^{3/2}M_{\rm
s}$. If, on the contrary, $\Om_r <1$, then reheating is only achieved once
the gauge-singlet fields, which carry the remainder of the total
energy density, have decayed into radiation. 

Such a reheating process generates an entropy density $s \sim
\r_r/T_r$, which is not sufficient, however, to dilute the unwanted relic
particles produced gravitationally during inflation, and also in
scatterings of the thermal bath for $t>t_1$. For the moduli, in
particular, the number-density-to-entropy-density ratio $Y_m
=n_m/s_m$ turns out to be \cite{Buo00} $Y_m \sim 0.3 /g_\ast$,
where $g_\ast$ is the number of degrees of freedom in the radiation
bath after thermalization.  For $g_\ast \sim 10^2$ -- $10^3$ this is well
above the limit imposed by nucleosynthesis \cite{Wei82}, which implies
$Y \laq 10^{-13}$. 

The same is true for gravitinos. Indeed, even if they are not
gravitationally produced during pre-big bang inflation because they
keep effectively massless \cite{BruHa00}, gravitinos are produced in
scatterings of the high-temperature thermal bath, and their number to
entropy ratio can be estimated (for $H_1 \sim M_{\rm s}$) as \cite{Buo00} 
$Y_{3/2} \sim g_1^{5/2} g_\ast ^{-7/4}$, which is still in excess of the
nucleosynthesis limit (see also \cite{Lemo99,Tsu01} for gravitino
production in a string cosmology context). 

The dilution of the unwanted relics possibly present at the beginning of
the post-big bang phase thus requires a strong entropy release, at the
level of about ten orders of magnitude for gravitationally produced
relics in the context of minimal pre-big bang models (so as to suppress
$Y$ from $10^{-3}$ to $10^{-13}$). Such an entropy production may be
due to a secondary reheating phase, as already discussed in Subsection
\ref{Sec6.3}, and a typical possibility corresponds to the oscillations
and decay of a scalar (or pseudoscalar) field, which gets a mass in the
post-big bang epoch through a symmetry- (or supersymmetry-) breaking
mechanism. 

A natural string-cosmology candidate for this effect is the axion (as
discussed in Subsection \ref{Sec7.5}), or even the dilaton, if it
is heavy enough. In particular, if the initial value of the oscillating field
is  at least of order $M_{\rm s}$, it is possible to dilute in this way moduli and
gravitinos, but not monopoles (possibly produced by GUT symmetry
breaking), while an initial value of order $M_{\rm P}$ is marginally
sufficient also for monopoles \cite{Buo00}. 

It is possible, however, that the monopole problem be independently
solved by an efficient monopole--antimonopole annihilation
\cite{Cop88}. We should keep in mind, also, that in more complicated,
non-minimal pre-big bang backgrounds (see Subsection \ref{Sec5.3}), the
energy distribution among the produced particles may be drastically
altered, with a resulting easier dilution of the relics component, and a
smaller required amount of entropy production (see \cite{Buo00} for a
detailed discussion). 

We note, finally, that the entropy-production process is independently
interesting by itself, in such a context, as it can naturally accommodate
baryogenesis if the oscillating scalar is identified with the Affleck--Dine
condensate, made of squarks and sleptons, whose decay generates the
baryon asymmetry \cite{AD85}. In particular, a ``mixed" reheating
phase with two oscillating fields (a modulus and a condensate), 
seems to be able to reproduce both the required dilution of relics and the
right baryon asymmetry  in the context of pre-big bang cosmology
\cite{Buo00}. The oscillations of the Affleck--Dine condensate could 
be sufficient, by themselves, to solve the moduli and gravitinos problem
in the context of non-minimal pre-big bang models, provided the
resulting baryon asymmetry were kept small enough by a very small
CP-violation parameter, or by a very efficient electroweak mechanism
of baryon-number erasure \cite{Gailla95}. 

\subsection{Outlook}
\label{Sec10.3}

Although it is always difficult to make forecasts, particularly in
theoretical physics, we are at least confident of one thing: research 
on the implications of string/M-theory on fundamental cosmological
questions is not just a momentary fashion. It is going to continue at an 
increasing rate as long as new cosmological data come in, and 
put more and more puzzles in front of theoretical cosmology.
Indeed, the last few years have witnessed a sudden jump of interest
in this field,  triggered partly by the new data on CMB anisotropy and
on evidence for an accelerating Universe, partly by the new theoretical
developments related to large extra dimensions, the 
possible lowering of the quantum-gravity scale, and the brane-world
ideas.

One line of research has been devoted to the modifications of the
standard cosmological equations for observers confined to a brane
immersed in  a higher-dimensional space-time \cite{Bine00}. Such
modifications are in principle sufficiently drastic to make us worry
about how to preserve the successes of the standard set-up, on the one
hand, and  about how to use the deviations in order to improve the
situation where the standard description may face difficulties  (e.g.
during inflation \cite{CoLiLi01} and early cosmology) on the other hand. 

A second development, which is even more relevant to this report, is
the proposal of new cosmologies \cite{KOST1,KOSST} that share many
properties with the pre-big bang scenario. As in the latter, they assume
that the big bang singularity is spurious, and that the Universe (and time)
had a long existence prior to it. Also, the initial state is assumed to be
very perturbative, although not as generic as in our approach. 
The most  important difference, however, is that such models are 
constructed around the brane-world idea (see Subsection \ref{Sec8.5}).
Thus, although some of the pictures look very similar, their meaning is
drastically different. As an example,  the plane-wave collision 
described  in Subsection \ref{Sec3.4}  superficially resembles the brane
collision of the ekpyrotic scenario; but while in the former our Universe
lies in the bulk, in the model of \cite{KOST1,KOSST} it lies  on one of the
two colliding branes. Also, the big bang is the moment of collision in
\cite{KOST1,KOSST} while, in the pre-big bang scenario, it emerges
sometime after the collision,  as a consequence of gravitational
collapse. 

The model in \cite{KOSST} is even closer to the pre-big bang idea,
 in the sense that, there, the extra dimension represents  the
dilaton  in the strong-coupling regime, according to the M-theory
conjecture \cite{Hor96}. It is thus possible to draw the
bouncing-Universe scenario of \cite{KOSST} in the same phase-space
plane as was used in Fig. \ref{f81}. The difference between the two
scenarios is very simple: while the pre-big bang Universe starts its
evolution in the upper-right quadrant of the diagram --and thus inflates,
in the  S-frame, as it evolves from weak to strong coupling-- in the
scenario of \cite{KOSST} the Universe starts in the lower-right quadrant
and thus contracts, even in the S-frame, while it evolves from strong to
weak coupling.

We have seen  in Subsections \ref{Sec1.3} and \ref{Sec2.4} that an
accelerated contraction is as  good as an accelerated expansion for
solving the horizon and flatness problems, hence the model of
\cite{KOSST} appears a priori as good as the pre-big bang model. 
It is important, however,  to stress some differences,
especially regarding the exit problem. If the exit is to  occur at
weak coupling, it should be possible to study it within the tree-level
effective action (although non-perturbatively in $\alpha'$). This was
precisely the framework described in Subsection
\ref{Sec8.2}, where it was shown that, at least order by order in the
$\alpha'$ expansion, a conserved quantity forbids a transition from
contraction to expansion in the S-frame. The possibility remains,
however,  of  non-perturbative (world-sheet instanton) effects for
achieving the bounce.

Still at the theoretical level, both the ekpyrotic and the pre-big bang
scenario suffer from the criticism of  \cite{DH00/1} about the
generic occurrence of a phase of chaotic oscillations. This is even more
crucial for the models of \cite{KOST1,KOSST}, since they need a highly
fine-tuned initial state \cite{KKL,KKLT} and, again,  cannot invoke
string-loop corrections for stopping BKL oscillations.

At the more phenomenological level the main difference between the
two classes of models concerns the spectrum of (adiabatic) scalar
metric perturbations. While these have  been claimed to have naturally
a blue spectrum in a pre-big bang context \cite{BruMu95} (see also
\cite{Hwang98,Tsu02}), the authors of  \cite{KOST1} have claimed to get
naturally (although through a certain hypothesis on the shape of a
potential) a scale-invariant spectrum, as favoured by observations.
Without taking sides, we will mention that this claim has been challenged
by several authors \cite{Ly02,BF01} (see however \cite{DuVe02}). On the
other hand, the curvaton mechanism of \cite{LyWa02} could still generate
scale-invariant adiabatic perturbations even if these were not
immediately generated during the pre-big bang (or pre-bounce) phase.

Since string/M-theory is proposed by its fans as a candidate unified
theory of all phenomena, it simply cannot allow itself to avoid tackling
the fundamental questions that classical and quantum gravity are 
proposing: What is the fate of the  classical singularities --that are so
ubiquitous in general relativity-- in the context of a consistent quantum
theory of gravity?

In this report the pre-big bang scenario was described in such a way
that the reader's attention was drawn  to the
big bang singularity of general relativity and to how   its  avoidance, in
string theory, could open up new  possibilities  for solving the
long-standing problems of classical, hot big bang cosmology. Amusingly,
in doing so, we were led to connect this question to the other big
puzzle of general relativity: the fate of black-hole singularities, with all
its ramifications into the information paradox, the loss of quantum
coherence, and the like.

To conclude, let us try to draw some generic lessons from the 
particular attempt at a new, string-based cosmology that we have
illustrated in the previous sections. 

\begin{itemize}
\item[(1)]
Neither our Universe, nor space and time themselves, have to emerge
from a singularity: the singularities of classical gravitational theories
should signal the lack of finite-size and quantum corrections, or the
need for new degrees of freedom in the description of physics at
very short distances. Also, the Universe did not have to start from the
very beginning as a hot and dense ``soup" of particles and radiation: a
hot Universe can emerge from a cold one, thanks to the parametric
amplification of the quantum fluctuations of the vacuum, during
inflation. 

\item[(2)]
Inflation does not need a scalar potential (or, more generally, an
effective cosmological constant), and can  naturally appear as the
result of the underlying duality symmetries of the cosmological
background. Also, inflation can be represented as a contraction (in the
appropriate frame), and the study of generic initial conditions can be
related, mathematically, to the study of gravitational collapse in
general relativity. 

\item[(3)]
The Planck scale does not provide any fundamental impenetrable
barrier, which would  limit our direct experimental information on
what happened before. On the contrary, the phenomenological imprint of
the pre-Planckian epoch can be encoded into a rich spectrum of
observable relics, reaching us today directly out of the pre-big bang
Universe. Pre-big bang physics can thus be the object of dedicated
experimental searches, which will be able to tell us, hopefully in a not 
too far future, whether or not there are chances for present
string-cosmology models to provide a successful description of our
primordial Universe.  

\end{itemize}

\vskip 5mm

\section*{Acknowledgements}
\addcontentsline{toc}{section}{Acknowledgements}

It is a pleasure to thank all our collaborators during more than a decade
of fun with string cosmology. In alphabetical order, they are:  Valerio
Bozza,  Ram Brustein, Alessandra Buonanno,  Cyril Cartier,  Edmund
Copeland, Thibault Damour,  Giuseppe De Risi,  Ruth Durrer,  Massimo
Giovannini, Amit Ghosh, Richard Madden, Michele Maggiore,  Jnan
Maharana, Krzysztov Meissner, Alessandro Melchiorri,  Slava
Mukhanov,  Federico Piazza, Giuseppe Pollifrone, Roberto Ricci, Mairi
Sakellariadou, Norma Sanchez, Carlo Ungarelli and Filippo Vernizzi. We
are also indebted to Suzy Vascotto for going through the pain of
proof-reading a preliminary version of this long paper.


\newpage 

\addcontentsline{toc}{section}{References}

\begin{thebibliography}{999}
\newcommand{\bb}{\bibitem}

\bb{Abb84}R. B. Abbott, S. 
M. Barr and S. D. Ellis, Phys. Rev.  {\bf D30}, 720 (1984). 

\bibitem{Abb86}R. B. Abbott, B. Bednarz and S. D. Ellis, Phys. Rev.
{\bf D33}, 2147 (1986).

\bb{AbbWi84}L.F. Abbot and M. B. Wise, Nucl. Phys. {\bf B244}, 541
(1984). 

\bb{Abramo92}A. Abramovici et al., Science {\bf 256}, 325 (1992).

\bb{AbSte}M. Abramowitz and  I. A. Stegun, {\em Handbook of
mathematical functions} (Dover, New York, 1972). 

\bb{AD85}I. Affleck and M. Dine, Nucl. Phys. {\bf B249}, 361 (1985). 

\bb{Ahn1}E. J. Ahn, M. Cavagli\`a and A. Olinto, hep-th/0201042.

\bb{Ahn2}E. J. Ahn and M. Cavagli\`a, hep-ph/0205168. 

\bb{AkaTera83}K. Akama and H. Terazawa, Gen. Relat. Gravit. {\bf 15}, 
201 (1983). 

\bb{ABE00}S. Alexander, R. Brandenberger and D. Easson,  {Phys. Rev.}
{\bf D62}, 103509 (2000).

\bb{ATU99}S. O. Alexeyev, A. V. Toporensky and V. O. Ustiansky, 
Class. Quant. Grav. {\bf 17}, 2243 (2000).  

\bb{Al88}B. Allen, {Phys. Rev.} {\bf D37}, 2078 (1988).

\bb{Al97}B. Allen, in {\em  
Astrophysical Sources of Gravitational waves}, Proc. Les Houches School,
1996, eds. J. A. Marck and J. P. Lasota ( University Press, Cambridge,
1997), p. 373.

\bb{AlBru97}B. Allen and R. Brustein, Phys. Rev. {\bf D55}, 3260 (1997). 

\bb{AlPapa99}B. Allen, E. E. Flanagan and M. A. Papa,   Phys. Rev. {\bf
D61}, 024024 (2000). 

\bb{Al94}B. Allen and S. Koranda, Phys. Rev. {\bf D50}, 3713 
(1994). 

\bb{AlRo99} B. Allen and J. D. Romano, Phys. Rev. {\bf D59}, 102001
(1999).

\bb{Alvarez85}E. Alvarez, Phys. Rev. {\bf D31}, 418 
(1985).

\bb{AlCon01}E. Alvarez and J. Conde, Mod. Phys. Lett. {\bf A17}, 413 
(2002).

\bb{AlvaOs89}E. Alvarez and M. Osorio, Int. J. Theor. Phys. {\bf 28}, 949
(1989). 

\bb{Ama91}U. Amaldi, W. de Boer and H. Furstenau, Phys. Lett. {\bf
B260}, 447 (1991). 

\bb{ACV1}D. Amati, M. Ciafaloni and G. Veneziano, Int. J. Mod. Phys. 
{\bf A3}, 1615 (1988). 

\bb{ACV2}D. Amati, M. Ciafaloni and G. Veneziano, Phys. Lett. 
{\bf B216}, 41 (1989). 

\bb{Amen00}L. Amendola, Phys. Rev. D {\bf 62}, 043511 (2000).

\bb{Am90}L. Amendola, M. Litterio and F. Occhionero,  Phys. Lett. {\bf
B237}, 348 (1990). 

\bb{AmenToc01}L. Amendola and D. Tocchini-Valentini,  Phys. Rev. D {\bf
64}, 043509 (2001).

\bb{AmenToc01a}L. Amendola and D. Tocchini-Valentini, 
 Phys. Rev. {\bf D65}, 063508 (2002). 

\bb{Ander94}S. Anderegg and V. N. Mukhanov, Phys.
Lett. {\bf B331}, 30 (1994). 

\bb{Angel95}C. Angelantonj, L. Amendola, M. Litterio and F. Occhionero, 
Phys. Rev. {\bf D51}, 1607 (1995).

\bb{Anto99}I. Antoniadis, in {\em 
 Gravitational Waves and Experimental 
Gravity}, Proc. XXXIV Rencontres de Moriond, Les Arcs, 1999,
eds. J. Tran Thanh Van et al. (World Scientific, Singapore, 2000). 

\bb{Anto98}I. Antoniadis,  N. Arkani-Hamed, S. Dimopoulos and G. Dvali, 
 Phys. Lett. {\bf B436}, 257 (1998). 

\bb{AnGav92}I. Antoniadis, E. Gava and K. S. Narain, Phys. Lett. {\bf
B283}, 209 (1992). 

\bb{AnGav92a}I. Antoniadis, E. Gava and K. S. Narain, Nucl. Phys. {\bf
B393}, 93 (1992). 

\bibitem{ART94}I. Antoniadis, J. Rizos, and K. Tamvakis,
Nucl. Phys. {\bf B415}, 497  (1994).

\bb{Arkani98}N. Arkani-Hamed, S. Dimopoulos and G. Dvali,  Phys. Lett.
{\bf B429}, 263 (1998). 

\bb{Ashte74}A. Ashtekar and R. Geroch, Rep. Prog. Phys. {\bf 37}, 1211
(1974). 

\bb{Asto98}P. Astone,  in Proc. {\em Second Edoardo Amaldi Conference
on Gravitational Waves}, CERN,  1997, eds. E. Coccia et al.  (World
Scientific, Singapore, 1998), p. 192.

\bb{Asto99}P. Astone et al.,  Astron. Astrophys. {\bf 351}, 811 (1999). 

\bb{Asto98a}P. Astone et al.,  in Proc. {\em Second Edoardo Amaldi 
Conference on Gravitational Waves}, CERN,  1997, eds. E. Coccia et al. 
(World Scientific, Singapore, 1998), p. 551.

\bb{Asto94}P. Astone, J. Lobo and B. Schutz, Classical Quant. Grav. {\bf
11}, 2093 (1994). 

\bb{Asto97a}P. Astone, G. Pallottino and G. Pizzella, Classical Quant. Grav.
{\bf 14}, 2019 (1997). 

\bb{Ax83}M. Axenides, R. H. Brandenberger and M. S. Turner,
Phys. Lett.  {\bf B126}, 178 (1983). 

\bb{BaBa01}D. Babusci, L. Baiotti, F. Fucito and A. Nagar,
Phys. Rev. {\bf D64}, 062001 (2001).
 
\bb{BaFoLosurdo}D. Babusci, S. Foffa, G. Losurdo, M. Maggiore, G.
Matone and R. Sturani, in  {\em Gravitational Waves}, Proc. SIGRAV
School, Centre A. Volta, Como, 1999,  eds. I. Ciufolini et al. (IOP Publishing,
Bristol, 2001), p. 179.

\bb{BabGio99}D. Babusci and M. Giovannini,  Int. J. Mod. Phys. {\bf D10},
477 (2001). 

\bb{BabGio99a}D. Babusci and M. Giovannini,   Phys. Rev. {\bf D60}, 
083511 (1999). 

\bb{BabGio99b}D. Babusci and M. Giovannini,  Classical Quant. Grav.  {\bf
17},  2621 (2000). 

\bibitem{Rey99}D. Bak  and S. Rey, Classical Quant. Grav. {\bf 17}, L1 
(2000).

\bb{Balbi01}A. Balbi, C. Baccigalupi, S. Matarrese, F. Perrotta and N.
Vittorio, Astrophys. J. {\bf 547}, L89 (2001). 

\bb{Banday97}A. J. Banday et al., Astrophys. J. {\bf 475}, 393 (1997).

\bb{BaFiMo99}T. Banks, W. Fischler and L. Motl, JHEP {\bf 9901}, 019
(1999). 

\bb{Banks93}T. Banks, D. B. Kaplan and A. E. Nelson, Phys.
Rev. {\bf D49}, 779 (1994).

\bb{Bar80}J. Bardeen, Phys. Rev. {\bf D22}, 1822 (1980).

\bb{Barkana96}R. Bar-Kana, Phys. Rev. {\bf D64}, 7138 (1996).

\bb{Bar86}J. D. Barrow, Phys. Lett. {\bf B180}, 335 (1986). 

\bb{Barrow93}J. D. Barrow, Phys. Rev. {\bf D48}, 3592 (1993). 

\bb{BaDa97}J. D. Barrow and M. P. Dabrowski, Phys. Rev. {\bf D55}, 630
(1997). 

\bb{BaDa98}J. D. Barrow and M. P. Dabrowski, Phys. Rev. {\bf D57}, 7204 
(1998). 

\bb{BaDa98a}J. D. Barrow and M. P. Dabrowski, Phys. Rev. {\bf D58},
103502  (1998). 

\bb{BaKu97}J. D. Barrow and K. E. Kunze, Phys. Rev. {\bf D55}, 623 
(1997). 

\bb{BaKu97a}J. D. Barrow and K. E. Kunze, Phys. Rev. {\bf D56},
741 (1997). 

\bb{BaKu97b}J. D. Barrow and K. E. Kunze, Phys. Rev. {\bf D57},
2255 (1998). 

\bb{Bar98}J. D. Barrow and K. E. Kunze,   Chaos, Soliton Frac. {\bf
10}, 257 (1999). 

\bb{Bar93}J. D. Barrow, J. P. Mimoso and M. R. de Garcia Maia, 
Phys. Rev. {\bf D48}, 3630 (1993).

\bb{BaLi02}N. Bartolo and A. R. Liddle, Phys. Rev. {\bf D65}, 121301
(2002). 

\bb{Bass97}B. A. Bassett, Phys. Rev. D {\bf 56}, 3439 (1997).

\bb{Bassett00}B. A. Bassett, C. Gordon, R. Maartens and D. I. Kaiser, 
Phys. Rev. {\bf D61}, 061302 (2000).

\bb{Shell96}R. A. Battye and E. P. S. Shellard, Phys. Rev. {\bf D53}, 1811
(1996). 

\bb{BeFor93}K. Behrndt and S. Forste, Phys. Lett. {\bf B320}, 253 (1993). 

\bb{BeFoSchwa97}K. Behrndt, S. Forste and S. Schwager, Nucl.  Phys. {\bf
B508}, 391 (1997). 

\bb{Bek73}J. D.  Bekenstein, Phys. Rev. {\bf D7}, 2333 (1973).

\bibitem{Bek1}J. D.  Bekenstein, Phys. Rev. {\bf D23}, 287 (1981).

\bibitem{Bek3}J. D.  Bekenstein, {Int. J. Theor. Phys.}  {\bf 28}, 967 
(1989).

\bibitem{Bek2}J. D.  Bekenstein, Phys. Rev. {\bf D49}, 1912 (1994).

\bibitem{GE} V. A. Belinskii and I. M. Khalatnikov, Sov. 
Phys. JETP {\bf 36}, 591 (1973). 

\bibitem{BKL} V. A. Belinskii, I. M. Khalatnikov and E. M. Lifshitz, Adv.
Phys. {\bf 19}, 525 (1970). 

\bb{Cobe94}C. L. Bennet et al., Astrophys. J. {\bf 430}, 423 (1994).

\bb{Ben96}C. L. Bennet et al., Astrophys. J.  {\bf 464}, L1 (1996). 

\bb{Bento95}M. G. Bento and O. Bertolami, Classical Quant. Grav. {\bf
12}, 1919 (1995). 

\bb{Bernard98}P. Bernard, G. Gemme, R. Parodi and E. Picasso,
Rev. Sci. Instrum. {\bf 72}, 2428 (2001). 

\bb{Berto99}O. Bertolami and D. F. Mota, Phys. Lett. {\bf B455}, 96
(1999). 

\bb{BiaCo98}M. Bianchi, M. Brunetti, E. Coccia, F. Fucito and J. A. Lobo,
Phys. Rev. D {\bf 57}, 4525 (1998).

\bb{BiaCo96}M. Bianchi, E. Coccia, C. Colacino, V. Fafone and F. Fucito,
Classical Quant. Grav. {\bf 13},  2865 (1996).

\bb{BiCoLid1}A. P. Billyard, A. A. Coley and  J. E. Lidsey,
Phys. Rev. {\bf D59}, 123505 (1999). 

\bb{BiCoLid2}A. P. Billyard, A. A. Coley and  J. E. Lidsey, 
J. Math. Phys. {\bf 40}, 5092 (1999). 

\bb{BiCoLid3}A. P. Billyard, A. A. Coley and  J. E. Lidsey, 
Classical Quant. Grav. {\bf 17}, 453 (2000). 

\bb{BiCoLid99}A. P. Billyard, A. A. Coley, J. E. Lidsey and U. S. Nilsson,
Phys. Rev. {\bf D61}, 043504 (2000). 

\bb{Bine00} P. Bin\'etruy, C. Deffayet and D. Langlois, Nucl. Phys. {\bf
B565}, 269 (2000).

\bb{BG86}P. Bin\'etruy and M. K. Gaillard, Phys. Rev. {\bf D34}, 3069 
(1986).

\bb{Bir82}N. D. Birrel and P. C. W. Davies, {\em Quantum fields in curved
spaces} (University Press, Cambridge, 1982). 

\bb{BisMaPra98}A. K. Bisvas, J. Maharana and R. K. Pradhan, 
Phys. Lett. {\bf B462}, 243 (1999). 

\bb{BP91}H. J. Blome and W. Priester, Astron. Astrophys. {\bf 250}, 43
(1991). 

\bibitem{Bon62} H. Bondi, M. G. J. van der Burg  and A. W.  K. Metzner,
Proc. Roy. Soc. Lond. {\bf A269}, 21 (1962). 

\bb{BoGuVil01}A. Borde, A. Guth and A. Vilenkin, gr-qc/0110012. 

\bb{BoVil94}A. Borde and A. Vilenkin, Phys. Rev.  Lett. {\bf  72}, 3305 
(1994).

\bibitem{Bousso99}R. Bousso,  JHEP {\bf 9906}, 028 (1999).

\bibitem{Bousso99a}R. Bousso,  JHEP {\bf 9907}, 004 (1999).

\bb{Bowick87}M. J. Bowick, L. Smolin and L. C. R. Wijewardhana, 
Gen. Relat. Gravit. {\bf 19}, 113 (1987). 

\bibitem{BaGasVe01}V. Bozza, M. Gasperini and G. Veneziano, Nucl.
Phys.  {\bf B619}, 191 (2001). 

\bb{BGGV02}V. Bozza, M. Gasperini, M. Giovannini and G. Veneziano,
hep-ph/0206131. 

\bibitem{BV} V. Bozza and G. Veneziano, JHEP {\bf 0010}, 35 (2000).

\bibitem{Bran01}R. Brandenberger, APCTP Bulletin {\bf 6} (2001).

\bb{BEK01} R. H. Brandenberger, D. Easson and D. Kimberly,
Nucl. Phys.  {\bf B623}, 421 (2002). 

\bibitem{BraEasMai98} R. H. Brandenberger, R. Easther and J. Maia,  
JHEP {\bf 9808}, 007 (1998).

\bb{BF01}R. Brandenberger and F. Finelli, JHEP {\bf 0111}, 056 (2001).  

\bibitem{Bran92}
R. Brandenberger, V. F. Mukhanov 
and T. Prokopec,  Phys. Rev. Lett. {\bf 69}, 3606 (1992).  

\bibitem{Bran93a}
R. Brandenberger, V. F. Mukhanov 
and T. Prokopec,   Phys. Rev.  {\bf D48}, 2443 (1993).

\bibitem{BraMuSor93}
R. Brandenberger, V. F. Mukhanov 
and A. Sornborger,   Phys. Rev.  {\bf D48}, 1629 (1993).

\bb{BraVa89}R. Brandenberger and C. Vafa, Nucl. Phys. {\bf B316}, 391
(1989). 

\bibitem{BraLuOv01}M. Brandle, A. Lukas and B. A. Ovrut, 
Phys. Rev.  {\bf D63}, 026003 (2001).

\bb{BruCo98}M. Brunetti,  E. Coccia,  V. Fafone and F. Fucito,  Phys. Rev.
{\bf D59}, 44027 (1999).

\bb{BruEl92}M. Bruni, G. F. R. Ellis and P. K. S. Dunsby, Classical Quant. 
Grav. {\bf 9}, 921 (1992).

\bb{Bru98}R. Brustein, Chaos, Soliton Frac. {\bf 10}, 283 (1999). 

\bibitem{BruFoStu99}R. Brustein, S. Foffa and R. Sturani, Phys.
Lett. {\bf B471}, 352 (2000). 

\bibitem{BruMu95}R. Brustein, M. Gasperini, M. Giovannini,
V. Mukhanov and G. Veneziano, Phys. Rev. {\bf D51}, 6744 (1995).

\bibitem{BruGas95}R. Brustein, M. Gasperini, M. Giovannini and G.
Veneziano, Phys. Lett. {\bf B361}, 45 (1995).

\bb{BruGas97}R. Brustein, M. Gasperini and G. Veneziano, Phys.  Rev 
{\bf  D55}, 3882 (1997).

\bibitem{BruGas98} R. Brustein, M. Gasperini and G. Veneziano, Phys.
Lett. {\bf B431}, 277 (1998).

\bibitem{BH98}R. Brustein  and M. Hadad, Phys. Rev. {\bf D57},  725
(1998).

\bb{BruHa00}R. Brustein and M. Hadad, Phys. Lett. {\bf B477}, 263
(2000).  

\bibitem{BruMad97}R. Brustein  and R. Madden,  Phys. Lett. {\bf B410},
110 (1997).

\bibitem{BruMad98}R. Brustein  and R. Madden,  Phys. Rev. {\bf D57},
712 (1998).

\bibitem{BruMad99}R. Brustein  and R. Madden,  JHEP {\bf 9907}, 006
(1999). 

\bb{BrusOa99}R. Brustein and D. H. Oaknin, Phys. Rev. {\bf D60}, 023508 
(1999).

\bibitem{BS93}R. Brustein and P. J. Steinhardt, Phys. Lett. {\bf B302},
196    (1993).

\bibitem{BruVe94}R. Brustein and G. Veneziano, Phys. Lett. {\bf B329},
429 (1994).

\bibitem{BruVe99}R.  Brustein and G. Veneziano,
Phys. Rev. Lett. {\bf 84}, 5695 (2000). 

\bibitem{BuoDa01}A. Buonanno and T. Damour, 
Phys. Rev. {\bf D64},  043501 (2001). 

\bibitem{BDV99}A. Buonanno, T. Damour and G. Veneziano, 
Nucl. Phys. {\bf B543}, 275 (1999). 

\bb{BuoGas97}A. Buonanno, M. Gasperini, M. Maggiore and C. Ungarelli,
Classical Quant. Grav.{\bf 14}, L97 (1997). 

\bb{BGU97}A. Buonanno, M. Gasperini and C. Ungarelli,
Mod. Phys. Lett. {\bf A12}, 1883 (1997). 

\bb{Buo00}A. Buonanno, M. Lemoine and K. A. Olive, 
Phys. Rev. {\bf D62}, 083513 (2000). 

\bibitem{BMU97}A. Buonanno, M. Maggiore and C. Ungarelli, Phys. Rev.
{\bf D55}, 3330 (1997).

\bb{BMUV98b}
A. Buonanno, K. A. Meissner, C. Ungarelli  and G. Veneziano, Phys. Rev.
{\bf D57}, 2543 (1998).

\bibitem{BMUV98a}
A. Buonanno, K. A. Meissner, C. Ungarelli  and G. Veneziano,
JHEP {\bf 9801}, 004 (1998). 

\bb{Caia91}E. R. Caianiello, M. Gasperini and G. Scarpetta, Classical 
Quant.  Grav. {\bf 8}, 659 (1991).

\bibitem{Cald92} R. R. Caldwell and B. Allen, Phys. Rev. {\bf D45}, 3447
(1992). 

\bibitem{CalDaSte98} 
R. R. Caldwell, R. Dave and P. J. Steinhardt,  Phys. Rev. Lett. {\bf 80}, 
582 (1998).  

\bibitem{CGHS92}C. Callan, S. Giddings, J. Harvey and A. Strominger,
Phys. Rev. {\bf D45}, R1005 (1992).

\bb{Cal88}C. G. Callan, R. C. Myers and M. J. Perry, Nucl. Phys. {\bf B311},
673 (1988). 

\bb{CAO90}B. A. Campbell, A. Linde and K. A. Olive, 
Nucl. Phys. {\bf B355}, 146 (1991). 

\bb{CapDeRi93}S. Capozziello and R. De Ritis, Int. J. Mod. Phys.  {\bf D2},
367 (1993). 

\bb{CapDeRi93a}S. Capozziello and R. De Ritis,  Int. J. Mod. Phys.  
 {\bf D2}, 373 (1993). 

\bb{CaDu02}C. Caprini and R. Durrer, Phys. Rev. {\bf D65}, 023517 (2002). 

\bb{Caron97}B. Caron et al., Classical Quant. Grav. {\bf 14}, 1461 (1997). 

\bibitem{Carr75}B. J.  Carr, Astrophys. J. {\bf 201}, 1 (1975). 

\bibitem{Carr74}B. J.  Carr  and S. W. Hawking, Mon. Not. Roy. Astron.
Soc. {\bf 168}, 399 (1974).

\bibitem{CaCoGas} C. Cartier, E. J. Copeland and M. Gasperini,  Nucl.
Phys. {\bf B607}, 406 (2001). 

\bibitem{CarCopMad00} C. Cartier, E. J. Copeland and R. Madden,
JHEP {\bf 0001}, 035 (2000).   

\bibitem{CaHwang01} C. Cartier, J. C. Hwang and  E. J. Copeland,  
Phys. Rev.  {\bf D64}, 103504 (2001). 
       
\bb{Cav01}M. Cavagli\`a, Classical Quant. Grav. {\bf 18}, 1355 (2001). 

\bb{CaDe97}M. Cavagli\`a and V. De Alfaro, Gen. Relat. Gravit. {\bf 29}, 773
(1997). 

\bb{CaMo01}M. Cavagli\`a and P. V. Moniz, Classical Quant. Grav. {\bf 18},
95 (2001). 

\bb{CaUnga99}M. Cavagli\`a and C. Ungarelli, Classical Quant. Grav. {\bf
16}, 1401 (1999). 

\bb{CaUnga00}M. Cavagli\`a and C. Ungarelli, 
Nucl. Phys. Proc. Suppl. {\bf B88}, 355 (2000). 

\bb{CCLOZ00}M. Cerdonio, L. Conti, J.A. Lobo, A. Ortolan and J. P. Zendri,
Phys. Rev. Lett. {\bf 87},  031101 (2001). 

\bb{Cheng94}B. Cheng and A. V. Olinto, Phys. Rev. {\bf D50}, 2421 (1994). 

\bibitem{Chiba99}T. Chiba, Phys. Rev. {\bf D59}, 083508 (1999).

\bb{Chib82}G. V. Chibisov  and V. N. Mukhanov, Mon. Not. Roy.  Astron.
Soc. {\bf 200}, 535 (1982)

\bb{Chinca02}A. Chincarini, G. Gemme, R. Parodi, P. Bernard and E.
Picasso, gr-qc/0203024. 

\bibitem{Chp93}M. W.  Choptuik, Phys. Rev. Lett. {\bf 70}, 9 (1993).

\bb{Chris92}N. Christensen, Phys. Rev. {\bf D46}, 5250 (1992). 

\bibitem{Chr86}D. Christodoulou, Commun. Math. Phys. {\bf 105}, 337
(1986).

\bibitem{Chr87}D. Christodoulou, Commun. Math. Phys. {\bf 109}, 613
(1987). 

\bibitem{Chr91}D. Christodoulou, Commun. Pure Appl. Math. {\bf 44}, 339  
(1991).

\bibitem{Chr93}D. Christodoulou, Commun. Pure Appl. Math. {\bf 46},
11331 (1993).

\bb{ClaFe99}D. Clancy, A. Feinstein, J. E. Lidsey and R. Tavakol,
Phys. Rev. {\bf D60}, 043503 (1999).  

\bb{CFLT99}D. Clancy, A. Feinstein, J. E. Lidsey and R. Tavakol, Phys. Lett.
{\bf B451}, 303 (1999). 

\bb{ClaLiTa98}D. Clancy, J. E. Lidsey and R. Tavakol,  Classical Quant. 
Grav. {\bf 15}, 257 (1998). 

\bb{CLT98}D. Clancy, J. E. Lidsey and R. Tavakol,  Phys. Rev.
{\bf D58}, 44017 (1998). 

\bb{CLT99}D. Clancy,  J. E. Lidsey and R. Tavakol, Phys. Rev.
{\bf D59}, 063511(1999). 

\bb{Coccia98}E. Coccia, V. Fafone, G. Frossati, J. Lobo and J. Ortega,
Phys. Rev. {\bf D57}, 2051 (1998). 

\bb{CoFaMo92}E. Coccia, V. Fafone and I. Modena, Rev. Sci. Instrum. {\bf
63}, 5432 (1992). 

\bb{CoGas02}E. Coccia, M. Gasperini and C. Ungarelli, 
Phys. Rev. {\bf D65}, 067101 (2002). 

\bb{Cop97}E. J. Copeland, R. Easther and D. Wands, Phys. Rev. 
 {\bf D56}, 874 (1997). 

\bb{Cop88}E. J. Copeland, D. Haws, T. W. Kibble, D. Mitchell and N. Turok, 
Nucl. Phys. {\bf B298}, 445 (1988). 

\bb{Cop94}E. J. Copeland, A. Lahiri and D. Wands, Phys. Rev. 
 {\bf D50}, 4868 (1994). 

\bb{Cop95}E. J. Copeland, A. Lahiri and D. Wands, Phys. Rev. 
 {\bf D51}, 1569 (1995). 

\bb{CoLiLi01}E. J. Copeland, A. R. Liddle and J. E. Lidsey, 
Phys. Rev. {\bf D64}, 023509 (2001).

\bb{Cop98} E. J. Copeland, A. R. Liddle, J. E. Lidsey and D. Wands,  Phys.
Rev.  {\bf D58}, 63508 (1998)

\bb{Cop98a} E. J. Copeland, A. R. Liddle, J. E. Lidsey and D. Wands,  Gen.
Relat. Gravit. {\bf 30}, 1711 (1998). 

\bb{Cop97a} E. J. Copeland, J. E. Lidsey and D. Wands,
Nucl. Phys. {\bf B506}, 407 (1997).

\bb{CoLiWa98} E. J. Copeland, J. E. Lidsey and D. Wands, Phys. Rev. {\bf
D57}, 625 (1998). 

\bb{Cop98b} E. J. Copeland, J. E. Lidsey and D. Wands, Phys. Lett. {\bf
B443}, 97 (1998). 

\bb{CopSaf00}E. J. Copeland, P. M. Saffin and O. Tornkvist, Phys.
Rev.  {\bf D61}, 105005 (2000). 

\bb{CJS78}E. Cremmer, B. Julia and J. Scherk, Phys. Lett. {\bf B76}, 409
(1978). 

\bb{Cruise00}A. M. Cruise, Classical Quant. Grav. {\bf 17}, 2525 (2000). 

\bb{Cuoco98}E. Cuoco, G. Curci and E. Beccaria, in Proc.  {\em Second
Edoardo Amaldi  Conference on Gravitational Waves}, CERN,  1997, eds. E.
Coccia et al.  (World Scientific, Singapore, 1998), p. 524. 

\bb{DaKie97}M. Dabrowski and C. Kiefer, Phys. Lett. {\bf B397}, 185
(1997). 

\bibitem{DH00/1}T. Damour and M. Henneaux, Phys. Rev. Lett. {\bf 85},
920  (2000). 

\bibitem{DH00/2} T. Damour and M. Henneaux, Gen. Relat. Gravit.  {\bf
32},  2339  (2000). 

\bibitem{DH00/3} T. Damour and M. Henneaux, Phys. Lett. {\bf B488}, 108
(2000). 

\bibitem{DH01} T. Damour and M. Henneaux, Phys. Rev. Lett. {\bf 86}, 
4749  (2001). 

\bibitem{DHJN01} T. Damour, M. Henneaux, B. Julia and H. Nicolai,
 Phys.  Lett. {\bf B509}, 323 (2001). 

\bb{DaPiaVe02} T. Damour, F. Piazza and G. Veneziano, gr-qc/0204094. 

\bb{DaPiaVe02a} T. Damour, F. Piazza and G. Veneziano,
hep-th/0205111. 

\bibitem{DP94a}T.  Damour  and A. M. Polyakov, Nucl. Phys. {\bf B423},
532 (1994). 

\bibitem{DP94b}T.  Damour and A. M. Polyakov, Gen. Relat. Gravit. {\bf 26},
1171 (1994).

\bibitem{DaVi96}T. Damour and A. Vilenkin, Phys. Rev. {\bf D53}, 2981 
(1996).

\bb{DaVi96a}T. Damour and A. Vilenkin, Phys. Rev. Lett. {\bf 78}, 2288 
(1997).

\bb{Davis00}A. Davis, K. Dimopoulos, T. Prokopec and O. Tornkvist,
 Phys. Lett.  {\bf B501}, 165 (2001). 

\bb{DeAlvis92}S. P. de Alwis, Phys. Lett. {\bf B289}, 278 (1992).

\bb{Debe00}P. de Bernardis et al.,  Nature {\bf 404}, 955 (2000). 

\bb{Debe02} P. de Bernardis et al.,  Astrophys. J. {\bf 564}, 559 (2002).

\bb{DeCa93}B. De Carlos, J. A. Casas, F. Quevedo and E. Roulet, Phys.
Lett. {\bf B318}, 447 (1993).

\bb{Ossa93}X. C. de la Ossa and F. Quevedo, Nucl. Phys. {\bf B403}, 377
(1993). 

\bb{Dem90}M. Demianski, in Proc. {\em It. Conf. on General
Relativity and Gravitational Physics}, Capri, 1990, eds. R. Cianci et al.
(World Scientific, Singapore, 1991), p. 19. 

\bb{Dem93}M. Demianski, A. G. Polnarev and P. Naselski, Phys. Rev. {\bf
D47}, 5275 (1993). 

\bb{DFKZ91}J. P. Derendinger, S. Ferrara, C. Kounnas and F. Zwirner, 
Phys. Lett. {\bf B271}, 307 (1991).

\bb{Tesi00}G. De Risi and M. Gasperini, Phys. Lett. {\bf B503}, 140 (2001).

\bb{DeRi01}G. De Risi and M. Gasperini, Phys. Lett. {\bf B521}, 335 (2001).

\bb{Deruelle92}N. Deruelle, C. Gundlach and D. Polarski, Classical Quant. 
Grav. {\bf 9}, 137 91992).

\bb{Deruelle92a}N. Deruelle, C. Gundlach and  D. Langlois, Phys. Rev. {\bf
D45}, R3301 (1992). 

\bibitem{DeMu95}N. Deruelle and V. F. Mukhanov, Phys. Rev. {\bf
D52}, 5549 (1995). 

\bb{VLS00}H. J. de Vega, A. L. Larsen and N. Sanchez, Phys. Rev. {\bf
D61}, 066003 (2000). 

\bibitem{Witt67}B. S. De Witt,  Phys. Rev. {\bf 160}, 1113 (1967). 

\bibitem{Dimo88}S. Dimopoulos et al., Astrophys. J. {\bf 330}, 545 (1988).

\bb{Dol93}A. D. Dolgov, Phys. Rev. {\bf D48}, 2499 (1993).

\bibitem{noncomm}M. R. Douglas, in {\em Nonperturbative Aspects of
Strings, Branes and Supersymmetry}, Proc. ICTP Spring School, Trieste,
1998  (World Scientific, Singapore, 1999), p. 131. 

\bb{Dur90}R. Durrer,  Phys. Rev. {\bf D42}, 2533 (1990).

\bb{Dur94}R. Durrer, Fundams.  Cosm. Phys., {\bf 15}, 209 (1994).

\bb{Dur01}R. Durrer, J. Phys. Stud. {\bf 5}, 177 (2001). 

\bb{DurGas98}R. Durrer, M. Gasperini, M. Sakellariadou and G.
Veneziano, Phys. Lett. {\bf B436}, 66 (1998).

\bb{DurGas99}R. Durrer, M. Gasperini, M. Sakellariadou and G.
Veneziano, Phys. Rev. {\bf D59}, 43511 (1999). 

\bb{DurLau96}R. Durrer and J. Laukenmann, Classical Quant. Grav. {\bf 13},
1069  (1996).

\bb{DurSak00}R. Durrer and M. Sakellariadou, 
Phys. Rev.  {\bf D62},  123504 (2000).

\bb{DuVe02}R. Durrer and F. Vernizzi, hep-ph/0203275. 

\bibitem{Giddings} D. M. Eardley and S. B. Giddings, gr-qc/0201034. 

\bb{Ea01}D. Easson, hep-th/0111055, in Proc. of the 
{\em COSMO-01 Workshop}, Rovaniemi,  2001.

\bb{EaBra99}D. A. Easson and R. Brandenberger, JHEP {\bf 9909}, 003
(1999). 

\bibitem{EL99}R. Easther  and D. A. Lowe,  Phys. Rev. Lett. {\bf 82}, 
4967 (1999).

\bibitem{EasMae96}
R. Easther and K. Maeda, Phys. Rev. {\bf D54}, 7572
(1996). 

\bibitem{Eas96}R. Easther, K. Maeda and D. Wands,   Phys. Rev.
 {\bf D53}, 4247 (1996).

\bb{ElBru89}G. F. R. Ellis and M. Bruni, Phys. Rev. {\bf D40}, 1804 
(1989). 

\bb{ERSD99}G. F. R. Ellis, D. C. Roberts, D. Solomons and P. K. S. Dunsby, 
Phys. Rev. {\bf D62}, 084004 (2000). 

\bb{EENQ86}J. Ellis, K. Enqvist, D. V. Nanopoulos and M. Quiros, Nucl.
Phys.  {\bf B277}, 231 (1986). 

\bb{Ellis89}J. Ellis, S. Kalara, K. A. Olive and C. Wetterich,  Phys.
Lett. {\bf B228}, 264 (1989).

\bb{ElNa86}J. Ellis, D. V. Nanopoulos and M. Quiros, Phys. Lett. {\bf
B174}, 176 (1986). 

\bb{ElNaSa85}J. Ellis, D. V. Nanopoulos and S. Sarkar, Nucl. Phys. {\bf
B259}, 175 (1985).

\bb{ElTsa86}J. Ellis, C. Tsamish and M. Voloshin, Phys. Lett. {\bf
B194}, 291 (1987). 

\bb{EnSlo01}K. Enqvist and M. S. Sloth, Nucl. Phys. {\bf B626}, 395
(2002). 

\bb{Tricomi}A. Erdelyi, W. Magnus, F. Oberhettinger and F. G. Tricomi, 
{\em Higher transcendental functions} (MacGraw-Hill, New York, 1955),
Vol. 3. 

\bb{Fab83}R. Fabbri and M. Pollock, Phys. Lett. {\bf B125}, 445 (1983).

\bb{Fen00}A. Feinstein, gr-qc/0106001, in {\em Fourth Mexican School
on Gravity and Mathematical Physics}, Huatulco,  2000. 

\bibitem{FKV} A. Feinstein, K. E. Kunze and M. A. V{\'a}zquez--Mozo,
Classical Quant. Grav. {\bf 17}, 3599 (2000). 

\bibitem{FeKuVa00} A. Feinstein, K. E. Kunze and M. A.
V{\'a}zquez--Mozo, Phys. Lett. {\bf  B491}, 190 (2000). 

\bb{FLV97}A. Feinstein, R. Lazkoz and M. A. V{\'a}zquez--Mozo,
Phys. Rev. {\bf D56}, 5166 (1997).

\bb{FeVa98}A. Feinstein and M. A. V{\'a}zquez--Mozo,  Phys. Lett. 
{\bf B441}, 40 (1998). 

\bb{FeVa00}A. Feinstein and M. A. V{\'a}zquez--Mozo, Nucl. Phys.
{\bf B568}, 405 (2000).

\bibitem{FKZ94}S. Ferrara, C. Kounnas and F. Zwirner, Nucl. Phys.
{\bf B429}, 589 (1994).

\bb{FiBra01}F. Finelli and R. Brandenberger, 
Phys. Rev. {\bf D65}, 103522 (2002).

\bb{Finelli00}F. Finelli and A. Gruppuso, 
Phys. Lett. {\bf B502}, 216 (2001).

\bb{Fis92} E. Fischbach and C. Talmadge, Nature {\bf 356}, 207 
(1992). 

\bibitem{FS98}W. Fischler and L. Susskind, hep-th/9806039.

\bb{Fla93}E. Flanagan, Phys. Rev. {\bf D48}, 2389 
(1993). 

\bibitem{Fla99}E. E. Flanagan, D. Marolf  and R. M. Wald, 
 Phys. Rev.  {\bf D62},  084035 (2000).

\bibitem{FoMa98} S. Foffa and M. Maggiore,  Phys. Rev.  {\bf D58},
023505 (1998). 

\bibitem{FoMaStu99} S. Foffa, M. Maggiore and R. Sturani, 
Phys. Rev. {\bf D59}, 043507 (1999). 

\bibitem{FofMagStu99} S. Foffa, M. Maggiore and R. Sturani, 
Nucl. Phys. {\bf B552}, 395 (1999).

\bb{Ford77}L. H. Ford and L. Parker, Phys. Rev. {\bf D16}, 1601 
(1977). 

\bb{Gailla95}M. Gaillard, H. Murayama and K. Olive, Phys. Lett. {\bf B355},
71 (1995).  

\bb{Occhio97}M. Galluccio, F. Litterio and F. 
Occhionero, Phys. Rev. Lett. {\bf 79}, 970 (1997). 

\bb{Garre92}W. D. Garretson, G. B. Field and S. M. Carrol, Phys. Rev.
{\bf D46}, 5346 (1992);

\bb{Gar89}J. Garriga and E. Verdaguer,  Phys. Rev. {\bf D39}, 1072
(1989). 

\bb{GasWeb}M. Gasperini's web site at
{\tt http://www.ba.infn.it/\~{}gasperin}

\bb{Gas85}M. Gasperini, Phys. Lett. {\bf B163}, 84 (1985). 

\bb{Gas86}M. Gasperini, Phys. Rev. Lett. {\bf 56}, 2873 (1986). 

\bb{MG91}M. Gasperini, Phys. Lett. {\bf B258}, 70 (1991). 

\bb{Gas91}M. Gasperini, in Proc. {\em Advances in Theoretical 
Physics}, Vietri, 1990, ed. E. R. Caianiello (World Scientific,
Singapore 1991), p. 77.

\bb{MG92}M. Gasperini, Gen. Relat. Gravit. {\bf 24}, 219 (1992). 

\bb{Gas94b}M. Gasperini, Phys. Lett. {\bf B327}, 314 (1994). 

\bb{Gas94}M. Gasperini, in  Proc. {\em Second Paris Cosmology
Colloquium}, Observatoire de Paris, 1994, eds. H. J. de Vega  and N.
Sanchez (World Scientific, Singapore, 1995), p. 429.

\bb{Gas94a}M. Gasperini, in  Proc. {\em Birth of the Universe and
Fundamental Physics}, Rome, 1994, ed.  F. Occhionero (Springer,
New York, 1995),  p. 71.

\bb{Gas96}M. Gasperini, in Proc. {\em Third Paris 
Cosmology Colloquium}, Observatoire de Paris,  1995, eds. H. J. de
Vega  and N. Sanchez (World Scientific, Singapore, 1996), p. 253.

\bb{Gas95}M. Gasperini, 
in {\em String Gravity and Physics at the Planck Energy Scale}, Proc. 
Erice School,  1995,
eds. N. Sanchez and A. Zichichi (Kluwer Acad. Publ., Dordrecht, 1996),
p. 305.

\bb{Gas97a}M. Gasperini, in Proc. {\em It. Conf. on 
General Relativity and Gravitational Physics}, Rome, 1996,
eds. M. Bassan et al. (World Scientific, Singapore, 1997), p. 181. 

\bb{Gas97}M. Gasperini, Phys. Rev.  {\bf D56}, 4815 (1997).

\bb{Gas98b}M. Gasperini, in {\em String theory in curved space times},
ed. N. Sanchez (World Scientific, Singapore 1998), p. 333.

\bb{Gas98d}M. Gasperini, in Proc. {\em Fourth Paris
Cosmology Colloquium}, Observatoire de Paris, 1997, eds. H. J. de
Vega  and N. Sanchez (World Scientific, Singapore, 1998), p. 85.

\bb{Gas98c}M. Gasperini, in  Proc. {\em Second Edoardo Amaldi
Conference on Gravitational Waves}, CERN, 1997, eds. E. Coccia et
al.  (World Scientific, Singapore, 1998), p. 62.

\bb{Gas98a}M. Gasperini, Gen. Relat. Gravit. {\bf 30}, 1703 
(1998).

\bb{Gas98}M. Gasperini, Int. J. Mod. Phys. {\bf A13}, 4779 (1998).

\bb{Gas99b}M. Gasperini, in  Proc. {\em Fifth Paris
Cosmology Colloquium}, Observatoire de Paris, 1998, eds. H. J. de
Vega  and N. Sanchez (Publications Observatoire de Paris, 1999), p. 317.

\bb{Gas99}M. Gasperini, Mod. Phys. Lett. {\bf A14}, 1059 (1999).

\bibitem {Gas99z}M. Gasperini, Phys. Lett. {\bf B470}, 67 (1999).

\bb{Gas00}M. Gasperini, Phys. Rev. {\bf D61}, 87301 (2000).

\bibitem {Gas00b}M. Gasperini, Phys. Lett. {\bf B477}, 242 (2000).

\bibitem {Gas00a}M. Gasperini, Classical Quant. Grav. {\bf 17}, R1
(2000). 

\bibitem {Gas99a}M. Gasperini, in  {\em Gravitational Waves},
Proc. SIGRAV School, Centre A. Volta, Como, 1999),  eds. I. Ciufolini et
al. (IOP Publishing, Bristol, 2001), p. 280.

\bb{Gas00d}M. Gasperini, Int. J. Mod. Phys. {\bf D10}, 15 (2001).

\bb{Gas00c}M. Gasperini, 
Phys. Rev.  {\bf D63},  047301 (2001).

\bb{Gas00e}M. Gasperini, in Proc.  {\em  IX Marcel Grossmann
Meeting}, Rome,  2000,  eds. V. G. Gurzadyan, R. Jantzen and R.
Ruffini (World Scientific, Singapore, 2002), in press. 

\bb{Gas01}M. Gasperini, Phys. Rev.  {\bf D64}, 043510  
(2001). 

\bibitem{GG92z}M. Gasperini and M. Giovannini, Phys. Lett. {\bf B287},
56 (1992). 

\bibitem{GG92}M. Gasperini and M. Giovannini, Phys. Lett. {\bf B282},
36 (1992). 

\bb{GG92a}M. Gasperini and M. Giovannini, Classical Quant. Grav. {\bf 9},
L137 (1992).

\bibitem{GG93a}M. Gasperini and M. Giovannini, Phys. Lett. {\bf B301},
334 (1993). 

\bibitem{GG93b}M. Gasperini and M. Giovannini,
Classical Quant. Grav. {\bf 10}, L133 (1993).

\bb{GG93}M. Gasperini and M. Giovannini, Phys. Rev.  {\bf D47}, 1519
(1993). 

\bb{GG97}M. Gasperini and M. Giovannini,  Classical Quant. 
Grav. {\bf 14}, 735 (1997).

\bb{GG98}M. Gasperini and M. Giovannini,  in {\em String theory in
curved space times}, ed. N. Sanchez (World Scientific, Singapore, 1998),
p. 249.

\bibitem{GGMV96} M. Gasperini, M. Giovannini, K. A. Meissner and G.
Veneziano, Nucl. Phys. (Proc. Suppl.) {\bf B49}, 70 (1996). 

\bb{GGMV98}M. Gasperini, M. Giovannini, K. A. Meissner and G.
Veneziano,  in {\em String theory in curved space times}, ed. N. Sanchez
(World Scientific, Singapore, 1998), p. 49.


\bb{GGV93}M. Gasperini, M. Giovannini and G. Veneziano, Phys. Rev. 
{\bf D48}, R439 (1993).

\bb{GGV95}M. Gasperini, M. Giovannini and G. Veneziano, Phys. Rev. Lett. 
{\bf 75}, 3796 (1995).

\bb{GGV95a}M. Gasperini, M. Giovannini and G. Veneziano, Phys. Rev. 
{\bf D52}, 6651 (1995).

\bibitem{GasMaVe97}M. Gasperini, M. Maggiore,
 and G. Veneziano, Nucl. Phys. {\bf B494}, 315  (1997).

\bibitem{GMV91}M.  Gasperini, J. Maharana and G. Veneziano, Phys.  
Lett. {\bf B272}, 277 (1991).

\bibitem{GMV92}M.  Gasperini, J. Maharana and G. Veneziano, Phys.  
Lett. {\bf B296}, 51 (1992).

\bibitem{GasMaVe96}
M. Gasperini, J. Maharana, and G. Veneziano, Nucl. Phys. 
{\bf B472}, 349  (1996). 

\bibitem{GPV01}
M. Gasperini, F. Piazza and G. Veneziano,   Phys. Rev. {\bf D65}, 023508
(2001).

\bb{GasRi95}M. Gasperini and R. Ricci, Classical Quant. Grav. {\bf 12}, 677
(1995). 

\bibitem{GRV93}M. Gasperini, R. Ricci and G. Veneziano, Phys. Lett. {\bf
B319}, 438 (1993).

\bb{GSV91a}M. Gasperini, N. Sanchez and G. Veneziano, Int. J. Mod.
Phys. {\bf A6}, 3853 (1991)

\bb{GSV91}M. Gasperini, N. Sanchez and G. Veneziano, Nucl. Phys. B 
{\bf 364}, 365 (1991).

\bb{GasUm91}M. Gasperini and G. Ummarino, Phys. Lett. {\bf
B266}, 275 (1991).

\bibitem{GasUn01} M. Gasperini and C. Ungarelli, 
Phys. Rev. {\bf D64}, 064009 (2001).

\bibitem{GasVe92}M. Gasperini and G. Veneziano, Phys. Lett. {\bf B277},
256 (1992).

\bibitem{GasVe93a}M. Gasperini and G. Veneziano, Astropart. Phys. 
{\bf 1}, 317 (1993).

\bibitem {GasVe93b}M. Gasperini and G. Veneziano, Mod. Phys. Lett. 
{\bf A8}, 3701 (1993). 

\bibitem{GasVe94a} M. Gasperini and G. Veneziano, 
Phys. Rev. {\bf D50}, 2519 (1994).

\bibitem{GasVe96a}M. Gasperini and G. Veneziano, Gen. Rel. Grav.
{\bf 28}, 1301 (1996). 

\bibitem {GasVe96b}M. Gasperini and G. Veneziano,  Phys. Lett. 
{\bf B387}, 715 (1996). 

\bb{GasVe99}M. Gasperini and G.
Veneziano, Phys. Rev. {\bf D59}, 43503 (1999). 

\bb{Gen01}U. Gen, A. Ishibashi and T. Tanaka, hep-th/0110286. 

\bb{Giov95}M. Giovannini, {\em Teoria quantistica delle perturbazioni in
cosmologia di stringa}, PhD Thesis (unpublished), University of Turin
(1995). 

\bibitem{Giov97}M. Giovannini, Phys. Rev. {\bf D55}, 595 (1997).

\bibitem{Giov97a}M. Giovannini, Phys. Rev. {\bf D56}, 631 (1997).

\bibitem{Giov97b}M. Giovannini, Phys. Rev. {\bf D56}, 3198 (1997).

\bibitem{Giov98}M. Giovannini, Phys. Rev. {\bf D57}, 7223 (1998).

\bibitem{Giov99a}M. Giovannini, Phys. Rev. {\bf D59}, 063503 (1999).

\bibitem{Giov99b}M. Giovannini, Phys. Rev. {\bf D59}, 083511 (1999).

\bibitem{Giov99c}M. Giovannini, Phys. Rev. {\bf D59}, 123518 (1999).

\bibitem{Giov99}M. Giovannini, Phys. Rev. {\bf D60}, 123511 (1999).

\bb{Giov00}M. Giovannini, Phys. Rev. {\bf D62}, 123505 (2000). 

\bb{Giov01}M. Giovannini, Phys. Rev. {\bf D64}, 064023 (2001). 

\bb{GioKu02}M. Giovannini, H. Kurki-Suonio and E. Sihvola,
astro-ph/0203430. 

\bb{GiovShap00}M. Giovannini and M. Shaposhnikov, 
 Phys. Rev. {\bf D62}, 103512 (2000). 

\bb{Giri00}G. Giribet and C. Simeone,  Mod. Phys. Lett. {\bf A16}, 19
(2001).

\bb{GiPoRa94}A. Giveon, M. Porrati and E. Rabinovici, Phys. Rep. {\bf 244},
77 (1994). 

\bb{Gleiser85}M. Gleiser and J. G. Taylor, Phys. Lett.  
{\bf B164}, 36 (1985).

\bb{Goncha84}A. S. Goncharov, A. D. Linde and M. I. Vysotsky, Phys. Lett.
{\bf B147}, 279 (1984).

\bibitem{GPV98}A. Ghosh, G. Pollifrone and G. Veneziano, Phys. Lett.  
{\bf B440}, 20 (1998).

\bibitem{Gosh00}A. Ghosh, R. Madden and G. Veneziano,
Nucl. Phys. {\bf B570}, 207 (2000).
 
\bb{Grasso00}D. Grasso and H. R. Rubinstein, Phys. Rep. {\bf 348},
163  (12001). 

\bb{GSW87}M. B. Green, J. Schwartz and E. Witten, {\em Superstring
theory} (University Press, Cambridge, 1987).

\bibitem{Greene99}B. Greene, {\em The Elegant Universe} (W. W. Norton 
and Co., New York \& London, 1999).

\bb{Gris75}L. P. Grishchuk, Sov. Phys. JETP {\bf 40}, 409 (1975). 

\bb{Gris88}L. P. Grishchuk, Sov. Phys. Usp. {\bf 31}, 490 (1988). 

\bb{Gris88a}L. P. Grishchuk, Sov. Sci. Rev. E. Astrophys. Space Phys. {\bf
7}, 267 (1988).

\bb{Gris92}L. P. Grishchuk, in {\em Squeezed states and uncertainty
relations}, eds. D. Han et al. (NASA Conf. Pub. N. 31353, 1992), p. 329. 

\bb{Gris92a}L. P. Grishchuk, in Proc.  {\em 6th Marcel Grossmann
Meeting}, Kyoto, 1991, ed. H. Sato (World Scientific, Singapore, 1992). 

\bb{Gris98}L. P. Grishchuk,  in {\em 
 Gravitational Waves and Experimental 
Gravity}, Proc. XXXIV Rencontres de Moriond, Les Arcs,  1999,
eds. J. Tran Thanh Van et al. (World Scientific, Singapore, 2000). 

\bb{GriSid89}L. P. Grishchuk and Y. V. Sidorov, Classical Quant. Grav. {\bf
6}, L161 (1989). 

\bb{GriSid90}L. P. Grishchuk and Y. V. Sidorov,  Phys. Rev. {\bf D42}, 3413 
(1990). 

\bb{GriSol91}L. P. Grishchuk and M. Solokhin, Phys. Rev. {\bf D43}, 2566
(1991). 

\bb{Guig88}M. McGuigan,  Phys. Rev. {\bf D38}, 3031 (1988).

\bb{Guig89}M. McGuigan,  Phys. Rev. {\bf D39}, 2229 (1989).

\bb{Guig90}M. McGuigan,  Phys. Rev. {\bf D41}, 418 (1990).

\bb{Gut81}A. Guth, Phys. Rev. {\bf D23}, 347 (1981). 

\bb{Ham98}W. Hamilton, in Proc. {\em Second Edoardo Amaldi Conference
on Gravitational Waves}, CERN,  1997, eds. E. Coccia et al.  (World
Scientific, Singapore, 1998), p. 115.

\bb{Bender98}A. Hammesfahr, Classical Quant. Grav. {\bf 18}, 4045
(2001). 

\bb{Han00}S. Hanany et al., Astrophys. J.  {\bf 545}, L5 (2000). 

\bb{Har70}E. R. Harrison, Phys. Rev. {\bf D1}, 2726 (1970). 

\bibitem{Harri73} E. R. Harrison, Phys. Rev. Lett. {\bf 30}, 188 (1973).

\bb{Hartle83}J. B. Hartle and S. W. Hawking, Phys. Rev.  {\bf D28}, 2960 
(1983). 

\bibitem{Has92} S. Hassan and A. Sen, Nucl. Phys. {\bf B375}, 103 (1992).

\bb{Haw66}S. W. Hawking, Astrophys. J. {\bf 145}, 544 (1966).

\bb{Haw75}S. W. Hawking, Commun. Math. Phys.  {\bf 43}, 199 (1975).

\bb{Haw84}S. W. Hawking, Nucl. Phys.  {\bf B239}, 257 (1984).

\bibitem{HawEl73}S. W.  Hawking  and G. F. R. Ellis, {\em The large scale
structure of spacetime} (University Press, Cambridge, 1973). 

\bibitem{HP70}S. W.  Hawking  and R. Penrose, Proc. Roy. Soc. Lond. {\bf
A314}, 529  (1970).

\bb{Wett00}A. Hebecker and C. Wetterich, 
Phys. Rev. Lett. {\bf 85}, 3339 (2000). 

\bb{Heng98}I. Heng, D. Blair, E. Ivanov and M. Tobar, 
 in  Proc. {\em Second Edoardo Amaldi Conference on
Gravitational Waves}, CERN, 1997,  eds. E. Coccia et al.  (World
Scientific, Singapore,  1998), p. 127.

\bibitem{Thooft93}G. 't Hooft,  in {\em Abdus Salam
Festschrift: A Collection of Talks}, eds. A. Ali , J. Ellis and
S. Randjbar-Daemi  (World Scientific, Singapore, 1993).  

\bibitem{Hor96}P. Horava and E. Witten, Nucl. Phys. {\bf B460}, 506
(1996). 

\bibitem{Hor96a}P. Horava and E. Witten, Nucl. Phys. {\bf B475}, 94 
(1996). 

\bb{Hotta97}K. Hotta, K. Kikkawa and H. Kunitomo, Prog. Theor. Phys.
{\bf 98}, 687 (1997). 

\bb{Hough98}J. Hough,  in  Proc. {\em  Second
Edoardo Amaldi Conference on Gravitational waves}, 
CERN, 1997, eds. E. Coccia et al. 
(World Scientific, Singapore,  1998), p. 97.

\bb{Hoyle01}C. D. Hoyle, U. Schmidt, B. R. Heckel, E. G. Adelberger, J. H.
Gundlach, D. J. Kapner and H. E. Swanson, Phys. Rev. Lett.  {\bf 86}, 1418
(2001). 

\bb{Hwang91}J. H. Hwang,  Astrophys. J. {\bf 375}, 443 (1991). 

\bb{Hwang98}J. H. Hwang,  Astropart. Phys. {\bf 8}, 201 (1998). 

\bb{HwaNo00}J. H. Hwang and H. Noh, Phys. Rev. {\bf D61}, 043511
(2000). 

\bb{HwaNo01}J. H. Hwang and H. Noh, Phys. Rev. {\bf D65}, 124010 
(2002). 

\bb{HwaVi91}J. H. Hwang and E. T. Vishniac,  Astrophys. J. {\bf 382}, 363
(1991). 

\bb{Israel66}W. Israel,  Nuovo Cimento {\bf B44}, 1 (1966). 

\bb{Tiga1}W. Johnson and S. Merkowitz, Phys. Rev. Lett. {\bf 70}, 2367
(1993). 

\bb{KaTo01}T. Kajita and Y. Totsuka, Rev. Mod. Phys. {\bf 73}, 85 (2001). 

\bb{KKL}R. Kallosh, L. Kofman and A. Linde, 
Phys. Rev. {\bf D64}, 123523 (2001).  

\bb{KKLT}R. Kallosh, L. Kofman, A. Linde  and A. Tseytlin,  
Phys. Rev. {\bf D64}, 123524 (2001). 

\bibitem{Kalo97}N. Kaloper,
Phys. Rev. {\bf D55}, 3394 (1997).

\bibitem{KaLin99}N. Kaloper and A. Linde,
Phys. Rev. {\bf D60}, 103509 (1999).

\bibitem{KaKoOl98}N. Kaloper, I. I. Kogan and K. A. Olive, Phys. Rev. {\bf
D57},  7340 (1998).

\bibitem{Lin99}N. Kaloper, A. Linde and R. Bousso, Phys. Rev. {\bf D59}, 
043508 (1999).

\bibitem{Kal95}N.  Kaloper,  R. Madden and K. A. Olive, Nucl. Phys. {\bf
B452}, 677 (1995).

\bibitem{Kal96}N.  
 Kaloper,  R. Madden  and K. A. Olive,  Phys. Lett. {\bf B371}, 34 (1996).

\bibitem{KalMeis97}N. Kaloper and K. A. Meissner,  Phys. Rev. {\bf D56},
7940 (1997).

\bibitem{KaOl93}N. Kaloper and K. A. Olive,  Astropart. Phys. {\bf 1}, 185 
(1993).

\bibitem{KaOl98}N. Kaloper and K. A. Olive,  Phys. Rev. {\bf D57}, 811
(1998).

\bb{KamJaf00}M. Kamionkowski and A. H. Jaffe, 
Int. J. Mod. Phys. {\bf A16}, 116 (2001).

\bb{KamKosSteb97}M. Kamionkowski, A. Kosowsky and A. Stebbins,
Phys. Rev. Lett. {\bf 78}, 2058 (1997). 

\bb{KRT98}P. Kanti, J. Rizos and K. Tamvakis, 
Phys. Rev. {\bf D59},  083512 (1999).

\bb{Kap85}V. Kaplunovsky, Phys. Rev. Lett. {\bf 55}, 1036 (1985). 

\bb{Kaspi94}V. Kaspi, J. Taylor and M. Ryba, Astrophys. J. {\bf 428}, 713
(1994).

\bb{Kawa97}K. Kawabe and the TAMA collaboration,  Classical Quant. 
Grav. {\bf 14}, 1477 (1997). 

\bb{KSS98}S. Kawai, M. Sakagami and J. Soda, Phys. Lett. {\bf  B437},
284  (1998). 

\bb{KS98}S. Kawai and J. Soda, Phys. Rev. {\bf D59},  063506 (1999).

\bb{KS99}S. Kawai and J. Soda,  Phys. Lett. {\bf B460}, 41 (1999). 

\bb{Kawa96}M. Kawasaki, N. Sugiyama and T. Tanagida, Phys Rev. {\bf
D54}, 	2442 (1996).

\bibitem{Ke96}A. A. Kehagias and A. Lukas,  Nucl. Phys. {\bf  B477}, 549 
(1996). 

\bibitem{Khleb97}S. Y. Khlebnikov and I. I. Tkachev, 
Phys. Rev. {\bf D56}, 653 (1997).  

\bibitem{KOST1}J. Khouri, B. A. Ovrut, P. J. Steinhardt and N. Turok, 
Phys. Rev. {\bf D64}, 123522 (2001).  

\bibitem{KOSST}J. Khouri, B. A. Ovrut, N. Seiberg, P. J. Steinhardt and N.
Turok,  Phys. Rev. {\bf D65}, 086007 (2002). 

\bb{Kibble95}T. W. Kibble and A. Vilenkin, Phys. Rev.  {\bf D52},
679 (1995). 

\bb{Kim79}K. Kim, Phys. Rev. Lett. {\bf 43}, 103 (1979).

\bibitem{KK94}E. Kiritsis and C. Kounnas, Phys. Lett. {\bf B331}, 51
(1994). 

\bibitem{KK95}E. Kiritsis and C. Kounnas, in  Proc. {\em Second Paris 
Cosmology Colloquium}, Observatoire de Paris,  1994, eds. H. J. De
Vega and N. Sanchez (World Scientific, Singapore, 1995), p. 500. 

\bibitem{Kohl} E. Kohlprath and G. Veneziano, gr-qc/0203093.

\bb{Kolb84}E. W. Kolb, D. Lindley and D. Seckel, Phys. Rev.  {\bf D30},
1205 (1984). 

\bibitem{KT90}E. W.  Kolb and M. S. Turner, {\em The Early Universe}  
(Addison-Wesley, Redwood City, CA, 1990).

\bb{Koz88}N. Kozimirov and I. I. Tkachev,  Mod. Phys. Lett. {\bf A4},
2377 (1988). 

\bb{Kra92}L. M. Krauss and M. White, Phys. Rev. Lett. {\bf 69}, 869
(1992). 

\bb{Kunze99}K. E. Kunze, Classical Quant. Grav. {\bf 16}, 3795 (1999).

\bb{KunDur00}K. E. Kunze and R. Durrer, Classical Quant. Grav. {\bf 17},
2597 (2000). 

\bibitem{LL62}L. Landau  and E. Lifshitz,  {\em The classical theory of
fields} (Pergamon Press, Oxford, 1962).

\bb{LarWil96}F. Larsen and F. Wilczek, Phys. Rev. {\bf D55}, 4591 (1997).

\bb{LaMar95}A. Lawrence and E. Martinec,  Classical Quant. Grav.  {\bf
13}, 63 (1996). 

\bb{Leblanc88}Y. Leblanc, Phys. Rev. {\bf D38}, 3087  
(1988).

\bb{Levin}J. Levin, Phys. Rev. {\bf D51}, 462 (1995). 

\bb{Lem95}D. Lemoine and M. Lemoine, Phys. Rev. {\bf D52}, 1955 
(1995).

\bb{Lemo99}M. Lemoine, Phys. Rev. {\bf D60}, 103522  (1999). 

\bb{Les99}J. Lesgourgues, D. Polarski, S. Prunet and A. A. Starobinski,
 Astron. Astrophys. {\bf 359}, 414 (2000). 

\bb{Lid93}A. R. Liddle and D. H. Lyth, Phys. Rep. {\bf 231}, 1 (1993). 

\bb{Lid95}J. E. Lidsey,  Phys. Rev. {\bf D52}, R5407 
(1995).

\bb{Lid97}J. E. Lidsey,  Phys. Rev. {\bf D55}, 3303  
(1997). 

\bb{Lid99}J. E. Lidsey,  Classical Quant. Grav. {\bf 15}, L77 (1999). 

\bibitem{Lidsey00}J. E. Lidsey, D. Wands and E. J. Copeland, 
 Phys. Rep. {\bf 337}, 343 (2000).

\bb{Lif63}E. M. Lifchitz and I. M. Kalatnikov, Adv.  Phys. {\bf 12}, 185
(1963). 

\bibitem{Lin83}A. Linde, Phys. Lett. {\bf B129}, 177 (1983).

\bb{Linde84} 
A. D. Linde, Sov. Phys. JETP {\bf 60}, 211 (1984). 

\bibitem{Lin90}A. D.  Linde, {\em Particle physics and inflationary
cosmology}  (Harwood, New York, 1990).

\bibitem{Lin95}A. D. Linde, Phys. Lett. {\bf B351}, 99 (1995).

\bb{Linde98}A. Linde, Phys. Rev. {\bf D58}, 083514 (1998). 

\bb{AdLigo98}LIGO MIT/Caltech Groups, {\em Research and
Development program for Advanced LIGO detectors},
LIGO-M970107-00-M (September 1997). 

\bb{Love84}C. Lovelace, Phys. Lett. {\bf B135}, 75 (1984).

\bb{LuMu97}H. Lu, S. Mukherji, C. N. Pope and K. W. Xu, Phys. Rev.
{\bf D55}, 7926 (1997).

\bb{Luc85} F. Lucchin and S. Matarrese, Phys. Lett.  {\bf B164}, 282 
(1985).  

\bb{Luck97}H. Luck and the GEO600 Team, Classical Quant. Grav. {\bf 14},
1741 (1997). 

\bb{LuOv98}A. Lukas and B. A. Ovrut, Phys. Lett. {\bf
B437}, 291 (1998).

\bb{LOSW99}A. Lukas, B. A. Ovrut, K. S. Stelle and D. Waldram, Phys.
Rev. {\bf D59}, 086001 (1999).

\bb{LOSW99a}A. Lukas, B. A. Ovrut, K. S. Stelle and D. Waldram, Nucl.
Phys. {\bf B552}, 246 (1999).

\bb{LuOWa97}A. Lukas, B. A. Ovrut and D. Waldram, Phys. Lett. {\bf
B393}, 65 (1997).

\bb{LuOWa97a}A. Lukas, B. A. Ovrut and D. Waldram, Nucl. Phys.  {\bf
B495}, 365 (1997).

\bb{LuPo97}A. Lukas and R. Poppe, Mod. Phys. Lett. {\bf A12}, 597 (1997). 

\bb{Luk80}V. N. Lukash, Sov. Phys. JETP {\bf 52}, 807 (1980).

\bb{Ly02}D. H. Lyth, Phys. Lett. {\bf B526}, 173 (2002). 

\bibitem{Lyt96}D. H. Lyth and D. Robert, 
Phys. Rev. {\bf D57}, 7120 (1998). 

\bibitem{LySte96}D. H. Lyth and E. Stewart, Phys. Rev. {\bf D53}, 1784 
(1996). 

\bibitem{LyWa02}D. H. Lyth and D. Wands, Phys. Lett. {\bf B524}, 5 
(2002). 

\bibitem{Madden01}R. Madden, JHEP {\bf 0101}, 23 (2001). 

\bb{MP86}K. Maeda and M. D. Pollock, Phys. Lett.  {\bf B173}, 251 
(1986).  

\bibitem{Mag98} M. Maggiore, Nucl. Phys. {\bf B525}, 
413 (1998). 

\bb{Mag00}M. Maggiore, Phys. Rep. {\bf 331}, 283 (2000).

\bb{MaNi99}M. Maggiore and A. Nicolis,  Phys. Rev. {\bf D62}, 024004 
(2000). 

\bibitem{MagRio98} M. Maggiore and A. Riotto, Nucl. Phys. {\bf B548}, 
427 (1999). 

\bb{MagTu97}M. Maggiore and R. Sturani, Phys. Lett. {\bf B415}, 335 
(1997). 

\bb{MaMu97}J. Maharana, S. Mukherji and S. Panda,  Mod. Phys. Lett.
{\bf A12}, 447 (1997). 

\bibitem{MOV98}J. Maharana, E. Onofri and G. Veneziano,  JHEP 
{\bf 9804}, 004  (1998). 

\bb{MahSch}J. Maharana and J. H. Schwartz, Nucl. Phys. {\bf B390}, 3
(1992). 

\bb{Map}MAP's web site at {\tt http://map.gsfc.nasa.gov}

\bb{Mar00}A. L. Maroto, Phys. Rev. {\bf D64},  083006 (2001). 

\bb{Mars95}M. Mars, Phys. Rev. {\bf D51}, 3899 (1995). 

\bb{MarVi96}X. Martin and A. Vilenkin, Phys. Rev. Lett. {\bf 77}, 2879
(1996). 

\bb{MaMo00}J. Marto and P. V. Moniz, in Proc. {\em  IX Marcel
Grossmann Meeting}, Rome, 2000,  eds. V. G. Gurzadyan, R. Jantzen
and R. Ruffini (World Scientific, Singapore, 2002), in press. 

\bb{Mau00}P. Mauskopf et al., Astrophys. J. {\bf 536}, L59 (2000). 

\bb{Meis97}K. A. Meissner, Phys. Lett. 
{\bf B392},  298 (1997). 

\bibitem{MeiVe91} K. A. Meissner and G. Veneziano, Mod. Phys. Lett. 
{\bf A6}, 3397 (1991). 

\bibitem{MeiVe91a} K. A. Meissner and G. Veneziano, Phys. Lett. 
{\bf B267},  33 (1991). 

\bb{Mel99}A. Melchiorri, F. Vernizzi, R. Durrer and G. Veneziano, Phys.
Rev. Lett. {\bf 83}, 4464 (1999). 

\bb{MenLid98}L. E. Mendes and A. R. Liddle, Phys. Rev. {\bf D60}, 063508
(1999). 


\bb{Tiga2}S. Merkowitz and W. Johnson, Phys. Rev. {\bf D51}, 2546 
(1995). 

\bb{MeTsey}R. R. Metsaev and A. A. Tseytlin, Nucl. Phys. {\bf B293}, 385 
(1987). 

\bb{Mil99}A. D. Miller et al., Astrophys. J.  {\bf 524}, L1 (1999). 

\bb{Vilko} A. G. Mirzabekian and G. A. Vilkovisky, 
 Classical Quant. Grav. {\bf 12}, 2173 (1995). 

\bb{MoTro95}R. Moessner and M. Trodden, Phys. Rev. {\bf D51},
2801 (1995). 

\bb{Mol90}S. Mollerach, Phys. Rev.  {\bf D42}, 313 (1990). 

\bb{Moroi}T. Moroi and T. Takahashi, Phys. Lett. {\bf B522}, 215 (2001). 

\bb{Mor02}T. Moroi and T. Takahashi, hep-ph/0206026.

\bb{Mul90}M. Mueller, Nucl. Phys. {\bf B337}, 
37 (1990). 

\bibitem{Mu88}V. F. Mukhanov, Sov. Phys. JETP {\bf 67}, 1297 (1988). 

\bibitem{MuBra92}V. F. Mukhanov and R. H.
Brandenberger, Phys. Rev.  Lett. {\bf 68}, 1969 (1992).

\bibitem{MFB92}V. F. Mukhanov, H. A. Feldman and R. H.
Brandenberger, Phys. Rep.  {\bf 215}, 203 (1992).

\bb{Nakao00}K. Nakao, T. Harada, M.  Shibata, S. Kawamura and  T.
Nakamura, Phys.  Rev. {\bf D63}, 082001 (2001). 

\bb{Nambu01}Y. Nambu, Phys.  Rev. {\bf D63}, 044013 (2001). 

\bb{Nambu02}Y. Nambu, Phys.  Rev. {\bf D65}, 104013 (2002). 

\bb{Nambu90}Y. Nambu and M. Sasaki,
Phys. Rev.  {\bf D42}, 3918 (1990).

\bb{NW92}C. R. Nappi and E. Witten, Phys. Lett. {\bf B293}, 309 (1992). 

\bb{Nara86}K. S. Narain, Phys. Lett. {\bf B169}, 41 (1986). 

\bb{Nara87}K. S. Narain, M. H. Sarmadi and E. Witten, Nucl. Phys. 
{\bf B279}, 369 (1987). 

\bb{Narita00}M. Narita, T. Torii and K. Maeda, Classical Quant. Grav. {\bf
17}, 4597 (2000). 

\bb{Nilles}H. B. Nilles, Phys. Rep. {\bf 110}, 1 (1984). 

\bb{NoRio}A. Notari and A. Riotto, hep-th/0205019. 

\bb{Novello93}M. Novello, L. A. R. Oliveira,  J. M. Salim and E. Elbaz, 
Int. J. Mod. Phys. {\bf D1}, 641 (1993). 

\bb{Novello78}M. Novello and J. M. Salim, 
Phys. Rev. {\bf D20}, 377 (1978). 

\bibitem{Nov80}I. D. Novikov and A. G. Polnarev, Astron. Zh. {\bf 57},
250 (1980) [Sov. Astron. {\bf 24}, 147 (1980)].

\bb{Pall97}G. Pallottino, in  Proc. {\em Gravitational Waves, Sources and
Detectors}, Cascina, 1996, eds. I Ciufolini and F. Fidecaro (World
Scientific, Singapore, 1997), p. 159. 

\bb{Pap51}A. Papapetrou, Proc. Roy. Soc. Lond. {\bf A209}, 248 (1951).

\bb{Parker79}E. N. Parker,
{\em Cosmical magnetic fields} (Clarendon Press, Oxford, 1979). 

\bb{Pea94}J. Peacock and S. Dodds, Mon. Not. Roy. Astron. Soc. {\bf 267},
1020 (1994). 

\bb{PeeVil99}P. J. E. Peebles and A. Vilenkin, Phys. Rev. {\bf D59},
063505 (1999). 

\bb{PiCo01}N. Pinto-Neto and R. Colistete Jr., Phys. Lett. {\bf A290}, 219 
(2001).

\bb{Peg78}F. Pegoraro, E. Picasso and L. A. Radicati, J. Phys. {\bf A11},
1949 (1978).

\bibitem{Pen65}R. Penrose, Proc.
Roy. Soc. Lond. {\bf A284}, 159 (1965).

\bibitem{Pen65a}R. Penrose, Phys. Rev. Lett. {\bf 14}, 57 (1965). 

\bibitem{Pen89}R. Penrose, {\em The emperor's new mind} 
(Oxford University Press, New York, 1989).

\bb{Perlmutter99}S. Perlmutter  et al., Astrophys. J. {\bf 517}, 565
(1999).  

\bb{Perl99}S. Perlmutter et al., Nature {\bf 391}, 51 (1999). 

\bb{PePi02}P. Peter and M. Pinto-Neto, hep-th/0203013. 

\bb{Pim99}L. O. Pimentel, Mod. Phys. Lett. {\bf A14}, 43 (1999). 

\bb{Pizz97}G. Pizzella, Classical Quant. Grav. {\bf 14}, 1481 (1997). 

\bb{Planck}PLANCK's web site at\\
 {\tt http://astro.estec.esa.nl/SA-general/Projects/Planck}

\bb{Star96}D. Polarski and A. A. Starobinski, Classical Quant. Grav. {\bf
13}, 377 (1996). 

\bb{Pol96}J. Polchinski,  Rev. Mod. Phys. {\bf 68}, 1245 (1996). 

\bibitem{Pol98}J. Polchinski, {\em String theory}, 
(University Press, Cambridge, 1998).

\bb{Pol89}M. D. Pollock, Nucl. Phys. {\bf B324}, 187 (1989). 

\bb{Pol92}M. D. Pollock, Int. J. Mod. Phys.  {\bf A7}, 4149 (1992). 

\bb{Poppe97}R. Poppe and S. Schwager, Phys. Lett. {\bf B393}, 51
(1997). 

\bb{Prodi98}G. A. Prodi et al., in Proc. {\em Second Edoardo Amaldi
Conference  on Gravitational Waves}, CERN,  1997, eds. E. Coccia et
al.  (World Scientific, Singapore, 1998), p. 148.

\bb{Rama97}K. S. Rama, Phys. Lett.  {\bf B408}, 91 (1997). 

\bb{RS2} L. Randall and R. Sundrum,  Phys. Rev. Lett. {\bf 83},
4690  (1999).

\bibitem{Ran95}L. Randall and S. Thomas,  Nucl. Phys. {\bf B449}, 229 
(1995).

\bibitem{Ratra93} B. Ratra, Astrophys. Lett. {\bf 391}, L1 (1992). 

\bb{Rey96}S. J. Rey, Phys. Rev. Lett. {\bf 77}, 1929 (1996).

\bb{Reece84}C. E. Reece, P. J. Reiner and A. C. Melissinos, Phys. Lett. 
{\bf A104}, 341 (1984). 

\bibitem{Riess98}A. G. Riess  et al., Astron. J. {\bf 116}, 1009 (1998). 

\bb{Rin56}W. Rindler, Mon. Not. Roy. Astron. Soc. {\bf 116}, 6 (1956).

\bb{Riotto00}A. Riotto, Phys. Rev. {\bf D61}, 123506 (2000).

\bb{Rit90}R. C. Ritter et al., Phys. Rev. {\bf D42}, 977 (1990).

\bb{RizTam94}J. Rizos and K. Tamvakis,  Phys. Lett. {\bf B326}, 57 
(1994).

\bb{Rosen85}N. Rosen, Astrophys. J. {\bf 297}, 347 (1985). 

\bb{Rub84}V. A. Rubakov, Phys. Lett. {\bf B148}, 280 (1984).

\bb{Rub88}V. A. Rubakov,  Phys. Lett. {\bf B214},  503 (1988).

\bb{Rub82}V. A. Rubakov, M. Sazhin and A. Veryaskin, Phys. Lett.
{\bf B115}, 189 (1982). 

\bibitem{Russo92}J. G. Russo, L. Susskind and L. 
Thorlacius, Phys. Rev. {\bf D46}, 3444 (1992). 

\bibitem{Sac62}R. K. Sachs,
Proc. Roy. Soc. Lond. {\bf A270}, 103 (1962). 

\bb{SW}R. K. Sachs and M. Wolfe, Astrophys. J. {\bf 147}, 73 (1967). 

\bb{Sahar97} A. A. Saharian, hep-th/9709188.

\bb{Sa90}V. Sahni, Phys. Rev. {\bf D42}, 453 (1990).

\bb{SaFeSte92}V. Sahni, H. Feldman and A. Stebbins, Astrophys. J. {\bf
385}, 1 (1992).

\bb{SaSaso01}V. Sahni, M. Sami and T. Souradeep, 
Phys. Rev.  {\bf D65}, 023518 (2002).

\bb{SV90}N. Sanchez and G. Veneziano, Nucl. Phys. {\bf B 333}, 253
(1990). 

\bb{SaMo01}A. K. Sanyal and B. Modak, Phys. Rev.  {\bf D63}, 
064021 (2001). 

\bb{Sas83}M. Sasaki, Prog. Theor. Phys. {\bf 70}, 394 (1980). 

\bb{Sas86}M. Sasaki, Prog. Theor. Phys. {\bf 76}, 1036 (1986). 

\bb{SatSi86}B. S. Sathyaprakash, P. Goswami and K. P. Sinha, 
Phys. Rev. {\bf D33}, 2196 (1986). 

\bb{Schwarz97}J. H. Schwarz,  Nucl. Phys. Proc. Suppl. {\bf 55B}, 1
(1997).

\bb{nucleo69}V. F. Schwarztmann, JETP Letters {\bf 9}, 184 (1969).

\bb{Sec85}D. Seckel and M. S. Turner,
Phys. Rev.  {\bf D32}, 3178 (1985).

\bb{SeZa96}U. Seljak and M. Zaldarriaga, Phys. Rev. Lett. {\bf 78}, 2054
(1997). 

\bibitem{Sen91} A. Sen, Phys. Lett. {\bf B271}, 295 (1991). 

\bb{Sen94}A. Sen,  Int. J. Mod. Phys. {\bf A9}, 3707 (1994).

\bb{Seno90}J. M. M. Senovilla, Phys. Rev. Lett. {\bf 64}, 2219 (1990). 

\bb{Sha84}D. Shadev, Phys. Lett.  {\bf B317}, 155 (1984). 

\bb{Shiba94}M. Shibata, K. Nakao and T. Nakamura,  Phys. Rev.  {\bf
D50}, 7304 (1994).

\bb{Schu86}B. L. Shumaker, Phys. Rep. {\bf 135}, 317 (1986). 

\bb{Siki94}P. Sikivie, Int. J. Mod. Phys.  {\bf D3} Suppl.,
1 (1994).

\bb{Cobe92}G. F. Smooth et al., Astrophys. J. {\bf 396}, 1 (1992).

\bb{SNO}SNO Collaboration, nucl-ex/0204008. 

\bb{SNOa}SNO Collaboration, nucl-ex/0204009. 

\bb{Star79}A. A. Starobinski, JETP Lett. {\bf 30}, 682 (1979).

\bb{Stew93}E. D. Stewart and D. H. Lyth, Phys. Lett. {\bf B302}, 171
(1993). 

\bb{Stein97}
P. Steinhardt, in {\em Critical problems in physics},  edis. V. L.
Fitch and D. R. Marlow (University Press, Princeton, NJ, 1997).

\bb{SteiTu01}P. J. Steinhardt and N. Turok, hep-th/0111030. 

\bibitem{Sus95}L. Susskind, J. Math. Phys. {\bf 36}, 6377 (1995).

\bb{Syn}J. L. Synge, {\em Relativity: the general theory} (North Holland, 
Amsterdam, 1960). 

\bibitem{Szekeres} P. Szekeres, J. Math. Phys. {\bf 13}, 286 (1972).

\bibitem{Tay88}T. R. Taylor  and G. Veneziano, Phys. Lett. {\bf B213}, 459
(1988).

\bibitem{Tol}R. C. Tolman, {\em Relativity, thermodynamics and
cosmology} (Dover Publications, New York, 1987). 

\bb{Tow95}P. Townsend, Phys. Lett. {\bf B350}, 184 (1995). 

\bibitem{Tse91}A. A. Tseytlin, Mod. Phys. Lett. {\bf A6}, 
1721 (1991).

\bibitem{Tse92}A. A. Tseytlin, Classical Quant. Grav. {\bf 9}, 403 (1992).

\bibitem{TseVa92}A. A.  Tseytlin  and C. Vafa, Nucl. Phys. {\bf  B372}, 443
(1992).

\bb{Tsu02}S. Tsujikawa,  Phys. Lett. {\bf B526}, 179 (2002). 

\bb{Tsu01}S. Tsujikawa and H. Yajima, Phys. Rev. {\bf D64}, 023519 
(2001). 

\bb{Tur97}M. S. Turner, Phys. Rev. {\bf D55}, 435 
(1997). 

\bb{Turner00}M. S. Turner, Phys. Rep. {\bf 333}, 619 (2000). 

\bb{Turner01} M. S. Turner and A. G. Riess, astro-ph/0106051. 

\bibitem{TW97}M. S. Turner  and E. J. Weinberg, Phys. Rev. {\bf D56}, 
4604 (1997). 

\bb{TurWi88}M. S. Turner and L. M. Widrow, Phys. Rev. {\bf D37}, 2473 
(1988). 

\bibitem{TurWil90}M. S. Turner and F. Wilczek, Phys. Rev. Lett. {\bf 65},
3080 (1990).

\bibitem{Tur88}N. Turok, Phys. Rev. Lett. {\bf 60},
549 (1988). 

\bb{TuHa98}N. Turok and S. W. Hawking, gr-qc/9802062. 

\bb{UVe0}C. Ungarelli and A. Vecchio,  in {\it Gravitational Waves},  
Proc. Third Edoardo Amaldi Conference, CALTECH, Pasadena, 1999, ed. S.
Meshkov (AIP Conference  Proceedings {\bf 523}  (2000)) p. 90. 

\bb{UVe1}C. Ungarelli and A. Vecchio, Phys. Rev. {\bf D63}, 064030
(2001).

\bb{Vacha91}T. Vachaspati, Phys. Lett. {\bf B265}, 258 (1991). 

\bb{Dorca00}C. van de Bruck, M. Dorca, R. H. Brandenberger and A.
Lukas, Phys. Rev. {\bf D62}, 123515 (2000).

\bibitem{GVFC86}G. Veneziano, Europhys. Lett. {\bf 2}, 133 (1986).

\bibitem{GVFC89}G. Veneziano, in  Proc.  {\em   School of
Subnuclear Physics}, Erice, 1989,  ed. A. Zichichi  (Plenum Press, New
York, 1990), p. 199.

\bibitem{GV91}G.  Veneziano, Phys. Lett. {\bf B265}, 287  (1991).

\bibitem{GVGvsG95}G. Veneziano, in Proc. {\em Second Paris Cosmology
Colloquium}, Observatoire de Paris, 1994, eds. H. J. de Vega  and N.
Sanchez (World Scientific, Singapore, 1995), p. 322.

\bb{Ve95}G. Veneziano
in {\em String Gravity and Physics at the Planck Energy Scale},
Proc. Erice School,  1995, eds. N. Sanchez and A. Zichichi
(Kluwer Acad. Publ., Dordrecht, 1996), p. 285. 

\bb{Ve97}G. Veneziano, in Proc. {\em Gravitational Waves: Sources and
Detectors}, Cascina, 1996, eds. I Ciufolini and F. Fidecaro (World
Scientific, Singapore, 1997), p.133.

\bibitem{GV97}G. Veneziano,  Phys. Lett. {\bf  B406}, 297 (1997). 

\bibitem{GV99}G. Veneziano,  Phys. Lett. {\bf  B454}, 22 (1999). 

\bibitem{Ve99a}G. Veneziano, in  {\em The primordial Universe},
Proc. LXXI Les Houches Summer School, 1999,  eds. P. Bin\'etruy, R.
Schaeffer, J. Silk and F. David (Springer-Verlag, Berlin, 2000), p. 581. 

\bb{Ven98}  G. Veneziano, in {\em The Geometric
Universe}, eds.   L. J. Mason, K. P. Tod, S. T. Tsou and N. M. J. Woodhouse  
(University Press, Oxford, 1998), p. 235.

\bb{Ven01}G. Veneziano, JHEP {\bf 0206}, 051 (2002). 

\bb{Vern00}F. Vernizzi, A. Melchiorri and R. Durrer,
Phys. Rev. {\bf D63}, 063501 (2001). 

\bb{Vil84}A. Vilenkin, Phys. Rev.  {\bf  D30}, 509 (1984).

\bb{Vil86}A. Vilenkin, Phys. Rev.  {\bf  D33}, 3650 (1986).

\bb{Vil88}A. Vilenkin, Phys. Rev.  {\bf  D37}, 888 (1988).

\bb{Vil92}A. Vilenkin, Phys. Rev.  {\bf  D46}, 2355 (1992).

\bb{Vil96}A. Vilenkin, in      
{\em String Gravity and Physics at the Planck Energy Scale}, Proc. Erice,
School,  1995, eds. N. Sanchez and A. Zichichi (Kluwer Acad.
Pub., Dordrecht, 1996),  p. 345.

\bb{Vit97}S. Vitale, M. Cerdonio, E. Coccia and A. Ortolan, Phys. Rev.  {\bf
D55}, 17414  (1997). 

\bb{Wald}R. M. Wald, {\em General relativity} (University Press,
Chicago, IL, 1984). 

\bb{Wands99}D. Wands, Phys. Rev. {\bf D60}, 023507 (1999).

\bb{Wang00}L. Wang, R. R. Caldwell, J. P. Ostriker and P. J. Steinhardt,
Astrophys. J. {\bf 530}, 17 (2000). 

\bb{Weinberg}S. Weinberg, {\em Gravitation and cosmology} (Wiley,
New York, 1971). 

\bb{Wein78}S. Weinberg, Phys. Rev. Lett. {\bf 40}, 223 (1978).

\bb{Wei82}S. Weinberg, Phys. Rev. Lett. {\bf 48}, 1303 (1982). 

\bb{Wesson85}P. S. Wesson, Astron. Astrophys. {\bf 151}, 276 (1985). 

\bb{Wheeler68}J. A. Wheeler,  in {\em Battelle Rencontres}, eds.  C.
De Witt  and  J. A. Wheeler  (Benjamin, New York, 1968).


\bb{Wilc78} F. Wilczeck, Phys. Rev. Lett. {\bf 40}, 279 (1978).

\bibitem{Wit84}E. Witten, Phys. Lett. {\bf B149}, 351 (1984).

\bibitem{Wit91}E. Witten, Phys. Rev {\bf D44}, 314 (1991). 

\bibitem{Wit95}E. Witten, Nucl. Phys. {\bf B443}, 85 (1995). 

\bibitem{Wit96}E. Witten, Nucl. Phys. {\bf B471}, 135 (1996). 

\bb{YMO99}H. Yajima, K. Maeda and H. Ohkubo, 
Phys. Rev.  {\bf D62}, 024020 (2000).

\bb{Yaza01}S. S. Yazadjiev, Phys. Rev. {\bf D63}, 063510 (2001). 

\bibitem{Yurtsever} U. Yurtsever, Phys. Rev. {\bf D37}, 2803 (1988). 

\bb{Zel72}Y. B. Zel'dovich, Mon. Not. Roy. Astron. Soc.  {\bf 160}, 1
(1972). 

\bb{Zeldo83}Y. B. Zeldovich, A. A. Ruzmaikin and D. D. Sokoloff,
{\em Magnetic fields in astrophysics} 
(Gordon and Breach, New York, 1983). 

\bb{Zeldo84}Y. Zel'dovich and A. A. Starobinski, Sov. 
Astron. Lett. {\bf 10}, 135 (1984).


\end{thebibliography}
\end {document}